\documentclass[12pt]{report}
\input epsf
\usepackage{hyperref}
\hypersetup{
    bookmarks=true,         
    unicode=false,          
    pdftoolbar=true,        
    pdfmenubar=true,        
    pdffitwindow=true,      
    pdfnewwindow=true,      
    colorlinks=true,       
    linkcolor=black,          
    citecolor=gray,        
    filecolor=magenta,      
    urlcolor=black          
}

\usepackage{amsmath}
\usepackage{amssymb}
\usepackage{stmaryrd}
\usepackage[dvips]{graphicx}
\usepackage{wrapfig}
\usepackage{cite}
\usepackage{verbatim}
\usepackage{setspace}
 \usepackage[usenames,dvipsnames]{pstricks}
 \usepackage{epsfig}
 \usepackage{pst-grad} 
 \usepackage{pst-plot} 

 \usepackage{crop}

 \usepackage{bbm}

    \def\be{\begin{eqnarray}}
    \def\ee{\end{eqnarray}}
    \def\no{\nonumber}

    \def\suml{\sum\limits}
    \def\prodl{\prod\limits}
    \def\intl{\int\limits}
    
    \def\intii{\int\limits_{-\infty}^{\infty}}

    \def\bn{\begin{enumerate}}
    \def\en{\end{enumerate}}
    \def\bi{\begin{itemize}}
    \def\ei{\end{itemize}}

    \def\({\left(\!}
    \def\){\!\right)}
    \def\<{\left\langle\,}
    \def\>{\, \right\rangle}
    \def\[{\left[}
    \def\]{\right]}

    \newcommand{\mtwo}[4]{\left(%
    \begin{array}{cc}
    #1 & #2 \\
    #3 & #4 \\
    \end{array}%
    \right)}

    \def\tilde{\widetilde}
    \def\bar{\overline}
    \def\hat{\widehat}

    \def\a{\alpha}
    \def\b{\beta}
    
    \def\g{\gamma}
    \def\G{\Gamma}
    \def\e{\epsilon}

    \def\m{\mu}

    \def\l{\lambda}
    \def\s{\sigma}
    
    \def\th{\theta}
    
    \def\o{\omega}
    \def\k{\kappa}
    \def\p{\phi}

    \def\CA{{\cal A}}
    
    \def\CC{{\cal C}}
    \def\CD{{\cal D}}
    
    \def\CF{{\cal F}}
    
    \def\CH{{\cal H}}
    
    \def\CJ{{\cal J}}
    \def\CK{{\cal K}}
    \def\CL{{\cal L}}
    
    \def\CN{{\cal N}}
    \def\CO{{\cal O}}
    \def\CP{{\cal P}}

    \def\CS{{\cal S}}
    \def\CT{{\cal T}}

    \def\CW{{\cal W}}

    \def\p{\partial}
    \def\pd{\partial}

    \def\MC{{\mathbb{C}}}

    \def\MP{{\mathbb{P}}}
    
    \def\MR{{\mathbb{R}}}

    \def\MZ{{\mathbb{Z}}}

    \def\sign{{\rm{sign}}\,}

    \def\Tr{{\rm Tr}\,}

    \def\Shrodinger{Schr$\ddot{\rm o}$dinger }

    \def\LQCD{{\Lambda}}
    \def\sigmacoupling{f}

    \def\xpm{ x^{\pm} }
    \def\xp{ x^{+} }
    \def\xm{ x^{-} }
    \def\ypm{ y^{\pm}}
    \def\yp{ y^{+}}
    \def\ym{ y^{-}}

    \def\BrP{B}
    \def\mass{{\rm m}}
    \def\scP{\CP}
    \def\chico{{\chi}}

   \def\xXX{{\begin{picture}(45,10)
\thicklines
\put(5.15,4){\line(1,0){6.85}}
\put(10.8,-0.28){\bf\Large $\times$}
\put(17.5,4){\circle{11.3}}
\put(23.3,4){\line(1,0){5.35}}
\put(27.8,-0.28){\bf\Large $\times$}
\put(34.5,4){\circle{11.3}}
\put(0,1.15){{$\mathbf{\times}$}}
\put(0,1.165){{$\mathbf{\times}$}}
\put(0,1.135){{$\mathbf{\times}$}}
\put(0.15,1.15){{$\mathbf{\times}$}}
\put(-0.15,1.15){{$\mathbf{\times}$}}
\end{picture}}}
 \def\xOX{{\begin{picture}(45,10)
\thicklines
\put(5.15,4){\line(1,0){6.85}}
\put(17.5,4){\circle{11.3}}
\put(23.3,4){\line(1,0){5.35}}
\put(27.8,-0.28){\bf\Large $\times$}
\put(34.5,4){\circle{11.3}}
\put(0,1.15){{$\mathbf{\times}$}}
\put(0,1.165){{$\mathbf{\times}$}}
\put(0,1.135){{$\mathbf{\times}$}}
\put(0.15,1.15){{$\mathbf{\times}$}}
\put(-0.15,1.15){{$\mathbf{\times}$}}
\end{picture}}}
 \def\xXO{{\begin{picture}(45,10)
\thicklines
\put(5.15,4){\line(1,0){6.85}}
\put(10.8,-0.28){\bf\Large $\times$}
\put(17.5,4){\circle{11.3}}
\put(23.3,4){\line(1,0){5.35}}
\put(34.5,4){\circle{11.3}}
\put(0,1.15){{$\mathbf{\times}$}}
\put(0,1.165){{$\mathbf{\times}$}}
\put(0,1.135){{$\mathbf{\times}$}}
\put(0.15,1.15){{$\mathbf{\times}$}}
\put(-0.15,1.15){{$\mathbf{\times}$}}
\end{picture}}}

\def\mI{\mathbbm{1}}

\begin{document}

\begin{titlepage}
\begin{center}
\begin{figure}[t]
\centering
\includegraphics[height=2cm]{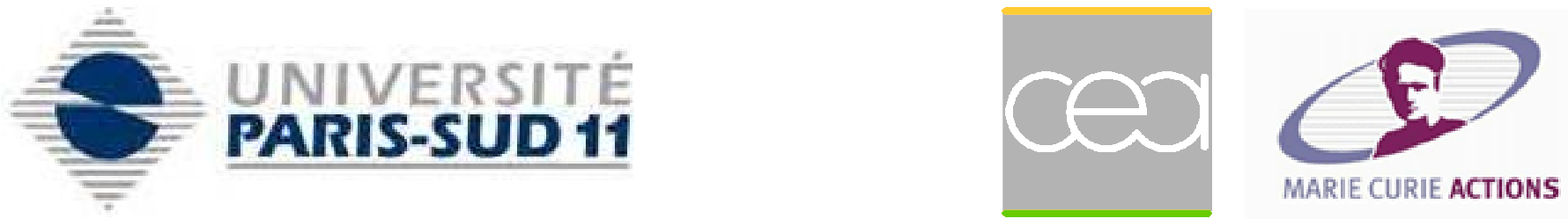}
\label{fig:logoparisXI}
\end{figure}

\vspace*{0.2in}

{\bf\Large Th\`ese de Doctorat}

\vspace*{0.1in}
{Pr\'esent\'e \`a l'Universit\'e Paris-XI}\\
{Sp\'ecialit\'e: Physique Th\'eorique}

\vspace*{0.6in}
{\LARGE Intégrabilit\'{e} quantique et \'{e}quations fonctionnelles.}
\par
{\large Application au probl\`{e}me spectral AdS/CFT et mod\`{e}les sigma bidimensionnels.}

\line(1,0){300}

\vspace{0.2in}
{\LARGE Quantum integrability and functional equations.}
\par
{\large Applications to the spectral problem of AdS/CFT and two-dimensional sigma models.}

\vspace{0.5in}
\Large
{Dmytro Volin}
\par
\large
{Institut de Physique Th\'eorique, CNRS-URA 2306}\\
{C.E.A. - Saclay,\ F-91191 Gif-sur-Yvette, France}
\par
\vspace{0.4in}
{\qquad Soutenu le 25 septembre 2009 devant le jury compos\'e de:}
\end{center}
\begin{flushleft}
\large
\begin{tabular}{ll}
  Gregory Korchemsky   \\
  Ivan Kostov & Directeur de These \\
  Joseph Minahan & Raporteur \\
  Didina Serban & Directeur de These \\
  Arkady Tseytlin & Raporteur \\
  Jean-Bernard Zuber
\end{tabular}
\end{flushleft}


\end{titlepage}
\ \pagestyle{empty}
\vspace{2in}
\begin{center}
{\it Dedicated to my grandmother}
\end{center}
\newpage
\begin{titlepage}
\begin{center}
    {\bf Acknowledgements}
\end{center}

\quad First of all I would like to thank my supervisors Ivan Kostov and Didina Serban, for all their help during the work on the dissertation. They introduced me into the subject, gave me many insights during our collaborative and my independent work, we had many days of interesting discussions. I appreciate the high standards that they introduced for the scientific research and the presentation of the results. Also I would like to thank for all the help I got from them during my stay in France.

I am grateful to my referees Joseph Minahan and Arkady Tseytlin, for their careful reading of the manuscript and a number of valuable suggestions, and to the members of the jury Gregory Korchemsky and Jean-Bernard Zuber for their interest in my work and for interesting remarks.

I thank  Zoltan Bajnok, Janos Balog, Benjamin Basso, Jean-Emile Bourgine,  Fran\c{c}ois David, Bertrand Eynard, Nikolay Gromov, Edmond Iancu, Nikolai Iorgov, Petro Holod, Romuald Janik, Vladimir Kazakov, Gregory Korchemsky, Peter Koroteev, Andrii Kozak, Sergey Lukyanov, Radu Roiban, Adam Rej, Hubert Saleur, Igor Shenderovich, Arkady Tseytlin, Pedro Vieira, Benoit Vicedo, Andrey Zayakin, Aleksander Zamolodchikov, Paul Zinn-Justin, Jean Zinn-Justin, and Stefan Zieme for many discussions that helped me a lot in my scientific research.

During my stay at IPhT I had interesting discussions in various topics in physics and beyond. I am greatful for these discussions to Alexander Alexandrov, Iosif Bena,  Riccardo Guida, David Kosower, Gregoire Misguich, Stephane Nonnenmacher, Jean-Yves Ollitrault, Vincent Pasquier,  Robi Peschanski, Pierre Vanhove and PhD students Alexey Andreanov, Adel Benlagra, Guillaume Beuf, Constantin Candu, Jerome Dubail, Yacine Ikhlef, Nicolas Orantin, Jean Parmentier, Clement Reuf, and Cristian Vergu. Special thanks to Jean-Emile Bourgine who was among my first teachers of French.

I would like to thank the administration of the IPhT, especially Henri Orland, Laure Sauboy, Bruno Savelli, and Sylvie Zaffanella. Their work made it possible not to worry about any organizational issues and completely concentrate on the research.

The work on the thesis was supported by the  European Union through ENRAGE network, contract MRTN-CT-2004-005616.  I am greatfull to Renate Loll for the perfect coordination of the network and to Fran\c{c}ois David who helped me with all the organization questions in the Saclay node of the network.

I would like to thank Vitaly Shadura and Nikolai Iorgov who organized science educational center in BITP, Kiev and would like to thank the ITEP mathematical physics group, especially Andrei Losev, Andrei Mironov, and Alexei Morozov.
BITP and ITEP were the places where I formed my scientific interests. There I studied mathematical physics together with Alexandr Gamayun, Peter Koroteev, Andrii Kozak, Ivan Levkivskii, Vyacheslav Lysov, Aleksandr Poplavsky, Sergey Slizovsky, Aleksander Viznyuk, Dmytro Iakubovskyi, and Alexander Zozula to whom I am greatful for many interesting seminars.

Last but not least, I would like to thank my parents, my grandmother, and my wife for all their support and encouragement throughout my studies.
\end{titlepage}
\newpage
\begin{titlepage} \begin{center}
    {\bf Resum\'{e}}
\end{center}
{Dans cette th\`{e}se, on d\'{e}crit une proc\'{e}dure permettant de repr\'{e}senter les \'{e}quations int\'{e}gra\-les de l'Ansatz de Bethe sous la forme du probl\`{e}me de Riemann-Hilbert. Cette proc\'{e}dure nous permet de simplifier l'\'{e}tude des cha\^{\i}nes de spins int\'{e}gra\-bles apparaissant dans la limite thermodynamique. A partir de ces \'{e}quations fonctionnelles, nous avons explicit\'{e} la m\'{e}thode qui permet de trouver l'ordre sous-dominant de la solution de diverses \'{e}quations int\'{e}grales, ces \'{e}quations \'{e}tant r\'{e}solues par la technique de Wiener-Hopf \`{a} l'ordre dominant.}

Ces \'{e}quations int\'{e}grales ont \'{e}t\'{e} \'{e}tudi\'{e}es dans le contextes de la correspondance AdS/CFT o\`{u} leur solution permet de v\'{e}rifier la conjecture d'int\'{e}grabilit\'{e} jusqu'\`{a} l'ordre de deux boucles du d\'{e}veloppement \`{a} fort couplage. Dans le contexte des mod\`{e}les sigma bidimensionnels, on analyse le comportement d'ordre \'{e}lev\'{e} du d\'{e}veloppement asymptotique perturbatif. L'exp\'{e}rience obtenue gr\^{a}ce \`{a} l'\'{e}tude des repr\'{e}sentations fonctionnelles des \'{e}quations int\'{e}grales nous a permis de r\'{e}soudre explicitement les \'{e}quations de crossing qui apparaissent dans le probl\`{e}me spectral d'AdS/CFT.

\ \\
\ \\

\begin{center}
    {\bf Abstract}
\end{center}
{In this thesis is given a general procedure to represent the integral Bethe Ansatz equations in the form of the Reimann-Hilbert problem. This allows us to study in simple way integrable spin chains in the thermodynamic limit. Based on the functional equations we give the procedure that allows finding the subleading orders in the solution of various integral equations solved to the leading order by the Wiener-Hopf techniques.

The integral equations are studied in the context of the AdS/CFT correspondence, where their solution allows verification of the integrability conjecture up to two loops of the strong coupling expansion. In the context of the two-dimensional sigma models we analyze the large-order behavior of the asymptotic perturbative expansion. Obtained experience with the functional representation of the integral equations allowed us also to solve explicitly the crossing equations that appear in the AdS/CFT spectral problem.}
\end{titlepage}
\pagestyle{plain}
\pagenumbering{roman}
\tableofcontents

\chapter*{Introduction}
\addcontentsline{toc}{chapter}{Introduction}
\pagenumbering{arabic}

\subsection*{Overview and motivation}

Quantum integrable systems play an important role in the theoretical physics. Many of these systems have direct physical applications. And also, since we can solve them exactly, the integrable systems are often considered as toy models and provide us with indispensable intuition for investigation of more complicated theories.

A class of integrable systems can be solved by means of the Bethe Ansatz. Bethe Ansatz was invented for the solution of the Heisenberg magnet in a seminal work \cite{Bethe:1931hc} in 1931. The solvability by means of the Bethe Ansatz essentially relies on the two-dimensionality of the considered system. Therefore the Bethe Ansatz works in the two-dimensional statistical models and $1+1$ dimensional quantum field theories. A two-dimensional integrable structure was also identified in a way explained below in the four-dimensional gauge theory: $\CN=4$ supersymmetric Yang-Mills (SYM). This theory became therefore the first example of a non-trivial four dimensional quantum field theory where some results were found exactly at arbitrary value of the coupling constant.

$\CN=4$ SYM is also famous due to its conjectured equivalence \cite{Maldacena:1997re,Gubser:1998bc,Witten:1998qj} to type IIB string theory on AdS$_5\times$S$^5$. This duality is the first explicit and the most studied example of the gauge-string duality known as the  AdS/CFT correspondence. This duality is usually studied in its weaker form, which states the equivalence between the 't Hooft planar limit of the gauge theory and the free string theory.

The AdS/CFT correspondence between gauge and string theories is the duality of weak/strong coupling type. It allows giving an adequate description of the strongly coupled gauge theory. On the other hand, the weak/strong coupling nature of the duality conjecture makes difficult to prove it. Except for the quantities protected by the symmetry and some special limiting regimes, comparison of gauge and string theory predictions requires essentially nonperturbative
calculations. This is where the integrability turns out to be extremely useful.

On the gauge side of the correspondence the integrability was initially discovered in the 1-loop calculation of anomalous dimensions of single trace local operators \cite{Minahan:2002ve}. Later it was conjectured to hold at all loops \cite{Beisert:2003tq}. According to the integrability conjecture, the single trace local operators correspond to the states of an integrable spin chain in which the dilatation operator plays the role of a Hamiltonian. At one-loop level this is a spin chain with the nearest neighbors' interactions. It can be diagonalized for example by algebraic Bethe Ansatz. The all-loop structure of the spin chain is much more complicated. In particular, the all-loop Hamiltonian is not known. Luckily, one of the beauties of the integrability is that it gives us a way to find the spectrum of the system even without knowledge of the exact form of the Hamiltonian. This solution can be obtained by application of the method which was initially developed in \cite{Zamolodchikov:1978xm} for the two dimensional relativistic integrable theories. We will now briefly recall this method.

Let us consider a two dimensional integrable relativistic quantum field theory which has massive particles as asymptotic states. Due to the existence of higher conserved charges, the number of particles is preserved under scattering and the scattering factorizes into $2\to 2$ processes. Therefore the dynamics of the system is determined by the two-particle $S$-matrix. This $S$-matrix is determined up to an overall scalar factor by the requirement of invariance under the symmetry group and by the Yang-Baxter equation (self consistency of two-particle factorization). The overall scalar factor is uniquely fixed by the unitarity and crossing conditions and the assumption about the particle content of the theory.

Let our field theory be defined on a cylinder of circumference $L$. The notion of asymptotic states and scattering can be defined only for $m L\gg 1$, where $m$ is the mass of the particles. If this condition is satisfied, the system of $N$ particles is completely described by the set of their momenta and additional quantum numbers. Once the quantum numbers are chosen, the momenta of the particles can be found from the periodicity conditions which lead to the Bethe equations. In the simplest case when particles do not bear additional quantum numbers the Bethe equations are written as:
\be\label{betheintroduction}
  e^{-ip_k L}=\prodl_{\substack{j=1\\j\neq k}}^{N}S[p_k,p_j],\ \ k=1,2,\ldots,N.
\ee
The energy of the system can be found from the dispersion relation:
\be
  E=\sum_{i=1}^N\varepsilon[p_i],\ \ \varepsilon[p]=\sqrt{m^2+p^2}.
\ee
Therefore the knowledge of the scattering matrix solves the spectral problem of the theory.

For simplicity we ignored the fact that the masses of the particles can be in principle different.

It turns out that the AdS/CFT integrable spin chain can be also described in terms of the factorized scattering. This idea was initially proposed by Staudacher in \cite{Staudacher:2004tk}. In \cite{Beisert:2006qh} Beisert showed that the scattering matrix can be fixed up to an overall scalar factor, known also as the dressing factor, already from the symmetry of the system. The Yang-Baxter equation is then satisfied automatically. The dispersion relation for the excitations is given by the expression
\be\label{dispersionintro}
    \varepsilon[p]=\sqrt{1+\frac{\l}{\pi^2}\sin^2{\frac p2}},
\ee
where $\l$ is the 't Hooft coupling constant. This dispersion relation was initially derived in \cite{Santambrogio:2002sb}.

In \cite{Beisert:2006qh} the Bethe Ansatz equations were derived from the knowledge of $S$-matrix using the nested Bethe Ansatz procedure. The equations coincided with the ones conjectured in \cite{Beisert:2005fw}. Note that the equations are defined up to the dressing factor which cannot be fixed from the symmetry.

Although the logic of derivation is similar to the one in the field theory, in AdS/CFT we are dealing with the spin chain. This is seen in particular in the dispersion relation which contains the sine function. As in field theory, the resulting Bethe Ansatz is valid only in the limit of large volume and is usually called the asymptotic Bethe Ansatz.

In spin chains the notion of the cross channel is not defined. Therefore the formulation of the crossing equations is not obvious and this prevents us from imposing constraints on the dressing factor. It seems that knowledge that the spin chain describes the spectral problem of $\CN=4$ SYM is insufficient to fix the dressing factor of the scattering matrix. However, we can try using the fact that due to the duality conjecture the AdS/CFT spin chain should solve also in some sense the string theory. We therefore turn to the discussion of the string side of the correspondence.

According to the duality conjecture, the conformal dimensions of local operators are equivalent to the energies of string states. The free string theory is described by the supersymmetric sigma model\footnote{with properly taken into account Virasoro constraints and remaining gauge freedom which come from the dynamical nature of the worldsheet metric in string theory.} \cite{Metsaev:1998it} which is classically integrable \cite{Bena:2003wd}\footnote{The integrable structures were discovered in the same time in gauge and string theories. Developments of the integrability ideas on both sides of the correspondence mutually used insights from each other}. Assuming its quantum integrability, we again can describe the system in terms of the factorized scattering matrix and construct the corresponding Bethe Ansatz from it. The AdS/CFT correspondence then implies that we will obtain the same Bethe Ansatz equations as for the solution of spin chain. In \cite{Hofman:2006xt} Hofman and Maldacena proposed to identify special one-particle excitations of the spin chain with special
string configurations known since then as giant magnons. The proposed identification conformed also the dispersion relation (\ref{dispersionintro}). The physical equivalence of scattering matrices for the sigma model and for the spin chain was shown in \cite{Arutyunov:2006yd}. Therefore we get an interesting phenomenon: a discrete spin chain solves a continuous field theory.

Due to this phenomenon we can expect the existence of the crossing equations for the scattering matrix of spin chain excitations. This is not granted, since relativistic invariance of the system is broken by the gauge fixing. In \cite{Janik:2006dc} Janik assumed that the crossing equations are however present. He derived these equations by purely algebraic means. We will explain his reasoning in details in subsection \ref{subsec:Janikderivation} of this thesis.

The existence of a non-trivial dressing factor was established before the crossing equations were formulated. The dressing factor was found at the leading \cite{Arutyunov:2004vx} and subleading \cite{Hernandez:2006tk} orders of strong coupling expansion by comparison of the algebraic curve solutions \cite{Kazakov:2004qf} for the sigma model and the asymptotic Bethe Ansatz. At first three orders of weak coupling expansion perturbative calculations of the gauge theory showed that the dressing factor equals $1$. If the duality conjecture is true the dressing factor should interpolate between its values at weak and strong coupling.

Beisert, Hernandez, and Lopez \cite{Beisert:2006ib} checked that the results \cite{Arutyunov:2004vx,Hernandez:2006tk} at strong coupling satisfy the crossing equations and proposed a class of asymptotic solutions to the crossing equations to all orders of perturbation theory. Based on this proposal and using a nontrivial ressumation trick, Beisert, Eden, and Staudacher \cite{Beisert:2006ez} conjectured a convergent weak coupling expansion for the dressing factor and found the nonperturbative expression that reproduced that expansion. Their proposal is known now as the BES/BHL dressing factor. Based on this conjecture they formulated the integral equation known as the BES equation. This equation was the Eden-Staudacher equation \cite{Eden:2006rx} modified because of the non-triviality of the dressing factor. The solution of the BES equation allows one to find the cusp anomalous dimension - the quantity that acquired attention both at gauge and string side of the correspondence. The BES/BHL proposal passed a non-trivial check. The cusp anomalous dimension calculated in \cite{Beisert:2006ez} up to fourth order in the weak coupling expansion from the BES equation coincided with the four-loop perturbative calculations in $\CN=4$ SYM \cite{Bern:2006ew}.

The $S$-matrix fixed by the symmetry and equipped with the BES/BHL dressing factor defines the AdS/CFT integrable system. The anomalous dimensions at arbitrary values of the coupling constant can be computed using the asymptotic Bethe Ansatz. Since integrability was not rigorously proven, the validity of the asymptotic Bethe Ansatz had to be checked.

\subsection*{Personal contribution}

 The verification of the AdS/CFT Bethe Ansatz at strong coupling was the subject which initiated my PhD research. This verification resulted in the papers \cite{KSV1,KSV2,V1}. In the papers \cite{KSV1,KSV2} we analyzed the strong coupling expansion of the BES equation, the paper \cite{V1} was devoted to the strong coupling solution of the generalization of the BES equation proposed in \cite{Freyhult:2007pz,Bombardelli:2008ah}. These problems proved to be rather difficult. The solution of these equations was a subject of interest of several theoretical groups \cite{Kotikov:2006ts,Alday:2007qf,Beccaria:2007tk,Benna:2006nd},\cite{Casteill:2007ct,Belitsky:2007kf},\cite{Basso:2007wd,Basso:2009gh},\cite{Basso:2008tx,Fioravanti:2008ak,Fioravanti:2008bh,Bajnok:2008it}.
  In our works we realized that it was particularly useful to rewrite the integral equation in the form of a specific Riemann-Hilbert problem. Such representation allowed further significant simplifications of the equations and finally allowed to solve them perturbatively.

The developed techniques turned out to be useful for the solution of similar integral equations that appeared in different integrable models. In \cite{V2} we applied this techniques to find the free energy of the $O(n)$ sigma model in the presence of magnetic field at first 26 orders of the perturbative expansion. The solution at leading and subleading orders allowed deriving analytically the exact value of mass gap, which was guessed previously from numerics in \cite{Hasenfratz:1990ab,Hasenfratz:1990zz}.  Higher orders of the perturbative expansion allowed testing the properties of Borel summability of the model.

My fifth paper \cite{V3} gave another evidence for the correctness of the AdS/CFT asymptotic Bethe Ansatz. The paper was devoted to the solution of the crossing equation for the dressing factor. Despite many checks of the BES/BHL proposal this dressing factor was never obtained directly from the solution of the crossing equation. In \cite{V3} we presented such kind of derivation. Moreover, we showed that the solution of the crossing equations is unique if to impose quite natural requirements on the analytical structure of the scattering matrix and to demand the proper structure of singularities that correspond to the physical bound states in the theory.

This thesis contains most of the results of the mentioned works \cite{KSV1,KSV2,V1,V2,V3}. The references to these works are listed at the end of the introductory chapter. However, as we will now explain, the discussed topics in the thesis are not enclosed with explanation of \cite{KSV1,KSV2,V1,V2,V3}.

\subsection*{About functional equations}
While we considered particular problems in the context of the AdS/CFT correspondence, it became clear that formulation following from the Bethe Ansatz integral equations in the functional form reveals resemblance between the AdS/CFT integrable system and rational integrable systems. Therefore we put a lot of attention in this thesis to reviewing of the simplest rational integrable models, such as XXX spin chain and Gross-Neveu and principal chiral field sigma models. Instead of using more usual language of integral equations or Fourier transform we perform the review in terms of the functional equations. Then we are able to present the AdS/CFT case as a generalization of rational integrable systems. Basically AdS/CFT spin chain requires for its formulation only one additional integral kernel $\tilde K$ with simple analytical properties.

In the first part of this thesis we also devote attention to such topics as Hirota relations and thermodynamic Bethe Ansatz (TBA). Also these subjects are not directly used by us in the AdS/CFT case, there are number of reasons to include them in the current work. First of all, TBA has a remarkable algebraic structure based on two deformed Cartan matrices. This algebraic structure is seen also on the level of (linear) functional equations. Second, the structure of the supersymmetric Bethe Ansatz is much better seen if Hirota relations are used for its derivation \cite{Kazakov:2007fy}. Third, the discussion of TBA in the rational case can be thought as a preparation for the subject of TBA in the AdS/CFT correspondence which has been recently studied in the literature \cite{Ambjorn:2005wa,Arutyunov:2007tc,Arutyunov:2009zu}, \cite{Gromov:2009tv,Arutyunov:2009ur,Bomb,GromovKKV}.

The name "functional equations" is mostly used in this thesis to denote the functional equations that were used for the asymptotic solution of integral Bethe Ansatz equations. These functional equations are linear.
This thesis deals also with two nonlinear functional equations. The first type is the Hirota equations. Since Hirota equations coincide with the TBA equations, the algebraic structure of Hirota equations can be also seen from the algebraic structure of linear functional equations. The second type is the crossing equations.

\subsection*{Original results presented in this thesis}
This thesis contains few minor original results that were not published before.

In chapter \ref{ch:sl2} we present the solution of the $SL(2)$ Heisenberg magnet in the logarithmic regime and at large values of $j$. This solution is based on the techniques developed in \cite{KSV2,V1,V2}. The solution gives a check of the two-loop strong coupling expansion of the generalized scaling function performed in \cite{Gromov:2008en},\cite{V1}.

In Sec.~\ref{sec:tdlimitfromHirota} we show for the case of the $SU(2)$ Gross-Neveu model and equally polarized excitations that the transfer matrices of the spin chain discretization become in the thermodynamic limit the $T$-functions which appear in the TBA system.

In Sec.~\ref{sec:stringhyp} we show that the integral equations for the densities of string configurations in the $gl(N|M)$ case fit the fat hook structure\footnote{We were informed that this result was also obtained by V.Kazakov, A.Kozak, and P.Vieira, however it has not been published.}. This is generalization of the known results for the $gl(N)$ and the $gl(2|2)$ cases.

In Sec.~\ref{sec:O6} we give an alternative to \cite{Basso:2008tx} derivation of the $O(6)$ sigma model from the BES/FRS equation. This an important check of the AdS/CFT integrable system originally proposed in \cite{Alday2007} and explicitly realized in \cite{Basso:2008tx}.

In appendix \ref{app:mirror} we solve the mirror crossing equations in a similar way as the physical crossing
equations were solved in \cite{V3}. This allows deriving the mirror integrable theory based only on the symmetries of the system, without performing analytical continuation from the physical theory.

\subsection*{Structure of the thesis}


The text of the thesis is divided into three parts. The parts were designed in a way that the reader familiar with basic aspects of integrability could read each part separately. Therefore some concepts are repeated throughout the text.

\paragraph{Part 1. Integrable systems with rational R-matrix.}

The main goal of this part is to give a pedagogical introduction to the subject of integrability.

In the first chapter we introduce the Bethe Ansatz for the integrable XXX spin chain. The XXX spin chain can be easily formulated and the Bethe Ansatz solution is the exact one. This Bethe Ansatz is also important since it describes the AdS/CFT integrable system at one-loop approximation on the gauge side.

In the second chapter we discuss the Bethe Ansatz solution of IQFT on the example of the principal chiral field (PCF) and the Gross-Neveu (GN) models. We exploit the constructions developed in the first chapter since the $S$-matrix in such models up to a scalar factor coincides with the $R$-matrix of the XXX spin chain.

The third chapter aims to show that the PCF and GN models can be viewed as a certain thermodynamic limit of the XXX spin chain. The derivation is based on the string hypothesis. We show also that the functional structure of the Y-system is basically dictated by the functional structure of the equations for the resolvents of string configurations.

The fourth chapter is devoted to the supersymmetric generalization of the ideas developed in the previous chapters.

\paragraph{Part 2. Integrable system of AdS/CFT.}

This part is devoted for reviewing of the integrability in AdS/CFT.

In the fifth chapter we make a general overview of the AdS/CFT correspondence, explain how the integrable system appeared in the spectral problem and discuss the main examples of the local operators/string states which were investigated in the literature.

In the sixth chapter we explain the main steps for the construction of the asymptotic Bethe Ansatz. The second part of this chapter is devoted to the solution of the crossing equations which is based on the author's work \cite{V3}.


\paragraph{Part 3. Integral Bethe Ansatz equations.}

This part is based on the original works \cite{KSV1,KSV2,V1,V2}. It is devoted to the perturbative solution of a certain class of integral Bethe Ansatz equations. The method of the solution is basically the same for all three cases that are considered (one chapter for one case).
This part  is provided with its own introduction.

\renewcommand{\bibname}{\large{List of author's works.}}

\renewcommand{\bibname}{Bibliography}

\part{Integrable systems with rational R-matrix}
\chapter{$SU(N)$ XXX spin chains}
\section{\label{sec:coordinate}Coordinate Bethe Ansatz}
The Bethe Ansatz is one of the most important tools for the solution of the quantum integrable models. It is applicable to all the integrable models considered in this thesis. In the first part of the thesis we would try to give a pedagogical review of the Bethe Ansatz and related topics with the perspective to the modern applications. For other pedagogical texts see \cite{Gaudin,Faddeev:1996iy,IntroBethe1,IntroBethe2,IntroBethe3,PZJ}.

The Bethe Ansatz was initially proposed in \cite{Bethe:1931hc} for the solution of the $SU(2)$ XXX spin chain which served as an approximation for the description of the one-dimensional metal. We will study a slightly more general $SU(N)$ case. The $SU(N)$ XXX spin chain is defined as follows. To each node of the chain  we associate the integer number with possible values from $0$ to $N-1$. In other words, each node carries a state in the fundamental representation of $SU(N)$.
The Hamiltonian is given by
\be\label{hamiltonianXXX}
  H=\sum_k \left(1-{\CP_{k,k+1}}\right),
\ee
where $\CP_{k,m}$ is the operator that permutes the $SU(N)$ states at sites $k$ and $m$.

The problem of the diagonalization of the Hamiltonian can be solved by the coordinate Bethe Ansatz. To show how it works, we should choose a pseudovacuum of the system. A possible choice is $|000...00000\rangle$ (the number zero is assigned to all the nodes). This is indeed a state with the lowest energy, although this energy level is highly degenerated due to the $SU(N)$ symmetry of the system. For the opposite sign of the Hamiltonian (antiferromagnetic spin chain) this state is no longer the vacuum but just a suitable state for the construction of the coordinate Bethe Ansatz.

 We consider subsequently the one-, two-, three- e.t.c particle excitations. By $n$-particle excitation we mean a state in which the values of exactly $n$ nodes are different from $0$.

From the one-particle excitations we can compose a plane wave excitation
\be
  \psi[\{p,a\}]=\sum_{k} e^{ipk}\phi[\{k,a\}],\;\;\;\phi[\{k,a\}]=|00...a(\textrm{$k$-th position})...000\rangle.
\ee
This plane wave is an eigenfunction of the Hamiltonian with the eigenvalue
\be
E[p]=2-2\cos[p].
\ee

For the investigation of the two-particle excitations let us first consider the following state:
\be
   \psi[\{p_1,a_1\},\{p_2,a_2\}]&=&\sum_{k_1<k_2} e^{ip_1k_1+ip_2k_2}\phi[\{k_1,a_1\},\{k_2,a_2\}],\\\phi[\{k_1,a_1\},\{k_2,a_2\}]&=&|00...a_1(\textrm{$k_1$-th position})\ldots a_2(\textrm{$k_2$-th position})\ldots 000>.\no
\ee
The Hamiltonian does not act diagonally on it. However, its action can be represented in the following form:
\be
  &&(H-E[p_1]-E[p_2])\psi[\{p_1,a_1\},\{p_2,a_2\}]=(1-e^{ip_2}+e^{i(p_1+p_2)})\chi_{a_1a_2}-e^{ip_2}\chi_{a_2a_1},\no\\
  &&\chi_{ab}=\sum_{k}e^{ik(p_1+p_2)}\phi[\{k,a\},\{k+1,b\}].
\ee
It is straightforward to check that the terms $\chi_{ab}$ cancel out if we consider the following linear combination
\be\label{twoparticle}
  \Psi_2=\psi[\{p_1,a_1\},\{p_2,a_2\}]+S_{a_1a_2}^{b_1b_2}[p_1,p_2]\psi[\{p_2,b_2\},\{p_1,b_1\}]
\ee
with
\be
  S_{a_1a_2}^{b_1b_2}[p_1,p_2]=\frac{e^{-ip_2}+e^{ip_1}}{1-2e^{ip_2}+e^{i(p_1+p_2)}}\delta^{b_1}_{a_1}\delta^{b_2}_{a_2}-
  \frac{\(e^{ip_2}-1\)\(e^{ip_1}-1\)}{1-2e^{ip_2}+e^{i(p_1+p_2)}}\delta^{b_1}_{a_2}\delta^{b_2}_{a_1}.
\ee
The matrix $S$ is called the scattering matrix. Note that indices $a_i$ and $b_i$ take values from $1$ to $N-1$. This is because the pseudovacuum is not invariant under the $SU(N)$ symmetry but only under the $SU(N-1)$ subgroup, therefore the excitations over the pseudovacuum transform under $SU(N-1)$.

It is useful to introduce the rapidity variable $\theta$ via the relation
\be
e^{ip}=\frac {\theta+\frac i2}{\theta-\frac i2}.
\ee
In terms of the rapidity variables the $S$ matrix has a simpler form:
\be\label{S-matrix}
  S[p_1,p_2]=S[\th_1-\th_2]=\frac{\th_1-\th_2-i\CP}{\th_1-\th_2+i},
\ee
Here the operator $\CP$ permutes the $SU(N-1)$ states of the particles. It is different from $\CP_{k,m}$ in (\ref{hamiltonianXXX}) where $SU(N)$ states were permuted.

For a sufficiently long spin chain we can give a physical interpretation for (\ref{twoparticle}) as the scattering process. Let us take $p_1>p_2$. The function $\psi[\{p_1,a_1\},\{p_2,a_2\}]$ is interpreted as the initial state: the particle with momenta $p_1$ is on the left side from the particle $p_2$. The function $\psi[\{p_2,b_2\},\{p_1,b_1\}]$ represents the out state. Scattering of the particles leads to the exchange of their flavors and to the appearance of the relative phase. The scattering is encoded in the $S$ matrix. The permutation operator $\CP$ in (\ref{S-matrix}) exchanges the flavors of the particles\footnote{We chose the notation in which the flavor is attached to the momenta of the particle. The other notation is also possible in which the flavor is attached to the relative position of the particle. In the latter notation the $S$ matrix differs by multiplication on the permutation operator.}.

The two particle wave function (\ref{twoparticle}) can be written in a condensed notation as
\be
  \Psi_2=\psi_{12}+S_{12}\psi_{21}.
\ee
It turns out that for the three-particle case the following linear combination
\be
  \Psi_3=\psi_{123}+S_{12}\psi_{213}+S_{23}\psi_{132}+S_{13}S_{12}\psi_{231}+S_{13}S_{23}\psi_{312}+S_{23}S_{13}S_{12}\psi_{321}
\ee
is an eigenfunction of the Hamiltonian with the eigenvalue $E[p_1]+E[p_2]+E[p_3]$. It can be interpreted as the combination of all possible configurations obtained from the \textit{pairwise} scattering of initial configuration given by $\psi_{123}$. The property that we encounter only the $2\to 2$ scattering process is an essential property of the integrability.

The last term in $\Psi_3$ can be written also as $S_{12}S_{13}S_{23}\psi_{321}$. This is because the process $(123)\to(321)$ can be obtained in two different ways. The independence of the order of individual scattering processes is encoded in the Yang-Baxter equation
\be\label{yang-baxter}
  S_{23}S_{13}S_{12}=S_{12}S_{13}S_{23}.
\ee
It is straightforward to check that this equation is satisfied by the $S$ matrix (\ref{S-matrix}). The scattering matrix also satisfies the unitarity condition
\be\label{unitarity}
  S_{12}S_{21}=1.
\ee

Let us consider now the states with larger number of particles. The solution of the diagonalization problem is obtained in a similar way. This solution is known as the coordinate Bethe Ansatz. The $K$-particle wave function is given by
\be\label{coordinate Bethe Ansatz}
  \Psi_K=\sum_{\pi}S_\pi\psi_{\pi[1]\pi[2]\ldots\pi[K]},
\ee
where the sum is taken over all permutations. The symbol $S_\pi$ means the following. We choose a realization of $\pi$ in terms of a product of the elementary permutations: $\pi=(a_1b_1)\ldots (a_kb_k)$. The transpositions of only the neighboring particles are allowed. Then $S_\pi=S_{(a_1b_1)...(a_kb_k)}\equiv S_{a_1b_1}S_{a_2b_2}\ldots S_{a_kb_k}$. The construction of the wave function is unambiguous since the $S$ matrix satisfies the Yang-Baxter equation (\ref{yang-baxter}) and the unitarity condition (\ref{unitarity}).

 The Ansatz (\ref{coordinate Bethe Ansatz}) gives the solution of the diagonalization problem for the case of the spin chain on infinite line. For the case of the periodic spin chain of length $L$ there are constraints on the possible values of momenta. To write down these constraints we first need to introduce an appropriate basis in the space of functions (\ref{coordinate Bethe Ansatz}).

 The scattering matrix is invariant under action of the $SU(N-1)$ group. Therefore, instead 
 $P_{Y}\Psi_K$ where $P_{Y}$ is the projector to a chosen representation $Y$ of the $SU(N-1)$ group.
 As we will show in the next section, it is possible to decompose the whole representation space $(\MC^{N-1})^{\otimes K}$ in a sum of irreps
 \be\label{decomposition}
  (\MC^{N-1})^{\otimes K}=\oplus_\a Y_{\a}
 \ee
  such that
 \be\label{115}
    \left[\prod_{j\neq k}^K S_{kj},P_{Y_\a}\right]=0,\ \ k=\overline{1,K}.
 \ee
 The ordering in the product is taken in such a way that $S_{kj}$ is to the right from $S_{kj'}$ if $(j-k)<(j'-k)$ mod $K$.

 For $K=1$ the decomposition (\ref{decomposition}) is a trivial property since the scattering matrix commutes with the symmetry generators. The nontrivial statement for $K\neq 1$ is in simultaneous validity of (\ref{115}) for different values of $k$.

  We chose a basis in such a way that each basis wave function lies inside some irrep $Y_\a$. The periodicity condition, which can be written as
  \be
  \phi[\{0,a\},\{k_2,a_2\},\ldots]=\phi[\{k_2,a_2\},\ldots,\{L,a\}],
   \ee
   implies the following equations on $p_k$, $k=\overline{1,K}$, for a chosen irrep $Y_\a$:
\be\label{prebethe}
  e^{-ip_kL}P_{Y_\a}=\prod_{j\neq k}^K S_{kj}\,P_{Y_\a}.
\ee
Once momenta satisfy (\ref{prebethe}), $P_{Y_\a}\Psi_k$ is the eigenfunction of the Hamiltonian on the periodic chain of length $L$ with the energy
\be\label{Energyformula}
    E=\sum_{k=1}^K(2-2\cos[p_k])=\sum_{k=1}^K\frac 1{\theta_k^2+\frac 14}\,.
\ee

Note that if the set of momenta $p_k$ is a solution of (\ref{prebethe}) for a given representation $Y_\a$, there is no guaranty that the same set of momenta would be a solution for a different representation. In fact, the general situation is that a set of momenta is a solution of (\ref{prebethe}) for only one representation $Y_\a$.

The equations (\ref{prebethe}) are best written in terms of the nested transfer matrix\footnote{We use the adjective "nested" to distinguish (\ref{TL}) with what is usually called the transfer matrix and which we introduce later.} which is defined as follows. We introduce an auxiliary particle $A$, with rapidity $\th$, in the fundamental representation of the $SU(N-1)$ group. The nested transfer matrix is given by
\be\label{TL}
  T[\th,\{\th_1,\ldots,\th_K\}]=\Tr_{\!A} \CT,\;\;\; \CT=S_{A K}....S_{A2}S_{A1},
\ee
where the trace is token over the representation space of the auxiliary particle. Since $S[0]=-\CP$, the periodicity conditions (\ref{prebethe}) can be written as
\be\label{period}
  e^{-ip_kL}P_{Y_\a}=-T[\th_k]P_{Y_\a}.
\ee
For the case $N=2$ the $S$ matrix is just a scalar function and the equations (\ref{period}) reduce to
\be\label{betheXXX}
  \(\frac{\th_k+\frac i2}{\th_k-\frac i2}\)^L=-\prod_{j=1}^K\frac{\th_k-\th_j+i}{\th_k-\th_j-i}
\ee
which are the Bethe Ansatz equations for the $SU(2)$ XXX spin chain.

\ \\

Let us now briefly describe how the solutions of the Bethe equations (\ref{betheXXX}) are mapped to the eigenstates of the Hamiltonian (\ref{hamiltonianXXX}) for the case $N=2$.

The solutions with at least two coinciding rapidities lead to a zero wave function. This is due to the fact that $S[0]=-1$. Each solution with pairwise different rapidities, $K\leq L/2$, and finite values of $\theta_k$ leads to an eigenstate of the Hamiltonian\footnote{There are few solutions of exceptional type of the Bethe equations, which require  the proper regularization to fit this picture. The example is the solution $\theta_1=-\theta_2=i/2$ for $K=2$ and $L=4$.}. This eigenstate is a highest weight vector in the irreducible multiplet of $SU(2)$. To obtain the other states of this multiplet we have to add subsequently the particles with infinite rapidities (zero momenta) to the solution. Adding the particle with infinite rapidity is equivalent to the action of the lowering generator of the $su(2)$ algebra on the wave function. The Bethe Ansatz solution is complete \cite{CompletenessSU2}. This means that all the eigenstates of the Hamiltonian can be obtained by the procedure that we described.

\section{Nested/Algebraic Bethe Ansatz}
For $N>2$ we have to diagonalize the nested transfer matrix (perform the decomposition (\ref{decomposition})) in order to solve the periodicity conditions (\ref{period}). This is the subject of the nested Bethe Ansatz. We consider first the case of the $SU(3)$ spin chain.
\begin{figure}[!h]
\centering
\includegraphics[height=4.5cm]{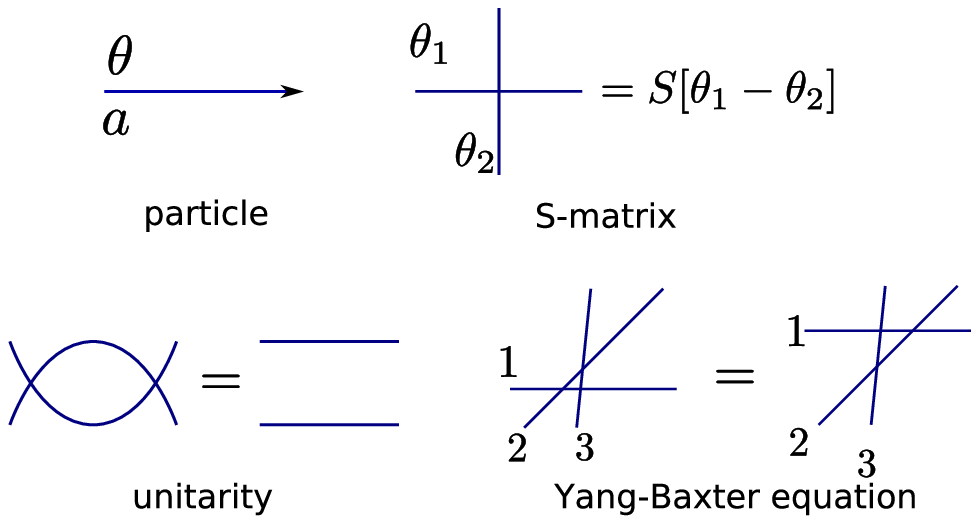}
\caption{\label{fig:definitions}Main definitions, unitarity and Yang-Baxter equations.}
\end{figure}In this case the excitations over the pseudovacuum can have only two labels, $1$ and $2$.

It is instructive to use the graphical representation (see Fig.~\ref{fig:definitions}) for the algebraic constructions that we will use. In the graphical representation each particle is represented by an arrow. For each arrow we assign the rapidity and the color.
If the direction is not shown explicitly we take by default that the particle propagates 1)from left to right and 2)from bottom to top. $S$ matrix is given as an intersection of two lines (scattering of the particles).

\begin{figure}[t]
\centering
\includegraphics[height=2.5cm]{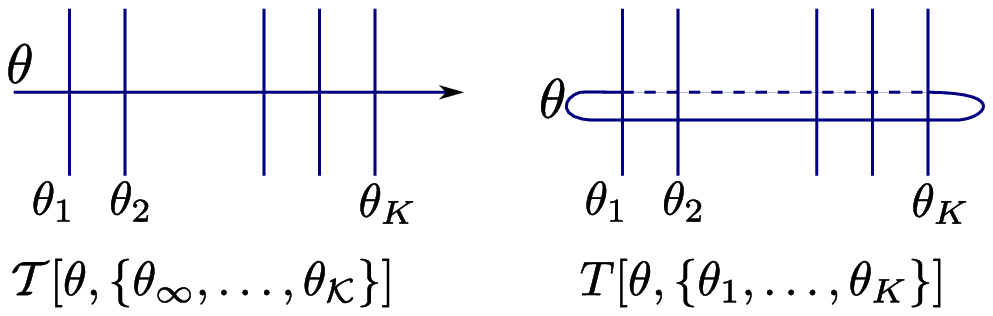}
\caption{\label{fig:transfermatrix}Nested monodromy and transfer matrices}
\end{figure}

The graphical representation of the nested monodromy matrix $\CT$ and the nested transfer matrix $T$ defined by (\ref{TL}) are shown in Fig.~(\ref{fig:transfermatrix}).
%
%

The nested monodromy matrix $\CT$ scatters the auxiliary particle through all the physical particles. Since in the considered case ($N=3$) particles can be in one of two states ($1$ or $2$), we can write the nested monodromy matrix as a two by two matrix with elements acting on the physical space only:
\be
  \CT=\left(
      \begin{array}{cc}
        A[\th] & B[\th] \\
        C[\th] & D[\th] \\
      \end{array}
    \right),\hspace{2.5EM}\ \parbox[c]{2.8cm}{\includegraphics[height=2cm]{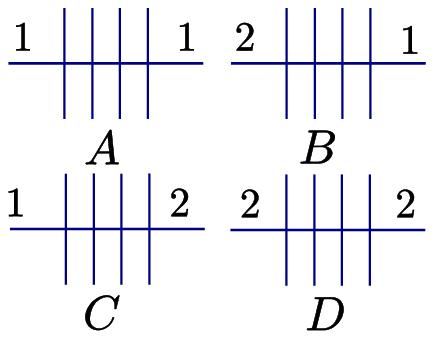}}.
\ee
Obviously, $T=A+D$.

The nested transfer matrices with different values of $\th$ commute with one another:
\be
[T[\theta],T[\theta']]=0.
\ee
Indeed, from the Yang-Baxter equation the following equation follows:
\be\label{eq:SLLLLS}
   \parbox[c]{2.8cm}{\includegraphics[width=2.8cm]{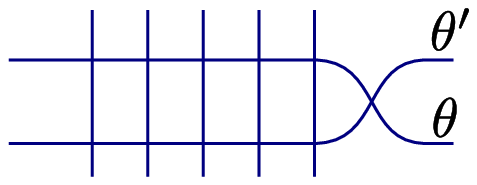}}&=&\parbox[c]{2.8cm}{\includegraphics[width=2.8cm]{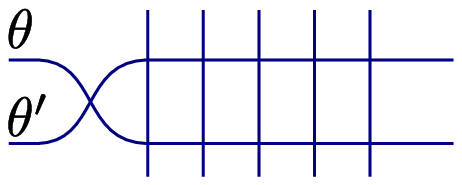}}\no\\
   S[\th-\th']\CT[\th]\CT[\th']&=&\CT[\th']\CT[\th]S[\th-\th'].
\ee
It can be also rewritten as
\be\label{STTSTT}
S\CT[\th]\CT[\th']S^{-1}=\CT[\th']\CT[\th].
\ee
Taking the trace over the representation spaces of both auxiliary particles in (\ref{STTSTT}) we prove the commutativity of the nested transfer matrices.

Due to the commutativity of the nested transfer matrices the simultaneous diagonalization of  $T[\theta_k]$ is possible. Since  $T[\theta]$ is invariant under the symmetry algebra, the condition (\ref{115}) can be also satisfied and the Bethe Ansatz equations (\ref{period}) can be constructed.

%

To solve the problem of diagonalization of $T[\theta]$ we first introduce a nested pseudovacuum. It consists of the particles with rapidities $\th_1,\ldots,\th_K$, all of them are of the color $1$. We will denote this new pseudovacuum as $|0_1\rangle$. 

The nested transfer matrix acts diagonally on $|0_1\rangle$:
\be
T[\theta]|0_1\rangle=\parbox[c]{4.1cm}{\includegraphics[width=4cm]{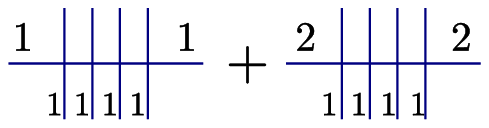}}=\(\frac{Q[\theta-i]}{Q[\theta+i]}+\frac{Q[\theta]}{Q[\theta+i]}\)|0_1\rangle,
\ee
where we introduced the Baxter polynomial
\be\label{BaxterPolynomial}
Q[\theta]=\prod\limits_{j=1}^K(\theta-\theta_j).
\ee

\begin{figure}[t]
\centering
\includegraphics[height=2.8cm]{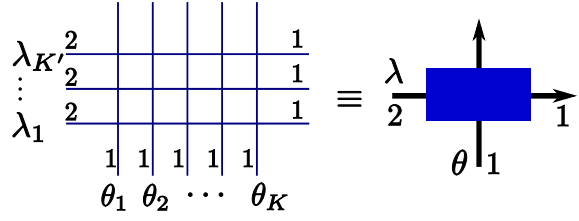}
\caption{\label{fig:fusion21}Construction of "exited" states (\ref{nesteeigenvector}) over the nested pseudovacuum.}
\end{figure}

Excitations over the nested pseudovacuum are generated by $B$ operators:
\be\label{nesteeigenvector}
  \Phi=B[\l_1]B[\l_2]\ldots B[\l_{K'}]|0_1\rangle.
\ee
These states are given by the diagram in Fig.~\ref{fig:fusion21}.

Generically, the action of the nested transfer matrix on the state $\Phi$ is not diagonal. However this is the case if the rapidities $\l$ satisfy the relation
\be\label{fusioninit}
  Q[\l_j-i]Q_\l[\l_j+i]+Q[\l_j]Q_{\lambda}[\l_j-i]=0,\ \ j=\overline{1,K'},
 \ \ Q_{\lambda}[\theta]\equiv\prod_{j=1}^{K'}(\theta-\l_j).
\ee

The eigenvalue of the transfer matrix is then given by
\be\label{eigenT}
  T[\th]\Phi=\(\frac{Q[\th-i]}{Q[\th+i]}\frac{Q_\l[\th+i]}{Q_\l[\th]}+\frac{Q[\th]}{Q[\th+i]}\frac{Q_\l[\th-i]}{Q_\l[\th]}\)\Phi.
\ee

Here we do not prove that the condition (\ref{fusioninit}) is necessary and sufficient to diagonalize action of $T$. See for example \cite{Faddeev:1996iy}.
However note that the relation (\ref{fusioninit}) can be read from the eigenvalue of the transfer matrix in the following way. From the form of the $S$ matrix we see that the nested transfer matrix cannot have poles at $\th=\l_j$. The application of this demand to (\ref{eigenT}) gives us (\ref{fusioninit}).

If we shift the variables $\l$ by $\frac i2$, the equation (\ref{fusioninit}) acquires the form of the nested Bethe Ansatz equation:
\be\label{nested1}
  1=-\frac{Q[\l_k-\frac i2]Q_\l[\l_k+i]}{Q[\l_k+\frac i2]Q_\l[\l_k-i]}=-\prod_{i=1}^K\frac{\l_k-\th_i-\frac i2}{\l_k-\th_i+\frac i2}\prod_{j=1}^{K'}\frac{\l_k-\l_j+i}{\l_k-\l_j-i}.
\ee
The periodicity condition (\ref{period}) leads to the following Bethe equation:
\be\label{nestedground}
  e^{-ip_kL}=-T[\th_k]=-\frac{Q[\th_k-i]}{Q[\th_k+i]}\frac{Q_\l[\th_k+\frac i2]}{Q_\l[\th_k-\frac i2]}.
\ee
From solution of (\ref{nestedground}) and (\ref{nested1}) we can construct the eigenstate for the $SU(3)$ XXX spin chain using (\ref{nesteeigenvector}) and (\ref{coordinate Bethe Ansatz}) and find its energy using (\ref{Energyformula}).

\subsubsection{Algebraic Bethe Ansatz for the $SU(2)$ XXX spin chain}
We can make another interpretation for the relations (\ref{fusioninit}). If to put all the $\th_k$ equal to $-i/2$, then (\ref{fusioninit}) transforms to
\be
  \(\frac{\l_k+\frac i2}{\l_k-\frac i2}\)^{K}=-\prod_{j=1}^{K'}\frac{\l_k-\l_j+i}{\l_k-\l_j-i}.
\ee
This is nothing but the Bethe Ansatz (\ref{betheXXX}) for the Heisenberg $SU(2)$ XXX spin chain with $L,K$ replaced by $K,K'$.

Of course, when all $\th_k$ are equal, we cannot interpret them as rapidities of excitations in the spin chain - the wave function would be just zero for them. Correspondingly, the periodicity condition (\ref{prebethe}) looses its sense. Instead, each particle with rapidity $\th=-\frac i2$ is interpreted as a node in the spin chain. Diagonalization of $T[\theta]$ automatically diagonalize the Hamiltonian due to the following equality:
\be\label{XXXhamiltonian}
  H_{\rm{XXX}}=-i\frac 1{T[\theta]}\frac {d T[\theta]}{d\theta}
\begin{picture}(5,20)
\put(2,-11){\line(0,1){20}}
\end{picture}
\raisebox{-0.35cm}[-0.7cm]{$\theta=-\frac i2$}\ .
\ee
The nested eigenvectors (\ref{nesteeigenvector}) are proportional to the corresponding
states (\ref{coordinate Bethe Ansatz}) of the coordinate Bethe Ansatz.

This approach of solving the XXX spin chain is called the algebraic Bethe Ansatz \cite{Faddeev:1979gh,Faddeev:1981ft}.
The matrix $T[\theta]$ in the algebraic Bethe Ansatz approach is called the transfer matrix of a spin chain. This explains why in previous section we used the notion of the "nested" transfer matrix: to distinguish between nested and algebraic Bethe Ansatz interpretations.

Together with the diagonalization of the Hamiltonian the algebraic Bethe Ansatz gives us the possibility to construct higher conserved charges. The local conserved charges are the coefficients of the expansion of the transfer matrix around a singular point $\theta=-\frac i2$:
\be
    T[-i/2+\e]=(-1)^K P e^{i(\e H_{XXX}+\e^2 H_3+\ldots)}.
\ee
Here $P$ is the operator of translation by one node of a spin chain.

Expansion around any other nonsingular point gives us other conserved charges that are not local. Of course the distinction between locality and no locality can be made only in the limit of infinite length.

\subsubsection{Nested Bethe Ansatz for $N>3$}
\begin{wrapfigure}{r}{0.5\textwidth}
 \vspace{-32pt}
  \begin{center}
    \includegraphics[width=0.35\textwidth]{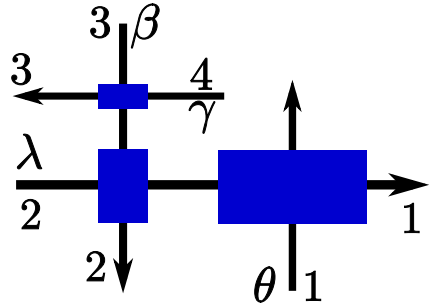}
  \end{center}
  \vspace{-15pt}
  \caption{\label{fig:highernested}Nesting procedure}
\end{wrapfigure}
The construction that we used to diagonalize $T[\theta]$ can be generalized to the case with arbitrary $N$ \cite{Kulish:1983rd}.

For example for $N=5$ any excited state can be built by the procedure shown in Fig.~\ref{fig:highernested}\footnote{We conjecture that Fig.~\ref{fig:highernested} is equivalent to the procedure in \cite{Kulish:1983rd}. Although we did not prove this explicitly, we checked on simple examples that the procedure in Fig.~\ref{fig:highernested} generates eigenstates of the transfer matrix once the rapidities satisfy nested Bethe Ansatz equations.}.

Again, the nested transfer matrix acts diagonally if and only if the rapidities $\l$, $\b$ and $\g$ satisfy relations which are exactly the nested Bethe Ansatz equations. The shortcut to write these relations can be read from the fact that the eigenvalue of the transfer matrix on the nested state  is given by
\begin{small}
\be\label{TLeigenvalue}
  \(\frac{Q[\th-i]}{Q[\th]}\frac{Q_\l[\th+i]}{Q_\l[\th]}
  +\frac{Q_\l[\th-i]}{Q_\l[\th]}\frac{Q_\b[\th+i]}{Q_\b[\th]}
  +\frac{Q_\b[\th-i]}{Q_\b[\th]}\frac{Q_\g[\th+i]}{Q_\g[\th]}+
  \frac{Q_\g[\th-i]}{Q_\g[\th]}\)\frac{Q[\theta]}{Q[\theta+i]}
\ee
\end{small}
and the requirement that the transfer matrix does not have poles except for $\th=\th_k+i$.

The nested Bethe Ansatz equations can be encoded in the following diagram:
\begin{figure}[!h]
\centering
\includegraphics[height=1cm]{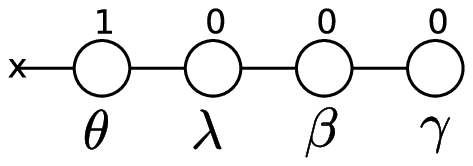}
\caption{\label{fig:dynkinnested}The Dynkin diagram for the nested Bethe Ansatz}
\end{figure}

Each node of the diagram corresponds to the one type of nested Bethe roots. For each node its left neighbor plays the role of the inhomogeneous spin chain. The cross corresponds to the initial homogeneous spin chain (each node of which can be interpreted as a particle with rapidity equal to $-i/2$).

 Actually, the diagram \ref{fig:dynkinnested} without cross is nothing but the Dynkin diagram. The cross corresponds to the fact that we consider particles in the fundamental representation of the $SU(N)$ group defined by the Dynkin labels $[1,0,0,0]$.

Each physically meaningful solution of the Bethe equations should not contain coinciding rapidities. The solutions with only finite Bethe roots and the numbers of Bethe roots that satisfy inequalities (\ref{inequalities}) give the highest weight vectors in the irreducible multiplet of the $SU(N)$ group. The highest weight states and the states obtained from them by action of the symmetry generators span the whole Hilbert space of the system \cite{BetheCompleteness}.

 If to put $\theta_k=-\frac i2$, we can interpret Fig.~\ref{fig:highernested} as the algebraic Bethe Ansatz for the $SU(4)$ spin chain. In this case the procedure shown in Fig.~\ref{fig:highernested} gives us also the wave function of the corresponding eigenstate.
  \ \\

It is also possible to construct the Bethe Ansatz for arbitrary simple Lie algebra and arbitrary irreducible representation (irrep). For a Lie algebra of rank $r$ defined by the Cartan matrix $c_{ab}$ and for the irrep given by the Dynkin labels $[\o_1,\ldots,\o_r]$ the Bethe Ansatz equations for a homogeneous spin chain of length $L$ are written as \cite{Ogievetsky:1986hu}:
 \be\label{Bethegeneralrepresentation}
    \(\frac{\theta_{a,j}+\frac i2\o_a}{\theta_{a,j}-\frac i2\o_a}\)^L=-\prod_{b=1}^{r}\prod_{k=1}^{K_{b}}\frac{\theta_{a,j}-\theta_{b,k}+\frac i2c_{ab}}{\theta_{a,j}-\theta_{b,k}-\frac i2c_{ab}},\ \ \ a=\overline{1,r},\ j=\overline{1,K_a}.
 \ee
%
\section{\label{sec:countingBethe}Counting of Bethe roots and string hypothesis}
For the simplest case of the $SU(2)$ magnet the number of the Bethe roots cannot be larger than the half of the length of the spin chain. This restriction can be explained by the representation theory. Each solution of the Bethe equations with all Bethe roots being finite corresponds to a highest weight vector. We cannot construct the highest weight vector for the number of excitations larger than the half of the length.

The same logic may be applied in principle for the $SU(N)$ magnet. However, it would be nice to see how the constraints on the number of the Bethe roots come directly from the Bethe equations. We will first do this derivation for the $SU(2)$ case and then generalize to arbitrary $N$.

It is useful to introduce the Baxter equation
\be\label{baxeq}
  \(u+\frac i2\)^L Q[u-i]+\(u-\frac i2\)^L Q[u+i]=T[u]Q[u].
\ee
Assuming that $Q[u]$ is a polynomial, it is easy to see that the set of the Bethe equations in the $SU(2)$ case (\ref{betheXXX}) is equivalent to the demand that the function $T[u]$ defined by (\ref{baxeq}) is an entire function (and therefore is a polynomial).

The zeroes of $Q[u]$ are the Bethe roots, therefore $Q[u]$ is the Baxter polynomial (\ref{BaxterPolynomial}). The reader can recognize in $T[u]$ the eigenvalue of the transfer matrix rescaled by an overall factor ({\it{cf.}} (\ref{eigenT})).

Let us take a solution of the Bethe equations which consists only from real roots and find all zeroes of the l.h.s. of (\ref{baxeq}). Among these zeroes there are Bethe roots (zeroes of $Q[u]$) and zeroes of $T[u]$.  The real zeroes of $T[u]$ we will call holes. The  zeroes of $T[u]$ with nonzero imaginary part will be called accompanying roots.

Accompanying roots are situated roughly on the distance $i$ above and below the Bethe roots. This fact can be understood in the large $L$ limit. Indeed, if we consider the region ${\rm Im}[u]>0$ for the large $L$, generically the second term in the l.h.s. of (\ref{baxeq}) is suppressed with respect to the first one. To estimate the magnitude of the suppression one can approximate
\be\label{eq:Bethesu2}
  \log \[\(\frac{u+\frac i2}{u-\frac i2}\)^L\frac{Q[u-i]}{Q[u+i]}\]\simeq \frac{iL}{Re[u]}-\sum_{k=1}^M \frac {2i}{u-u_k}+L\frac{{ \rm Im}[u]}{{\rm Re}[u]^2}-\sum_{k=1}^M\frac {2{\rm Im}[u-u_k]}{{\rm Re}[u-u_k]^2}.
\ee
We see that suppression is strong at large $L$ if $u$ is sufficiently close to origin.

The first term in the l.h.s. of (\ref{baxeq}) is suppressed for ${\rm Im}[u]<0$. Therefore in the large $L$ limit we can write approximately
\be
  \(u+\frac i2\)^LQ[u-i]&=&T[u]Q[u],\ \ {\rm Im}[u]>0,\no\\
  \(u-\frac i2\)^LQ[u+i]&=&T[u]Q[u],\ \ {\rm Im}[u]<0.
\ee
We see that the accompanying roots are given in the first approximation by zeroes of $Q[u-i]$ and $Q[u+i]$. Therefore each Baxter root gives 2 accompanying zeroes in $T[u]$. Counting the number of zeroes on both sides of the Baxter equation, we obtain
\be
  L+n=n+2n+n_h,\ \to\ \ n=\frac {L-n_h}{2},
\ee
where $n$ is the number of Bethe roots and $n_h$ is the number of holes.

\begin{figure}[t]
\centering
\includegraphics[width=12.52cm]{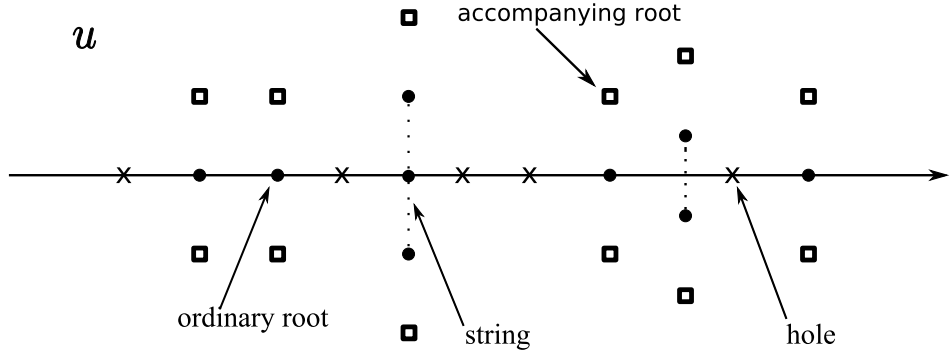}
\caption{\label{fig:polesforsimplerootbaxter}Structure of zeroes of the l.h.s. of (\ref{baxeq}) (the complex Bethe roots are allowed).}
\end{figure}
The maximally filled state does not contain holes and we obtain the known restriction on the number of the Bethe roots.

This analysis is simply generalized to the case of $SU(N)$ magnet. In this case we have $N-1$ types of Bethe roots. Let us denote the number of roots of each type by $n_1,n_2,\ldots, n_{N-1}$. For the $k$-th type of the Bethe roots we can write the Baxter equation:
\be
  Q_{k-1}\[u+\frac i2\]Q_{k+1}\[u+\frac i2\]Q_{k}\[u-i\]+{\rm c.c}=Q_k[u]T_k[u].
\ee
Now the $Q_{k-1}\[u\pm\frac i2\]Q_{k+1}\[u\pm\frac i2\]$ play the same role as $(u\pm\frac i2)^L$ in (\ref{baxeq}) and we obtain $2n_k\leq n_{k-1}+n_{k+1}$.

The complete set of inequalities can be written as
\be\label{inequalities}
  2n_1\leq L+n_2,\ \ 2n_2\leq n_1+n_3,\ \ldots \ 2n_{N-2}\leq n_{N-3}+n_{N-1},\ \ 2n_{N-1}\leq n_{N-2}.
\ee
From here it follows that a maximally saturated state (antiferromagnetic vacuum), which corresponds to the trivial representation, is constructed from the following number of Bethe roots:
\be
  n_k=\frac {N-k}{N}L,\ \ k=\overline{1,N-1}.
\ee

\subsubsection{String hypothesis}

The accompanying roots could not be the roots of $Q[u]$ since we had restricted ourselves for the case when all the Bethe roots are real. If we allow the Bethe roots to be complex, then an accompanying root can become the complex root of $Q[u]$. But in this case this complex root of $Q[u]$ will have its own accompanying root. In turn, we can allow this second-level accompanying root to enter the Baxter polynomial or not. In such a way we construct a so called string.

The string of length $s$, or $s$-string, is the following set of the Bethe roots:
\be
  \theta_{0}+in,\ \ \  -\frac{s-1}2\geq n\geq \frac{s-1}2,
\ee
where, depending on $s$, $n$ is integer or half-integer.
The $s$-string is completely defined by the position of its center $\theta_{0}$.

The string hypothesis states that in the limit $L\to\infty$ all the Bethe roots are organized in strings. It was shown by counterexamples that the string hypothesis is strictly
speaking wrong. Although, as we can see from (\ref{eq:Bethesu2}), stringy configurations dominate for $\theta_0$ being close to the origin, for $\theta_0\sim L$ the configuration resembles string only qualitatively. In this regime the imaginary distance between roots scales as $\sqrt{L}$ and the positions of roots belonging to string significantly deviate from lying on a straight line. There are also more sophisticated examples of solutions which do not satisfy string hypothesis even qualitatively\footnote{For numerical and analytical studies related to the string hypothesis see \cite{Isler:1993fc,Ilakovac:1999pe,Antipov2006,Bargheer:2008kj} and references therein.}.

Although the string hypothesis is wrong, the stringy configurations describe the low energy excitations in some regimes that we will be interested in. Therefore it is reasonable to study Bethe equations as if the string hypothesis was correct.


What is a string from the point of view of the coordinate Bethe Ansatz? Let us consider a string of length two which is composed from two rapidities $\theta_{\pm}=\theta_0\mp i/2$. The wave function for such string is given by
\be
  \Psi&=&\sum_{k_1<k_2}\((\theta_--\theta_++i)e^{ip_-k_1+ip_+k_2}+(\theta_--\theta_+-i)e^{ip_+k_1+ip_-k_2}\)\phi[k_1,k_2]=\no\\
  &=&2i\sum_{k_1<k_2}e^{ip_-k_1+ip_+k_2}\phi[k_1,k_2]
\ee
Since $\rm{Im}[p_\pm]\gtrless 0$, this wave function describes the propagation of the bound state with momenta $p=p_++p_-=-i\log\[\frac{\theta_0+i}{\theta_0-i}\]$. Therefore strings correspond to the bound states.

\subsubsection{Interaction of strings}
Each nested level of the $SU(N)$ Bethe Ansatz has its own string solutions. Let us assume that the string hypothesis is valid and write down Bethe equations explicitly for the center of strings.

The Bethe equations are constructed from the following building block:
\be\label{buildblock}
  \ldots \prod_{j}\frac{u_{a;k}-u_{a';j}+\frac i2}{u_{a;k}-u_{a';j}-\frac i2}\ \ldots\ .
\ee
The indices $a,a'=a\pm 1$ label the nested level of the Bethe roots (the case $a=a'$ is considered below), $k,j$ enumerate Bethe roots at each level.

Let us introduce the shift operator
\be
  D\equiv e^{\frac i2\p_u}
\ee
and the following notation
\be\label{Dexpnotation}
  (f[u])^{D}\equiv e^{D\log f[u]}.
\ee
In this notation the function $\frac{u-v+i/2}{u-v-i/2}$ will be written as
\be
  (u-v)^{D-D^{-1}}.
\ee
To write the Bethe equation for the center of a given string we have to multiply the Bethe equations for each root which constitutes the string. For the string of length $s$ and with the center at $u_0$ we will get the get a factor of the type
\be
  \ldots\prod_{n=-\frac {s-1}2}^{\frac {s-1}2}\(u_0-v+in\)^{(D-D^{-1})}\ \ldots=\ldots(u_0-v)^{D^{s}-D^{-s}}\ \ldots\ .
\ee
From (\ref{buildblock}) we conclude that the interaction of the strings of length $s$ and $s'$ with centers at $u_0$ and $v_0$ are written as
\be\label{stringintdif}
  (u_0-v_0)^{\frac{(D^{s}-D^{-s})(D^{s'}-D^{-s'})}{D-D^{-1}}}\equiv (u_0-v_0)^{\CL_{ss'}}\ .
\ee
We have to understand the denominator of $\CL_{ss'}$ as such power series that $\CL_{ss'}$ represents a finite linear combination of shift operators. More precisely:
\be
  \CL_{ss'}&\equiv&-D\frac{D^{s+s'}-D^{|s-s'|}}{1-D^{2}}+D^{-1}\frac{D^{-s-s'}-D^{-|s-s'|}}{1-D^{-2}}=
  \no\\&=& D^{s+s'-1}+D^{s+s'-3}+\ldots +D^{|s-s'|+1}-\no\\&&-D^{-|s-s'|-1}-D^{-|s-s'|-3}-\ldots-D^{-s-s'+1}.
\ee
The expression (\ref{stringintdif}) describes the interaction of strings from different nested levels ($a\neq a'$). In the case when strings belong to the same nested level, the interaction will look like
\be
  (u_0-v_0)^{(D+D^{-1})\CL_{ss'}}.
\ee
Now we are ready to write down the set of Bethe equations for the centers of strings. We introduce the following notations.

First, the centers of strings are marked by $u_{a,s;j}$, where $a$ labels the nested level (node of the Dynkin diagram), $s$ labels the length of the string and $j$ enumerates different strings with the same $a,s$.

Then we will also need the "$D$-deformed" Cartan matrix of the Dynkin diagram:
\be\label{CartanMatrix}
  C_{aa'}=(D+D^{-1})\delta_{aa'}-A_{aa'},
\ee
where $A$ is the adjacency matrix of the Dynkin diagram. For $A$ series which we consider $A_{aa'}=\delta_{a,a'+1}+\delta_{a,a'-1}$. For example, for $A_3$ the deformed Cartan matrix is given by
\be
  \left(
    \begin{array}{ccc}
      D+D^{-1} & -1 & 0  \\
      -1 & D+D^{-1} & -1   \\
      0 & -1 & D+D^{-1} \\
    \end{array}
  \right).
\ee
Using these notations, the Bethe equations for centers of strings that follow from (\ref{Bethegeneralrepresentation}) can be written as
\be\label{completeBetheAnsatz}
  u_{a,s;k}^{\CL_{s,\mu_a}}=-\prod_{a'=1}^{N-1}\prod_{s'=1}^{\infty}Q_{a',s'}[u_{a,s;k}]^{C_{aa'}\CL_{ss'}},\ \ Q_{a,s}[u]\equiv\prod_{j}(u-u_{a,s;j}).
\ee

\section{\label{sec:Fusion procedure}Fusion procedure and Hirota equations}
The Bethe Ansatz equations can be also derived by the procedure different from the one presented above. This procedure includes derivation of the functional (Hirota) equations (\ref{fusioncomplete}) and then solution of them via the chain of Backlund transforms
(\ref{Backlundsequence}). This procedure is interesting in particular because the functional equations (\ref{fusioncomplete}) reflect the symmetry algebra of the system. The analytic peculiarities of the system appear then by imposing proper boundary conditions when solving (\ref{fusioncomplete}). In principle, it is possible to choose different boundary conditions and therefore obtain different integrable systems based on the same symmetry group.

In this section we will explain the meaning of the functional equations (\ref{fusioncomplete}). In the next section we will show how to solve them. We will consider only the rational case which leads to the Bethe equations of the XXX spin chain.

First, let us make a simplification. The $S$ matrix (\ref{S-matrix}) can be represented as the ratio of the $R$-matrix
\be\label{R-matrix}
  R[\th]=\th-i\CP
\ee
and the scalar factor $\th-i$. The $R$-matrix satisfies the Yang-Baxter equation. However, the unitarity condition is replaced by $R[\th]R[-\th]=-(\th^2+1)$.

In the following we will assign to the scattering of the particles the $R$-matrix (\ref{R-matrix}) instead of the $S$ matrix (\ref{S-matrix}). This is a reasonable since the common scalar factors, like $\theta-i$, do not play the role in the problem of the diagonalization of the transfer matrix. Due to this we will also define the transfer matrix through the $R$-matrices:
\be
  T=\Tr_{\!A}\(R_{AK}\ldots R_{A1}\).
\ee
Starting from now, we will also understand the transfer matrix in the sense of its definition for the algebraic Bethe Ansatz procedure.

Until now we studied only scattering of the particles in the fundamental representation. It turns out useful to introduce the particles in different representations.  We can introduce them by a so called fusion procedure. Let us first understand how the fusion procedure works for the construction of the particles in the symmetric and the antisymmetric representations.

We define the particle in the symmetric/antisymmetric representation as a composite of two fundamental particles with symmetrization/antisymmetrization of the color:
\be\label{fusionsymasym}
    \parbox[c]{4.5cm}{\includegraphics{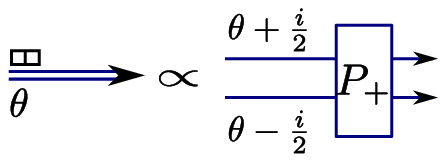}},\ \ \parbox[c]{4.5cm}{\includegraphics{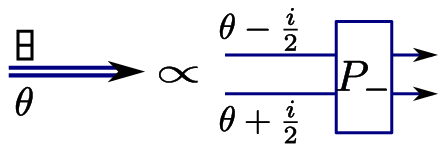}},
\ee
where the projectors $P_{\pm}$ are defined as:
\be
    P_+=\frac{1+\CP}2,\ \ P_-=\frac {1-\CP}2.
\ee
This definition of the composite particle makes sense only if the projection to the symmetric or antisymmetric representation survives under scattering with other particles:
\be\label{surviving}
    \parbox[c]{6.1cm}{\includegraphics[width=6cm]{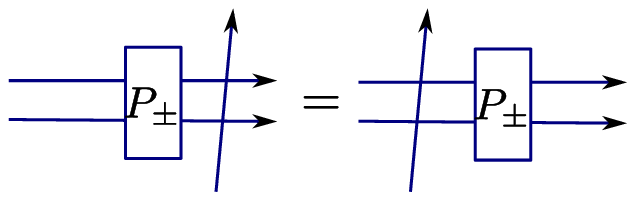}}.
\ee
This requirement is satisfied if to choose the relative rapidities of the constituent fundamental particles as shown in (\ref{fusionsymasym}). Indeed, let us use the fact that the operator $\check R[\theta]\equiv R[\theta]\CP$ has the following property:
\be\label{projectors}
\check R[\pm i]=\raisebox{0.1cm}{\parbox[c]{2.4cm}{\includegraphics[width=2.4cm]{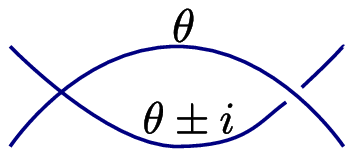}}}=\pm 2iP_{\mp}.
\ee
Then the property (\ref{surviving}), for example for the particle in the symmetric representation, is a simple consequence of
\be
\parbox[c]{13cm}{\includegraphics{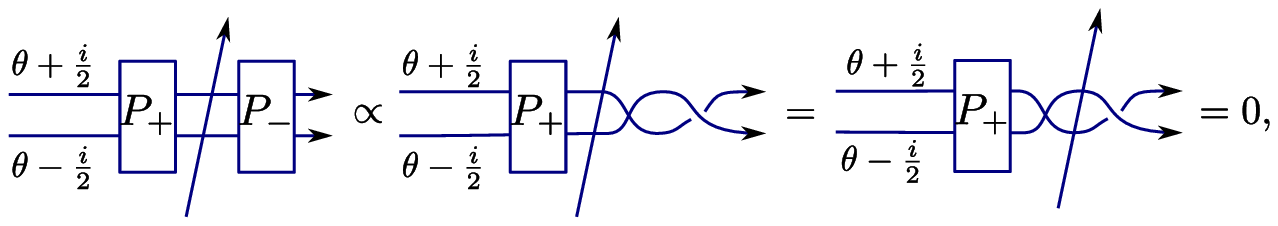}}
\ee
where we used the Yang-Baxter equation and the property that $P_+P_-=0$.

 Let us take the particle in the symmetric representation with the color $\{ab\}$\footnote{ curly brackets means symmetrization} and scatter it with the particle in the fundamental representation which carries the color $c$. Direct calculation shows that the scattering process is the following:
 \be
  \{ab\}\otimes c\rightarrow \(\th-\frac i2\)\(\(\th+\frac i2\) \{ab\}\otimes c-i(\{cb\}\cdot a+\{ac\}\otimes b)\).
 \ee
 We will drop the overall scalar factor $\th-i/2$. Then the $R$-matrix of this process is given by
 \be
  R_{\footnotesize{\Box\!\Box,\Box}}[\theta]=\theta+\frac i2-i\CP,
 \ee
 where $\CP$ now means a generalized permutation: $\CP:\{ab\}\otimes c\mapsto \{ac\}\otimes b+\{cb\}\otimes a $.

 \

 The particle in any representation given by the Young table with $n$ boxes can be constructed as a composite particle of $n$ fundamental particles with a corresponding symmetrization of color indices. The relative rapidities of the fundamental particles are chosen such that the symmetrization commutes with the scattering process. The scattering of composite particles satisfies the Yang-Baxter equation since the scattering of the fundamental particles does. The $R$-matrix can be calculated in the way analogous to the presented above calculation of $R_{\footnotesize{\Box\!\Box,\Box}}[\theta]$. For the explicit formulas and more detailed discussion see \cite{Zabrodin:1996vm}.

In the following we will be interested only in the rectangular representations - the representations $Y^{a,s}$ given by the rectangular Young tables with $a$ rows and $s$ columns\footnote{$a$ for antisymmetrization, $s$ for symmetrization}.
For the scattering of the rectangular representation with the fundamental one the R-matrix has a simple form\footnote{The $R$-matrix for the scattering of two arbitrary representations is in general complicated. It is given as a polynomial over the generalized permutation operator with coefficients that depend on $\theta$.}: 
\be\label{Reeboks}
  R_{Y^{a,s},\Box}=\theta+\frac {is}2-\frac {ia}2-i\CP,
\ee
where $\CP$ is a generalized permutation operator.

The generalized permutation operator $\CP$ acting  on the tensor product of two arbitrary representations $Y_1\otimes Y_2$ is defined as follows. The color of the particle in the representation $Y_i$ is given by $P_{Y_i}[a_1\ldots a_{n_i}]$, where $n_i$ is a number of boxes in the Young table and $P_{Y_i}$ is a projector. The operator $\CP_{Y_1,Y_2}$ is a sum over all possible pairwise permutations of the indices $a_j$ of  $P_{Y_1}[a_1\ldots a_{n_1}]$ with the indices $b_j$ of $P_{Y_2}[b_1\ldots b_{n_2}]$. For example, the generalized permutation operator acts on the tensor product of symmetric and antisymmetric representations in the following way:
\be
    \CP\!:\{a,b\}\otimes\[c,d\]\!\mapsto\!\{c,b\}\otimes\[a,d\]+\{a,c\}\otimes\[b,d\]+\{a,d\}\otimes\[c,b\]+\{d,b\}\otimes\[c,a\]\!.
\ee

Now we are ready to introduce the transfer matrix in a given representation. The $T$-matrix in the representation $Y$ is defined as follows:
 \be
  T_Y[\th;\{\th_1,\ldots\th_k\}]=Tr_Y\(R_{Yk}[\th-\th_K]...R_{Y1}[\th-\th_1]\).
 \ee
Here the auxiliary particle is in the representation $Y$. All the physical particles\footnote{In fact these "physical particles" are the nodes of a spin chain. Each node carries an inhomogeneity $\theta_i$.} are in the fundamental representation. For the rectangular representations we will additionally use the notation
\be
    T^{a,s}\equiv T_{Y^{a,s}}.
\ee

Using the equation (\ref{eq:SLLLLS}), which is valid for two arbitrary representations, we can prove that
\be
  [T_{Y_1}[\th],T_{Y_2}[\th']]=0.
\ee
Therefore we can diagonalize simultaneously all the transfer matrices.

The transfer matrices in different rectangular representations are not independent but satisfy  the so called fusion or Hirota equations. In the simplest case of the fusion of two transfer matrices in the fundamental representation the Hirota equation reads
\be\label{hirotasimple}
  T_{\raisebox{0cm}{\tiny$\Box$}}[\theta+i/2]T_{\raisebox{0cm}{\tiny$\Box$}}[\theta-i/2]=T_{\raisebox{0cm}{\tiny$\Box\!\Box$}}[\theta]
  Q[\theta-i/2]+T_{\raisebox{-0.075cm}{\tiny$\Box$}\hspace{-0.46EM}\raisebox{0.075cm}{\tiny$\Box$}}Q[\theta+i/2].
\ee
The proof of this relation is most easily done graphically:
\be\label{derimage}
    T_{\raisebox{0cm}{\tiny$\Box$}}[\theta+i/2]T_{\raisebox{0cm}{\tiny$\Box$}}[\theta-i/2]&=&
    \parbox{5.5cm}{\includegraphics{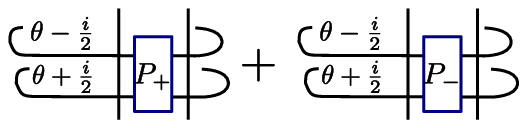}}=\no\\
    &=&Q[\theta-i/2]\parbox{2cm}{\includegraphics[width=2cm]{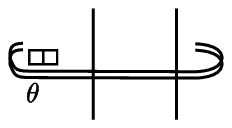}}+Q[\theta+i/2]\parbox{2cm}{\includegraphics[width=2cm]{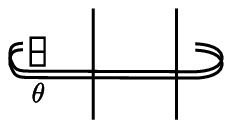}}.
\ee
To obtain the transfer matrix in the antisymmetric representation in derivation (\ref{derimage}) we should use the representation (\ref{projectors}) for the projector $P_-$, Yang-Baxter equation, and cyclicity of the trace.

If we would scatter the fundamental representation with the trivial one, there would be no interchange of color. Therefore it is natural to define the corresponding $R$-matrix as
\be\label{RBOX}
    R_{\Box,\bullet}=\theta.
\ee
Then the Baxter polynomial $Q[\theta]$ can be considered as a transfer matrix in the trivial representation:
\be
    Q[\theta]=T^{0,0}[\theta].
\ee
Moreover, using fusion procedure and (\ref{RBOX}) we conclude that
\be
  R_{Y^{a,s},\bullet}=\theta+\frac {is}2-\frac {ia}2
\ee
and thus
\be\label{BaxterasT}
  Q\[\theta-\frac {ia}2\]=T^{a,0}[\theta],\ \ Q\[\theta+\frac {is}2\]=T^{0,s}[\theta].
\ee
Therefore equation (\ref{hirotasimple}) can be rewritten as
\be
    T^{1,1}[\theta+i/2]T^{1,1}[\theta-i/2]=T^{1,2}[\th]T^{1,0}[\th]+T^{2,1}[\th]T^{0,1}[\th].
\ee

This equation is generalizable for arbitrary rectangular representations:
\be\label{fusioncomplete}
  T^{a,s}[\th+i/2]T^{a,s}[\th-i/2]=T^{a,s+1}[\th]T^{a,s-1}[\th]+T^{a+1,s}[\th]T^{a-1,s}[\th].
\ee
Equation (\ref{fusioncomplete}) is known as the Hirota equation.

\begin{figure}[t]
\centering
\includegraphics[height=4.14cm]{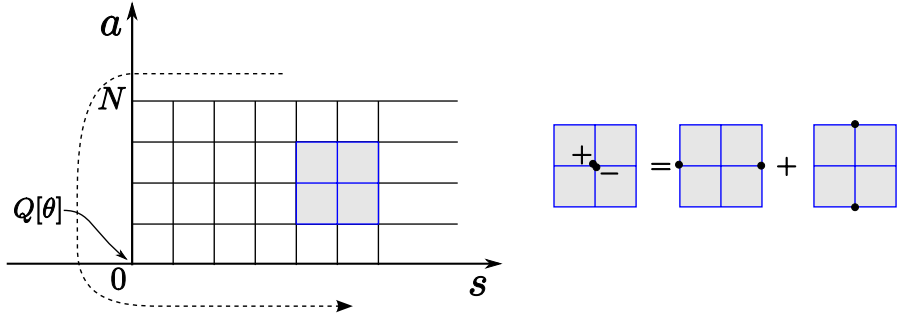}
\caption{\label{fig:hook} Transfer matrices are nonzero on the shown shape (strip) on the left and zero outside it. The Hirota equation (\ref{fusioncomplete}) is represented graphically by the picture on the right. The boundary of the strip carries the wave which propagates in the direction shown by the dashed arrow. The value of the wave at the origin is the Baxter polynomial $Q[\theta]$.}
\end{figure}

The transfer matrices are defined on the lattice bounded by the rectangle $0\leq a\leq N, s\geq 0$ which any meaningful Young diagram should fit. To make the Hirota equations valid for any integer values of $a$ and $s$ we define $T^{a,s}$ on the lines $a=0$ and $a\geq 0,s=0$ by the relations (\ref{BaxterasT}). Outside the rectangle and these two lines $T^{a,s}\equiv 0$. Therefore, the fusion relations (\ref{fusioncomplete}) are nontrivial on the shape shown in Fig.~\ref{fig:hook}.

The Hirota equation simplifies on the boundary of the rectangle. Let us take for example the lower boundary ($a=0$). The fusion relation on it is given by
\be
  T^{0,s}\[\th+\frac i2\]T^{0,s}\[\th-\frac i2\]=T^{0,s+1}[\th]T^{0,s-1}[\th].
\ee
It is solved as a product of left and right moving waves. On the other hand $T^{0,s}[\theta]=Q[\theta+is/2]$. Therefore we have only the left-moving wave.

The same situation occurs on the left boundary. On the upper boundary we also have the transfer matrix in the trivial representation. However now the corresponding $R$-matrix is given by
\be\label{RNs}
  R^{N,s}=\theta+\frac {is}2-\frac {iN}2-i\CP.
\ee
The generalized permutation is not zero, as it was on the upper and the left boundaries, but is equal to $s\times Id$. For example: $\CP:[a_1\ldots a_N]\cdot c\mapsto [a_1\ldots a_N]\cdot c$ due to the fact that $[a_1\ldots a_N]$ is proportional to the completely antisymmetric tensor.

Due to (\ref{RNs}) we have $T^{N,s}=Q[\theta-iN/2-is/2]$. Combining three boundaries together we see that there is only the left-moving wave on the boundary generated at the origin by $Q[\theta]$.

\section{\label{sec:nestedBABacklund}Nested Bethe Ansatz via Backlund transform}
The Hirota equation (\ref{fusioncomplete}) is a discrete integrable system in the sense that it can be obtained as a compatibility condition of the system of the linear equations \cite{Kazakov:2007fy}:
\be\label{Backlund}
  T^{a,s}[\theta]F^{a-1,s}[\theta-i/2]=T^{a-1,s}[\theta-i/2]F^{a,s}[\theta]+T^{a,s-1}[\theta-i/2]F^{a-1,s+1}[\theta],\no\\
  T^{a,s+1}[\theta-i/2]F^{a,s}[\theta]=T^{a,s}[\theta]F^{a,s+1}[\theta-i/2]+T^{a+1,s}[\theta-i/2]F^{a-1,s+1}[\theta]
\ee
that can be also represented graphically as
\be\label{Backlundpicture}
    \parbox{5cm}{\includegraphics[width=5cm]{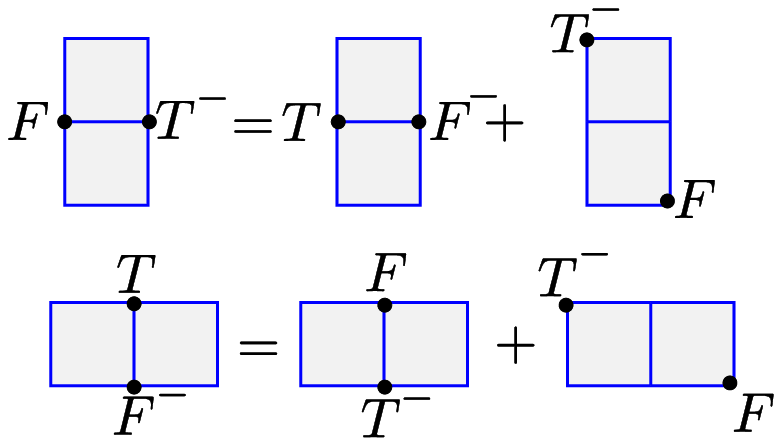}}.
\ee
An interesting fact is that if we consider (\ref{Backlund}) as the equations on $T$ then the consistency condition gives us fusion relations (\ref{fusioncomplete}) on $F$! Therefore $F$ can be thought as the transfer matrix of some integrable system. The transfer matrix $F$ is called the Backlund transform of $T$.

There are two possible and different solutions to (\ref{Backlund}) which generate respectively the first-type (BT1) and the second-type (BT2) Backlund transformations. BT2 is relevant for the study of supersymmetric groups and is discussed in chapter \ref{ch:susyspch}. Here we discuss BT1.

For BT1 the strip on which the functions $F$ are nonzero is given by the constraints $0\leq a\leq N-1,s\geq 0$, that is the number of rows is
diminished by 1. Thus $F^{a,s}$ can be viewed as the transfer matrices for an $SU(N-1)$ integrable system. From (\ref{Backlund}) one can see that the boundary of the strip of $F$ carries left-moving wave induced by the left-moving wave of $T$. To obtain a rational integrable system we have to choose the value of $F^{0,0}$ to be a polynomial.
\begin{figure}[t]
\centering
\includegraphics[height=3.14cm]{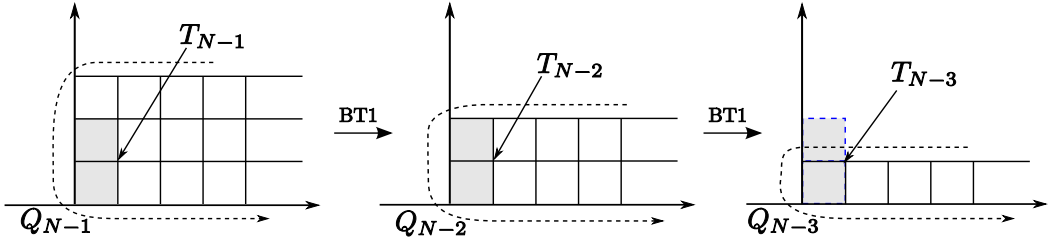}
\caption{\label{fig:Backlundchain}A chain of Backlund transformations.}
\end{figure}

The idea of solving the Hirota equations is to perform a sufficient number of Backlund transforms so that we are eventually left with a strip with only one row. To make the notation systematic we define by $T^{a,s}_{N}$ the transfer matrices associated with the $SU(N)$ group. The boundary condition is given by $T^{0,0}_{N}[\theta]=Q_{N}[\theta]$. For the rational integrable system $Q_{N}$ is a polynomial. The zeroes of $Q_{N}$ are denoted by $\l_{N,k}$. Subsequent application of Backlund transformations generates the sequence
\be\label{Backlundsequence}
\begin{array}{c}
    T_{N}  \\
    Q_{N}  \\
    \end{array}\rightarrow
\begin{array}{c}
    T_{N-1}  \\
    Q_{N-1}  \\
\end{array}\rightarrow\ldots\rightarrow
\begin{array}{c}
    T_{1}  \\
    Q_{1}  \\
    \end{array}.
\ee
The linear equation (\ref{Backlundpicture}) evaluated for the gray rectangle in Fig.~\ref{fig:Backlundchain} has a particular interest for us. This equation gives a relation between $T^{1,1}_{M}$, $T^{1,1}_{M+ 1}$ and the Baxter polynomials $Q_M$, $Q_{M+1}$. Indeed, this equation reads as
\be\label{Baxter}
  Q_{M}T^{1,1}_{M+1}=Q_{M+1}T^{1,1}_{M}+Q_{M+1}^{--}Q_{M}^{++},
\ee
where the notation $f^{\pm\pm}[\theta]\equiv f[\theta\pm i]$ is used.

Since $T^{1,1}_1=Q_1^{--}$, the recursive relation (\ref{Baxter}), known also as a Baxter equation, gives us the possibility to express $T^{1,1}_{N}$ in terms of Baxter polynomials only:
\be\label{TaL}
  \frac{T^{1,1}_{N}}{Q_{N}}=\frac{Q^{--}_{N}Q^{++}_{N-1}}{Q_{N}Q_{N-1}}+\frac{Q^{--}_{N-1}Q^{++}_{N-2}}{Q_{N-1}Q_{N-2}}
  +\ldots+\frac{Q^{--}_{2}Q^{++}_{1}}{Q_{2}Q_{1}}+\frac{Q_1^{--}}{Q_1}.
\ee
This is enough to solve the Hirota system since all the $T^{a,s}_{N}$ can be found from the knowledge of $T^{1,1}_{N}$ and $Q_{N}$.

 In (\ref{TaL}) we recognize up to an overall factor the expression (\ref{TLeigenvalue}). The Bethe equations can be read from (\ref{TaL}) as the condition that $T^{1,1}_{N}$ is a polynomial by construction and therefore does not have poles.

The Hirota equations were obtained as a relation between transfer matrices. They give a set of Bethe equations via the sequence of Backlund transformations. The essential point in this derivation of the Bethe equations is that the boundary conditions for the transfer matrices and the transfer matrices themselves are required to be polynomials. It is also possible to impose different requirements on the analytical structure of the transfer matrices. Then $T^{a,s}$ will be considered as a transfer matrices based on an $R$-matrix different from (\ref{R-matrix}). In this way we can obtain for example trigonometric and elliptic integrable systems. The transfer-matrices of the Hirota system that was proposed for the AdS/CFT \cite{Gromov:2009tv} contain square root branch points.

\chapter{\label{ch:twodimqft}Two-dimensional integrable field theories}
\section{Two-dimensional sigma-models}
The integrable 1+1 dimensional quantum field theories (IQFT) give us another example of the systems that can be solved by the Bethe Ansatz techniques. A remarkable development in this direction started in the late 70's with the realization of the fact that the scattering matrix in these theories can be found exactly \cite{Zamolodchikov:1978xm}.

In this and the next chapter we will consider the following two examples of the integrable models:
$SU(N)$ chiral Gross-Neveu model (GN) and $SU(N)\times SU(N)$ principal chiral field model (PCF). Our choice is dictated by the simplicity of these models. In chapter \ref{ch:massgap} we also study the $O(N)$ vector model. The actions for these three models are given by:
\be\label{actionsofsigmamodels}
   S_{GN}&=&\frac 1{\sigmacoupling}\int d^2 x\ \overline\psi_a i\pd\!\!\!\slash\psi^a+\frac{1}2\(\(\ \overline\psi_a\psi^a\ \)^2-\(\ \overline\psi_a\gamma^5\psi^a\ \)^2\),\ \ a=\overline{1,N},\no\\
   S_{PCF}&=&\frac 1{2\sigmacoupling}\int d^2 x\ \Tr(\mathfrak{g}^{-1}\partial_\mu \mathfrak{g})(\mathfrak{g}^{-1}\partial^\mu \mathfrak{g}),\ \ \ \mathfrak{g}[x]\in SU(N),\no\\
   S_{O(N)}&=&\frac 1{2\sigmacoupling}\int d^2x\ (\pd_\mu \overrightarrow{n}\pd^\mu\overrightarrow{n}),\ \ \overrightarrow{n}^2=n_1^2+\ldots+n_N^2.
\ee

All three theories can be treated on a similar footing. They are determined by the coupling constant $\sigmacoupling$ and the parameter $N$ - size of the matrix of a symmetry group. These theories are asymptotically free. Therefore, at large energy scales they can be studied perturbatively. 

The infrared catastrophe makes the perturbative description inappropriate at low energies. It is believed that the spectrum of these theories develops a mass gap. The mass scale comes through the mechanism of the dimensional transmutation and is given through the beta-function $\beta[g]$:
\be
  \Lambda=\mu e^{-\int^\infty_{g[\mu]} \frac{dg}{\beta[g]} }.
\ee

The chiral Gross-Neveu model\footnote{It is also known as a two-dimensional Nambu-Jona-Lasinio model or a massless Thirring model.} is a model with a four-fermion interaction and continuous chiral symmetry $\psi\to e^{i\gamma_5\theta}\psi$. It was discussed in \cite{Gross:1974jv} together with the other fermion field theories with quartic interaction. In particular it was shown in the large $N$ limit that the operator $(\overline\psi\psi)^2+(i\overline\psi\g^5\psi)^2$ acquires on the quantum level a nonzero average proportional to $\Lambda^2$. 
In the large $N$ limit it is possible to identify the particle content of the model and calculate the masses of the particles. The theory contains one massless particle which is invariant under the $SU(N)$ group but transforms under the action of the $U(1)$ chiral symmetry. There are also massive particles which are blind to the chiral symmetry and do not interact with the massless particle. There are $N-1$ different types of massive particles. The $k$-th type transforms under an antisymmetric representation $[k]$ of the $SU(N)$ group.  Since the massless particle is completely decoupled from the massive ones, we will not consider it in the following.

The PCF for finite values of $N$ was solved by Polyakov and Wiegmann \cite{Polyakov:1983tt,Polyakov:1984et}. In \cite{Fateev:1994ai} the large $N$ solution of this model using different means was given. There are also $N-1$ different types of massive particles. The $k$-th type transforms under the $[k]\times[k]$ representation of the $SU(N)\times SU(N)$ symmetry group of the system.

The $O(N)$ vector sigma model was solved at large $N$ \cite{Bardeen:1976zh,Brezin:1976qa}. This solution shows the presence of the only one particle multiplet in the vector representation of the $O(N)$ group.

\section{\label{sec:scatteringmatrix}Scattering matrix}

As we see, the sigma models can be exactly solved at large values of $N$. The theories can be also exactly solved at finite values of $N$ if to assume theirs integrability and make an assumption about the particle content of the theory\footnote{Using integrability, the masses of the particles could be exactly found at finite $N$ \cite{Forgacs:1991nk} as we discuss in chapter \ref{ch:massgap}.}. More precisely, we can exactly find the scattering matrix \cite{Zamolodchikov:1978xm}. The arguments go as follows.

First, the infinite number of the conserved charges implies conservation of the number of particles. The reason for this is that the $n$-th conserved charge $Q_n$ acting on a free particle with momenta $p$ gives roughly speaking $p^n$. So $Q_1$ is the total momentum of the system and $Q_2$ is the total energy of the system. Therefore we have an infinite set of conservation laws
\be
  \sum \(\ p_{i}^{in}\ \)^n=\sum \(\ p_i^{out}\ \)^n
\ee
that can be satisfied only if the number of particles is conserved and the momenta interchange.

Second, the scattering process factorizes into $2\to 2$ processes. The argument of why it happens is the following\footnote{For a rigorous treatment see \cite{Iagolnitzer:1977sw}}. The particle is described by a wave packet
\be
  \psi[x,t]\propto \int dp\ e^{ip(x-x_0)-i\varepsilon[p](t-t_0)-\a (p-p_0)^2}.
\ee
Action of $e^{i Q_1}$ and $e^{i Q_2}$ generates the translation in space and time respectively (as it should). Let us consider the action of $e^{iQ_3}$:
\be
  e^{iQ_3}\psi[x,t]\propto \int dp\ e^{ip^3+ip(x-x_0)-i\varepsilon[p](t-t_0)-\a (p-p_0)^2}.
\ee
Expanding the term $p^3$ around a saddle point value $p_0$ we see that the action of $e^{iQ_3}$ shifts the wave packet by the value that depends on the momentum $p_0$ of the particle. Therefore the action of $e^{iQ_3}$ on a system of particles will shift each particle by a different distance. So, using this operator we can always represent any scattering as a combination of $2\to 2$ scattered processes. Therefore we should know only $2\to 2$ scattering matrix to define the system.

\begin{wrapfigure}{r}{0.3\textwidth}
 \vspace{-18pt}
  \begin{center}
    \includegraphics[width=0.20\textwidth]{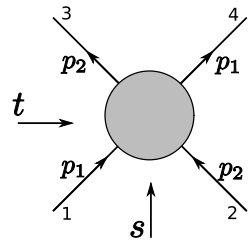}
  \end{center}
  \vspace{-5pt}
  \caption{\label{fig:4point}4-point amplitude}
  \vspace{-5pt}
\end{wrapfigure}
The same reasoning with application of $e^{iQ_3}$ leads to the Yang-Baxter equation (\ref{yang-baxter}) which is depicted in Fig.~\ref{fig:definitions}.

To impose further constraints on the structure of the $S$-matrix we will use the fact that we are dealing with the relativistic quantum field theory\footnote{The pedagogical discussion of the analytical structure of the scattering matrix can be found for example in \cite{Chew}.}. The two to two scattering process is defined by the $4$-point function shown in Fig.~\ref{fig:4point}. For simplicity we consider the scattering of particles with equal masses. Since in two dimensions the momenta are only interchanged after scattering, the scattering matrix depends on only the one invariant. For this invariant we can take $$s=(p_1+p_2)^2$$ or the difference of rapidities $\theta$ which is related to $s$ through:
\be\label{srapidity}
  p_k&=&m\cosh\theta_k,\ \ \theta=\theta_1-\theta_2,\no\\
  s&=&(p_1+p_2)^2=2m^2(1+\cosh\theta).
\ee
The invariant of the $t$-channel is given by
\be
  t=(p_1-p_2)^2=2m^2(1-\cosh\theta)=2m^2(1+\cosh[i\pi-\theta]).
\ee
The amplitude of the reverse process can be obtained simply by replacing $\theta$ with $-\theta$. Therefore the unitarity condition reads
\be\label{Sunitraity}
  S[\theta]S[-\theta]=1.
\ee
We can pass from $(12)\to(34)$ process to $(\overline 31)\to(4\overline2)$ process (overline means antiparticles and charge conjugation) by simple change of the sign of $p_2$. This leads to the crossing equations
\be\label{Scrossing}
  S_{\overline{3}1}^{4\overline 2}[-\theta]=S_{12}^{34}[i\pi+\theta].
\ee

The $S$-matrix as a function of the $s$ variable has square root branch points at $s=4m^2$ and $s=0$ which correspond to two-particle and particle-antiparticle thresholds. The on-shell two-particle scattering is given by $S[s+i0]$ for $s>4m^2$, the on-shell particle-antiparticle scattering is given by $S[s-i0]$ for $s<0$. The square root cuts are resolved after introduction of the rapidity variable $\theta$ via (\ref{srapidity}).
\begin{figure}[t]
\centering
\includegraphics[width=10cm]{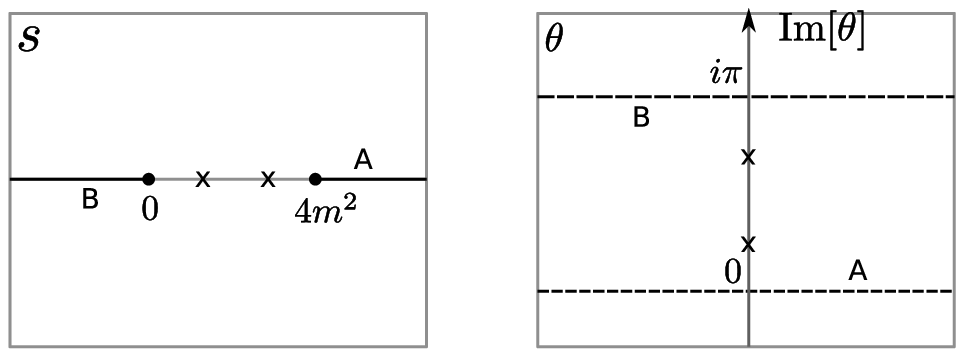}
\caption{\label{fig:smatrixanalytics}Analytical structure of the $S$-matrix. The square root cuts $A$ and $B$ in the s-plane are resolved by introducing the rapidity variable. These cuts are mapped to the lines $A$ and $B$ in the $\theta$-plane. The physical strip is inside these lines.}
\end{figure}

The $S$-matrix is a meromorphic function of $\theta$. The physical $s$-sheet is mapped into the strip $0<{\rm Im[\theta]}<i\pi$. The poles on the imaginary axes of $\theta$-plane and inside this strip correspond to the physical particles in the theory. These particles can be thought as bound states of the scattered particles. We can understand this correspondence from tree level diagrams:
\be
\parbox{9cm}{\includegraphics{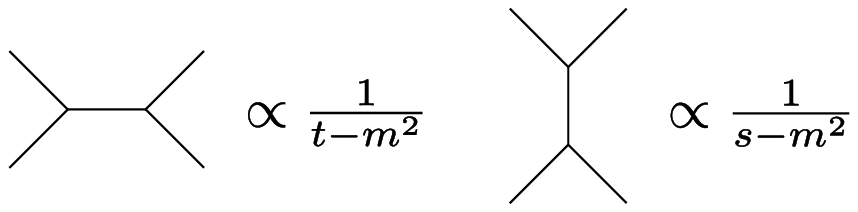}}.
\ee

Using the trick with $e^{iQ_3}$ one can calculate the scattering which includes these bound particles in the following way:
\be
\parbox{6cm}{\includegraphics{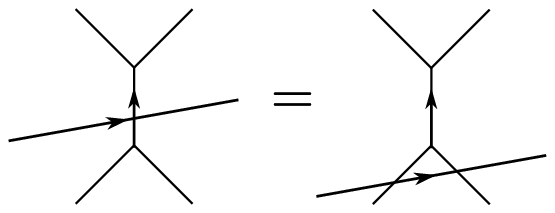}}.
\ee

The invariance under the symmetry of the system, the Yang-Baxter equation (\ref{yang-baxter}), the unitarity (\ref{Sunitraity}) and crossing (\ref{Scrossing}) conditions, and the assumption on the pole structure inside the physical strip $0<{\rm Im[\theta]}<\pi$ uniquely fix the two-particle scattering matrix and therefore completely determine the system.

The Yang-Baxter equation and invariance under the symmetry fixes the $S$-matrix up to an overall scalar factor. The scattering matrix discussed in the first chapter also satisfies these two conditions. Therefore the algebraic structure of the $S$-matrices in two cases is the same and we can apply the Bethe Ansatz techniques developed in the first chapter.

The unitarity, crossing, and the pole structure are the physical constraints on the $S$-matrix. The assumption on the pole structure is a consequence of an assumption on the particle content of the theory. This assumption is hard to be proven exactly. It is usually confirmed by the exact large $N$ solution. Another possibility to verify this assumption is to study the renorm-group behavior and Borel summability properties of the system which we will discuss in chapter \ref{ch:massgap}.

The knowledge of the $S$-matrix allows solving the theory at large volume with the help of the Bethe Ansatz as we will discuss now on the example of the PCF and GN models.

\section{PCF and Gross-Neveu model}\label{sec:PCFGNbound}
The particle content of GN model includes in particular particles in the fundamental representation. The PCF model contains in particular particles in the fundamental$\times$fundamental representation. We choose the following normalization of the rapidities of these particles
\be
  p=m\sinh\[\frac{2\pi\theta}{N}\],\no\\
  E=m\cosh\[\frac{2\pi\theta}{N}\].
\ee
In this normalization the Bethe equations will resemble most the Bethe equations (\ref{betheXXX}) of the XXX spin chain.

The scattering matrix of the fundamental particles was determined to have the following form \cite{Polyakov:1983tt}:
\be\label{S0GN}
   S_{GN}[\theta]&=&\(P_++\frac{\theta+i}{\theta-i}P_-\)S_{0,GN}=\frac{\theta-i\CP}{\theta-i}S_{0,GN},\no\\
  S_{0,GN}&=&-\frac{\Gamma[1-\frac \theta{Ni}]\Gamma[1-\frac 1N+\frac \theta{Ni}]}{\Gamma[1+\frac \theta{Ni}]\Gamma[1-\frac 1N-\frac \theta{Ni}]},
\ee
\be\label{S0PCF}
  S_{PCF}&=&\(\frac{\theta-i\CP}{\theta-i}\otimes\frac{\theta-i\CP}{\theta-i}\)S_{0,PCF},\no\\
  S_{0,PCF}&=&-\(\frac{\Gamma[1-\frac \theta{Ni}]}{\Gamma[1+\frac \theta{Ni}]}\)^2\frac{\Gamma[1-\frac 1N+\frac \theta{Ni}]\Gamma[\frac 1N+\frac \theta{Ni}]}{\Gamma[1-\frac 1N-\frac \theta{Ni}]\Gamma[\frac 1N-\frac \theta{Ni}]}.
\ee
There is a relation between $S_{PCF}$ and $S_{GN}$:
\be
 S_{PCF}[\theta]&=&\left(S_{GN}[\theta]\otimes S_{GN}[\theta]\right)X[-\theta],\ \
  X[\theta]=-\frac{\Gamma[1-\frac 1N+\frac \theta{Ni}]\Gamma[\frac 1N-\frac \theta{Ni}]}{\Gamma[1-\frac 1N-\frac \theta{Ni}]\Gamma[\frac 1N+\frac \theta{Ni}]}.
\ee

The scalar factors can be also rewritten in terms of the shift operators. For this let us use the following \textit{definition} (see also definition (\ref{Dexpnotation})):
\be\label{gamma}
  \theta^{-\frac {D^{2N}}{1-D^{2N}}}\equiv \Gamma\[1+\frac {\theta}{iN}\],\ \ D\equiv e^{\frac i2\pd_\theta},
\ee
and similar expressions for other gamma functions. Naively, the l.h.s. of (\ref{gamma}) should be understood as product of poles
\be\label{gammadef}
 \theta^{-\frac {D^{2N}}{1-D^{2N}}}\simeq \frac 1{\theta+iN}\frac 1{\theta+2iN}\ldots\ .
\ee
Of course, this product should be regularized. We will always use regularization which leads to the gamma function. For more details see appendix \ref{app:shift}.

Using the representation (\ref{gamma}) for the gamma functions one can write:
\be\label{s0gnpcf}
  S_{0,GN}[\theta]\!&=&\!\! \theta^{\frac{1-D^{2N-2}}{1-D^{2N}}-\frac{1-D^{-2N+2}}{1-D^{-2N}}}\simeq \theta^{\frac{-(D-D^{-1})(D^{N-1}-D^{1-N})}{D^{N}-D^{-N}}},\no\\
  S_{0,PCF}[\theta]\!&=&\!\!\theta^{\frac{(1-D^2)(1-D^{2N-2})}{1-D^{2N}}-\frac{(1-D^{-2})(1-D^{-2N+2})}{1-D^{-2N}}}\simeq
  \theta^{-2\frac{(D-D^{-1})(D^{N-1}-D^{1-N})}{D^{N}-D^{-N}}}.
\ee
The first equalities for the expressions of $S_{0,GN}$ and $S_{0,PCF}$ are exact in view of the definition (\ref{gamma}). It is worth to recall that the asymptotic behavior of these scalar factors is given by
\be
  S_{0,GN}&\to&-e^{\pm \frac{i\pi}{N}},\ \ \theta\to \pm\infty,\no\\
  S_{0,PCF}&\to& 1,\ \ \theta\to \infty.
\ee
This asymptotic behavior is not evident from the representation (\ref{s0gnpcf}) but can be found if to use (\ref{gamma}).

The second equalities in (\ref{s0gnpcf}) have no precise meaning and are given for heuristic reasons. In particular, the second equalities formally suggest that $S_{0,PCF}=-S_{0,GN}^2$. This is not true of course, however this suggestion means that $S_{0,PCF}$ and $S_{0,GN}^2$ differ only by a CDD factor. The CDD factor, which is in our case $X[\theta]$, is a factor multiplication by which leaves a scattering matrix obeying the crossing equations. It is possible to fix this factor by the requirement of a precise structure of poles in the physical strip.

\subsubsection{Bound states}
The physical strip is given by $0\leq Im[\theta]<\frac {N}2$. The fundamental particles can form bound states if the $S$-matrix has poles inside this strip.


The $S$-matrices (\ref{S0GN}) and (\ref{S0PCF}) have a pole at $\theta=i$ in the purely antisymmetric channel. This means that bound states are particles in the antisymmetric representations: $[2]$ for GN and $[2]\times [2]$ for PCF. The mass of the obtained particle is easily calculated from the conservation law:
\be
  E_{[2]}=m \cosh\[\frac{2\pi}{N}(\theta+i/2)\]+m\cosh\[\frac{2\pi}N(\theta-i/2)\]=2m\cos\[\frac{\pi}{M}\]\cosh\[\frac{2\pi}{N}\theta\].
\ee
Therefore
\be
m_{[2]}=2m\cos\left[\frac{\pi}{N}\right]
\ee

This bound state (fused particle) can be represented by two particles with rapidity difference equal to $i$. Due to the existence of the higher conserved charges any scattering with such bound state is equivalent to successive scattering with its constituents.  One can build therefore the correspondent $S$-matrix. The procedure is almost the same as for the fusion procedure discussed in Sec.~\ref{sec:Fusion procedure}. The only difference is that the fused $S$-matrix should satisfy physical requirements of unitarity and crossing and therefore has a different from the fused $R$-matrix overall scalar factor.

It is instructive to compare the bound state with the strings in the XXX spin chain. The string configurations in the XXX spin chain can be also considered as bound states, which however reflect the analytical structure of the $R$-matrix (\ref{R-matrix}).  Since for the XXX spin chain $e^{ip}=\frac{\theta+\frac i2}{\theta-\frac i2}$, Im$[p]$ has opposite sign with Im$[\theta]$. Therefore strings correspond to zeroes of $R[\theta]$ for $\rm{Im}[\theta]>0$. These zeroes appear in the symmetric channel of scattering.

 Since strings in the XXX spin chain are bound states in the symmetric channel we can form them at any nested level of the Bethe Ansatz. In the case of the $SU(N)$ GN and PCF, the bound states appear in the antisymmetric channel. Thus these bound states should correspond to collection of Bethe roots on different levels of the nested Bethe Ansatz. This collection of roots will be constructed in the next subsection.

One can perform scattering of the bound state with a fundamental particle. The scattering matrix of this process again has a pole inside the physical strip. The easy way to see this is to notice that this pole should be due to a pole of the scattering with one of the two constituents of the bound state. The constituents of the bound state have rapidities
$\theta_0\pm \frac i2$. Therefore the pole occurs at $\theta-\theta_0=\frac {3i}2$.

Continuing in a similar way we find that the considered sigma models contain particles in the representations $[k]$ ($[k]\times [k]$) for $k=\overline{1,N-1}$. The masses of the particles come from the conservation law:
\be\label{massesGN}
  m_k=m\frac{\cosh[\frac{2\pi}{N}(\theta+\frac {ik}2)]+\cosh[\frac{2\pi}{N}(\theta+\frac {i(k-2)}2)]+\ldots\cosh[\frac{2\pi}{N}(\theta-\frac {ik}{2})]}{\cosh[\frac{2\pi}{N}\theta]}=m\frac{\sin{\frac{\pi k}{N}}}{\sin{\frac{\pi}{N}}}.
\ee


\subsubsection{Asymptotic Bethe Ansatz for GN.}
Once the $S$-matrix of the system is known, one can build the wave function for each state in the same way as it was done for the XXX spin chain. If we consider the theory on a circle, the periodicity of the wave function leads to a nested Bethe Ansatz. Its construction is completely parallel to that of the XXX spin chain. The answer for the Gross-Neveu model is the following. There are momentum carrying roots $\theta_k$ and $N-1$ types of nested roots $\lambda_{a;j}$, where $a$ denotes the nested level. The Bethe Ansatz equations read
\be\label{BA-12}
e^{-imL\sinh[\frac{2\pi}N\theta_k]}&=&-\prod_{k'=1}^{M}S_{0,GN}[\theta_k-\theta_{k'}]\prod_{j=1}^{K_1}\frac{\theta_k-\lambda_{1;j}+\frac i2}{\theta_k-\lambda_{1,j}-\frac i2},\no\\
\prod_{k=1}^M\frac{\l_{a;j}-\theta_k+\delta_{a,1}\frac i2}{\l_{a;j}-\theta_k-\delta_{a,1}\frac i2}&=&-\prod_{a'=1}^{N-1}\prod_{j'=1}^{K_{a'}}\frac{\l_{a;j}-\l_{a';j'}+\frac{ic_{aa'}}2}{\l_{a;j}-\l_{a';j'}+\frac{ic_{aa'}}2},
\ee
where $c_{aa'}$ is the usual Cartan matrix of the Dynkin diagram for the $A_{N-1}$ Lie algebra.

\begin{figure}[t]
\centering
\includegraphics[width=4cm]{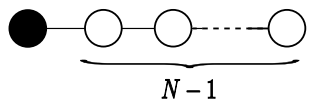}
\caption{\label{fig:DynkinGN}"Dynkin diagram" for Bethe equations (\ref{BA-12}). The black node symbolizes equations for $\theta_k$, white nodes symbolize equations for nested Bethe roots.}
\end{figure}

Sometimes this set of Bethe Ansatz equations is depicted by a kind of Dynkin diagram shown in Fig.~\ref{fig:DynkinGN}.

An important difference with the XXX spin chain is that the Bethe Ansatz (\ref{BA-12}) is asymptotic. It is valid only when the volume $L$ of the system is large. Otherwise the notion of scattering and free particles would be impossible. The exact expression for the energy of the system in the finite volume is different from the answer given by the asymptotic Bethe Ansatz (ABA) by the correction of order $e^{-m L}$.

Although the Bethe Ansatz above is formally for the particles in the fundamental representation, it contains all the other particles as well. They appear as special string-type configurations.

 To recover the configurations which correspond to the bound states we first note that from the fusion procedure we know that the $[k]$-particle should contain $k$ rapidities $\theta$ which form a $k$-string. Let us take a rapidity $\tilde\theta$ which belongs to this string and which has a positive imaginary part . For  $L\to\infty$ we see that $e^{-imL\sinh[\pi\tilde\theta]}\to \infty$. Since $S_{0,GN}$ has no poles in the physical strip, we need to have a nested root $\l_{1}\simeq \tilde\theta-i/2$. Equivalently, for each rapidity $\tilde\theta'$ with negative imaginary part we have $\l_{1}\simeq \tilde\theta'+i/2$. Therefore the existence of the $k$-string of $\theta$-s requires presence of the $(k-1)$-string of $\lambda_1$-s with the same center.

\begin{figure}[t]
\centering
\includegraphics[width=6cm]{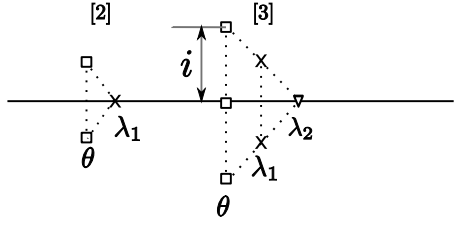}
\caption{\label{fig:trianglesGN}Bound states for the Gross-Neveu model, example of particles in irrep $[2]$ and $[3]$. The real part of the Bethe roots is the same and is depicted to be different only for convenience.}
\end{figure}

 Following a similar logic one can show that $k$-string of $\theta$-s and $(k-1)$-string of $\lambda_1$-s imply the $(k-2)$-string of $\lambda_2$-s. The iteration procedure is performed until we reach the string of the length $1$.
Therefore the bound particles in GN are described by stacks shown in Fig.~\ref{fig:trianglesGN}.

If we denote by $\theta_{a;k}$  the center of the stack that contains $a$ $\theta_k$ roots ($a$-stack) (it is associated with the $a-th$ bound state) then the Bethe Ansatz equations for the centers of stacks will have the following form:
\be\label{GN1}
  \prod_{k=1}^{M_a}\frac{\l_{a;j}-\theta_{a;k}+\frac i2}{\l_{a;j}-\theta_{a;k}-\frac i2}&=&-
  \prod_{a'=1}^{N-1}\prod_{j'=1}^{K_{a'}}\frac{\l_{a;j}-\l_{a';j'}+\frac{ic_{aa'}}2}{\l_{a;j}-\l_{a';j'}-\frac{ic_{aa'}}2},\no\\
  e^{-im_aL\sinh[\theta_{a;k}]}&=&-\left(\prod_{a'=1}^{N-1}\prod_{k'=1}^{M_{a'}}S_{0,GN}^{aa'}[\theta_{a;k}-\theta_{a';k'}]\right)\prod_{j=1}^{K_a}\frac{\theta_{a;k}-\l_{a;j}+\frac i2}{\theta_{a;k}-\l_{a;j}-\frac i2},
\ee
\be\label{SGN1aa}
  S_{0,GN}^{aa'}[\theta]&=&\theta^{\frac{\left(1-D^{2{\rm min}[a,a']}\right)\left(D^{|a-a'|}-D^{2N-a-a'}\right)}{\left(1-D^2\right)\left(1-D^{2N}\right)}-(D\ \to\  D^{-1})}\simeq\no\\&\simeq&\theta^{\frac{\left(D^{N-{\rm max}[a,a']}-D^{{\rm max}[a,a']-N}\right)\left(D^{{\rm min}[a,a']}-D^{-{\rm min}[a,a']}\right)}{D^N-D^{-N}}}.
\ee

 Remarkably, each bound state  interacts with only one level of nested Bethe roots. And this interaction is the same as the interaction between different nested levels (see also Fig.~\ref{fig:GNandXXX}).

\subsubsection{Asymptotic Bethe Ansatz for PCF.}
The asymptotic Bethe Ansatz for PCF contains two wings of nested levels. We will denote the nested Bethe roots by $\l_{R,a;j}$ and $\l_{L,a;j}$, where $R$ and $L$ stand for right and left wings.

The Bethe equations are the following:
\be\label{BethePCF}
\prod_{k=1}^M\frac{\l_{L,a;j}-\theta_k+\delta_{a,1}\frac i2}{\l_{L,a;j}-\theta_k-\delta_{a,1}\frac i2}&=&-\prod_{a'=1}^{N-1}\prod_{j'=1}^{K_{L,a'}}\frac{\l_{L,a;j}-\l_{L,a';j'}+\frac{ic_{aa'}}2}{\l_{L,a;j}-\l_{L,a';j'}-\frac{ic_{aa'}}2},\no\\
e^{-imL\sinh[\frac{2\pi}N\theta_k]}&=&-\prod_{k'=1}^{M}S_{0,PCF}[\theta_k-\theta_{k'}]\prod_{j=1}^{K_{L,1}}\frac{\theta_k-\lambda_{L,1;j}+\frac i2}{\theta_k-\lambda_{L,1,j}-\frac i2}\prod_{j=1}^{K_{R,1}}\frac{\theta_k-\lambda_{R,1;j}+\frac i2}{\theta_k-\lambda_{R,1,j}-\frac i2},\no\\
\prod_{k=1}^M\frac{\l_{R,a;j}-\theta_k+\delta_{a,1}\frac i2}{\l_{R,a;j}-\theta_k-\delta_{a,1}\frac i2}&=&-\prod_{a'=1}^{N-1}\prod_{j'=1}^{K_{R,a'}}\frac{\l_{R,a;j}-\l_{R,a';j'}+\frac{ic_{aa'}}2}{\l_{R,a;j}-\l_{R,a';j'}-\frac{ic_{aa'}}2}.
\ee

The identification of the bound states is similar to the case of GN model. The only modification is that since $S_{0,PCF}[i]=0$, a $k$-string of $\theta$-s induces $(k-1)$-strings on both left and right wings. As a result we have the structure of stacks shown in Fig.~\ref{fig:rombesPCF}.

If we denote by $\theta_{a;k}$  the center of the stack that contains $a$ $\theta_k$ roots then the Bethe Ansatz equations will have the following form:
\be\label{PCF1}
  \prod_{k=1}^{M_a}\frac{\l_{L,a;j}-\theta_{a;k}+\frac i2}{\l_{L,a;j}-\theta_{a;k}-\frac i2}&=&-
  \prod_{a'=1}^{N-1}\prod_{j'=1}^{K_{L,a'}}\frac{\l_{L,a;j}-\l_{L,a';j'}+\frac{ic_{aa'}}2}{\l_{L,a;j}-\l_{L,a';j'}-\frac{ic_{aa'}}2},\no\\
  e^{-im_aL\sinh[\theta_{a;k}]}&=&-\left(\prod_{a'=1}^{N-1}\prod_{k'=1}^{M_{a'}}S_{0,PCF}^{aa'}[\theta_{a;k}-\theta_{a';k'}]\right)\times\no\\&&\times\prod_{j=1}^{K_{L,a}}\frac{\theta_{a;k}-\l_{L,a;j}+\frac i2}{\theta_{a;k}-\l_{L,a;j}-\frac i2}\prod_{j=1}^{K_{R,a}}\frac{\theta_{a;k}-\l_{R,a;j}+\frac i2}{\theta_{a;k}-\l_{R,a;j}-\frac i2},\no\\
  \prod_{k=1}^{M_a}\frac{\l_{R,a;j}-\theta_{a;k}+\frac i2}{\l_{R,a;j}-\theta_{a;k}-\frac i2}&=&-
  \prod_{a'=1}^{N-1}\prod_{j'=1}^{K_{R,a'}}\frac{\l_{R,a;j}-\l_{R,a';j'}+\frac{ic_{aa'}}2}{\l_{R,a;j}-\l_{R,a';j'}-\frac{ic_{aa'}}2},
\ee
\be
  S_{0,PCF}^{aa'}[\theta]&=&\theta^{\frac{\left(1-D^{2{\rm min}[a,a']}\right)\left(D^{|a-a'|}-D^{2N-a-a'}\right)}{\left(1-D^{2N}\right)}-(D\ \to\  D^{-1})}\simeq\no\\&\simeq&\theta^{2\frac{\left(D^{N-{\rm max}[a,a']}-D^{{\rm max}[a,a']-N}\right)\left(D^{{\rm min}[a,a']}-D^{-{\rm min}[a,a']}\right)}{D^N-D^{-N}}}.
\ee

\begin{figure}[t]
\centering
\includegraphics[width=8cm]{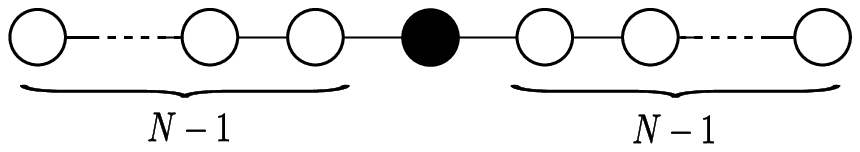}
\caption{\label{fig:DynkinPCF}"Dynkin diagram" for Bethe equations (\ref{BethePCF}).}
\end{figure}
\begin{figure}[b]
\centering
\includegraphics[width=6cm]{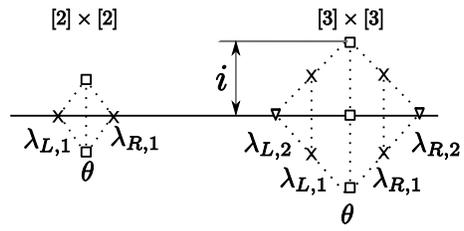}
\caption{\label{fig:rombesPCF}Bound states for the PCF model.}
\end{figure}

\chapter{\label{ch:tdlimit}IQFT as a continuous limit of integrable spin chains}

An important problem to study is what field theories can be obtained in the continuous limit of integrable spin chains. Assuming that integrability and symmetry of the system is preserved in the continuous limit we may expect to recover from the $SU(N)$ symmetric spin chains the $SU(N)$ GN model, just based on the argument of the universality of low energy effective theories.

In this chapter we study excitations over the antiferromagnetic vacuum of the $SU(N)$  XXX spin chain\footnote{This study for the $SU(2)$ case was first time correctly done by Faddeev and Takhtajan \cite{Faddeev:1981ip,Takhtajan:1982zz,Faddeev:1984ft}. These authors also found the scattering matrix of the excitations using the method of \cite{Korepin:1979qq}. The general case was studied in \cite{Ogievetsky:1987vv}.}. This vacuum is interesting since it is invariant under the $SU(N)$ group. The scattering of the excitations, which are also called spinons, depends on the representation in which the spin chain was defined. For the fundamental representation the scattering of low energetic spinons in the infinite volume is described by the same scattering matrix as the one for the $SU(N)$ GN model. In the limit of the infinite spin representation the low energy scattering matrix is identical with the one of the PCF model \cite{Polyakov:1983tt,Faddeev:1985qu}. Interestingly, the PCF model has a larger $SU(N)\times SU(N)$ symmetry.

A nontrivial scalar factor in the scattering matrix, such as $S_{0,GN}$ in (\ref{S0GN}), is recovered from the algebraic Bethe Ansatz equations as an effective interaction of spinons. From the point of view of the Bethe Ansatz, spinons are not elementary particles but are holes in a Dirac sea of magnons - excitations over ferromagnetic vacuum. Curiously, the scattering matrix of spinons obtained in this purely algebraic way is crossing-invariant, although the XXX spin chain does not have such discrete symmetry.

In opposite to the GN and PCF models, the spinon spectrum is gapless. To obtain massive particles we may consider an inhomogeneous spin chain which is also known as a light cone spin chain\footnote{Such spin chain was proposed in \cite{Faddeev:1985qu}. The Bethe equations for it were used before in \cite{Polyakov:1983tt} were the PCF model was solved in terms of a certain fermionic model.}.

The described above scattering picture is applicable when we consider sigma models in the infinite volume. In sections \ref{sec:TBA} and \ref{sec:tdlimitfromHirota}  we discuss applicability of the spin chain discretization for the finite volume case. We recall a thermodynamic Bethe Ansatz (TBA) \cite{Yang:1968rm,Zamolodchikov:1989cf} and suggest that  the $T$-functions which appear in the context of TBA should be identified with the transfer matrices of the light cone spin chain. We show that this is indeed the case on the simplest example of equally polarized spinons of the $SU(2)$ GN model.

%
%

\section{\label{excitationsintheantiferro}Excitations in the antiferromagnetic $SU(2)$ XXX spin chain}
We will first consider in details the $SU(2)$ case, which will be our guiding example for a more complicated cases of $SU(N)$ models. The Bethe equations for the $SU(2)$ XXX spin chain,
\be\label{BetheXXX}
  \(\frac{u_k+\frac i2}{u_k-\frac i2}\)^L=\prod_{\substack{j=1 \\j\neq k}}^{M}\frac{u_k-u_j+i}{u_k-u_j-i},
\ee
were derived from the scattering theory of the excitations over the ferromagnetic vacuum. 
An interesting physics arise if to consider the Hamiltonian with opposite sign and therefore excitations over the antiferromagnetic vacuum. An important property of the antiferromagnetic vacuum is that it is a trivial representation of the symmetry group\footnote{for the spin chain of even length which we will consider in the following. For the antiferromagnetic spin chain of the odd length the lowest energy state is doubly degenerated.}.

We have at our disposal excitations over the ferromagnetic vacuum (magnons).
The question is to build an effective theory in which the vacuum is a Dirac sea of magnons. For general length of the spin chain this problem was not solved. However, in the thermodynamic limit $L\to\infty$ the Bethe equations are considerably simplified and the problem can be solved. The usual way to solve it is to introduce the integral equation for the density of Bethe roots, as it is done for example in \cite{Faddeev:1996iy}. We will use a slightly different description, in which we deal with the resolvents and functional equations on them. The advantage is that all the integral kernels which appear are replaced with the rational functions of the shift operator\footnote{Fourier transform also leads to these rational functions. The functional approach has an advantage of keeping explicitly the analytical structure of the resolvents and therefore of the density functions. This is especially useful for the problems discussed in the third part of this work.}.

\subsubsection{Holomorphic projection}
To derive the functional equations for the resolvents let us start from a simple configuration which contains only real Bethe roots. The main assumption which goes through all derivations is that in the large $L$ limit one or other terms in the l.h.s. of the Baxter equation (\ref{baxeq})
\be
  \left(u+\frac i2\right)^L Q[u-i]+\left(u-\frac i2\right)^L Q[u+i]=T[u]Q[u]
\ee
 can be neglected, depending on whether we are in the upper or in the lower half plane of the rapidity variable. As we discussed in section \ref{sec:countingBethe}, in this approximation the transfer-matrix is represented as
\be
 T[u]=Q^*[u]Q[u-i]Q[u+i],
\ee
 where $Q^*$ stands for the Baxter polynomial of holes
\be
  Q^*[u]=\prod_{k}(u-\theta_{h,k}).
\ee
Therefore the Baxter equation reduces to
\be\label{bQequation}
  \(u\pm\frac i2\)^L=Q^*[u]Q[u]Q[u\pm i],\ \ {\rm Im}[u]\gtrless 0.
\ee
If we introduce the resolvents
\be
  R[u]=\frac d{du}\log Q[u],\ \ \ R^*[u]=\frac d{du}\log Q^*[u],
\ee
and take derivative of the logarithm of (\ref{bQequation}), we will get
\be\label{firstR}
  (1+D^2)R+R^*=L\ D\frac 1u,\ \ {\rm Im}[u]>0,\no\\
  (1+D^{-2})R+R^*=L\ D^{-1}\frac 1u.\ \ {\rm Im}[u]<0,
\ee
where $D\equiv e^{\frac i2\pd_u}$.

At first glance this derivation seems to be not rigorous. However, 
the equations (\ref{firstR}) are equivalent to the linear integral equation which is derived in the canonical approach (used for example in the lectures \cite{Faddeev:1996iy}). Indeed, taking the difference of the functional equations (\ref{firstR}) on the real axis we get
\be\label{diff epxplicit}
  R[u+i0]+R[u+i]-R[u-i0]-R[u-i]+R^*[u+i0]-R^*[u-i0]=\frac {-i L}{u^2+\frac 14}.
\ee
Using the fact that
\be
  R[u]=\int dv\frac{\rho[v]}{u-v},\ \ \rho[u]=-\frac 1{2\pi i}\left(R[u+i0]-R[u-i0]\right),
\ee
it is easy to see that (\ref{diff epxplicit}) is equivalent to
\be\label{xxxintexplicit}
  \rho[u]+\rho_h[u]+\int\frac{dv}{\pi}\frac {1}{(u-v)^2+1}\rho[v]=\frac{L}{2\pi}\frac 1{u^2+\frac 14},
\ee
which is the integral Bethe Ansatz equation for the XXX spin chain.

To go back from (\ref{xxxintexplicit}) to (\ref{firstR}) we integrate (\ref{xxxintexplicit}) with a Cauchy kernel. We call this operation the holomorphic projection. The word holomorphic stands for passing from the density which is the function defined only on the real axis to the resolvent which is an analytical function everywhere in the complex plane except on the support of the density. The word projection stands for the fact that the expression (\ref{diff epxplicit}) can be split into two equations (\ref{firstR}) defined on the upper and the lower half planes.

\subsubsection{Bethe equations for holes}
The true excitations over the antiferromagnetic vacuum are holes. The equations (\ref{firstR}) can be rewritten as\footnote{We omit the equation in the l.h.p. since it is just conjugated to the one in the u.h.p.}
\be\label{forholes}
  \(1+\frac{-D^2}{1+D^2}\)R^*+R=L\frac {D}{1+D^2}\frac 1u.
\ee
Remarkably, such an equation can be obtained in the large $L$ limit from the following Bethe Ansatz:

\be\label{AsBA}
    e^{-ip[\theta_k]L}=-\prod_{j}S_{0,GN}[\theta_k-\theta_j],
\ee
where $p$ is defined through
\be
  -i \frac{dp}{du}=\(\frac {D}{1+D^2}-\frac {D^{-1}}{1+D^{-2}}\)\frac 1u=-\frac{i\pi}{\cosh\pi u}.
\ee
$S_{0,GN}$ is nothing but the $S$-matrix of the $SU(2)$ GN model for scattering of equally polarized particles. The rapidities $\theta_k$ describe holes from the point of view of the ferromagnetic spin chain. In this derivation we did not require that the number of holes should be infinite. So the equation (\ref{forholes}) can describe few excitations over the antiferromagnetic vacuum as well.

What is missing for matching (\ref{AsBA}) with the Asymptotic Bethe Ansatz of the GN model is the nested level of the Bethe equations and the correct expression for $p$.

We will recover the correct nested structure if we will take into account also all string configurations in the XXX spin chain. The functional equation that describes the thermodynamic limit of the complete set of Bethe equations (\ref{completeBetheAnsatz}) is the following:
\be\label{cool}
R_{s}^*+\frac{D+D^{-1}}{D-D^{-1}}\sum_{s'=1}^\infty (D^{s+s'}-D^{|s-s'|})R_{s'}=L D^s\frac 1{u},\ \ s\geq 1.
\ee
The second term on the l.h.s. comes from the operator $\CL_{ab}$. We can split this operator in the following way:
\be\label{313}
  \CL_{ab}=\frac {\left(D^{a}-D^{-a}\right)\left(D^b-D^{-b}\right)}{D-D^{-1}}=\frac{D^{a+b}-D^{|a-b|}}{D-D^{-1}}+\frac{D^{-a-b}-D^{-|a-b|}}{D-D^{-1}}.
\ee
Each of two terms in the r.h.s. of (\ref{313}) is a sum of finite number of shift operators. The first term contains only shifts in the positive imaginary direction. Only this term survives under the holomorphic projection in the upper half plane.

A remarkable property of the operator $\frac{D^{a+b}-D^{|a-b|}}{D-D^{-1}}$ is that it is an element of the inverse deformed Cartan matrix (\ref{CartanMatrix}) of the $A_\infty$ algebra\footnote{We are grateful to Paul Zinn-Justin for pointing out this fact. We would like also to refer to his PhD thesis \cite{PZJ} where many ideas of this chapter are also discussed.}. Therefore the equation (\ref{cool}) can be also rewritten as
\be\label{trescool}
  C_{ss'}^{\infty}R_{s'}^*+(D+D^{-1})R_s=\delta_{s,1}\frac {L}{u}.
\ee
In the last formula we used the operator $D^{-1}$ to put the expression into a more symmetric form. To avoid ambiguities of how the action of this shift operator is defined, we should consider (\ref{trescool}) in the region ${\rm Im}[u]>1/2$, not ${\rm Im}[u]>0$. Then this equation can be analytically continued. 

To make the particle-hole transformation\footnote{Passing to functional equations that allow to interpret holes as the roots of a certain Bethe Ansatz equation.} more explicit we introduce
\be
  \tilde R=R^*_1,\ \ \tilde R^*=R_1.
\ee
The equations (\ref{trescool}) for $s=1$ can be written also as
\be
  \(1+\frac {-D^2}{1+D^2}\)\tilde R+\tilde R^*+\sum_{s'=2}^\infty D^{s-1} R_{s'}=L\frac D{1+D^2}\frac 1u.
\ee
It is easy to see that it leads to the Bethe equation of the $SU(2)$ GN for the momentum-carrying particle (but with different $p[u]$).

The equations (\ref{trescool}) for $s\geq 2$ are written as
\be
  \sum_{s'=1}^{\infty}C_{ss'}R_{s'+1}^*+(D+D^{-1})R_{s+1}=\delta_{s,1}\tilde R,\ \ s\geq 1.
\ee
These equations have the same form as (\ref{trescool}) except for the replacing the source term $\frac Lu$ by $\tilde R$. The same equations can be obtained from the nested level of the Bethe Ansatz for $SU(2)$ GN. Interestingly, the centers of the strings of the length two play the role of the real roots of the nested Bethe equations.

\subsubsection{A model with the gap in the spectrum}
Still, the obtained equations do not have correct expression for the momentum. In fact, the energy of a hole is given by
\be
  E[\theta]=\frac 1{\pi}\frac {dp}{du}=\frac {1}{\cosh[\pi\theta]}.
\ee
The spectrum of the theory does not contain a gap, therefore we can obtain only a massless field theory.

To cure the situation we follow the approach of Faddeev and Reshetikhin \cite{Faddeev:1985qu} who suggested to consider the XXX spin chain with alternating inhomogeneities $\omega$ such that the Bethe Ansatz is written as
\be\label{FRBA}
  \(\frac{u_k+\omega+\frac i2}{u_k+\omega-\frac i2}\)^{L/2}\(\frac{u_k-\omega+\frac i2}{u_k-\omega-\frac i2}\)^{L/2}=-\prod_{k'}\frac{u_k-u_{k'}+i}{u_k-u_{k'}-i}.
\ee
The same equations appeared in the solution of PCF by Polyakov and Wiegmann \cite{Polyakov:1983tt} before \cite{Faddeev:1985qu}. However the equations (\ref{FRBA}) had a different meaning; they described a fermionic model with a four-fermion interaction. The reason of equivalence is that the first-quantized version of the fermionic model are the particles interacting through the $\delta$-potential. Such system is solved by the same Bethe Ansatz equations as the ones for the XXX spin chain. In fact, exactly the Polyakov-Wiegmann approach was used in \cite{Ogievetsky:1987vv} for the solution of the PCF model for an arbitrary simple group. Translation of this result to the Faddeev-Reshetikhin picture is straightforward and for the $SU(N)$ case is given in the next sections.

Let us introduce the spacing $\ell$ between the nodes of the spin chain. The energy for the excitations over the antiferromagnetic vacuum can be written as
:
\be\label{FRdispersion}
  E[\theta]&=&\frac 1{\ell}\frac 12\(\frac{1}{\cosh[\pi(\theta-\omega)]}+\frac{1}{\cosh[\pi(\theta+\omega)]}\).
\ee
There are low energy excitations for small values of $\theta$.
In the limit $\omega\to\infty$ we obtain for them:
\be
  E[\theta]=m \cosh\pi\theta,\ \ m=\frac {2}{\ell}e^{-\pi\omega}.
\ee
To obtain a finite value for the mass $m$ we have to consider a proper limit
\be\label{FRlightcone}
  \ell\to 0,\ \ L\to\infty,\ \ \omega\to\infty,\no\\
  \ell L=L_{\rm phys}\ \ {\rm fixed},\ \ m=\frac {2}{\ell}e^{-\pi\omega}\ \ \ {\rm fixed}.
\ee

In this limit we recover exactly the integral equations which follow from the asymptotic Bethe Ansatz of  $SU(2)$ GN.

Now we will generalize the obtained results to the case of $SU(N)$ group and  then to the case of PCF.

\section{Generalization to the case of $SU(N)$ GN.}
The generalization is straightforward. Let us label the resolvent for strings in the position $\{a,s\}$ as $R_{a,s}$. The label $a$ corresponds to the nested level of the Bethe root. The label $s$ labels the length of the string. The general Bethe equations are written as
\be\label{almostrem}
  R_{a,s}^*&+&\frac{D+D^{-1}}{D-D^{-1}}\sum_{s'}(D^{s+s'}-D^{|s-s'|})R_{a,s'}-\no\\&-&\frac 1{D-D^{-1}}\sum_{s'}(D^{s+s'}-D^{|s-s'|})\(R_{a+1,s'}+R_{a-1,s'}\)=\delta_{a,1}D^s R_p,
\ee
where $R_p$ is the resolvent for the source term. It equals to $L/u$ in the homogeneous case and to $\frac L2(\frac 1{u-\omega}+\frac 1{u+\omega})$ in the case of the light-cone spin chain.

Remarkably, the equation (\ref{almostrem}) within the usage of the deformed Cartan matrix (\ref{CartanMatrix}) can be rewritten as:
\be\label{vcool}
  \sum_{s'=1}^\infty C^\infty_{s,s'}R_{a,s'}^*+\sum_{a'=1}^{N-1}C^{N-1}_{a,a'}R_{a',s}=\delta_{a,1}\delta_{s,1}R_p.
\ee
We use two different Cartan matrices. One is of $A_\infty$ Dynkin diagram, another one is of $A_{N-1}$ Dynkin diagram. We reflect these distinctions in the superscripted indexes of $C_{ab}$.

\begin{figure}[t]
\centering
\includegraphics[width=14cm]{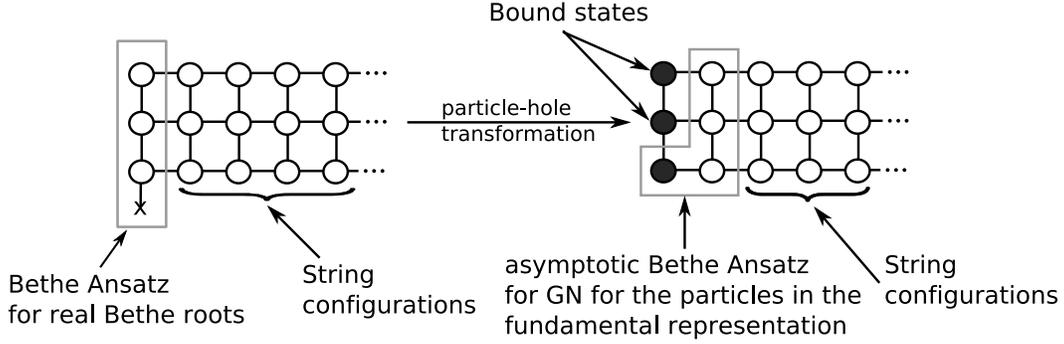}
\caption{\label{fig:GNandXXX} Particle-hole transformation. In the proper thermodynamic limit the excitations of the inhomogeneous XXX spin chain scatter as the particles in the GN model. The picture reflects the structure of the functional equation (\ref{vcool}).}
\end{figure}

This form of the functional equation is usually described in terms of the Dynkin-type graph in Fig.~\ref{fig:GNandXXX} with $A_\infty$ graph in the horizontal direction and with $A_{N-1}$ in the vertical direction.

To obtain the $SU(N)$ GN model we perform a particle-hole transformation on the roots of type $\{a,1\}$. For this we introduce the resolvents
\be
\tilde R_{a}=R_{a,1}^*,\ \ \tilde R_{a}^*=R_{a,1}.
\ee

The equations (\ref{vcool}) for $s\geq 2$ describe the nested levels of the $SU(N)$ GN. As in the $SU(2)$ case, the roots $\{a,1\}$ play the role of the source terms.

The equations (\ref{vcool}) for $s=1$ lead to the following equations:
\be\label{FRSUNGN}
  D^{-1}\(C^{N-1}\)^{-1}_{aa'}\tilde R_{a'}+\tilde R_{a}^*+\sum_{s'=2}^\infty D^{s'-1}R_{a,s'}=\(C^{N-1}\)^{-1}_{a1}R_p.
\ee
We see that interaction between the states is determined by the inverse Cartan matrix. It is straightforward to check that we get a correct expression for $S_{0,GN}^{aa'}$ given in (\ref{SGN1aa}).

The inverse deformed Cartan matrix can be written explicitly as\footnote{The generalization of this expression for the distinguished Dynkin diagram of $gl(N|M)$ algebra was given in \cite{SKR}.}
\be
  (C^{N-1})^{-1}_{aa'}=\left\{\begin{array}{c}
                         \frac{[N-a]_{D}[a']_{D}}{[N]_D},\ a\geq a' \\
                         \frac{[N-a']_D[a]_D}{[N]_{D}},\ a'>a
                       \end{array}\right.,\ \ [x]_D\equiv \frac{D^x-D^{-x}}{D-D^{-1}},
\ee
where we used the notations $[x]_q$ for $q$-numbers. This inverse Cartan matrix should be understood as a series over positive powers of $D$ for the equations written in the u.h.p. and as a series over negative powers for the equations in the l.h.p.

The momenta $p_k$ of a $k$-particle of the GN model should enter into the integral equation for $\tilde \rho_k$ (density associated with $\tilde R_k$) as $-\frac{dp_{k}}{du}$. We can calculate this expression from the explicit form of the inverse Cartan matrix:
\be
  -i\frac {d\tilde p_k}{du}=D\frac{1-D^{2N-2k}}{1-D^{2N}}R_p-D^{-1}\frac{1-D^{-2N+2k}}{1-D^{-2N}}R_p.
\ee
Explicit calculation with the help of inverse Laplace transform gives
\be
  \(D\frac{1-D^{2N-2k}}{1-D^{2N}}-D^{-1}\frac{1-D^{-2N+2k}}{1-D^{-2N}}\)\frac 1{u}=2\pi i\sum_{r=1}^\infty(-1)^r\sin\[\frac{N-k}{N}\pi r\]e^{-\frac{2\pi}{N} |u| r}.
\ee
Using the parity properties of $\frac {d\tilde p_k}{du}$ we obtain in the limit analogous to (\ref{FRlightcone})
\be
  \frac {d\tilde p_k}{du}=2\pi\sin\[\frac{\pi k}{N}\]\cosh\[\frac{2\pi}{N}u\]L e^{-\frac{2\pi}N\omega},
\ee
which leads to the following effective mass parameter:
\be
  m=\frac {2}{\ell\sin\frac{\pi}N}e^{-\frac{2\pi}{N}\omega},\ \ \ m_{[k]}=\frac{\sin\left[\frac{\pi k}{n}\right]}{\sin\left[\frac\pi n\right]}.
\ee
We see that masses of the particles obtained from a light-cone spin chain are the same as it should be in the $SU(N)$ GN model (see (\ref{massesGN})).

\section{\label{sec:PCFgeneralization}Generalization to the case of $SU(N)$ PCF.}
If we consider the XXX spin chain in the fundamental representation of the $SU(N)$ group, the dynamics of low energy excitations over the antiferromagnetic vacuum is described by the GN model. To obtain in continuous limit the PCF model we have to consider the spin chain in a different representation \cite{Faddeev:1985qu}. The general Bethe Ansatz for arbitrary representation of $SU(N)$ group is given by (\ref{Bethegeneralrepresentation}). The version of the Bethe Ansatz which includes a string hypothesis is given by the equation (\ref{completeBetheAnsatz}). Remarkably, the Bethe Ansatz which corresponds to a rectangular representation with Young table given by $A$ rows and $S$ columns leads to a simple functional equation on the resolvents:
\be\label{vvcool}
  \sum_{s'=1}^\infty C^\infty_{s,s'}R_{a,s'}^*+\sum_{a'=1}^{N-1}C^{N-1}_{a,a'}R_{a',s}=\delta_{a,A}\delta_{s,S}R_p.
\ee
To obtain a PCF we have to consider the case $A=1$ and $S\to\infty$. The antiferromagnetic vacuum corresponds to the maximal filling of strings of type $\{a,S\}$, $a=\overline{1,A}$.

\begin{figure}[b]
\centering
\includegraphics[width=10cm]{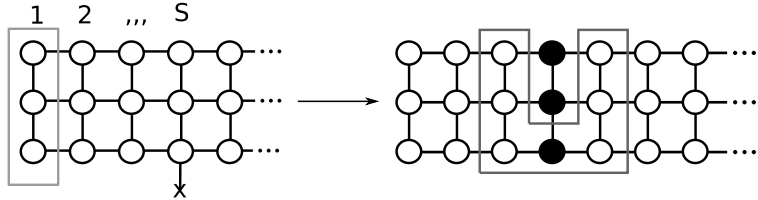}
\caption{\label{fig:PCFandXXX} The particle-hole transformation of a spin chain in the representation with spin $S/2$. In the limit $S\to\infty$ the global symmetry of the system becomes $SU(N)\times SU(N)$, and the thermodynamic limit leads to the PCF model.}
\end{figure}
As usual, we perform a particle-hole transformation for the nodes $\{a,S\}$:
\be
  \tilde R_a=R_{a,S}^*,\ \ \tilde R_a^*=R_{a,S}.
\ee
Since the resolvent decreases at infinity one can approximate in the large $S$ limit
\be
  \(C^\infty\)^{-1}_{S,S+s'}\equiv\frac{D^{2S+s'}-D^{|s'|}}{D-D^{-1}}\simeq -\frac{D^{|s'|}}{D-D^{-1}}.
\ee
Using this approximation, we can write for the central nodes\footnote{the nodes that a subject of a particle-hole transformation.}:
\be
  -(D-D^{-1})\left(C^{N-1}\right)^{-1}_{aa'}\tilde R_{a'}+\tilde R_a^*+\sum_{s'=1}^\infty D^{s'}\(R_{a,S+s'}+R_{a,S-s'}\)=\left(C^{N-1}\right)^{-1}_{a1}R_p.
\ee
These are functional Bethe Ansatz equations that can be also derived from the asymptotic Bethe Ansatz of $SU(N)$ PCF. To obtain the massive theory we use the limit (\ref{FRlightcone}) which is the same as in the case of the GN model.

The nested Bethe equations come from the nodes that are on the left and on the right from the central nodes. We see that in the limit $S\to\infty$ two, left and right, $SU(N)$ symmetries appear.


\section{\label{sec:TBA}Thermodynamic Bethe Ansatz}
In the previous sections we saw that the functional Bethe equations for the excitations over the antiferromagnetic vacuum coincide with the equations that can be obtained from the asymptotic Bethe Ansatz for the sigma-models. It is therefore natural to ask the question whether light-cone spin chains provide a correct lattice regularization of the sigma model. If it is the case we will be able to describe sigma-models in a finite volume. A possible way to describe a system in a finite volume known as a thermodynamic Bethe Ansatz (TBA) \cite{Yang:1968rm,Zamolodchikov:1989cf}. Our goal would be to compare the description given by TBA with the description which is suggested by the spin chain discretization. In this section the TBA approach is reviewed. In the next section we argue that the spin chain leads to the same results and support this suggestion by the simplest example.

\paragraph{TBA for calculation of the ground state energy.} Let us consider the theory on the torus with periods $L$ and $R$. The sigma-model in the finite volume $L$ is recovered in the large $R$ limit. The basic idea of the TBA is to write the partition function in two different ways:
\be
  Z[R,L]=\Tr\ e^{-R H[L]}=\Tr\ e^{-L H[R]}.
\ee
Here $H[L]$ ($H[R]$) is the Hamiltonian that generates translation in the $L$ ($R$) direction. Since the theory is relativistically invariant these Hamiltonians are the same. The energy of the ground state of the sigma-model at finite $L$  can be found as
\be\label{ZamGrounstate}
  E_0[L]=-\lim_{R\to\infty}\frac{\log Z}{R}.
\ee
We can calculate partition function since the energy spectrum at large $R$ is described by the asymptotic Bethe Ansatz. In the large $R$ limit a saddle point approximation is used. Then $-\log Z$ is given by a free energy that can be found by the minimization of
\be\label{CFLHS}
  \CF=L H[R]-S,
\ee
where $S$ is the entropy of the macroscopic state with given energy.

We use an assumption that in the thermodynamic limit the Bethe roots organize themselves into strings. Then the macroscopic state in the thermodynamic limit is described by densities of the strings. We already know that the functional equations that describe sigma model can be written in a much simpler form if we interchange the role of particles and holes for the momentum carrying roots. Therefore let us denote by $\rho_{A}$ ($\rho_{A}^*$) the densities of strings (corresponding holes) of type $A$
in the case when they are not momentum carrying. The momentum carrying strings will be denoted in inverse way - $\rho_{A}$ will stand for densities of holes and $\rho_A^*$ for densities of particles.

The densities satisfy the equation\footnote{In the literature the notation $1+\tilde K_{AB}$ instead of $K_{AB}$ is often used. Our choice of notation is motivated by the fact that $K_{AB}$ is given by the inverse Cartan matrix (up to multiplication by $D^{-1}$ or $(D-D^{-1})$). The kernel $\tilde K_{AB}=1-K_{AB}$ does not have such clear algebraic interpretation.}
\be
  \sum_{B}\int_{-\infty}^\infty K_{AB}[u,v]\rho_{B}^*[v]dv+\rho_A=J_A,
\ee
where  $J_A$ is a source term. For GN or PCF $A\equiv\{a,s\}$ and  $J_{a,s}[u]=m_a \cosh\[{\frac{2\pi}{N}u}\]\delta_{s,0}$.

The free energy $\CF$ is given by the following equation
\be
  \CF=\sum_A\intii dv L \rho_A^* J_A-(\rho_A+\rho_A^*)\log(\rho_A+\rho_A^*)+\rho_A\log\rho_A+\rho_{A^*}\log\rho_{A^*},
\ee
where the first term is $L H[R]$ of (\ref{CFLHS}) and the second one is the entropy of the system.

Minimization of the free energy leads to the TBA equations:
\be\label{integralTBA}
  \log\[1+\frac{\rho_A[v]}{\rho_A^*[v]}\]=L\ J_A[v]+\intii du K_{BA}[u,v]\log\[1+\frac{\rho_B^*[u]}{\rho_B[u]}\].
\ee
The ground state energy given by (\ref{ZamGrounstate}) then can be evaluated as
\be\label{energygroundstate}
  E_0[L]=-\sum_A\intii dv J_A[v]\log\[1+\frac{\rho_{A}^*[v]}{\rho_{A}[v]}\].
\ee
In the case of the GN and PCF models one can further simplify the expressions. First, the
kernels $K_{AB}[u,v]$ can be most generally represented as
\be
  K_{AB}[u,v]=\sum_{n=0}^\infty \left(c_n[A,B]D^{n}_u\frac 1{u-v+i0}-c_n[A,B]D^{-n}_u\frac 1{u-v-i0}\right),
\ee
where coefficients $c_n[A,B]$ are symmetric with respect to interchange of $A$ and $B$.
We conclude therefore that
\be
  K_{AB}[u,v]=K_{BA}[u,v]=K_{BA}[v,u].
\ee
Let us introduce the functions $Y$:
\be
  Y_{a,s}=\frac{\rho^*_{a,s}}{\rho_{a,s}}.
\ee
Then the TBA equations for $s\geq 1$ will be written as\footnote{The sum over $s'$ is from $0$ to $\infty$ for
GN and from $-\infty$ to $\infty$ for PCF. We always assign $s=0$ to the momentum-carrying nodes.}
\be\label{TBAY}
  \log\[1+Y_{a,s}^{-1}\]=L m_a\delta_{s,0}\cosh\[\frac{2\pi}{N}u\]+\sum_{s'}\sum_{a'=1}^{N-1} \(\hat C^{N-1}\)^{-1}_{a,a'}\hat C_{s,s'}*\log\[1+Y_{a',s'}\] \ee
Here $\hat C_{s,s'}$ denotes the integral kernel whose holomorphic projection is the deformed Cartan matrix $C_{s,s'}$. Let
us evaluate this kernel. If the function $g$ of integration is analytic in the strip $-1/2<{\rm Im}[u]<1/2$, then
\be
  \sum_{s'}\hat C_{s,s'}*g_{s'}&=&-g_{s+1}-g_{s-1}-\frac 1{2\pi i}\((D+D^{-1})G_s[u+i0]-(D+D^{-1})G_s[u-i0]\)=\no\\&&=g_s[u+i/2]-g_s[u-i/2]-g_{s+1}[u]-g_{s-1}[u],
\ee
where $G_s$ is the resolvent of $g_s$ and $D^{-1}G_s[u+i0]$ means analytical continuation of $G_s$ to the point $u-i/2$ from the upper half plane. We see that
for the functions analytic in the strip ${\rm Im}[u]<1/2$, $\hat C_{s,s'}$ coincides with $C_{s,s'}$.

Let us therefore act with $C_{a,a'}^{N-1}$ on the equation (\ref{TBAY}). It is straightforward
to check that $J_{a,0}$ is a zero mode of this action. Therefore we will get
\be\label{prefY}
  \sum_{a'=1}^{N-1}C_{a,a'}\log\[1+Y_{a',s}^{-1}\]=\sum_{s'}C_{s,s'}\log\[1+Y_{a,s'}\],
\ee
or more explicitly
\be\label{fY}
  \frac{Y_{a,s}^+Y_{a,s}^-}{Y_{a+1,s}Y_{a-1,s}}=\frac{(1+Y_{a,s+1})(1+Y_{a,s-1})}{(1+Y_{a+1,s})(1+Y_{a-1,s})}.
\ee
We have obtained the functional $Y$-system. The algebraic structure of (\ref{prefY}) is determined by the equations (\ref{vcool}). Therefore the structure of $Y$-system reflects the strings hypothesis
and the interaction between strings.


\paragraph{TBA in the case of excited states.}
The set of equations (\ref{fY}) requires appropriate boundary conditions to have a unique solution. Boundary conditions which follow  from (\ref{integralTBA}) lead to the solution which gives the energy of the ground state.
The results of \cite{Bazhanov:1996aq,Dorey:1996re} suggest that changing of the boundary conditions allows describing a certain class of excited states. The energy of such excited states is given by generalization of (\ref{energygroundstate}):
\be\label{energyexcitedstate}
  E_0[L]=-\sum_A\intii dv J_A[v]\log\[1+\frac{\rho_{A}^*[v]}{\rho_{A}[v]}\]+\sum_A\sum_{k_A}m_A\cosh\left[\frac{2\pi}{N}\theta_{k_A}\right],
\ee
where $\theta_{k_A}$ are solutions of
\be\label{YA}
    Y_{A}[\theta_{k_A}\pm i/2]=-1.
\ee

It is often believed that all the excited states of the sigma model can be described as certain solutions of the $Y$-system. The energies of these states are given by (\ref{energyexcitedstate}). As was shown in \cite{Gromov:2008gj} for a wide class of excited states of the $O(4)$ sigma model, equations (\ref{YA}) reduce in the large volume limit to the Bethe Ansatz equations. Therefore (\ref{YA}) are thought as exact Bethe Ansatz equations for the systems in finite volume.

The conjecture about applicability of the $Y$-system for the description of the excited states is supported by the spin chain point of view on the problem as we will now discuss.
%

\section{\label{sec:tdlimitfromHirota}Transfer matrices of spin chain as $T$-functions of TBA.}
The functional $Y$-system (\ref{fY}) can be rewritten as the Hirota system (\ref{fusioncomplete})
\be\label{hirotacomplete2}
  T^{a,s}[\th+i/2]T^{a,s}[\th-i/2]=T^{a,s+1}[\th]T^{a,s-1}[\th]+T^{a+1,s}[\th]T^{a-1,s}[\th]
\ee
if we introduce the $T$-functions by
\be\label{Ydefinition}
  Y_{a,s}=\frac{T^{a,s+1}T^{a,s-1}}{T^{a+1,s}T^{a-1,s}}.
\ee
Relation (\ref{Ydefinition}) defines the functions $T$ modulo the gauge transformation
\be\label{gaugetr}
  {T^{a,s}_g[\theta]}=g_1\!\!\[\theta+\frac {i(a+s)}2\]\!\!g_2\!\!\[\theta+\frac {i(a-s)}2\]\!\!g_3\!\!\[\theta-\frac {i(a-s)}2\]\!\!g_4\!\!\[\theta-\frac {i(a+s)}2\]\!\!T^{a,s}[\theta].
\ee
The Hirota system (\ref{hirotacomplete2}) is also invariant under (\ref{gaugetr}), therefore there is an equivalence between $Y$-system and Hirota system modulo gauge transformations.

{\it A priori} $T$-functions defined by (\ref{Ydefinition}) do not have meaning of transfer matrices. However (\ref{hirotacomplete2}) has the same form as the fusion relations for the transfer matrices of the spin chain. Based on this observation we suggest that $T$ functions obtained from (\ref{Ydefinition}) coincide with transfer matrices of the spin chain discretization of the sigma model.

 Here we will show this equivalence on the simplest case - $SU(2)$ GN model with the excitations polarized in the same direction. A more general case would be the subject of our future work.

\begin{figure}[t]
\centering
\includegraphics[height=5cm]{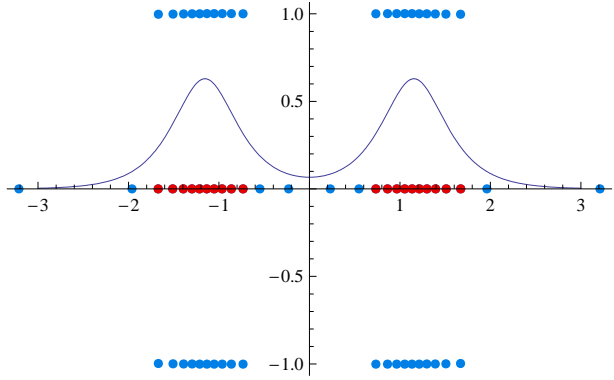}
\caption{\label{fig:FRsolution}A particular solution of (\ref{FRBA}) close to the antiferromagnetic vacuum. $N$=48, $M$=20. The inhomogeneity parameter $\omega=\pi^{-1}\log\[\frac{2\pi L}8\]$ corresponds to $m L=8$.}
\end{figure}

 Since all particles are equally polarized, the spin chain is described by the equations (\ref{FRBA}). In Fig.~\ref{fig:FRsolution} we show a typical solution of this equation. The red dots represent the zeroes of the Baxter polynomial and the blue dots represent the zeroes of $T$. The complex zeroes of $T$ are accompanying\footnote{This term was introduced in Sec.\ref{sec:stringhyp}.} roots and do not have physical meaning. The real zeroes of $T$ are holes. They play the role of physical excitations. Their energy is proportional to $\cosh[\pi(u+\o)]^{-1}+\cosh[\pi(u-\o))]^{-1}$ and is also plotted in Fig.~\ref{fig:FRsolution} (as a potential with a well).

In the Faddeev-Reshetikhin limit (\ref{FRlightcone}) the holes inside the well in Fig.~\ref{fig:FRsolution} describe the massive excitations of the GN model. The holes outside the well should describe the massless field theory. However in the limit (\ref{FRlightcone}) massless excitations do not interact with the massive modes as we will see below.

Let us use (\ref{Ydefinition}) to define $Y$ functions based on the transfer matrices of the spin chain. These $Y$ functions will be denoted as $\hat Y_{a,s}$ to distinguish them from the $Y$-functions of TBA.  From (\ref{Ydefinition}) we have
\be\label{Y1132}
  \hat Y_{1,1}[\theta]=\frac{T^{1,1}[\theta+\frac i2]T^{1,1}[\theta-\frac i2]}{T^{0,1}[\theta]T^{2,1}[\theta]}-1=\frac{T^+T^-}{Q_s^{++}Q_s^{--}}-1,
\ee
where $Q_s[\theta]=(\theta-\o)^{L/2}(\theta+\o)^{L/2}$ and $T\equiv T^{1,1}$.

The function $\hat Y_{1,1}$ satisfies the equations
\be\label{Y1111}
  \hat Y_{1,1}\left[\theta_h\pm\frac i2\right]=-1.
\ee
Since $\theta_h$ are the rapidities of spinon excitations which are true particles in the antiferromagnetic case, we should identify (\ref{Y1111}) with (\ref{YA}) and therefore we expect that $\hat Y_{1,1}=Y_{1,0}$.
From this identification it also follows that all excited states should satisfy (\ref{Y1111}) which cannot be shown from the TBA point of view.

The transfer-matrix $T$ satisfies the Baxter equation
\be
  Q^+_sQ^{--}+Q^-_sQ^{++}=Q\ T.
\ee
We have already discussed (see Sec.~\ref{sec:countingBethe}) that $T$ is represented in the form
\be
  T=\tilde{Q}^{--}\tilde{Q}^{++}Q_{h},
\ee
where $\tilde{Q}\to Q$ in the large $L$ limit. In the large $L$ limit we can also approximate the Baxter equation by:
\be\label{Baxtersimplified}
  Q^{\pm}_s=QQ^{\pm\pm} Q_{h},\ \ {\rm Im}[\theta]\gtrless 0.
\ee
Let us verify if the approximation (\ref{Baxtersimplified}) is valid in the limit (\ref{FRlightcone}). We are interested in the dynamics of excitations inside the well, therefore $|\theta|\lesssim \log L$. As it follows from (\ref{eq:Bethesu2}), approximation (\ref{Baxtersimplified}) can be trusted\footnote{We consider the Bethe roots inside the well. For these roots the last term in (\ref{eq:Bethesu2}) is small} if $\theta$ is at the distance from the source origin smaller then $\sqrt{L}$. This condition is perfectly satisfied since inside the well all the distances are maximally of order $\log L$. We see that approximation (\ref{Baxtersimplified}) is perfectly satisfied. This also means the validity of the string hypothesis.

In view of (\ref{Baxtersimplified})
\be\label{Y11}
  \hat Y_{1,1}[\theta]=\frac{Q_s Q^{+++}}{Q_s^{++}Q^{-}}+\frac{Q_s Q^{---}}{Q_s^{--}Q^{+}}+...=\frac{Q_s}{Q^+Q^-}\(\frac 1{Q_{h}^+}+\frac 1{Q_{h}^-}\),
\ee
where the dots stand for the term that can be neglected in the large $L$ limit.

The equation (\ref{Baxtersimplified}) for $Q$ is solved explicitly by\footnote{We require $Q$ to be analytic outside the real axis.}
\be
  Q&=&\(\mp\frac i2\)^{-M}\(\frac{\Gamma[\mp \frac i2(\theta-\o)+\frac 34]\Gamma[\mp \frac i2(\theta+\o)+\frac 34]}{\Gamma[\mp \frac i2(\theta-\o)+\frac 14]\Gamma[\mp \frac i2(\theta+\o)+\frac 14]}\)^{N/2}\prod_{k}\frac{\Gamma[\mp\frac i2(\theta-\theta_{h,k})]}{\Gamma[\mp\frac i2(\theta-\theta_{h,k})+\frac 12]}\equiv\no\\&\equiv&\(\mp\frac i2\)^{- M}Q_s^{\frac{D^{\pm 1}}{1+D^{\pm 2}}}Q_h^{\frac 1{1+D^{\pm 2}}},\ \ {\rm Im}[\theta]\gtrless 0.
\ee
Inserting this solution into (\ref{Y11}) we obtain
\be
  \hat Y_{1,1}=4^{-M}Q_s^{1-\frac {D^2}{1+D^2}-\frac {D^{-2}}{1+D^{-2}}} Q_{h}^{\frac{D}{1+D^{2}}+\frac{D^{-1}}{1+D^{-2}}}\(\frac 1{Q_{h}^+}+\frac 1{Q_{h}^-}\).
\ee
The factor containing $Q_s$ can be calculated explicitly in the limit (\ref{FRlightcone}). If we take logarithm and then derivative of this factor we will get:
\be
  \partial_\theta\sum_{n}(-1)^n \log[Q_s[\theta+in]]=\frac {L}{2}\sum_{n}(-1)^n\(\frac 1{\theta-\o+in}+\frac 1{\theta+\o-in}\)=\no\\
  =\frac {\pi L}{2}\(\frac 1{\sinh[\pi(\theta-\o)]}+\frac 1{\sinh[\pi(\theta+i\o)]}\)\to -\pi m L_{\rm phys}\sinh\pi\theta.
\ee
The overall constant of integration can be found from the fact that at $\theta=0$ the considered factor in the limit (\ref{FRlightcone}) should be equal to $(-4)^{N/2}$.

Therefore in the limit (\ref{FRlightcone})
\be\label{Yderived}
  \hat Y_{1,1}=(-1)^{L/2+n_r}e^{-m L_{\rm phys}\cosh\pi\theta}\(\frac 1{Q_{h}^+}+\frac 1{Q_{h}^-}\)\prod_{k}2\frac{\Gamma[\frac{i(\theta-\theta_{h,k})}2+\frac 34]\Gamma[-\frac{i(\theta-\theta_{h,k})}2+\frac 34]}{\Gamma[\frac{i(\theta-\theta_{h,k})}2+\frac 14]\Gamma[-\frac{i(\theta-\theta_{h,k})}2+\frac 14]}.
\ee
Here the product is taken only over the interior holes (inside the well in Fig.~\ref{fig:FRsolution}). The residual factor from the exterior holes is $(-1)^{n_r}$, where $n_r$ is the number of exterior holes to the right from the well. Since the holes are excited by pairs, the value of the multiplier $(-1)^{n_r}$ does not depend on the type of excitation we consider. In the following we will consider the case when $L$ is divisible by $4$, therefore $(-1)^{L/2+n_r}=1$.

The representation (\ref{Yderived}) for $\hat Y_{1,1}$ is correct in the strip $-1/2<{\rm Im}[\theta]<1/2$, where we can use approximation (\ref{Baxtersimplified}) for both $Q^+$ and $Q^-$. From (\ref{Y1132}) we do not expect singularities on the lines ${\rm Im}[\theta]=\pm 1/2$, at least inside the well. Therefore let us evaluate l.h.s. of (\ref{Y1111}) using (\ref{Yderived}). From this evaluation we see that (\ref{Y1111}) gives asymptotic Bethe Ansatz equations for the $SU(2)$ GN model:
\be
  e^{-imL_{\rm phys} \sinh\pi\l_h}=-\prod_{\l_h'}(\l_h-\l_h')^{-\frac{D^{2}}{1+D^{2}}+\frac{D^{-2}}{1+D^{-2}}}.
\ee
A given asymptotic behavior of $\hat Y$ functions in the large $L_{\rm phys}$ limit is sufficient to solve the $Y$ system in the finite volume. It follows for example from the method developed in \cite{Gromov:2008gj}. Since we reproduce asymptotic Bethe Ansatz in the large volume limit, we therefore can identify $\hat Y_{a,s}$ and $Y_{a,s-1}$.

\begin{figure}[t]
\centering
\includegraphics[height=2cm]{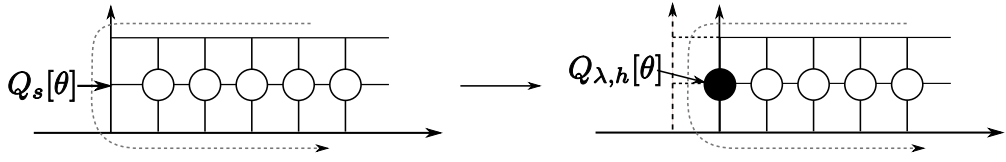}
\caption{\label{fig:HirotaTD}Reduction of the hook in the large volume limit. Boundary condition are given by $i/2$-shifted Baxter polynomials. On the left this is inhomogeneity polynomial $Q_s[\theta]$. On the right this is a  Baxter polynomial of holes (nontrivial term of the transfer-matrix $T^{1,1}$).}
\end{figure}

Let us now see how the transfer matrices are identified. We expect to find the gauge transform which will give the following boundary conditions in the large $L_{\rm phys}$ limit:
\be\label{bn1}
  T^{0,s}_g&=&Q_{\l,h}[\theta+\frac {is}2],\ \ s\geq 1, \no\\
  T^{2,s}_g&=&Q_{\l,h}[\theta-\frac {is}2],\ \ s\geq 1
\ee
and
\be\label{bn2}
  T^{1,1}_g&=&Q_{\l,h}[\theta].
\ee
In this gauge $T^{a,0}$ should be of order $e^{-mL_{\rm phys}\cosh[\theta]}$ and therefore the first row in the strip were Hirota equations are nontrivial is decoupled.
Therefore (\ref{bn1}) and (\ref{bn2}) define boundary conditions for the Hirota system defined on a strip $s\geq 1,a=0,1,2$, see Fig.~\ref{fig:HirotaTD}.  $T^{1,2}$ is the nested transfer matrix. The condition of absence of poles of $T^{1,2}$ leads to the nested Bethe equations of $GN$ sigma model.
%

The gauge transformation which leads to (\ref{bn1}) and (\ref{bn2}) should satisfy the following equations
\be
g_1\!\[\theta+\frac {ia}2\]\!g_2\!\[\theta+\frac {ia}2\]\!g_3\!\[\theta-\frac {ia}2\]\!g_4\!\[\theta-\frac {ia}2\]&=&\frac 1{Q_{\l}[\theta+\frac {ia}2]Q_{\l}[\theta+\frac {ia}2+i]},\no\\
g_1\!\[\theta+\frac {ia}2+i\]\!g_2\!\[\theta+\frac {ia}2-i\]\!g_3\!\[\theta-\frac {ia}2+i\]\!g_4\!\[\theta-\frac {ia}2-i\]&=&\frac 1{Q_{\l}[\theta-\frac {ia}2]Q_{\l}[\theta-\frac {ia}2-i]},\no\\
  g_1\!\[\theta+i\]\!g_2\!\[\theta\]\!g_3\!\[\theta\]\!g_4\!\[\theta-i\]&=&\frac 1{Q_{\l}[\theta-i]Q_{\l}[\theta+i]}.
\ee
There is indeed a solution of such system.
\be
  g_1[\theta]=g_4[\theta]=\frac 1{g_3[\theta]}=\prod_{k}\frac 1{\G[1+i(\theta-\l_k)]},\ \ g_2[\theta]=-\prod_{\l_k}\Gamma[i(\theta-\l_k)-1]
\ee
Therefore we see that the Hirota equations for the discretized spin chain have a meaningful thermodynamic limit if a proper gauge is chosen.

\chapter{\label{ch:susyspch}Supersymmetric spin chains}
The current achievements in the field of supersymmetric integrable systems can be summarized as follows\footnote{We will restrict to the most studied case of $gl(N|M)$ algebras.}. Most of the algebraic constructions based on the $gl(N)$ algebra have a direct generalization to the supersymmetric case. However, these generalizations do not cover the all variety of properties of supersymmetric systems. Also, the physics of the excitations over the antiferromagnetic vacuum is much more complicated and was studied to some extent only for few simple cases.

Generalizing $gl(N)$ case construction, all finite-dimensional irreducible representations (irreps) of $gl(N|M)$ algebra were constructed by Kac \cite{Kac:1977em,Kac:1977qb} as highest weight representations. Equivalently these representations can be labeled by Young tableaux \cite{Balantekin:1980qy,Balantekin:1980pp,Bars:1982se}. The determinant Weyl formula for characters and its parameter dependent generalization - Bazhanov-Reshetikhin formula \cite{Bazhanov:1989yk} has a straightforward generalization \cite{Tsuboi:1997iq,Tsuboi:1998ne,Kazakov:2007na}. This allows, in a complete analogy with $gl(N)$ case, to construct the nested Bethe Ansatz for a supersymmetric group from Hirota dynamics via a chain of Backlund transforms \cite{Kazakov:2007fy}. The Bethe Ansatz equations can be also derived using analytical Bethe Ansatz techniques \cite{Ragoucy:2007kg}.


What makes the $gl(N|M)$ case more difficult than the $gl(N)$ case is that Lie superalgebras allow also reducible but indecomposable representations. A good example is the $sl(2|1)$ algebra representations of which were studied in \cite{Scheunert:1976wj,Marcu:1979se}.  Although all irreps of $gl(N|M)$ were classified, the complete classification of indecomposable representations for arbitrary supersymmetric simple algebra is not known\footnote{This is mainly due to the representations of zigzag type. See \cite{Gotz:2005jz} and references therein.}.


Even if we consider only irreducible representations, their tensor product may contain an indecomposable representation. It is possible to restrict to the class of the representations tensor product of which is given by a direct sum of irreps. Unfortunately, such restriction does not allow construction of a spin chain which contains a singlet state (antiferromagnetic vacuum). Therefore, whenever an antiferromagnetic vacuum is present in a spin chain, studying of excitations around this vacuum encounters the problem of presence of irreducible representations; this leads to significant complications in the construction of a physical theory \cite{Essler:2005ag,Saleur:2006tf}.

Here we aim to discuss the techniques mostly for further applications to the integrable system that appears in the context of the AdS/CFT correspondence. It turns out that, at least for the purposes of this text, it is sufficient to consider the class of irreps that do not contain indecomposable parts in their tensor product. These representations can be obtained from tensoring of the fundamental representation.

In this chapter we will first recall basic properties of the supersymmetric algebras and theirs representations on the example of the $sl(2|1)$ algebra. Then we will discuss the structure of the representations that appear as a result of the fusion procedure for the $gl(N|M)$ case. We will see that, as in the nonsupersymmetric case, the string-like solutions are in the one-to-one correspondence with rectangular irreps. The corresponding integral equations will be naturally written on a fat hook shape.

The integral equations can be written also in the case of the Bethe Ansatz in the representation of type $[0\ldots010\ldots0]$. In this case the equations are naturally written on a so called T-hook shape. This raises a question about representation theory behind T-hook. This question has not been solved.

The fusion relations for a fat hook shape were discussed in \cite{Tsuboi:1997iq,Tsuboi:1998ne,Kazakov:2007fy}. In the context of the AdS/CFT correspondence the T-hook shape was first proposed in \cite{Gromov:2009tv} and then in \cite{Bomb,GromovKKV} built starting from the string hypothesis of \cite{Takahashi:1972,Arutyunov:2009zu}. The string hypotheses and TBA systems for various cases were also known in the literature before \cite{Saleur:1999cx}.

In Sec.~\ref{sec:stringhyp} we present a general construction: the string hypothesis and corresponding functional equations for the resolvents\footnote{Equivalently, we can write integral equations for the density functions.} for $gl(N|M)$ algebra and for Bethe equations with one arbitrary source term. The Bethe equations may be based on any Kac-Dynkin diagram. Although this generalization of \cite{Takahashi:1972,Saleur:1999cx,Gromov:2009tv} is quite obvious, it was not discussed in the literature before.

\section{\label{sec:sl21algebra}$Sl(2|1)$ algebra and its representations}

The defining representation of the algebra is given by the following $3\times 3$ matrices
\begin{small}
\be
  e=\left(
      \begin{array}{ccc}
        0 & 1 & 0 \\
        0 & 0 & 0 \\
        0 & 0 & 0 \\
      \end{array}
    \right)\!,\ f=\left(
      \begin{array}{ccc}
        0 & 0 & 0 \\
        1 & 0 & 0 \\
        0 & 0 & 0 \\
      \end{array}
    \right)\!,\ h=\left(
      \begin{array}{ccc}
        1/2 & 0 & 0 \\
        0 & -1/2 & 0 \\
        0 & 0 & 0 \\
      \end{array}\right)\!,\
      B=\left(\begin{array}{ccc}
        1/2 & 0 & 0 \\
        0 & 1/2 & 0 \\
        0 & 0 & 1 \\
      \end{array}\right)\!,&& \no\\
      Q_+=\left(\begin{array}{ccc}
        0 & 0 & 1 \\
        0 & 0 & 0 \\
        0 & 0 & 0 \\
      \end{array}
    \right)\!,\
    Q_-=\left(\begin{array}{ccc}
        0 & 0 & 0 \\
        0 & 0 & 1 \\
        0 & 0 & 0 \\
      \end{array}
    \right)\!,\ S_+=\left(\begin{array}{ccc}
        0 & 0 & 0 \\
        0 & 0 & 0 \\
        0 & -1 & 0 \\
      \end{array}
    \right)\!,\ S_-=\left(\begin{array}{ccc}
        0 & 0 & 0 \\
        0 & 0 & 0 \\
        1 & 0 & 0 \\
      \end{array}
    \right)\!.\hspace{1EM}&&
\ee
\end{small}
\noindent The generators $e,f,h$ of $sl(2)$ bosonic subalgebra obey canonical commutation relations
\be
  [h,e]=e,\ \ [h,f]=-f,\ \ [e,f]=2h.
\ee
The fermionic generators $S_\pm$ and $Q_\pm$ transform under adjoint action of $e,f,h$ as a spin $1/2$ representation. Nonzero commutators are given by
\be
  [h,Q_\pm]&=&\pm\frac 12 Q_\pm,\ \ [h,S_\pm]=\pm\frac 12 S_\pm,\no\\ \
  [e,Q_-]&=&Q_+,\ \ [f,Q_+]=Q_-,\no\\ \
  [e,S_-]&=&S_+,\ \ [f,S_+]=S_-.
\ee

The Cartan subalgebra of $sl(2|1)$ is spanned by $h$ and $B$. The generator $B$ acts as zero on the  bosonic generators. On the fermionic generators it acts as follows
\be
  [B,S_\pm]=\frac 12S_\pm, \ \ [B,Q_\pm]=-\frac 12 Q_\pm.
\ee
Finally, the fermionic commutators\footnote{For the case of superalgebras we understand any commutator as a graded commutator, that is anticommutator when both generators are fermionic and usual commutator otherwise.} are given by
\be\label{sl21commutrelations}
  [S_+,Q_+]=-e,\ \ [S_-,Q_-]=f,\ \
  \[Q_\pm,S_\mp\]= h\pm B.
\ee

\begin{figure}[t]
\centering
\includegraphics[height=6cm]{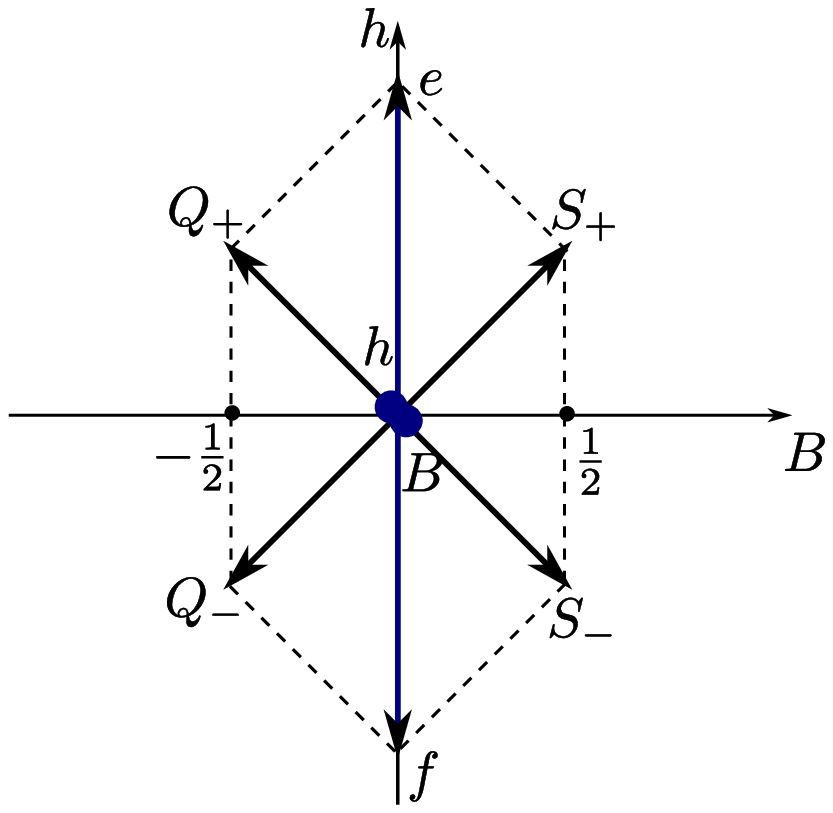}\qquad
\includegraphics[height=6cm]{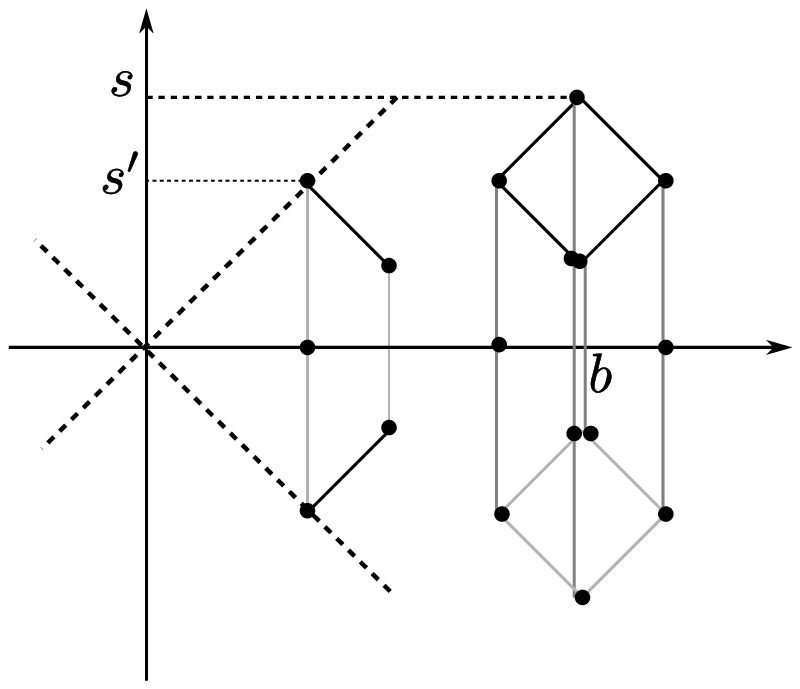}

\parbox[t]{0.45\textwidth}{\caption{\label{fig:weightSL21}The weight diagram for the generators of $sl(2|1)$ algebra.}}\qquad
\parbox[t]{0.45\textwidth}{\caption{\label{fig:irrepSL21}Irreducible representations of $sl(2|1)$. The right one - typical irrep $\langle s,b\rangle$ , the left one - atypical irrep $|s',s'\rangle$. }}
\end{figure}

The weight diagram for the generators is shown in Fig.~\ref{fig:weightSL21}.

A standard method to build an irreducible representation of the $sl(2|1)$ algebra is to introduce the highest weight vector $|s,b\rangle$ defined by
\be
  S_+|s,b\rangle =0,\ \ Q_+|s,b\rangle=0,\ \ e|s,b\rangle=0,\ \
  B|s,b\rangle = b|s,b\rangle,\ \ h|s,b\rangle=s|s,b\rangle.
\ee
The other states of the representation are obtained by action of lowering operators $Q_-,S_-$, and $f$. Due to the fermionic nature of $Q_-$ and $S_-$, each of these operators is applied consecutively at most once. For the case when they applied exactly once, we get a typical representation, which we will denote as $\langle s,b\rangle$. Under the action of the bosonic $su(2)\oplus u(1)$ subalgebra this representation decomposes into four $su(2)$ irreps: ($s$,$b$), ($s-1/2,b+1/2$), ($s-1/2,b-1/2$), ($s-1,b$), where the first argument in the brackets is the spin of the $su(2)$ irrep and the second argument is the eigenvalue of the operator $B$. The dimension of $\langle s,b\rangle$ is the following:
\be
  {\rm dim}\langle s,b\rangle=8s.
\ee
The parameter $b$ can acquire arbitrary values except $\pm s$.

An atypical representation is the representation for which either $Q_-|s,b\rangle =0$ or $S_-|s,b\rangle=0$. Due to the commutation relations (\ref{sl21commutrelations}) the first condition can be realized only for $b=s$. The second condition can be realized only for $b=-s$. Both conditions are satisfied simultaneously only when $b=s=0$, \textit{i.e.} for the trivial representation. We will denote the atypical representation with $b=s$  as $|s\rangle$ and the atypical representation with $b=-s$ as $\langle s|$. The reason for such notation is clear from the structure of the representation shown in Fig.~\ref{fig:irrepSL21}. The dimension of atypical representation is given by
\be
  {\rm dim}|s\rangle={\rm dim}\langle s|=4s+1.
\ee
Since the dimension of atypical representation is smaller than the dimension of typical representation with the same spin, atypical representation is often called short representation.

For $s=1/2$ we have ${\rm dim}|1/2\rangle=3$. This is nothing but the fundamental representation.

The structure of reducible but indecomposable representations is quite involved. Here we would like to point out only the simplest type of indecomposable representation which can be obtained as continuation of the typical representation $\langle b,s\rangle$ to the point $b=s$ (or $b=-s$). The irreducible submodule in such representation is $|s-1/2\rangle$ ($\langle s-1/2|$). 
For more information about indecomposable representations we refer to \cite{Gotz:2005jz}.

\begin{figure}[t]
\centering
\includegraphics[height=4.5cm]{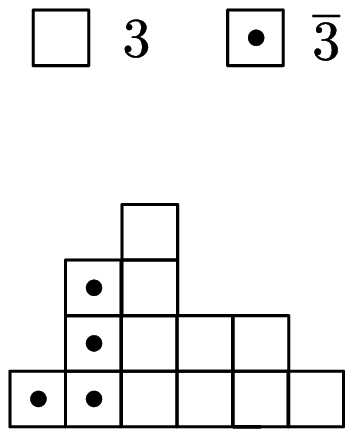}\qquad\qquad\qquad\qquad
\includegraphics[height=4.5cm]{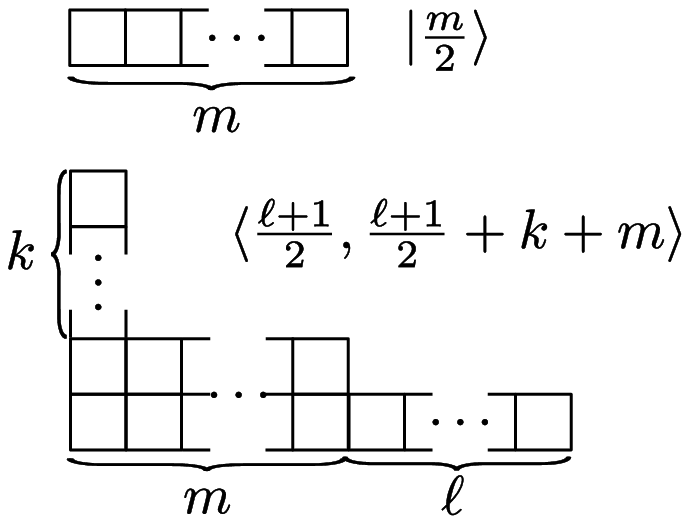}

\parbox[t]{0.46\textwidth}{\caption{\label{fig:YoungSL21left}At top: fundamental and antifundamental representations. At bottom: an example \cite{Balantekin:1980pp} of irrep that is built from tensor products of fundamental and antifundamental representations.}}\ \ \ \
\parbox[t]{0.46\textwidth}{\caption{\label{fig:YoungSL21}Irreducible representations of $sl(2|1)$ built from tensoring of fundamental representation. At top: atypical irrep $|s'\rangle$, at bottom - typical irrep $\langle s,b\rangle$.}}
\end{figure}

It is useful to identify irreps with the Young tableaux. There are two 3-dimensional representations: $|1/2\rangle$ and $\langle 1/2|$. We will refer to the first one as fundamental ($3$) and to the second one as antifundamental ($\overline 3$). Following \cite{Balantekin:1980pp}, we label $3$ with a box and $\overline 3$ with a dotted box (see Fig.~\ref{fig:YoungSL21left}). In general, to obtain all irreps one should consider tensor products of $3$ and $\overline 3$. A general irrep corresponds to a union of two Young tableaux \cite{Balantekin:1980pp} shown in Fig.~\ref{fig:YoungSL21left}. It is tempting to answer the question whether it is meaningful to fit this union into the T-hook domain that is introduced in section \ref{sec:thookth}.

An important difference with the $gl(n)$ case is that the antifundamental representation never appears in the tensor product of the fundamental representation. This is also true for a trivial representation, atypical representations with negative $b$ and any typical representations with $b<s$. A natural way to construct antiferromagnetic vacuum is to consider a spin chain with half of the nodes in the fundamental and half of the nodes in the antifundamental representations \cite{Essler:2005ag}. Since tensor products of $3$ and $\overline 3$ contain also indecomposable representations, we will encounter such representations when dealing with the excitations over the antiferromagnetic vacuum.

In the following we will consider a simpler particular case - the representations with $b\geq s$.
These representations can be obtained from the tensor product of fundamental representation only. Such tensor product always decomposes into irreps.

To fix grading of the representation module, we take the vector space with two bosonic and one fermionic components for the fundamental representation. A straightforward combinatorial analysis gives the identification of Young tableaux and irreps shown in Fig.~\ref{fig:YoungSL21}. We used inverse notation for Young tableaux with the largest line below. Such notation was used in \cite{Kazakov:2007fy} to describe Hirota dynamics in integrable systems. Horizontal lines in the Young tableaux mean supersymmetrization: symmetrization of bosonic elements with other elements and antisymmetrization of fermionic elements. The vertical lines in the Young tableaux imply superantisymmetrization: antisymmetrization of bosonic elements with other elements and symmetrization of fermionic elements. Note that Young tableaux give typical irreps with only discrete half-integer values of $b$.

\section{Fusion procedure in the $gl(N|M)$ case.}
The procedure of construction of irreps of $gl(N|M)$ algebra is a straightforward generalization of such procedure for $gl(2|1)$ case. Again, we have a  fundamental representation defined by Dynkin labels $[100\ldots]$ and its conjugate. Arbitrary irrep can be obtained only from tensor product of both of these representations. Appearance of the reducible indecomposable representations in the tensor product is controlled by the value of the continuous parameter $b$. The parameter $b$ is a Cartan weight with respect to the generator $B$, where generator $B$ is given by the following matrix in the defining representation:
\be
  B=\left(
    \begin{array}{c|c}
      N^{-1} \mI_N& 0 \\
      \hline \\
      0 & M^{-1} \mI_M
    \end{array}
  \right)\!.
\ee

In the following we consider only the representations that can be obtained from the tensor product only of the fundamental representation $[100\ldots]$ itself. Such irreps are described by Young tableaux which should be inside a fat hook structure shown in Fig.~\ref{fig:susyhirota}. If the border of a Young table does not touch the internal border\footnote{By the internal border of fat hook we mean the border which contains the corner point $\{M,N\}$.} of the fat hook then such Young table defines atypical representation. Otherwise, such Young table defines typical representation.

We can derive the Hirota dynamics on a fat hook using a straightforward generalization of discussion in Sec.~\ref{sec:Fusion procedure} \cite{Kazakov:2007fy}. Namely, we consider an integrable system based on the $R$-matrix
\be
  R=\theta-i\scP,
\ee
where $\scP$ is the graded permutation.

Starting with this $R$-matrix one can construct the transfer matrix in a fundamental representation and use a fusion procedure to obtain the transfer matrices in other representations. The transfer matrices $T^{a,s}[\theta]$ in the rectangular irrep satisfy the Hirota equation
\be\label{fusionblabla}
  T^{a,s}\[\theta+\frac i2\]T^{a,s}\[\theta-\frac i2\]=T^{a,s+1}[\theta]T^{a,s-1}[\theta]+T^{a+1,s}[\theta]T^{a-1,s}[\theta].
\ee
The transfer matrices are non-zero if $a=0$ or $(a,s)$ is inside or on the border of the shape (fat hook) in Fig.~\ref{fig:susyhirota}.

  The representations are nontrivial on the internal border of the fat hook. They are different from the representation in the corner $\{M,N\}$ only by the value of the parameter $b$.

\section{\label{sec:Thook}Bethe Ansatz}

\begin{figure}[t]
\centering
\includegraphics[height=5cm]{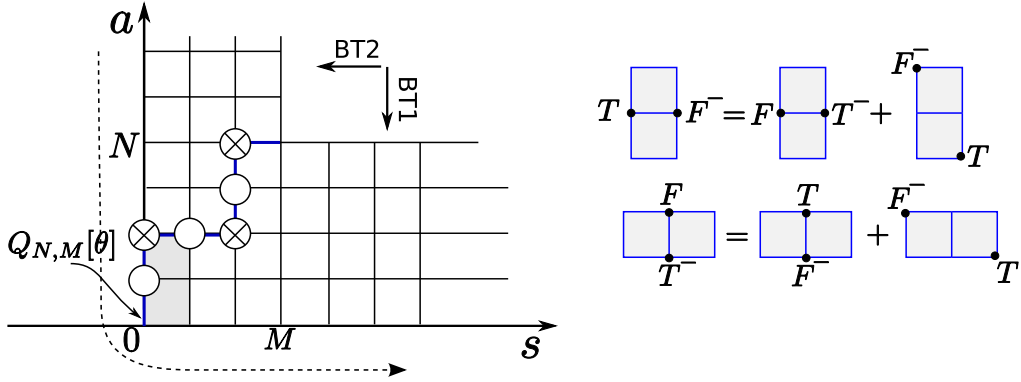}
\caption{\label{fig:susyhirota}On the left: fat hook for $gl(N|M)$ algebra. On the right: functional equations on $T$ and $F$ for BT2 transform.}
\end{figure}

The Bethe Ansatz equations can be derived from the fusion relations (\ref{fusionblabla}) via a chain composed from two different types of Backlund transforms. The two different types of Backlund transforms appear in the following way. The bilinear equations (\ref{Backlund}) on $T$ and $F$ considered for a fat hook configuration allow two solutions. For the first one $F$ is nonzero on a fat hook $(N-1,M)$. This is the same Backlund transform which was discussed in section \ref{sec:nestedBABacklund}. It is called the Backlund transform of the first type (BT1).

For the second solution  $F$ is nonzero on a fat hook $(N,M+1)$. Exchanging the roles of $T$ and $F$ for such Backlund transform we obtain the Backlund transform of the second type (BT2) which reduces $M$ by one. The functional equations for BT2 are shown in Fig.~\ref{fig:susyhirota}.


The set of the Bethe Ansatz equations is obtained after successive application of Backlund transforms that reduce the fat hook to a single strip. This can be done by application of different chains of Backlund transforms \cite{Kazakov:2007fy}. A possible chain is shown in Fig.~\ref{fig:susyhirota}. The shape of the chain corresponds to a particular Kac-Dynkin diagram of the $gl(N|M)$ algebra. Each turning of the path corresponds to the fermionic node. Each straight pass corresponds to the bosonic node.

The Bethe Ansatz equations are determined by the Cartan matrix of a given Kac-Dynkin diagram and have the formal expression
\be\label{susyBAE}
  \frac{\l_a+\frac i2\delta_{a,1}}{\l_a-\frac i2\delta_{a,1}}=\left.\prod_{\l_b'}\right.^\prime\frac{\l_a-\l_b'+\frac i2 c_{ab}}{\l_a-\l_b'-\frac i2 c_{ab}}.
\ee
The Cartan matrix $c_{ab}$ for a given Kac-Dynkin diagram up to an overall sign is given by the following identification:
\begin{figure}[!h]
\centering
\includegraphics[height=1.5cm]{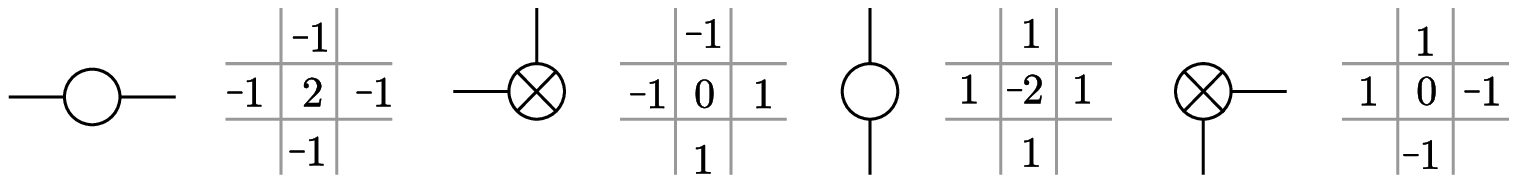}
\end{figure}

The overall sign is chosen such that the sum of the elements in the first row of $c_{ab}$ is positive.

For example, the Cartan matrix for the Kac-Dynkin diagram in Fig.~\ref{fig:susyhirota} is the following:
$$
\left(
  \begin{array}{cccccc}
    2 & -1 & 0 & 0 & 0 & 0 \\
    -1 & 0 & 1 & 0 & 0 & 0 \\
    0 & 1 & -2 & 1 & 0 & 0 \\
    0 & 0 & 1 & 0 & -1 & 0 \\
    0 & 0 & 0 & -1 & 2 & -1 \\
    0 & 0 & 0 & 0 & -1 & 0 \\
  \end{array}
\right).
$$

The Bethe equations (\ref{susyBAE}) can be also written for an arbitrary representation \cite{Ragoucy:2007kg}. It is not clear how to obtain the Bethe equations for an arbitrary representation from a fat hook construction.

\section{\label{sec:DualityTransformation}Duality transformations.}
The system of the Bethe equations is different depending on what underlying Kac-Dynkin diagram was chosen. But the physical properties of the system should not depend on the choice of the Kac-Dynkin diagram. And indeed, there exists a so called duality transformation \cite{Woynarovich,tJmodel},\cite{Tsuboi:1998ne},\cite{GohmannSeel} which allows to pass from one system of equations to another.

\begin{figure}[t]
\centering
\includegraphics[height=3cm]{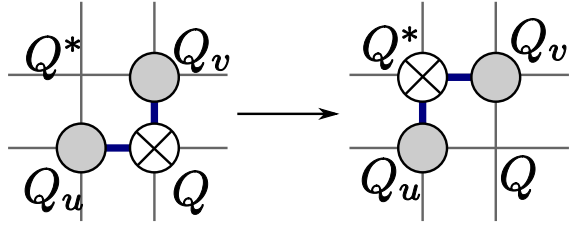}
\caption{\label{fig:duality}Duality transformation. The nodes denoted by gray can be of any statistics. Obviously, the duality transformation changes the statistics of these nodes.}
\end{figure}

Let us consider a fermionic node. We will denote the Baxter polynomial for this node as $Q=\prod (\theta-\theta_k)$. The fermionic node is coupled to two other nodes. We will denote the Baxter polynomials that correspond to these nodes as
\be
  Q_u=\prod (\theta-u_k),\ \ Q_v=\prod (\theta-v_k).
\ee
The Bethe equation for the fermionic node
\be\label{BAEf}
  1=\frac{Q_u[\theta_k+\frac i2]}{Q_u[\theta_k-\frac i2]}\frac{Q_v[\theta_k-\frac i2]}{Q_v[\theta_k+\frac i2]}
\ee
can be derived from the Baxter equation
\be\label{Bf}
  Q_u^+Q_v^--Q_u^-Q_v^+=QQ^*.
\ee
From the point of view of a fat hook picture the Baxter polynomials $Q_u,Q_v,Q,Q^*$ are just some $Q_{N'|M'}$ polynomials\footnote{$Q_{N'|M'}$ is the polynomial that defines a boundary condition for the $(N',M')$ fat hook. This is a generalization of the $Q_{N'}$ polynomial in Sec.~\ref{sec:nestedBABacklund}. The polynomial $Q_{N|M}$ gives a source term for the Bethe equations. For the homogeneous spin chain of length $L$ we have $Q_{N|M}=(\theta-\frac i2)^L$.} with appropriate shifts of the arguments on $i/2$. These polynomials are associated to the nodes which form a square shown in Fig.~\ref{fig:duality}. The Baxter equation (\ref{Bf}) is a $QQ$ relation on these polynomials.

Duality transformation is a passage from the Bethe equations which contain $\theta_k$ as Bethe roots to the Bethe equations which contain $\theta_k^*$ (zeroes of $Q^*$) as Bethe roots. We see that $Q$ and $Q^*$ enter (\ref{Bf}) in a symmetric way. Therefore $\theta_k^*$ are also fermionic roots. They obey the same equation as (\ref{BAEf}), with $\theta_k$ replaced by $\theta_k^*$.

The roots $\theta_j$ enter to the Bethe equations for $u_k$ as the following ratio
\be
  \frac{Q\[u_k-\frac i2\]}{Q\[u_k+\frac i2\]}.
\ee
An analogous ratio appears in the Bethe equations for $v_k$.

Using (\ref{Bf}) it is easy to see that
\be
  \frac{Q\[u_k-\frac i2\]}{Q\[u_k+\frac i2\]}=\frac{Q_u\[u_k-i\]}{Q_u\[u_k+i\]}\frac{Q^*\[u_k+\frac i2\]}{Q^*\[u_k-\frac i2\]}.
\ee
Therefore the Bethe equations for $u_k$ ($v_k$) change under duality transformations in accordance with the change of the Kac-Dynkin diagram shown in Fig.~\ref{fig:duality}.

Let us also consider Bethe equations which correspond to the representation\\
$[0\ldots 01(k{\rm -th place})0\ldots 0]$. Such Bethe equations are written as
\be\label{01}
 \frac{\l_a+\frac i2\delta_{a,k}}{\l_a-\frac i2\delta_{a,k}}=\prod_{\l_b'}\frac{\l_a-\l_b'+\frac i2 c_{ab}}{\l_a-\l_b'-\frac i2 c_{ab}}.
\ee
A special attention should be devoted to the momentum carrying node in these equations if $k\neq 1$. If the momentum carrying node is fermionic, the Bethe equations for this node are written as
\be
  \(\frac{\theta_k+\frac i2}{\theta_k-\frac i2}\)^L=\frac{Q_u\[\theta_k+\frac i2\]}{Q_u\[\theta_k-\frac i2\]}\frac{Q_v\[\theta_k-\frac i2\]}{Q_v\[\theta_k+\frac i2\]}.
\ee

If we perform a duality transformation for the momentum carrying node then the building block of the Bethe equations for $v_k$ roots will be the following:
\be
  \frac{Q\[v_k-\frac i2\]}{Q\[v_k+\frac i2\]}=\(\frac{\theta_k-i}{\theta_k+i}\)^L\frac{Q_v\[v_k-i\]}{Q_v\[v_k+i\]}\frac{Q^*\[v_k+\frac i2\]}{Q^*\[v_k-\frac i2\]}.
\ee
Therefore the $v_k$ node becomes also momentum carrying. Although the duality transformation of the momentum-carrying node is possible, the resulting equations bring the Bethe Ansatz equations in a nonstandard form. Due to this we should forbid to perform the duality transformation of the momentum-carrying node if it is situated inside the Kac-Dynkin diagram. This restriction has interesting consequences. For example, if each wing (on the left and on the right from the momentum-carrying node) of the Kac-Dynkin diagram contains at least one fermionic node, one cannot rewrite the Bethe equations in a form which corresponds to the distinguished diagram (the one with a single fermionic node).

\begin{figure}[t]
\centering
\includegraphics[height=2.5cm]{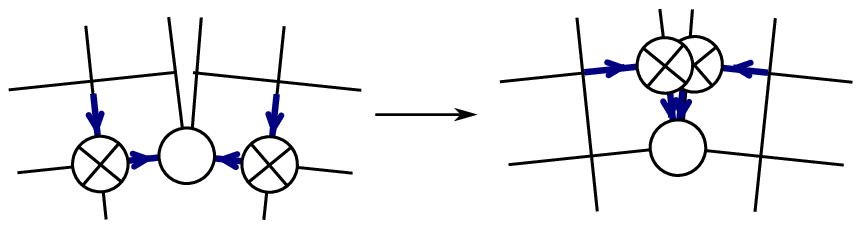}
\caption{\label{fig:cnodenoncompact}Duality transformation which reveals noncompactness of in supersymmetric spin chains.}
\end{figure}

The momentum-carrying node has another important feature. If we start from the Bethe equations
\be\label{cncomp}
  \(\frac{\theta_k+\frac i2}{\theta_k-\frac i2}\)^L=-\frac{Q\[\theta_k+i\]}{Q\[\theta_k-i\]}\frac{Q_u\[\theta_k-\frac i2\]}{Q_u\[\theta_k+\frac i2\]}\frac{Q_v\[\theta_k-\frac i2\]}{Q_v\[\theta_k+\frac i2\]},
\ee
which correspond to the left diagram in Fig.~\ref{fig:cnodenoncompact}, and perform duality transform both on $u$ and $v$ nodes, then the Bethe equations for the momentum carrying node will become
\be\label{cnnoncomp}
  \(\frac{\theta_k+\frac i2}{\theta_k-\frac i2}\)^L=-\frac{Q\[\theta_k-i\]}{Q\[\theta_k+i\]}\frac{Q_u^*\[\theta_k+\frac i2\]}{Q_u^*\[\theta_k-\frac i2\]}\frac{Q_v^*\[\theta_k+\frac i2\]}{Q_v^*\[\theta_k-\frac i2\]}.
\ee
If we put $Q_u=1$ and $Q_v=1$ in equation (\ref{cncomp}) then we will get the equation for a compact (SU(2)) XXX spin chain. On the opposite, if we put  $Q_u^*=1$ and $Q_v^*=1$ in (\ref{cnnoncomp}) then we will get the equation for a noncompact (Sl(2)) XXX spin chain.

The second description reveals noncompactness of the spin chain described by (\ref{01}). Indeed, let us find the constraints on the possible number of the Bethe roots. For simplicity we suppose that there are only three types of the Bethe roots. Bosonic Bethe roots $\theta_k$ which satisfy (\ref{cncomp}) (or, equivalently (\ref{cnnoncomp})) and fermionic Bethe roots $u_k,v_k$ which satisfy the Baxter equations
\be
  Q^+-Q^-=Q_uQ_u^*,\ \ \ Q^+-Q^-=Q_vQ_v^*.
\ee
From the Baxter equation we read that
\be\label{numberuv}
  n_\theta=n_u+n_u^*=n_v+n_v^*.
\ee

In Sec.~\ref{sec:countingBethe} we obtained the constraints on the number of bosonic Bethe roots. In our case we have
\be\label{numbertheta}
  n_\theta\leq \frac{L+n_u+n_v}2.
\ee
Combining (\ref{numberuv}) and (\ref{numbertheta}) we see that the number of the Bethe roots is unconstrained.

\section{\label{sec:stringhyp}String hypothesis and integral equations}
To formulate a set of integral equations in the thermodynamic limit we will establish first the string hypothesis for the Bethe Ansatz based on the $gl(N|M)$ algebra.

The string hypothesis for bosonic roots was discussed in Sec.~\ref{sec:countingBethe} and in Sec.~\ref{sec:PCFGNbound}. Bosonic Bethe roots can form string configurations alone. Also they can participate in the formation of stacks as intermediate nodes. The mechanism how the bosonic roots enter the stacks was discussed in Sec.~\ref{sec:PCFGNbound} in the context of GN and PCF models.

To formulate a string hypothesis for fermionic roots we will consider the Baxter equation for fermionic roots
\be\label{Bf2}
  Q_u[\theta+i/2]Q_v[\theta-i/2]-Q_u[\theta-i/2]Q_v[\theta+i/2]=Q[\theta]Q^*[\theta],
\ee
where the notation is the same as for (\ref{Bf}).

Let us assume for simplicity the following ordering of Bethe roots: $n_u\geq n_\theta\geq n_v$. Then in the thermodynamic limit the roots of $u$ type will determine the dominant term in the l.h.s. of (\ref{Bf2}). For ${\rm Im}[\theta]>0$ the first term will be dominant.

Fermionic Bethe roots do not form string configurations themselves. Instead, they participate in the formation of stacks as the first or the final node. Indeed, if $\theta$-roots organize in a string of length $k$ then (\ref{Bf2}) implies the string configuration with the same center either for $u$-roots of length $k+1$, or for $v$-roots of length $k-1$.

\begin{figure}[!h]
\centering
\includegraphics[height=4cm]{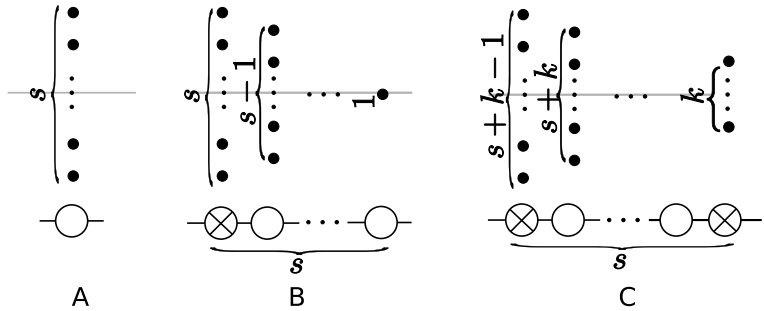}
\caption{\label{fig:stringstacks}Possible string/stack configurations. A) an $s$- string of arbitrary length formed of bosonic roots of the same type. B) an $s$-stack with the longest string of length $s$ formed of fermionic roots. All the other roots are bosonic. The total number of root types is $s$. This stack should terminate with a string of length 1. C) an $(s+k-1)$-stack with the first and the last nodes being fermionic. All other nodes are bosonic. The length of the initiating string is arbitrary.\newline\ \ \
All the centers of strings are situated at the same position. They are depicted in different positions for convenience.\newline\hspace{1EM}
The ordering of number of Bethe roots in the nested Bethe Ansatz is the following: the node with the largest amount of roots is on the left.}
\end{figure}

All possible string/stack configurations that can be built from bosonic and fermionic nodes\footnote{For the rational Bethe Ansatz in any fundamental representation.} are shown in Fig.~\ref{fig:stringstacks}

\subsubsection{Integral equations on a fat hook}
The procedure of construction of integral equations is analogous to the one described for the bosonic case. In appendix \ref{app:susy} we prove the following statement:

The set of the functional equations derived from the Bethe Ansatz for a given Kac-Dynkin diagram of $gl(N|M)$ algebra and under the assumption that the string hypothesis is valid is equivalent to the following set of equations:
\be\label{fathookie}
  \delta_{a,1}\delta_{s,1}R_p=\sum_{a'}C_{aa'}R_{a',s}+\sum_{s'}C_{ss'}\overline{R}_{a,s'},
\ee
for $\{a,s\}$ being a coordinate of a point situated strictly inside the fat hook, and, for $\{a=M,s=N\}$,
\be\label{TMN}
  R_{M,N}+{\overline R}_{M,N}=T_{M,N},
\ee
where $T_{M,N}$ is different for different Kac-Dynkin diagrams; the expression for of $T_{M,N}$ is given in (\ref{TNM1}) and (\ref{TNM2}).

$C_{aa'}$ is a "D-deformed" Cartan matrix defined by
\be\label{cddcdd}
  C_{aa'}=(D+D^{-1})\delta_{aa'}-\delta_{a,a'+1}-\delta_{a,a'-1}.
\ee
The resolvents $R_{a,s}$ and ${\overline R}_{a,s}$ are zero on the boundary and outside the fat hook except for $R_{N,M}, \overline{R}_{N,M}$.

There is a string configuration associated with each $\{a,s\}$ node of a fat hook. Depending on the node, either $R_{a,s}$ or ${\overline R}_{a,s}$ is the resolvent of the density function of strings. Then correspondingly
either ${\overline R}_{a,s}$ or $R_{a,s}$ is the resolvent of the correspondent density of holes.
\begin{figure}[t]
\centering
\includegraphics[height=6cm]{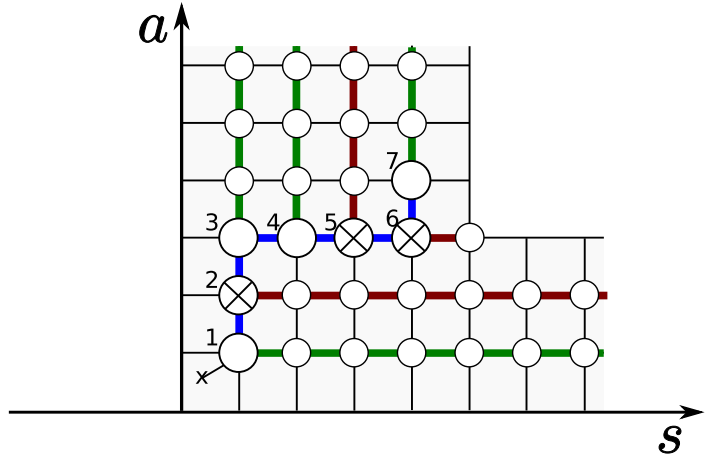}
\caption{\label{fig:stringfathook}Organization of strings into fat hook. The blue line shows the path of embedding of the Kac-Dynkin diagram. Green lines show the string configurations based on the bosonic nodes. Red lines show stack configurations based on the fermionic nodes.}
\end{figure}

The exact identification of $R_{a,s}$ with holes or particles is given by the following procedure. First we draw the Kac-Dynkin diagram on the nodes of the fat hook. The momentum-carrying node corresponds to the node \{1,1\}. We are allowed then to move only to the east or to the north. The initial direction is to the east(north) if the last movement in the chain of the Backlund transforms was to the west(south). For example for the fat hook construction in Fig.~\ref{fig:susyhirota} the initial direction is to the north.

One should turn at each node next to the fermionic one. 

Each node of the fat hook occupied by the Kac-Dynkin diagram is associated with the corresponding type of the Bethe roots. The string/stack configurations with the longest string formed from this type of Bethe roots are situated on the straight line which starts from the considered node. There is the only way to assign straight lines to the nodes of the Kac-Dynkin diagram such that all nodes of the fat hook will be occupied.

The example of the whole construction is shown in Fig.~\ref{fig:stringfathook}.

The correspondence between holes/particles and $R/{\overline R}$ is chosen in such a way that interaction (in the sense of equation (\ref{fathookie})) inside the same line should be through the holes and the interaction between different lines should be through the particles.



Although formally the equations (\ref{fathookie}) are the same for different Kac-Dynkin diagrams, these equations describe different regimes of a spin chain. As well, in general there is no relations between the resolvents used in (\ref{fathookie}) for different Kac-Dynkin diagrams. We will illustrate now this issue on the example of $sl(2|1)$ spin chain

\section{Thermodynamic limit of the Bethe equations in the $sl(2|1)$ case}
In the $gl(n)$ case we can consider the thermodynamic limit for configurations close either to the ferromagnetic or to the antiferromagnetic vacuum. In the supersymmetric case there are more possibilities.

Let us consider the Bethe equations for $sl(2|1)$ spin chain in a fundamental representation.
Each site of the spin chain is occupied by one particle which can be a boson with spin up or down or a fermion.

From the representation theory we conclude that the total number of fermions can be changed only by $\pm 1$ by the action of the symmetry algebra. Indeed, the number of fermions is changed under action of $Q_\pm$ and $S_\pm$ generators. The square of these generators is zero.

Therefore the state with large amount of fermions never lies in the same symmetry multiplet with the state with small amount of fermions. Thus we have four possible regimes in the thermodynamic limit. If we choose as the pseudovacuum a state with each site occupied with bosons with spin down then these regimes are described as follows:
\bn
  \item Large spin, small amount of fermions.
  \item Small spin, small amount of fermions.
  \item Large spin, large amount of fermions.
  \item Small spin, large amount of fermions.
\en
These regimes can be shown on the following weight diagram:
\be
    \parbox{5cm}{\includegraphics[width=0.25\textwidth]{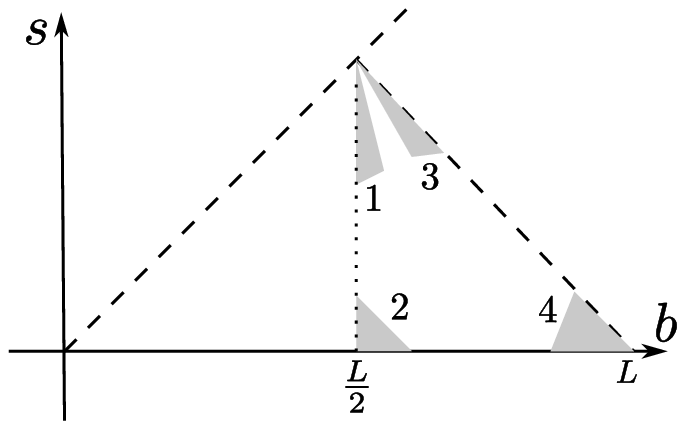}}.
\ee

It turns out that different types of the thermodynamic limit, except the second one, are naturally related to different dual configurations of the Bethe equations. We will consider each configuration separately.

\paragraph{}{\bf\xOX case.\hspace{1EM}}
In this case the Bethe equations are written as
\be\label{someBethe2}
  \(\frac{\theta+\frac i2}{\theta-\frac i2}\)^L&=&-\prod_{\theta'}\frac{\theta-\theta'+i}{\theta-\theta'-i}\prod_{u}\frac{\theta-u-\frac i2}{\theta-u+\frac i2},\no\\
  1&=&\prod_\theta\frac{u-\theta+\frac i2}{u-\theta-\frac i2}.
\ee
The inequalities for the number of Bethe roots are written as follows
\be\label{ineq1}
  n_u\leq n_\theta,\  \ n_\theta\leq\frac{L+n_u}{2}\leq L.
\ee
Naturally, the thermodynamic limit is possible in the case when we replace all inequalities $\leq$ by stronger demands $\ll$. This corresponds to the first type of the thermodynamic limit.

The bosonic Bethe roots $\theta$ can form string configurations while the fermionic roots are real. Introducing the resolvents $R_s$ for the $s$-strings and $F$ for fermions we obtain the integral equations
\be
  \delta_{s,1}R_p&=&C^\infty_{ss'}R_{s'}^*+(D+D^{-1})R_s-\delta_{s,1}F,\no\\
  0&=&D^{s'}R_{s'}-F-F^*
\ee
One can try to perform a particle-hole transformation of $R_1$ as we did in the $su(2)$ case. In this way we will realize the second type of the limit. The obtained integral equations seem do not correspond to the asymptotic Bethe Ansatz of any known integrable field theory.

\paragraph{}{\bf\xXX case.\hspace{1EM}}
This case is obtained from the previous one by performing the dualization of the root $u$. The Bethe equations are written as
\be\label{someBethe1}
  \(\frac{\theta+\frac i2}{\theta-\frac i2}\)^L&=&\prod_{{u^*}}\frac{\theta-{u^*}+\frac i2}{\theta-{u^*}-\frac i2},\no\\
  1&=&\prod_\theta\frac{{u^*}-\theta+\frac i2}{{u^*}-\theta-\frac i2}.
\ee
The inequalities for number of Bethe roots are written as follows
\be
  n_{u^*}\leq n_\theta\leq\frac{L}{2},\  \ n_\theta\leq {L-n_{u^*}}.
\ee
The thermodynamic limit corresponds to the case when $n_u\lesssim n_\theta$, which is a third type of thermodynamic limit.

Since the resolvents $R_s$ remain the same as in the previous case, all the integral equations which do not include $F$ should remain the same. This is indeed the case.  Note that the underlying $s$-string configurations in the \xOX case correspond to stack configurations of two fermionic nodes in the \xXX case.

The complete set of integral equations is written as
\be
\delta_{s,1} R_p&=&C^\infty_{ss'}R_{s'}^*+(D+D^{-1})R_s-\delta_{s,1}F_*^*,\no\\
0&=&D^{s'}R_{s'}-F_*-F_*^*,
\ee
where by $F_*$ we denoted the resolvent for dual variables and by $F_*^*$ the resolvent of holes for the dual variables. Note that by construction $F_*^*\neq F$. 

\paragraph{}{\bf\xXO case.\hspace{1EM}}
This case is obtained after duality transformation on the $\theta$-variable of the \xXX case. The Bethe equations are written as
\be\label{someBethe1}
  \(\frac{\theta^*+\frac i2}{\theta^*-\frac i2}\)^L&=&\prod_{{u^*}}\frac{\theta^*-{u^*}+\frac i2}{\theta^*-{u^*}-\frac i2},\no\\
  -1&=&\prod_{{u^*}'}\frac{{u^*}-{u^*}'+i}{{u^*}-{u^*}'-i}\prod_\theta^*\frac{{u^*}-\theta^*-\frac i2}{{u^*}-\theta^*+\frac i2}.
\ee
This case is the best suited for description of the low-spin regime with most of the excitations being fermions (case 4). The $s$-string of $\theta$-s in the \xOX case transforms into $s-1$ string of $u^*$-s in the \xXO case. The bosonic excitations of the \xOX case are now described by the stack composed of $\theta^*$ 2-string and one $u^*$ Bethe root. We will denote the resolvent of $(s-1)$-string configuration, $s\geq 2$, by $R_s$ (it should be the same resolvents as in previous case), the resolvent for free $\theta^*$ roots by $R_1$, and the resolvent for the stack by $G$.
Then the integral equations will be written as
\be
  \delta_{s,1} R_p&=&C^\infty_{ss'}R_{s'}^*+(D+D^{-1})R_s-\delta_{s,1}G^*,\no\\
0&=&\sum_{s'=2}^\infty D^{s'}R_{s'}-D R_1^*-G-G^*.
\ee

\begin{figure}[t]
\centering
\includegraphics[height=3cm]{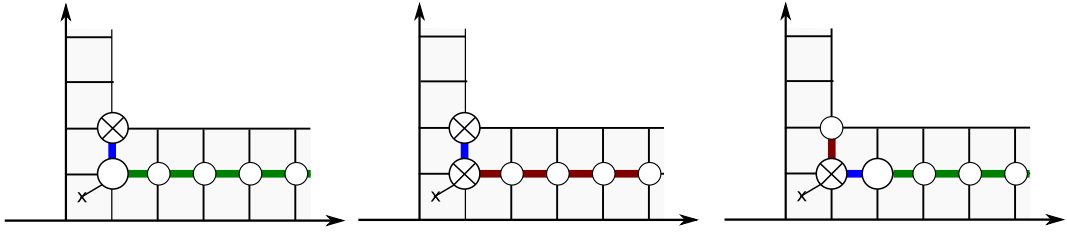}
\caption{\label{fig:tdsl21}Organization of integral equations for different Kac-Dynkin diagrams of $sl(2|1)$.}
\end{figure}

\section{\label{sec:thookth}T-hook}
In the previous two chapters we discussed an algebraic structure of the functional equations for the case of the irrep $[100\ldots]$ of the $gl(N|M)$ algebra. This structure has a natural generalization for the case of other fundamental representation. Let us consider an integrable system which is described by the following Kac-Dynkin diagram:
\be\label{KDinside}
    \parbox[c]{6.1cm}{\includegraphics[width=6cm]{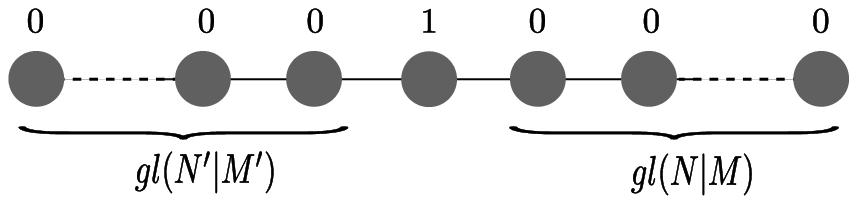}}.
\ee
This is a diagram of the $gl(N+N'|M+M')$ algebra\footnote{If a grading convention is not specified, it could be also a $gl(N+M'|M+N')$ algebra. However, we can always define a self-consistent notation to obtain only $gl(N+N'|M+M')$ case.}.  Gray nodes stay either for bosonic or fermionic nodes. Therefore we do not require the Kac-Dynkin diagram to be distinguished. Numbers above the nodes are the Dynkin labels of the representation.

\begin{figure}[t]
\centering
\includegraphics[height=4cm]{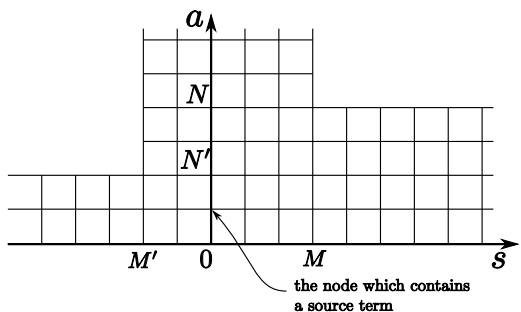}
\caption{\label{fig:thookstrings}T-hook shape on which (\ref{fathookie}) is defined.}
\end{figure}

By a direct generalization of the results of Sec.~\ref{sec:stringhyp} we deduce that the interactions between string configurations in the thermodynamic limit of the Bethe Ansatz based on (\ref{KDinside}) are described by (\ref{fathookie}) which is now defined on a T-hook shape shown in Fig.~\ref{fig:thookstrings}. The nodes $(a,0)$ correspond to string or stack configurations based on the momentum carrying node. The nodes to the right(left) of the central line $s=0$ denote string or stack configurations constructed from the Bethe roots that correspond to the Dynkin nodes to the right(left) of the momentum carrying node.

Based on the functional equations (\ref{fathookie}), a $T$-system defined on the T-hook shape can be derived from the TBA. This is how a T-hook was initially proposed in \cite{Gromov:2009tv,Bomb,GromovKKV} for $N=N'=M=M'=2$.

In the case of the $gl(n)$ algebra and of the representation $[100\ldots]$ of the $gl(n|m)$ algebra there is a strange coincidence that the fusion relations for the transfer  matrices of these system are defined exactly on the same shape as the $T$-system following from TBA. Based on this coincidence we even suggested in section \ref{sec:tdlimitfromHirota} an equivalence between $T$-functions and transfer matrices of the spin chain discretization. It is therefore natural to ask a question whether a T-hook shape has a transfer matrix interpretation.

Unfortunately such kind of interpretation is not known in the literature. Probably one of the obstacles is that consideration of fusing relations that include representation of type $[00010\ldots]$ leads to indecomposable representations and therefore is complicated. Still, in a lack of the transfer matrix description we can ask the question whether T-hook $T$-system can be solved by the Bethe Ansatz equations, in the same fashion as was discussed in sections \ref{sec:nestedBABacklund} and \ref{sec:Thook}. Such kind of analysis was partially done in \cite{Hegedus:2009ky}. In this work indeed a solution of $T$-system was given in terms of the boundary $Q$-functions. If to assume that $Q$ are polynomials, they can be considered as the Baxter polynomials that define positions of the Bethe roots. In \cite{Hegedus:2009ky} however only the set of $QQ$ relations which leads to the Bethe equations on the fermionic Bethe roots was given. $QT$ relations which lead to the bosonic Bethe equations were not given. It would be interesting to find such kind of relations. If it is possible then it would be natural to ask what is the meaning of $T$-functions from the point of view of the Bethe Ansatz. This should be a nontrivial question in view of the complicated representation theory of superalgebras.

\part{Integrable system of AdS/CFT}
\chapter{\label{ch:overview}Overview}
In this chapter  we are going to review the subject of integrability in the AdS/CFT correspondence. For other reviews on this  subject we refer to \cite{Klebanov:2000me,Maldacena:2003nj,Aharony:1999ti,D'Hoker:2002aw} (correspondence itself), \cite{Beisert:2004ry,Rej:2009je,SerbanMemoire} (integrability, with the stress on the gauge side), \cite{Tseytlin:2003ii,Vicedo:2008jk,GromovThese,Arutyunov:2009ga} (integrability, with the stress on the string side), \cite{Okamura:2008jm,VieiraThese,DoreyReview} (both aspects). In this text we do not discuss scattering amplitudes in the context of the AdS/CFT correspondence. For reviews of this subject we refer to \cite{Alday:2008yw,Alday:2008zz}.

\section{Essentials of the AdS/CFT correspondence.}
String theory initially appeared as an attempt to describe strong interactions. However, this attempt was not very successful and with the discovery of nonabelian gauge theories it receded into the background. A new relation of strings, now to the gauge theory,
was proposed by 't Hooft \cite{tHooft:1973jz}. In his work he considered the large $N$ limit of $SU(N)$ gauge theories:
\be\label{thooftlimit}
  g_{YM}^{-1},N\to \infty\ \ \  {\rm with}\ \ \ \l\equiv g_{YM}^2N\ \ {\rm fixed}.
\ee
In this limit the sum of Feynman diagrams can be organized as the $1/N^{2}$ expansion. Interestingly, this expansion is nothing but the expansion over the topologies. For example, the partition function of the system is expanded as
\be
  \log Z=N^2 \CF_0[\l]+\CF_1[\l]+N^{-2} \CF_2[\l]+\ldots\ ,
\ee
where $\CF_n[\l]$ is a sum of Feynman diagrams which can be drawn on a Riemann surface with $n$ handles.

The topological expansion is reminiscent to the perturbative expansion in a string theory. In this analogy we interpret the Feynman diagram (more precisely, the dual graph) as the discretization of a string worldsheet and $1/N$ should be then proportional to the string coupling $g_s$.

The idea to interpret Feynman diagrams as a discretization of the two-dimensional surface was fruitfully used to define the two-dimensional gravity through the matrix models \cite{Kazakov:1985ds,David:1985nj,Ambjorn:1985az,Kazakov:1985ea} (for a review see \cite{DiFrancesco:2004qj}). In the case of the matrix models the sum over the planar graphs $Z_0[\l]$ can be calculated explicitly. This sum is convergent with a finite radius of convergence $\l_0$. Close to a critical value $\l=\l_0$ the sum is dominated by the diagrams with large number of vertices. Therefore the continuous surfaces are recovered in the proper double scaling limit which includes $\l\to\l_0$.

The identification of the string theory which corresponds to a 't Hooft construction in the gauge theories is a highly nontrivial problem. For a long time there were no explicit examples except the ones for the two-dimensional gauge theories \cite{Gross:1992tu,Gross:1993hu,Minahan:1992sk}. The discovery of the AdS/CFT correspondence \cite{Maldacena:1997re,Gubser:1998bc,Witten:1998qj} gave us explicit examples of a gauge/string duality in four dimensions. Often it is suggested that this duality realizes the 't Hooft idea, however this was not explicitly shown.

The first example of the AdS/CFT correspondence is the equivalence between
$\CN=4$ supersymmetric Yang-Mills theory (SYM) and the type IIB string theory with AdS$_5\times$S$^5$ manifold as a target space. This equivalence was first proposed in \cite{Maldacena:1997re} developing the study of string theories in the presence of $D$-branes.

 The $Dp$-branes are $p+1$ dimensional objects on which strings can terminate \cite{Polchinski:1996na} ($D$ stands for Dirichlet boundary conditions). The low-energy modes of the $D$-branes are described by the maximally supersymmetric gauge theory living on $D$-brane. A stack of $N$ branes leads to the $SU(N)$ gauge theory \cite{Witten:1995im}. For $p=3$ this is $\CN=4$ SYM. This is how SYM theory appears in the context of the string theory.

 The AdS/CFT conjecture is based on the observation that string theory in the presence of $D3$-branes can be effectively described, at least when $N$ is large, as a closed string theory in a nontrivial gravitational background. We will now explain this description in more details. Let  us consider a type IIB string theory. Its action is the action for the supersymmetric sigma model, with a worldsheet metric being dynamic. The bosonic part of the action can be written as
 \be\label{Sbos}
    S_{bos}=\frac 1{4\pi\a'}\int d^2\sigma \sqrt{\eta}\(\ \eta^{\mu\nu}G_{mn}[X]+\e^{\mu\nu}B_{mn}[X]\ \)\pd
     _\mu X^m\pd_\nu X^m+\a'R_\eta\Phi[X],
 \ee
 where $G_{mn}$ is a target space metric, $B_{mn}$ is an antisymmetric tensor and $\Phi$ is the dilaton.

 $G_{mn}[X],B_{mn}[X],\Phi[X]$ can be considered as an infinite set of coupling constants in the quantum field theory which are subjects of renormalization. This is how strings govern the geometry of the target space. The demand that the $\b$-function for these coupling constants equals zero leads to the supergravity equations of motion.

 The type IIB string theory contains also the selfdual 4-form $A[X]$. There exists a solution of supergravity EOM with a flux of a field strength $F=dA$ equal to $N$, $\int_{S^5}F=N$:
 \be\label{Maldacenaspace}
     ds^2&=&\(1+\frac{R^4}{r^4}\)^{-1/2}(-dt^2+dx^2+dy^2+dz^2)+\(1+\frac{R^4}{r^4}\)^{1/2}(dr^2+r^2d\Omega_5^2),\no\\
     &&\hspace{2EM}R^4=4\pi g_s N \a'^2.
 \ee
 This is a background which is suggested to be effectively generated by a stack of $N$ $D3$-branes.

Due to the red shift the energies for the observer at infinity are not the same as the energies for the observer at finite $r$. They are related as
 \be\label{EEr}
  E_\infty=\(1+\frac{R^4}{r^4}\)^{-1/4}E_r\simeq \frac{r}{\sqrt{\a'}}E_r,\ \ r\ll R.
 \ee
 One can consider the following limit introduced by Maldacena \cite{Maldacena:1997re}. We take $\a'\to 0$ with $g_s N$ fixed. In this limit the near horizon dynamics at distances $r\sim\a'$ decouples from the rest. In terms of the variable $u=R^2/r$ the near horizon geometry is approximated by
 \be\label{adsmetric}
   ds^2\simeq R^2\(\frac{-dt^2+dx^2+dy^2+dz^2+du^2}{u^2}+d\Omega_5^2\),
 \ee
which is the AdS$_5\times$S$^5$ geometry. Therefore we obtain the type IIB strings propagating in the AdS$_5\times$S$^5$ space-time.

%
%
%
Due to (\ref{EEr}), the near horizon excitations are viewed by the observer at infinity as low energy excitations. As we already discussed, the low energy excitations are described in a $D3$-brane picture by the gauge theory.

The comparison of these two different descriptions of the near horizon physics leads to the AdS/CFT correspondence conjecture. The conjecture includes also the relation between $g_{YM}$ and $g_s$:
\be
  4\pi g_s=g_{YM}^2.
\ee
From (\ref{adsmetric}) and (\ref{Sbos}) we conclude that the string sigma model depends not on $R^2$ and $\a'$ separately but on their dimensionless ratio which is a 't Hooft coupling constant:
\be
  \frac{R^2}{\a'}=\sqrt{4\pi g_sN}=\sqrt{\l}\,.
\ee
The AdS/CFT correspondence is often trusted only in the 't Hooft limit (\ref{thooftlimit}), when the stack of $D$-branes is a heavy object capable to generate a nontrivial gravity background. In the 't Hooft limit $g_s\to 0$, so the first quantization of the string theory is enough. This is the limit which we consider in this text.

The discussed construction of the duality picture can be generalized. The possible variations of the string target spaces and brane configurations are discussed in a great detail in the review \cite{Aharony:1999ti}.

Recently an AdS$_4$/CFT$_3$ 
version of the duality was proposed by Aharony, Bergman, Jafferis,
and Maldacena (ABJM) \cite{Aharony:2008ug}. 
The ABJM conjecture relates $\CN=6$ $U(N)\times U(N)$ superconformal Chern-Simons theory at level $k$ with M-theory on AdS$_4\times$S$^7/\MZ_k$. The quotient with respect to $\MZ_k$ should be understood in the following way: S$^7$ can be viewed as the S$^1$ Hopf bundle over $\MC\MP_3$. The discrete group $\MZ_k$ acts on the S$^1$ fiber. For $k\neq 1,2$ it breaks the complete $\CN=8$ supersymmetry to $\CN=6$ one. In the 't Hooft limit ($k,N\to\infty$ with $\l=N/k$ fixed) the dimension reduction from S$^7$ to $\MC\MP_3$ takes place and we obtain the correspondence between the planar Chern-Simons gauge theory and the first quantized type IIA string theory on AdS$_4\times\MC\MP_3$.


Since in the limit (\ref{thooftlimit}) the duality is of weak/strong coupling type, it is especially useful for the description of the gauge theories at strong coupling. It would be very interesting to find a non-supersymmetric version of the duality which could describe QCD. Of course it is not obvious whether such generalization is possible at all. For a discussion of possible restrictions for this generalization and to what extent they can be overcome see \cite{Erlich:2009me}.
\ \\

Let us come back to the 't Hooft attempt to describe the worldsheet discretization of a string with the help of Feynman diagrams.
If to follow the analogy with matrix models, the continuous description of the worldsheet appears when the coupling constant approach some critical value $\l_0$. The finiteness of $\l_0$ for $\CN=4$ SYM\footnote{or other gauge theory conjectured by the duality.} would signal that the 't Hooft description is inappropriate for the AdS/CFT correspondence. In the presence of the supersymmetry one could believe that by virtue of some cancellations $\l_0=\infty$. However, the finite radius of convergence for small 't Hooft coupling expansion of various observables in $\CN=4$ SYM suggests that $\l_0$ is finite. 

A huge progress in the understanding of the AdS/CFT correspondence during the last decade was possible in particular due to the discovery of integrability. It is believed that integrability will allow to find the spectrum of the AdS/CFT system at arbitrary value of the coupling constant and to give a direct proof of the duality. On the string side the integrability appears in a standard way - as integrability of a coset sigma model. On the gauge side the integrability structures are identified in quite unusual way. We will now discuss the integrability structures from the gauge point of view, then pass to the string side, and at the end review the main tests of the AdS/CFT correspondence and the integrability conjecture.

\section{Gauge side of the correspondence and underlying integrable system.}
The gauge side of the AdS/CFT correspondence is the $\CN=4$ super Yang-Mills (SYM) theory in $3+1$ dimensions. This is a conformal field theory. The vanishing of the beta function in it was shown in \cite{Brink:1982wv,Novikov:1983uc,Howe:1983sr}. The field content of $\CN=4$ SYM is a gauge field, four left and four right Majorana spinors, and 6 real scalars:
\be
  A_\mu,&&\ \ \mu=\overline{1,4},\no\\
  \psi_{a,\a},\overline\psi_{\dot\a}^a,&&\ \ a=\overline{1,4},\ \a,\dot\a=1,2,\no\\
  \Phi_i,&&\ \ i=\overline{1,6}.
\ee
All fields are in the adjoint representation of the $SU(N)$ gauge group.

The global symmetry algebra of the system is $psu(2,2|4)$.  The bosonic subalgebra of $psu(2,2|4)$ is given by the direct sum $so(2,4)\oplus so(6)$. Its first term is the conformal symmetry in $3+1$ dimensions, the second term is the $R$ symmetry. The $R$ symmetry trivially acts on $A_\mu$, spinors transform in its spinorial representation ($so(6)\simeq su(4)$ acts on the index $a$), scalars transform in its vector representation ($so(6)$ acts on the index $i$).

The lagrangian of the theory is given by
\be
  \CL&=&\frac 1{g_{YM}^2}\Tr\(\frac 14 F_{\mu\nu}F^{\mu\nu}+\frac 12D_\mu\Phi_iD^\mu\Phi^i+{\bar\psi}^a\sigma^\mu D_\mu\psi_a-\right.\no\\ &&\left.-\frac 14[\Phi_i,\Phi_j][\Phi^i,\Phi^j]-\frac i2 \s_{i}^{ab}\psi_a[\Phi^i,\psi_b]-\frac i2\s^i_{ab}\overline \psi^a[\Phi_i,\overline\psi^b]\).
\ee
The coefficients $\s^{i}_{ab},\s_i^{ab}$ are the Clebsch-Gordan coefficients which provide the invariance of the lagrangian under action of the $R$ symmetry.

We are interested in the 't Hooft planar limit (\ref{thooftlimit}) of the theory. Instead of the 't Hooft coupling constant $\l$ the coupling constant $g$ is often used. It is defined by
\be\label{gdefinition}
  g^2=\frac{\l}{16\pi^2}.
\ee

The objects to study in the planar limit are single trace local operators. 
These operators can be viewed as the words built from the letters corresponding to the elementary fields of the theory. A possible example of such operator is
\be\label{word}
  \CO=\Tr[\Phi_1D_3\Phi_2\psi_{1\dot1}\ldots],\ \ D_\mu\cdot\equiv\pd_\mu+[A_\mu,\cdot].
\ee
Single trace operators (\ref{word}) are organized into multiplets of the symmetry algebra $psu(2,2|4)$. Its representations are parameterized by a set of numbers
\be\label{numbersd}
(s_1,s_2),\ \Delta,\ [r_1,r_2,r_3],
\ee
where $(s_1,s_2)$ are spin labels of the representation of the Lorentz group $so(3,1)\subset so(2,4)$, $[r_1,r_2,r_3]$ are Dynkin labels of $so(6)$ and $\Delta$ is the conformal dimension or, what is the same, the eigenvalue of the dilatation operator. We will denote the dilatation operator by $\hat\Delta$.

The symmetry algebra is preserved on the quantum level. However, certain symmetry generators depend on the coupling constant due to their renormalization.

The unitary representations of $psu(2,2|4)$ were classified in \cite{Dobrev:1985qv,Dobrev:1985vh}. They include three series of the so called BPS representations. These representations are analogs of atypical irreps discussed in Sec.~\ref{sec:sl21algebra}. The value of $\Delta$ in the BPS multiplets can acquire only half-integer or integer values, therefore it does not change on the quantum level.

There is also one series of irreps for which the value of $\Delta$ can change continuously. For these irreps the dilatation operator acquires in general situation a quantum correction:
\be\label{definition of H}
  \hat\Delta[g]=\hat\Delta_0+g^2\CH[g].
\ee

The quantum correction to the dilatation operator $\CH[g]$, unlike the classical part $\hat\Delta_0$, acts nondiagonally on the single trace local operators:
\be
  \hat\CH[g]:\,\CO^a\mapsto \CH[g]^a_b \CO^b.
\ee
According to the duality conjecture the eigenvector of $\hat\Delta_0+g^2\CH$ corresponds to a certain string state. The corresponding eigenvalue (anomalous dimension of the operator) is equal to the energy of this state. Finding the spectrum of $\hat\Delta[g]$ is the subject on which we concentrate in this thesis.

If we are interested in finding the spectrum of very long operators, the diagonalization of the matrix $\CH[g]^a_b$ seems to be a complicated problem since this matrix is very large. For short operators a direct diagonalization is possible, however $\CH[g]$ is known explicitly only at first few orders of small $g$ expansion. The integrability methods turned out to be a very powerful tool which allowed to overcome these problems. We will explain how integrability appears in the spectral problem of AdS/CFT and then discuss it in details. To start, we should first explain some details on the structure of the single trace operators and operator $\CH$.

One can think about a single trace operator as being a state of a spin chain. Each occurrence of a given field (but not covariant derivative) corresponds to one site of this chain with the corresponding color index.

As an example we can consider the operator which consists of the fields only of two types, $X$ and $Z$:
\be\label{forspan}
  \CO=\Tr(XZZZXXZ\ldots),\ \ \ X=\Phi_1+i\Phi_2,\ Z=\Phi_5+i\Phi_6.
\ee
The vector space spanned by these operators is invariant under action of the quantum dilatation operator. $X$ and $Z$ form a fundamental multiplet of the $su(2)$ subalgebra of $psu(2,2|4)$. Therefore the space spanned by operators (\ref{forspan}) is usually called the $SU(2)$ subsector. The mapping to the spin chain is the following: we identify $Z$ field with a spin down $\downarrow$ state of the node and $X$ field with a spin up $\uparrow$ state of the node, so the operator (\ref{forspan}) is mapped to
\be
  |\uparrow\downarrow\downarrow\downarrow\uparrow\uparrow\downarrow\ldots\rangle.
\ee

Another important example is the operators built only from the $Z$ field and a light-cone covariant derivative $D=D_0-D_1$:
\be\label{twistJ}
  \CO=\Tr(ZD^2ZD^4ZZDZ\ldots).
\ee
Such operators are called twist $J$ spin $S$ operators, where $J$ is a number of $Z$-fields and $S$ is a number of covariant derivatives.

This subsector is also invariant under the action of the dilatation operator and is called the $Sl(2)$ subsector. Here $Z$ is identified with a node of the spin chain and the number of covariant derivatives acting on $Z$ gives a number of excitations at that particular node. This subsector is noncompact in the sense that the number of excitations at each node can be arbitrary.

Although the symmetry generators $J_A$ are renormalized, their conformal dimensions are preserved since the commutation relations are preserved. In other words, $\CH[g]$ commutes with symmetry generators:
\be\label{comut1}
  [\CH,J_A]=0.
\ee
Therefore, if one thinks of $\CH$ as of the Hamiltonian of the described above spin chain, $psu(2,2|4)$ is the symmetry of this spin chain. 

Note that $\CH$ enters as a central extension of $psu(2,2|4)$ algebra which is different from for example such spin chain as Heisenberg ferromagnet.
Instead, this property resembles to the case of relativistic theories in the sense that in both cases the Hamiltonian is a part of a symmetry algebra.

The Hamiltonian $\CH$ can be expanded in the powers of the coupling constant as
\be\label{Hexpansion}
  \CH[g]=\CH_2+g^2\CH_3+g^{4}\CH_4+\ldots,
\ee
where $\CH_n$ is the $(n-1)$-loop contribution of the perturbative theory. The perturbation theory in the 't Hooft planar limit is such that $\CH_n$ includes only the interactions between $n$ nearest neighbors. For example, for the $SU(2)$ subsector $\CH_2$ and $\CH_3$ are given by \cite{Beisert:2003tq}:
\be
    \CH_2=2\sum_{i}(1-\scP_{i,i+1}),\ \
    \CH_3=-4\sum_{i}(1-\scP_{i,i+1})+\sum_{i}(1-\scP_{i,i+2}),
\ee
where $\scP$ is a permutation operator.

The integrability in $\CN=4$ SYM was first observed by Minahan and Zarembo in \cite{Minahan:2002ve} for the $SO(6)$ subsector of the spin chain at one loop\footnote{Prior to this work, integrability had been already discovered in one loop QCD. It first appeared in a study of the Regge asymptotics of scattering amplitudes \cite{Lipatov:1993yb,Faddeev:1994zg}. The one-loop dilatation operator in the $Sl(2)$ subsector of QCD was identified with the Hamiltonian of integrable spin chain in \cite{Braun:1998id}. The $Sl(2)$ subsector then was further studied in \cite{Braun:1999te,Belitsky:1999ru,Belitsky:1999bf,Derkachov:1999ze}}. This subsector is built only from the scalar operators $\Phi_i$ and is invariant under action of $\CH_2$. Minahan and Zarembo explicitly calculated $\CH_2$ in this sector and found that it was the Hamiltonian of the integrable $SO(6)$ spin chain with rational $R$-matrix.

\begin{figure}[t]
\centering
\includegraphics[width=8cm]{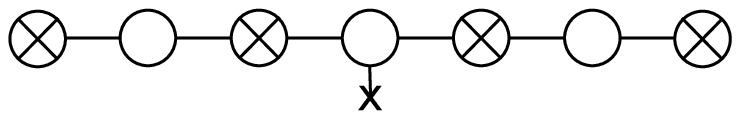}
\caption{\label{fig:Dynkin1l}A particular Kac-Dynkin diagram for the 1-loop Bethe Ansatz.}
\end{figure}
In \cite{Beisert:2003jj} the one-loop integrability for the whole symmetry algebra was checked. The underlying integrable spin chain is based on the rational $gl(4|4)$ $R$-matrix. Therefore it can be solved by standard means described in details in the first part of this text. In particular, we have to choose a pseudovacuum to write down the Bethe Ansatz equations. The standard choice for the pseudovacuum is the BPS state
\be\label{BPSvac}
  \Tr Z^L.
\ee
The pseudovacuum is not invariant under the whole $psu(2,2|4)$ symmetry, but only under $psu(2|2)^2\rtimes \MR$. The Bethe Ansatz equations for the excitations over this pseudovacuum are encoded into the Kac-Dynkin diagram\footnote{A particular choice of a Kac-Dynkin diagram is dictated by further generalization of integrability to higher loops.} shown in Fig.~\ref{fig:Dynkin1l}.

The first evidence that integrability might hold below one loop was obtained by Beisert, Kristjansen, and Staudacher in \cite{Beisert:2003tq}, where the two loop dilatation operator ($\CH_3$) was calculated and a degeneracy of the spectrum characteristic for integrable system was found. In this work they conjectured also that integrability holds at all loops. In \cite{Beisert:2003ys} Beisert showed that the dilatation operator, at least in $SU(2|3)$ subsector, can be fixed up to three loops by supersymmetry, an input from perturbative field theory, and additional requirement - the proper BMN scaling \cite{Berenstein:2003gb,Gross:2002su,Santambrogio:2002sb}. This gave a check of integrability up to three loops. The three loop Hamiltonian was shown to be related to the Inozemtsev model \cite{Serban:2004jf} and derivable from the Hubbard spin chain \cite{Rej:2005qt}.

Such properties as the BMN scaling of the spectrum and the relation of the Hamiltonian to Inozemtsev and Hubbard models, do not hold starting from four loops. Starting from this order, the checks of integrability are more sophisticated.
So far there is no proof of the all-loop integrability conjecture, however there are nontrivial checks at four \cite{Beisert:2006ez,Bern:2006ew,Beisert:2007hz},\cite{Bajnok:2008bm,Bajnok:2008qj} and even at five loops \cite{Bajnok:2009vm,Fiamberti:2009jw,Lukowski:2009ce}\footnote{The four loop tests \cite{Bajnok:2008bm,Bajnok:2008qj} and five loop tests consider short operators which are sensible to wrapping corrections. These are the verifications of the thermodynamic Bethe Ansatz which is next level after construction of the asymptotic Bethe Ansatz that we discuss in this chapter.}. Of course, an important argument for the all-loop integrability is that at strong coupling we should reproduce the string sigma model, which is integrable at least classically (we discuss the integrability on the string side of the correspondence in the next section).

The identification of integrability structures at higher loops is a nontrivial procedure. As we have already mentioned, the range of interaction increases by one with each order of the perturbation theory. Soon the structure of the Hamiltonian becomes very involved. The exact Hamiltonian of the system is not known\footnote{For the $SU(2)$ subsector a construction that allows one to generate the series of $\CH$ in $g^2$ was proposed \cite{Bargheer:2008jt,Bargheer:2009xy}. Still, this construction does not give the Hamiltonian in a close form. Also it does not include wrapping interactions.}. Also we do not know the complete $R$-matrix of the system.

For the finite length spin chain at a certain order of the perturbation theory the range of interaction spreads to the whole spin chain. At this order new types of Feynman diagrams appear that contribute to the Hamiltonian. They lead to the so called wrapping interactions. In \cite{Kotikov:2007cy} it was shown that the Bethe Ansatz approach, at least in the known form, does not correctly describe the system with wrapping interactions. Therefore, to apply Bethe Ansatz one should generically restrict to the case of the infinitely long spin chain. More precisely, for the spin chain of the length $L$, the Bethe Ansatz can be trusted up to the order $g^{2L-4}$ in the expansion (\ref{Hexpansion}). Due to this reason the Bethe Ansatz is called asymptotic.

This is reminiscent to the asymptotic Bethe Ansatz in the integrable sigma-models. The latter is built based on the factorized $S$-matrix which describes the scattering of particles. The notion of a scattering is possible only in the case when $m L\gg 1$, where $m$ is the mass of the particle and $L$ is the size of the system. For $m L\simeq  1$ or smaller the asymptotic Bethe Ansatz should be corrected.

The scattering matrix in integrable sigma-models is constructed by the bootstrap approach discussed in Sec.~\ref{sec:scatteringmatrix}. This approach turned out to be very useful to describe the spectrum of $\CH[g]$. The importance of the factorized $S$ matrix in the AdS/CFT case was recognized by Staudacher in \cite{Staudacher:2004tk}. As was shown by Beisert \cite{Beisert:2005tm}, the $S$-matrix which describes the scattering over the BPS vacuum (\ref{BPSvac}) can be fixed up to an overall scalar factor from the symmetry requirements. The overall scalar factor is restricted by the crossing equations \cite{Janik:2006dc}\footnote{The existence of the crossing equations is expected from the fact that the integrable system should describe also the string sigma model (see Sec.~\ref{sec:crossingequations}).}. The solution of the crossing equations is also known \cite{Beisert:2006ib,Beisert:2006ez}. This solution is uniquely fixed by the reasonable physical requirements \cite{V3}.

The knowledge of the $S$-matrix allows to build the all-loop Bethe Ansatz equations \cite{Beisert:2006qh,Martins:2007hb} initially proposed by Beisert and Staudacher \cite{Beisert:2005fw}. The $S$-matrix  is not a rational one, except for $g=0$ where we recover the one-loop integrable system. The overall scalar factor, the logarithm of which is also known as the dressing phase, enters to the central node equations.
\begin{figure}[t]
\centering
\includegraphics[width=8cm]{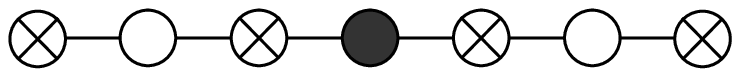}
\caption{\label{fig:Dynkinalll}Formal Kac-Dynkin diagram for all-loop AdS/CFT asymptotic Bethe Ansatz. We show this diagram to show the resemblance of the structure of the asymptotic Bethe Ansatz with the one of PCF, {\it cf.} Fig.~\ref{fig:DynkinPCF}. The AdS/CFT Bethe Ansatz behind this diagram is not based on rational R-matrix. More details are given in Sec.~\ref{Sec:BSasBA}.}
\end{figure}

The construction of the asymptotic Bethe Ansatz will be discussed in details in chapter \ref{ch:bootstrap}. To give a brief idea of how it looks like we write down an explicit form of the Bethe Ansatz for the $SU(2)$ subsector:
\be\label{bethesu2sector}
  \(\frac{x[u_k+i/2]}{x[u_k-i/2]}\)^L=-\prod_{j=1}^M\frac{u_k-u_j+i}{u_k-u_j-i}\sigma[u_k,u_j]^2.
\ee
Here $x$ is the Zhukovsky map defined by (\ref{Juk}) and $\sigma[u,v]$ is the dressing factor. For $g=0$ we have $x[u]=u$ and $\sigma[u,v]=1$. Therefore we recover the Bethe Ansatz equations for the XXX Heisenberg ferromagnet (\ref{BetheXXX}). The Bethe Ansatz equations in the absence of the dressing factor were first proposed by Beisert, Dippel and Staudacher \cite{Beisert:2004hm} and are known as the BDS Bethe Ansatz. The necessity of the nontrivial dressing factor became clear from the perturbative calculations on the string theory side \cite{Arutyunov:2004vx}.

The logic of construction and the structure of all-loop Bethe equations are much closer to the integrable field theories then to the integrable spin chains. Therefore, we depict the Dynkin diagram for the all-loop Bethe Ansatz as shown in Fig.~\ref{fig:Dynkinalll}\footnote{(\ref{bethesu2sector}) corresponds to the black node of the diagram in Fig.~\ref{fig:Dynkinalll} with nested Bethe roots being turned off.}, in analogy with Fig.~\ref{fig:DynkinPCF} for PCF. This analogy should not be very surprising since at strong coupling the Bethe Ansatz should reproduce the spectrum of the string theory, which is a two-dimensional field theory.

\section{String side of the correspondence}
The first-quantized string is described by a Green-Schwarz-Metsaev-Tseytlin action \cite{Metsaev:1998it}, whose bosonic part has a nonlinear sigma model form
\be\label{sbos}
  S_{bos}=\frac 1{4\pi\a'}\int_{0}^{2\pi}\!\!d\sigma\!\int\! d\tau\ G_{MN}\pd X^M\pd X^N.
\ee
The metric $G_{MN}$ is the metric of AdS$_5\times$S$^5$ (\ref{adsmetric}). The target space is not AdS$_5\times$S$^5$ but its universal covering. As we already mentioned, the theory depends on the unique parameter - 't Hooft coupling constant given by $\sqrt{\l}=R^2/\a'$, where $R$ is the radius of $AdS_5$ and $S^5$ spaces.

This theory can be formulated as a coset sigma model. Similarly to bosonic sigma models on a coset manifold \cite{Zakharov:1973pp}, it can be shown to be classically integrable \cite{Bena:2003wd}. Namely, it is possible to construct the flat current $j$ in terms of which the equations of motion are written as
\be\label{someEOM}
  d*j=0.
\ee
In the simplest cases $j=g^{-1}dg$. For the string sigma model its definition is slightly more complicated.

One can introduce the Lax connection $J$
\be
  J=\frac 1{1-x^2}j+\frac{x}{1-x^2}*j
\ee
which is flat as it can be easily checked from (\ref{someEOM}) and the flatness of $j$.

\begin{wrapfigure}{r}{0.30\textwidth}
 \vspace{-12pt}
  \begin{center}
    \includegraphics[width=0.20\textwidth]{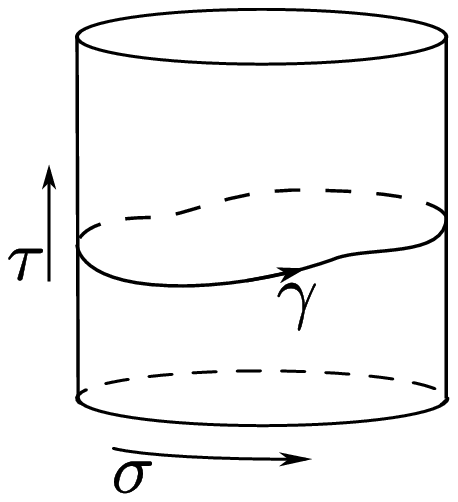}
  \end{center}
  \vspace{-5pt}
\end{wrapfigure}
Using the Lax connection, one can build the monodromy  $\Omega$
\be\label{mOmega}
  \Omega[x]=Pe^{\int_\gamma J},
\ee
where the contour $\gamma$ is a closed contour on the string worldsheet shown in figure on the right.

Due to the flatness of $J$ the function $\Tr\Omega[x]$ does not change if the contour $\g$ is continuously deformed, including translation in the $\tau$ direction. Therefore $\Tr\Omega[x]$ does not depend on time $\tau$. Since $\Tr\Omega[x]$ depends on the additional parameter $x$ it generates an infinite set of integrals of motion. As usual for integrable systems, the local integrals of motion appear if to expand $\Tr\Omega[x]$ near its singular points which are $x=\pm 1$ (compare to the formula (\ref{XXXhamiltonian}) and the discussion after it).

For the AdS/CFT sigma model this construction was first realized in \cite{Bena:2003wd}. However, the existence of infinite set of integrals of motion is not sufficient for proving classical integrability. One has to show that these integrals of motion are in convolution.
This was done first for the $SU(2)$ subsector in \cite{Dorey:2006mx}, then for the bosonic string in \cite{Kluson:2007vw,Kluson:2007md}, and finally for the whole string in \cite{Magro:2008dv}.

Each string state is characterized, not uniquely in general, by the values of the Cartan generators of $psu(2,2|4)$ algebra acting on it. These are three angular momenta $(J_1,J_2,J_3)$ for the rotation of string in $S^5$, two angular momenta $(S_1,S_2)$ for the rotation of string in $AdS_5$ and the energy $E$ which corresponds to the translations in time direction. These parameters are related to the parameters (\ref{numbersd}) which define the representation in the following way:
\be\label{ChargesDefinition}
  [r_1,r_2,r_3]=[J_2-J_1,J_3-J_2,J_2+J_1],\ \ (s_1,s_2)=\frac 12(S_1+S_2,S_1-S_2),\ \ E=\Delta.
\ee

Initial interest for particular solutions to the string equations of motion and their quasiclassical quantization was motivated by the discovery of the so called BMN scaling \cite{Berenstein:2003gb}. The conjecture of the BMN scaling states that in the limit $J_3\gg 1$, $\l\lesssim J_3$ the energy is effectively expanded in terms of the BMN coupling $\l'=\l/J_3^2$. This scaling gave a window for a direct comparison of strong and weak coupling results since smallness of $\l'$ can be achieved both in perturbative gauge and string theories since $J_3$ is large. For some of the references concerning study of string configurations in this scaling and its generalizations see \cite{Gubser:2002tv,Frolov:2002av,Minahan:2002rc,Frolov:2003qc,Arutyunov:2003uj,Beisert:2003ea,Frolov:2003xy}. We will discuss some particular cases in the next section.

These initial investigations played an important role for the development of integrability techniques. Later it became evident that the BMN scaling is violated at higher orders. But soon the development of the integrability gave
precise predictions for certain anomalous dimensions at strong coupling without relying on the BMN regime.

Solutions of string equations of motion can be described by a general approach known as a finite gap method. Its application for AdS/CFT was developed in a series of papers \nocite{Kazakov:2004qf}\cite{Beisert:2005bm,Beisert:2004ag,Dorey:2006zj,Gromov:2007aq,Vicedo:2008jy} initiated by \cite{Kazakov:2004qf}. Based on the finite gap method, it was possible to perform a general test for the validity of the asymptotic Bethe Ansatz on the string side at 1-loop approximation \cite{Gromov:2007ky}.

There are also few two-loop calculations in string theory \cite{Roiban:2007dq,Roiban:2007ju,Roiban:2009aa}. The two-loop check of the integrability seems to be more profound than the one loop one. The one-loop results could be obtained by the quasiclassical quantization. Instead, a two-loop calculation requires computing of Feynman diagrams.  We discuss comparison of the Bethe Ansatz prediction and the string two-loop predictions  in subsection \ref{sec:cusp} and chapter \ref{ch:BESFRS}. So far, at two loops there is an agreement for the value of the cusp anomalous dimension \cite{Roiban:2007dq}. However, for more involved examples: generalized scaling function \cite{Roiban:2007ju} and Konishi operator \cite{Roiban:2009aa} there are disagreements. The reason for these disagreements is still not understood.

Before going to the explicit examples let us mention that there is an important difference between the sigma-model and the string theory. In the latter the world-sheet metric is a dynamical quantity. Although we can always choose a reference frame in which the worldsheet metric is flat, the dynamical nature of the metric leads to Virasoro constraints.
%
 By fixing the Virasoro constraints we typically introduce the mass scale into the system, as it happens for example in a light-cone gauge. Therefore, finally we consider not the initial conformal sigma model (\ref{sbos}) but a field theory with massive excitations. The $\beta$-function for the coupling constant remains zero of course.

\section{Verifications of the asymptotic Bethe Ansatz}

All the examples considered below are based on the string states which have at most two nonzero angular momenta. The formulation of the perturbative string theory is known only when one of the angular momenta is large. This requirement is also needed to be able to use the asymptotic Bethe Ansatz.

\subsection{BMN particles and giant magnons.}
In the seminal paper of Berenstein, Maldacena, and Nastase \cite{Berenstein:2003gb} the string theory was considered in a type of the Penrose plane wave limit \cite{Penros} which described perturbations around fast moving point-like string solution. The limit in \cite{Berenstein:2003gb} included
\be
  J\sim N\to\infty,\ \ g_{YM}\ \ {\rm fixed}.
\ee
Here we will rather use the 't Hooft limit (\ref{thooftlimit}) with additional demand that the BMN coupling
\be
  \l'=\frac{\l}{J^2}
\ee
is small. The excitations we are interested in are the same in two cases. Therefore we will keep the name of BMN excitations (particles) also for our case. Our reasoning follows closely to \cite{Gubser:2002tv}.

To understand the nature of the BMN excitations it is useful to replace string by a particle on $S^2$. The lagrangian for such a particle is given by
\be
  \CL=\frac{mr^2}{2}\(\(\frac{d\theta}{dt}\)^2+\sin^2[\theta]\(\frac{d\phi}{dt}\)^2\).
\ee
One has the conserved angular momentum
\be
  J_\phi=\frac{\pd{\CL}}{\pd\dot\phi}=mr^2\sin^2[\theta]\frac{d\phi}{dt}.
\ee
Fixing its value one can find the effective lagrangian for the coordinate $\theta$:
\be
  \CL_{\theta}=\CL-J_\phi\dot\phi=\frac{mr^2}{2}\(\frac{d\theta}{dt}\)^2-V[\theta],\ \ V[\theta]=\frac{J_\phi^2}{2mr^2\sin^2[\theta]}.
\ee
We got a particle which moves in the effective potential with minimum at $\theta=\pi/2$. For large values of $J_\phi$ it is enough to use quadratic approximation:
\be
  V[\theta]=\frac{J_\phi^2}{2mr^2}\(1+\(\theta-\frac\pi 2\)^2\),
\ee
which leads to oscillations with frequency $\omega=\frac{J}{mr^2}$. Note that in this example $mr^2$ plays the role of the coupling constant.

The picture in the string theory is completely similar. The energy of the particle is the analog of the string \textit{worldsheet} energy. The coupling constant in the string case is $\sqrt{\l}$. The fast rotating string with momenta $J$ leads to a massive relativistic field theory with $m=J/\sqrt{\l}$. The worldsheet energy is given by
\be\label{eq4}
  \delta=\frac 12\frac{J^2}{\l}-\frac 12\frac{\Delta^2}{\l}+\sum_{n}N_n\sqrt{n^2+\frac{J^2}{\l}}.
\ee
The spectrum is discrete since we consider closed strings.

The written expression (\ref{eq4}) for $\delta$ looks like if we have only one type of excitations. In fact, there are 16 of them - 8 bosonic and 8 fermionic. However, all their masses are equal as was shown by Metsaev in \cite{Metsaev:2001bj} using symmetry.

To find the target space energy $\Delta$ we should use the on-shell condition which states that $\delta=-\ell+1$, where $\ell$ is a number of worldsheet derivatives in the vertex operator corresponding to the considered state. If we consider states with finite values of $\ell$, the on-shell condition and (\ref{eq4}) lead to

\be\label{disp1}
  \Delta=J+\sum_{n}N_n\sqrt{\frac{\l n^2}{J^2}+1}+\CO(\sqrt{\l}/J).
\ee
Now we understand the meaning of the BMN coupling $\l'=\l/J^2$. This is a square of the inverse mass of the BMN particle. If we assume that spectrum of the theory is described by BMN particles, it is natural to expect that the energy of the state is expanded in terms of $\l'$. 
In reality the BMN scaling is not an exact property. The reason is that the harmonic oscillator approximation works only for the oscillations with small mode numbers $n$.

It will be convenient to introduce the worldsheet momentum
\be
  p=\frac{2\pi n}{J},
\ee
which will be a natural variable if we rescale the worldsheet volume such that it is equal to $J$.

The momentum $p$ is continuous in the large $J$ limit. The harmonic oscillator approximation works for $p\ll 1$. 
For $p\gtrsim 1$ one may expect a different from (\ref{disp1}) dispersion relation. In fact, the exact energy which is associated to each particle is equal to
\be\label{dispersionrelation}
  E[p]=\sqrt{1+16g^2\sin^2\frac p2}-1.
\ee
This expression was initially derived in \cite{Santambrogio:2002sb}. 

The excitations whose momentum scales at large $g$ as $p\sim 1$ are called giant magnons.

To understand better the giant magnons let us consider the bosonic string sigma model on $\MR\times S^2$ and choose the temporal gauge $X^0=\tau$. In this gauge the theory reduces to the $O(3)$ sigma model. The equations of motion for the $O(3)$ sigma model {\it via} Pohlmeyer reduction \cite{Pohlmeyer:1975nb} reduce to the equations of motion for the sine-Gordon model. It is well-known that the sine-Gordon model has kink solutions.  Giant magnons are equivalent to these kink solutions. For more details we refer to the original paper of Hofman and Maldacena \cite{Hofman:2006xt} and to the review of Dorey \cite{DoreyReview}.

There is also an intermediate regime between the one of BMN particles and giant magnons - near plane wave regime \cite{Maldacena:2006rv}.  The momentum of excitations in this regime scales as $p\sim g^{-1/2}$. From the dispersion relation (\ref{dispersionrelation}) it follows that $E\sim g^{1/2}$. Interestingly, this scaling is characteristic not only for the near plane wave excitations but for the operators with finite value of $J$ \cite{Gubser:1998bc,Witten:1998qj}.

Let us now identify the considered excitations from the point of view of the gauge theory.

On the gauge theory side the string moving at the speed of light corresponds to the BPS vacuum (\ref{BPSvac}). The transversal fluctuations around this vacuum correspond to insertion of waves of some other operator $\CW$ different from $Z$:
\be
  |p\rangle=\sum_{n}e^{ip n}Tr(ZZZZ\CW ZZZZ).
\ee
The dispersion relation (\ref{dispersionrelation}) is the consequence of invariance under supersymmetry algebra \cite{Beisert:2006qh} as we will explain in the next chapter.

In the case of the $SU(2)$ sector $\CW=X$. At weak coupling the expression for the energy (\ref{dispersionrelation}) can be approximated by
\be
  E=8g^2\sin^2 p/2=2g^2(2-2\cos p).
\ee
In this expression we recognize, up to an overall factor $2g^2$, the dispersion relation for the XXX ferromagnet.

Let us summarize. At finite coupling there is one type of excitation - spin wave in the integrable spin chain. At strong coupling three possible scalings of this excitation are possible\footnote{numerical analysis shows that for multipartite solutions a more general scaling of the type $p\sim 1/g^{a},\ \ 0<a<1$ is possible. However there is no macroscopically significant amount of such particles.}: giant magnon ($p\sim 1$), near plane wave ($p\sim g^{-1/2})$), and BMN particle ($p\sim 1/g$).

\subsection{\label{sec:foldedandspinningstrings}Folded and spinning strings.}
A further possible generalization of the single particle excitations discussed above is to consider a string with two large angular momenta. If two of them, say $J_1$ and $J_3$, are on the sphere, we obtain the folded string rotating in $S^5$. In $AdS$ this is a point-like string situated at the center of $AdS$. If one momentum $J=J_3$ is on $S^5$ and another $S=S_1$ is on $AdS_5$, we get the folded string rotating in $AdS$. In $S^5$ this is a point-like string spinning around equator.

On a gauge side the folded string on $S^5$ corresponds to the multiparticle state in the $SU(2)$ sector, the folded string on $AdS_5$ corresponds to the multiparticle state in the $SL(2)$ sector. The $SU(2)$ case was initially analyzed in \cite{Beisert:2003xu}, then both $Sl(2)$ and $SU(2)$ cases were treated on the common footing in \cite{Beisert:2003ea}.

Let us discuss in more details these solutions in the one-loop approximation on the gauge side of the correspondence.

The Bethe equations are written as
\be
  \(\frac{u_k+\frac i2}{u_k-\frac i2}\)^{J}=-\prod_{j=1}^M\(\frac{u_k-u_j+i}{u_k-u_j-i}\)^{\eta},
\ee
where the notation is the following. For the $SU(2)$ case: $\eta=1,\ J=J_1+J_3,\ M=J_1$. For the $SL(2)$ case: $\eta=-1,\ J=J_3,\ M=S_1$.

The energy is given by
\be
  E=2g^2\sum_k\frac 1{u_k^2+\frac 14}.
\ee

It will be useful to introduce the ratio
\be\label{alphadef}
  \a=-\eta M/J.
\ee
The structure of the Bethe Ansatz restricts $\a$ to be larger than $-1/2$. The positive values of $\a$ can be arbitrary, however the solution that we discuss in this subsection requires $\a$ to be finite in the large $J$ limit.

Taking the logarithm of the Bethe equations we get
\be\label{logbethe}
  J\log\frac{u_k+\frac i2}{u_k-\frac i2}=2\pi i n_k+\eta\sum_{j\neq k}\log\frac{u_k-u_j+i}{u_k-u_j-i}.
\ee
For finite values of $\a$, positive or negative, and large values of $J$ the Bethe roots scale as $J$. Therefore we can approximate (\ref{logbethe}) by
\be\label{bethesimplified}
  \frac J{2u_k}=\pi n_k+\eta\sum_{j\neq k}\frac 1{u_k-u_j}.
\ee
We are interested in the solution with minimal energy and for simplicity we take $M$ to be even. Such a solution is symmetric and is defined by taking half of $n_k$ equal to 1 and other half equal to $-1$ \cite{Beisert:2003xu,Beisert:2003ea}.

\begin{figure}[t]
\centering
\includegraphics[width=6cm]{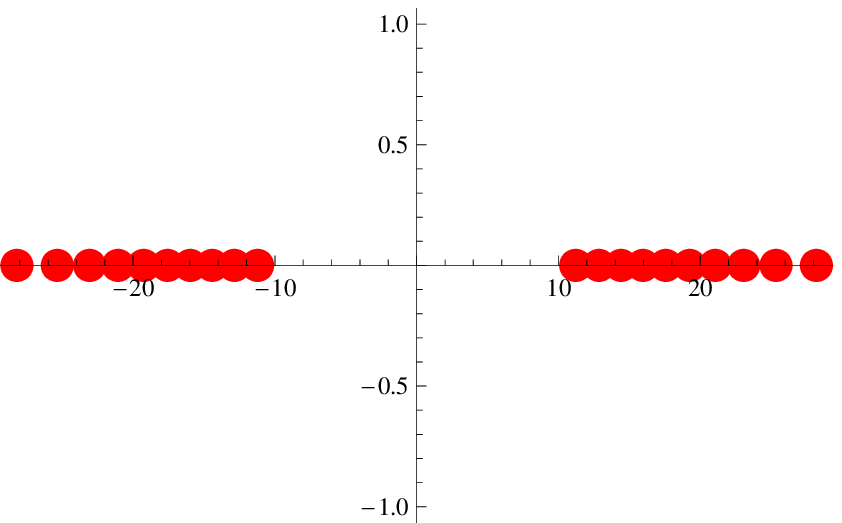}\hspace{3EM}
\includegraphics[width=6cm]{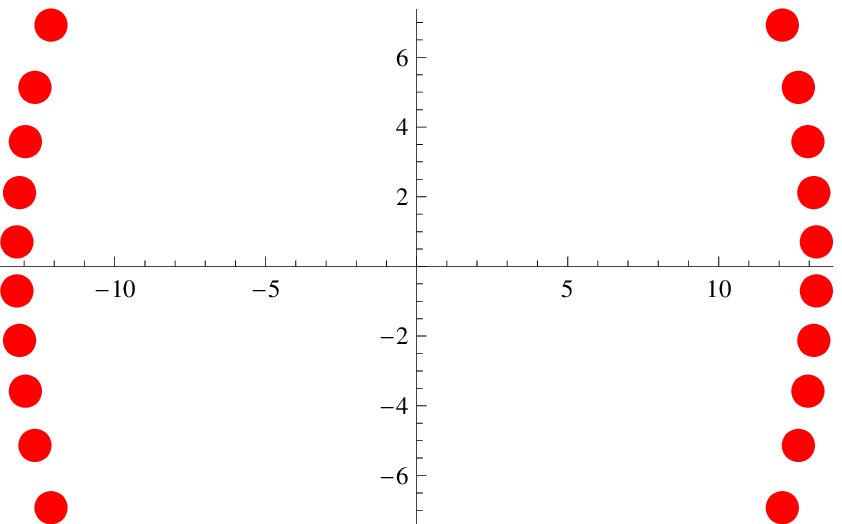}
\caption{\label{fig:sl2su2roots}Numerical solution to Bethe equations with $L=100$ and $M=20$. To the left: $Sl(2)$ case, to the right: $SU(2)$ case}
\end{figure}

One can interpret (\ref{bethesimplified}) as the force equilibrium equation in the classical mechanics. If $\eta$ is negative, then the particles repel along the real axis. This leads to their distribution on the real axis. If $\eta$ is positive, then the particles attract along the real axis. Instead, they repel in the imaginary direction. This leads to the fact that for the $SU(2)$ subsector the lowest energy solution is given by two symmetrically situated string-like configurations (see Fig.~\ref{fig:sl2su2roots}).

In the large $J$ limit (\ref{bethesimplified}) equation is solved by the techniques developed in the context of the $O(n)$ matrix model \cite{Brezin:1977sv,Kostov:1988fy}. Namely, we rely on the fact that the roots condensate on two cuts in the complex plane (see Fig.~\ref{fig:sl2su2roots}). To see condensation one should rescale $\hat u= u/J$.

Then it is useful to introduce the resolvent $G$
\be
  G=\frac 1J\sum_k\frac 1{\hat u-\hat u_k}.
\ee
This function is analytic everywhere except on the cuts. In the large $J$ limit (\ref{bethesimplified}) becomes
\be\label{eq5}
  \frac 1{2\hat u}=\pi n_k+\frac 12(G[\hat u+0]+G[\hat u-0]),
\ee
where $G[\hat u+0]$ and $G[\hat u-0]$ are the values of the resolvent from the right and from the left of the cut respectively. If we differentiate the last equation, then we obtain the equation for $dG/d\hat u$ which has the solution
\be\label{eqgu}
  \frac{dG[\hat u]}{d\hat u}=-\frac 1{J\hat u^2}+\frac 1{J^2}\frac 1{\sqrt{(\hat u^2-a^2)(\hat u^2-b^2)}}\(\frac{ab}{2\hat u^2}+J-M\).
\ee
To fix the solution we used the fact that the resolvent should be analytic outside the cuts and decrease as $M/(J\hat u)$ at infinity.

The branch points of the resolvent are $\pm a$ and $\pm b$. $a$ and $b$ are real (and are defined to be positive) in the $SL(2)$ case and complex conjugated (and are defined to have positive real part) in the $SU(2)$ case. The values of $a$ and $b$ are fixed from
the additional condition
\be\label{eq6}
  \int_\infty^{\pm a} d\(G[\hat u]-\frac 1{\hat u}\)=\int_\infty^{\pm b} d\(G[\hat u]-\frac 1{\hat u}\)=\mp\pi\eta J,
\ee
which follows from (\ref{eq5}).

The energy found from the solution (\ref{eqgu}) with $a$ and $b$ fixed by (\ref{eq6}) is given by \cite{Beisert:2003ea}:
\be\label{Egauge}
E=-\frac{8\pi^2\eta}{J}{\rm K}[q]((2-q){\rm K}[q]-2{\rm E}[q]),\ \ \a=\frac 1{2\sqrt{1-q}}\frac{{\rm E}[q]}{{\rm K}[q]}-\frac 12,
\ee
where ${\rm E}[q]$ is the elliptic integral of the first kind and ${\rm K}[q]$ is the elliptic integral of the second kind:
\be
  {\rm E}[q]\equiv\int_0^{\frac\pi 2}d\phi\sqrt{1-q\sin^2\phi},\ \ {\rm K}[q]\equiv\frac {d\phi}{\sqrt{1-q\sin^2\phi}}.
\ee

Interestingly, the energy in two different sectors is defined by a single analytic function ($\eta E$) of the parameter $\a$.

To obtain the result (\ref{Egauge}) one considers the large $J$ limit while the coupling constant is small (zero in fact). Of course, string perturbative solution is valid only for large values of the coupling constant. Therefore the large $J$ limit on a string theory side should be formulated differently. The first step which is used for the calculations on the string theory side is to consider the limit
\be\label{stringlimit2}
  \sqrt{\l}\sim J_i\sim S_i\to\infty\ \ {\rm with}\ \ \CJ_i=J_i/\sqrt{\l},\ \ \CS_i=S_i/{\sqrt{\l}}\ \ {\rm fixed}.
\ee
This limit can be thought as introduction of the BMN couplings $\l'_i\sim 1/\CJ_i^2$. Some of the charges may be zero. Then the string movement in the corresponding  direction is switched off and we do not require the scaling $J_i(S_i)\sim\sqrt{\l}$ for such charges.

In the limit (\ref{stringlimit2}) the classical solution for the equations of motion is known exactly \cite{Frolov:2003qc,Beisert:2003ea}. The energy for such classical solution can be represented as
\be
  E=\sqrt{\l}E[\CJ_i,\CS_i].
\ee
To be precise, let us consider the $SU(2)$ case. The large $\CJ=\CJ_1+\CJ_2$ limit with fixed $\a=-\CJ_1/\CJ$ leads to the following expansion of the energy
\be
  E=\sqrt\l\(\CJ+\frac{E_1[\a]}{\CJ}+\frac{E_2[\a]}{\CJ^3}+\frac{E_3[\a]}{\CJ^5}+\ldots\).
\ee
The term containing $E_1$ indeed coincides with (\ref{Egauge}) \cite{Beisert:2003ea}. $\sqrt{\l}E_2/\CJ^3$ agrees with two loop prediction from the gauge theory side \cite{Beisert:2003ea}. However, $\sqrt{\l}E_3/\CJ^5$ does not match with the three loop prediction from the gauge theory side. This does not mean the violation of the AdS/CFT correspondence but that the BMN scaling is not valid at this order of perturbation theory.

Let us take a look once more on the solution (\ref{eqgu}). $dG[u]$ can be viewed as a meromorphic differential which defines a Riemann surface. Therefore the solution of the Bethe Ansatz in the thermodynamic (large $J$) limit is characterized by the algebraic curve \cite{Kazakov:2004qf}.

The algebraic curve appears also on the string side. Indeed, let us consider the differential $d\log\Omega[x]$, where $\Omega$ is defined by (\ref{mOmega}). This is a $8\times 8$ supermatrix. Let us take its particular eigenvalue $dp[x_0]$ at some point $x=x_0$. In general there are nontrivial cycles starting and ending at $x_0$, analytical continuation over which brings $dp[x]$ to another eigenvalue of $d\log\Omega[x]$. Therefore $dp[x]$ is the differential defined in general on the 8-sheeted Riemann surface. 

What we can do is not to compare the energies given from string theory and asymptotic Bethe Ansatz, but to compare the algebraic curves which with additional constraints are in one to one correspondence with given solutions. This is indeed can be done \cite{Kazakov:2004qf},\cite{Beisert:2005bm,Beisert:2004ag,Dorey:2006zj,Gromov:2007aq,Vicedo:2008jy}.
This approach is known as the finite gap method. It gives a more systematic way to describe possible string solutions. However, it is not evident how to apply it beyond the quasiclassical approximation.

Of course, it is not compulsory to look only for the solutions only in the BMN window. We can directly demand that the (all-loop)asymptotic Bethe Ansatz gives the same results at strong coupling as the perturbative results of the string sigma model. Using this demand it was possible to fix the dressing factor at the tree  \cite{Arutyunov:2004vx} and one loop \cite{Beisert:2005cw,Freyhult:2006vr,Hernandez:2006tk,Gromov:2007cd} levels of the strong coupling expansion.

\subsection{\label{sec:cusp}Operators with logarithmic scaling of energy}
Large $J$ ($\CJ$) scaling with fixed $\a$ is not the only possible regime for the operators that are described by two nonzero angular momenta. Another interesting regime in the $SL(2)$ case is when $S\to\infty$ and $J$ scales in way that $\a\to 0$.

The possibility of using this regime to test the AdS/CFT correspondence was initially recognized by Belitsky, Gorsky, and Korchemsky in \cite{Belitsky:2006en}. They showed that the spectrum in this regime depends on the value of the parameter\footnote{In \cite{Belitsky:2006en} the inverse parameter $\xi=j^{-1}$ was used. We use the notations of \cite{Freyhult:2007pz}}
\be
    j=\frac{J}{\log S}.
\ee
In particular they discussed the case $j\ll 1$ and showed that the one-loop correction to the anomalous dimension scales in this case as $\log S$:
\be\label{sud1}
  \Delta-S=f[g]\log S+\CO(1).
\ee
 Appearance of the parameter $j$ is not accidental. At sufficiently small $j$ the distance between the two cuts in the two-cut solution (like in Fig.~\ref{fig:sl2su2roots} on the right) can be approximated by the quantity proportional to $j$. This distance remains finite for finite $j$. Therefore $j$ can be thought as measure of distance between two cuts.

  When the cuts are situated at the finite distance from each other, the solution (\ref{eqgu}) becomes inappropriate, since due to the required rescaling $u\to S u$ (\ref{eqgu}) develops a singularity at the origin.

The logarithmic scaling regime acquired a lot of attention. In \cite{Alday2007,Freyhult:2007pz} it was shown that the logarithmic scaling (\ref{sud1}) also takes place in the case of finite $j$:
\be
   \Delta-S=f[g,j]\log S+\CO(1).
\ee
The quantity $f[g,j]$ is called the generalized scaling function.

\subsubsection{Cusp anomalous dimension}
The situation $j=0$ is realized in particular in the case when we are dealing with finite twist operators (\ref{twistJ}) with large number $S$ of covariant derivatives.
The composite operators with large number of covariant derivatives were investigated for arbitrary four-dimensional gauge theory. All of them obey the logarithmic scaling (\ref{sud1}) known also as a Sudakov scaling. 

For the case of twist two operators the scaling function $f[g]$ is equal to twice the cusp anomalous dimension \cite{Korchemsky:1988si,Korchemsky:1992xv}:
\be\label{fgammacusp}
    f[g]=2\Gamma_{cusp}[g].
\ee
The same equality holds for the ground state of arbitrary finite twist operator.

The cusp anomalous dimension was introduced in \cite{Korchemsky:1985xj,Korchemsky:1987wg}. This quantity depends on the gauge theory that we consider and it is important for calculation of the scattering amplitudes. For $\CN=4$ SYM the cusp anomalous dimension allows one to find exactly the four- and five-point gluon amplitudes \cite{Bern:2005iz}. The cusp anomalous dimension was found at weak coupling analytically up to three loops \cite{Vogt:2004gi,Kotikov:2004er} and numerically up to four loops, after an impressive effort \cite{Bern:2006ew,Cachazo:2006az}.

On a string theory side the finite twist operator correspond to the folded string rotating in $AdS$ \cite{Gubser:2002tv}. The logarithmic scaling appears due to the approaching of the ends of string to the boundaries of $AdS$. The factor of two in (\ref{fgammacusp}) corresponds to the fact that the folded string has two ends\footnote{A more general situation of the so called spiky strings that have $n$ cusps, was discussed in \cite{Dorey:2008vp}. For that string configurations the coefficient of proportionality in \ref{fgammacusp} was found to be $n$.}. A nice physical arguments for the origin of logarithmic scaling were given in \cite{Alday2007}. The cusp anomalous dimension was found at strong coupling  at tree level \cite{Gubser:2002tv,Frolov:2002av}, one \cite{Frolov:2006qe} and two \cite{Roiban:2007dq} loops.

Beisert, Eden, and Staudacher \cite{Beisert:2006ez} derived from the asymptotic Bethe Ansatz the integral equation, known as the BES equation, solution of which allows to find the scaling function $f[g]$ at finite values of the coupling constant. Their proposal was based on the work of Eden and Staudacher \cite{Eden:2006rx} and was different by introduction of the nontrivial dressing factor $\sigma[u,v]$. This was the first work where an exact nonperturbative proposal for the dressing factor, built on an earlier work \cite{Beisert:2006ib}, appeared.

The calculation of the scaling function $f[g]$ through the BES equation gave the first nontrivial test for the AdS/CFT integrability conjecture at four loops of weak coupling expansion \cite{Beisert:2006ez}. The strong coupling solution of the BES equation proved to more complicated. In \cite{Kotikov:2006ts,Alday:2007qf},\cite{KSV1,Beccaria:2007tk} the leading  order of the strong coupling expansion was found analytically. In \cite{Benna:2006nd} the first three orders were found numerically. The subleading order was obtained in \cite{Casteill:2007ct,Belitsky:2007kf} by means different from solving the BES equation (see section \ref{sec:perturbativeregime}). In \cite{Basso:2007wd},\cite{KSV2} a recursive procedure for analytical expansion to any desired order was given. The obtained results reproduced the string theory calculations at tree level \cite{Gubser:2002tv,Frolov:2002av}, one \cite{Frolov:2006qe}, and two \cite{Roiban:2007dq} and gave a strong evidence of the correctness of the asymptotic Bethe Ansatz. We present the results of \cite{KSV2} in section \ref{sec:cuspanomaly} and in appendix \ref{app:cusp}.

\subsubsection{Generalized scaling function}
As we already mentioned, the logarithmic scaling also takes place for finite values of the parameter $j$ \cite{Alday2007,Freyhult:2007pz}. A generalization of the BES equation which allows finding the generalized scaling function $f[g,j]$ was given in \cite{Freyhult:2007pz,Bombardelli:2008ah}. This generalization is known as the BES/FRS equation.  Since we have an additional parameter $j$, by solving this equation at strong coupling we can test the integrability on the functional level. The solution of the BES and BES/FRS equations was a subject of our work. We discussion these equations in chapter \ref{ch:BESFRS}.

\subsubsection{\label{sec:O6physical}$O(6)$ sigma model}
One important regime proposed by Alday and Maldacena in \cite{Alday2007} needs to be mentioned.
This is strong coupling limit in which $j$ is non-zero but is exponentially small with respect to the coupling constant.

To understand what happens in this case let us discuss the spectrum of excitations in the string sigma model. In the same way as it was for the circular string in S$^5$, the folded string rotating in AdS$_5$ creates a centrifugal force. However, this force leads to the nonzero mass only for bosonic fluctuations on AdS$_5$ and for fermionic fluctuations. The bosonic fluctuations on S$^5$ are left massless on the classical level. Therefore, if we consider low energetic fluctuations of the folded string in AdS, they will be described by the fluctuations on S$^5$, \textit{i.e.} by the O(6) sigma model. The other massive fluctuations serve as a cutoff for such low energy description.

On the quantum level the O(6) sigma model acquires a new mass scale via the mechanism of the dimensional transmutation. The mass of the particles can be given through the 't Hooft coupling constant which in the quasiclassical approximation coincides with the coupling constant of O(6) sigma model:
\be\label{mo6}
  m_{O(6)}=k g^{1/4}e^{-\pi g}(1+\CO(g^2)).
\ee
The power ${1/4}$ and the exponential factor are defined through the beta function of the sigma model (see chapter \ref{ch:massgap}). The constant $k$ is not universal and depends on how the O(6) sigma model is embedded in the whole string sigma model.

The strong coupling and the large $S$ regime with $j\sim m$ is described by the $O(6)$ sigma model in the presence of the chemical potential which creates a finite density of particles (proportional to $j$) with rapidities on an interval $[-B,B]$. The boundary of the interval $B$ is a function $j/m$:
\be
  j/m=\sqrt{B}e^{B}+...\ .
\ee
When $B$ becomes of order of $\pi g$, we reach the cutoff where the $O(6)$ sigma model is no more appropriate. At larger energies we have to deal with the full spectrum of the string sigma model.

The $O(6)$ sigma model is the theory with a non-zero beta function which allows in fact to have the parameter of dimensional transmutation (\ref{mo6}). From the other side, the parameter (\ref{mo6}) is defined in terms of the 't Hooft coupling constant which has a vanishing beta function. This seeming contradiction is resolved in the following way. We consider a system which depends on three parameters: $g, J$, and $\log S$. In the regime when the $O(6)$ sigma model is applicable, it sufficient to have two parameters: $M=m\log S$ and $J=j\log S$. The reason why we multiplied by $\log S$ is the following: if we normalize the worldsheet volume to some constant, the mass of AdS excitations will scale as $\log S$ at large $S$. Therefore $\log S$ serves effectively as a mass scale. If we simultaneously change $g$ and $\log S$ (keeping $J$ constant) in a way that $M$ do not change, then the physics of the $O(6)$ sigma model will not change. We see that effectively $g$ runs with $\log S$, and  (\ref{mo6}) reflects the nontrivial beta-function of this dependence.

The BES/FRS equation in the Alday-Maldacena regime was investigated in \cite{Basso:2008tx,Fioravanti:2008ak,Fioravanti:2008bh}. In \cite{Basso:2008tx}
Basso and Korchemsky derived the integral equation for the $O(6)$ sigma model from the BES/FRS equation. They also derived the proper expression (\ref{mo6}) for the mass scale and the explicit value of the coefficient $k$. In \cite{Fioravanti:2008rv},\cite{Fioravanti:2008ak,Fioravanti:2008bh} a numerical evidence for the presence of the mass scale (\ref{mo6}) was given. In subsection \ref{sec:O6} we will give an alternative derivation of the integral equation for the $O(6)$ sigma model however without the derivation of the coefficient $k$.


\section{Summary}
The AdS/CFT correspondence gives an explicit realization of long standing attempts to formulate duality between gauge and string theories. Despite its strong/weak coupling type, the AdS/CFT correspondence can be directly verified, at least on the level of the spectral problem, using the integrability technique. The problem of diagonalization of the dilatation operator is shown to be equivalent to the problem of diagonalization of the Hamiltonian of a certain spin chain conjectured to be integrable. This problem is solved using the bootstrap approach analogical to the one proposed by Zamolodchikov and Zamolodchikov. The approach is based on the assumption of the factorized scattering which allows to express all the processes in terms of the two-particle S-matrix. The two-particle S-matrix can be uniquely fixed by symmetry, unitarity, crossing, and an assumption about the particle content\footnote{As we discuss in the next chapter, a strange feature of the $su(2|2)$ algebra allows to fix the S-matrix even without using of Yang-Baxter equation.}. The asymptotic Bethe Ansatz which can be derived once the scattering matrix is known gives a possibility to explicitly verify the integrability conjecture by comparison of the Bethe Ansatz results with perturbative calculations on both sides of the AdS/CFT correspondence. In this chapter we discussed checks that are based on the consideration of such objects as single-particle excitations (magnons), folded and spinning strings, operators with logarithmic scaling of energy. The verifications of the asymptotic Bethe Ansatz are done today up to four loops on the gauge side and up to two loops on the string side. Based on the thermodynamic Bethe Ansatz, which we did not discuss here, recently even five-loop check of integrability was done. These results provide a strong evidence that the AdS/CFT system is indeed integrable.

The properties of operators with logarithmic scaling of energy was a subject of our work. We will present the details of our calculations in chapters \ref{ch:sl2} and \ref{ch:BESFRS}. Before this, in chapter \ref{ch:bootstrap} we discuss in details the derivation and main properties of the asymptotic Bethe Ansatz.

\chapter{\label{ch:bootstrap}Bootstrap approach and asymptotic Bethe Ansatz}
In this chapter we discuss the derivation of the asymptotic Bethe Ansatz of the AdS/CFT integrable system
. The derivation is based on determination of the scattering matrix using the symmetry constraints and then application of the nested or algebraic Bethe Ansatz procedure. Technically the scattering matrix is found based on the arguments similar to those of Zamolodchikov and Zamolodchikov \cite{Zamolodchikov:1978xm} for relativistic systems. However, the AdS/CFT scattering matrix has a different physical interpretation. While the approach of \cite{Zamolodchikov:1978xm} deals with the excitations over the true vacuum of the theory, in the AdS/CFT case we consider excitations over the BMN state $\Tr Z^J$ which is more likely as a pseudovacuum of the XXX spin chain.

The BMN state $\Tr Z^J$ is invariant under the psu(2$|$2)$^2\rtimes\MR$ subalgebra of the whole psu(2,2$|$4) symmetry of the system\cite{Beisert:2004ry}. The generator of the central extension of this subalgebra is given by
\be\label{Definition of C}
    C=\frac 12(\hat\Delta-J).
\ee
%

In the Bethe Ansatz description an excitation of $\Tr Z^J$ is seen as the state composed from a fixed number of magnons (see section \ref{sec:coordinate}). Far from the collision point a single magnon is given by a plane wave:
\be\label{magnon1}
  &&\hspace{7.5EM} n{\rm-th\ position}\no\\
  |\Upsilon^A[p]\rangle&=&\sum_n e^{i p n}|\ldots ZZZ\Upsilon^A ZZ\ldots\rangle.
\ee
Here $A$ stands for a type of a magnon.

Although an excitation over the BMN state should be covariant under the action of psu(2$|$2)$^2\rtimes\MR$, a single magnon does not have this property. The symmetry algebra closes when acting on the one magnon configuration only for the case $p \in 2\pi\MZ$. 

It was suggested in \cite{Beisert:2006qh} that we should extend the symmetry algebra with two additional central charges to get a closed action on (\ref{magnon1}). Then a symmetry algebra which allows us to study a magnon scattering is  psu(2$|$2)$^2\rtimes\MR^3$. The appearance of two additional central charges was confirmed from the string theory perspective \cite{Arutyunov:2006ak}. From the point of view of the gauge theory such analysis was not done. Though, the idea of central extension leads to simple and self-consistent explanations for many properties of the AdS/CFT system. This makes it hard to doubt in the correctness of the suggestion.

We see that it is logical to start discussion by reviewing the basic properties of the centrally extended $psu(2|2)$ algebra and its representations. After that we come back to the discussion of the magnons and their scattering and give the expression for the scattering matrix up to an overall scalar factor. These topics are the subject of the section \ref{sec:BS1} in which we follow closely the ideas of \cite{Staudacher:2004tk,Beisert:2006qh,Beisert:2005tm}\footnote{Mostly a gauge theory point of view on the scattering is discussed here. A complementary approach based on a Zamolodchikov-Faddeev algebra and more natural for the string theory is given in \cite{Arutyunov:2006yd}.}.
In section \ref{Sec:BSasBA} we formulate the asymptotic Bethe Ansatz proposed in \cite{Beisert:2005fw} and derived in \cite{Beisert:2006qh,Martins:2007hb,deLeeuw:2007uf}. Then in section \ref{sec:crossingequations} we formulate the crossing equations (based on \cite{Janik:2006dc}, see also \cite{Arutyunov:2006yd}) on the overall scalar factor of the S-matrix, explicitly solve them, and formulate a sufficient set of conditions that uniquely fix the solution. The presented solution of the crossing equations is based on \cite{V3}. The resulting expression for the scalar factor was initially proposed in \cite{Beisert:2006ib,Eden:2006rx} and then checked to solve the crossing equations in \cite{Arutyunov:2009kf}. We finish this chapter with section \ref{sec:Kernels} where is given, based on Cauchy type integral operators, a shorthand notation for the Bethe equations and following from them in a thermodynamic limit functional equations.

\section{\label{sec:BS1}Scattering matrix from symmetry constraints}
\subsection{$su(2|2)$ algebra}
We will start from the $psu(2|2)$ algebra with one central charge. This is just an $su(2|2)$ algebra.

To describe the structure of $su(2|2)$ algebra it is instructive to consider its four-dimensional representation. The vector space of this representation is a direct sum $V_0\oplus V_1$ of two-dimensional spaces: $V_0$ with even grading and $V_1$ with odd grading. The $su(2|2)$ generators can be combined in the following block matrix:
\be
\left(%
   \begin{array}{c|c}
    L^{a}_{b} & Q^{a}_{\b} \\
    \hline
    S^{\a}_{b} & \CL^{\a}_{\b} \\
    \end{array}%
    \right)\left(%
   \begin{array}{c}
    V_0 \\
    \hline
    V_1 \\
    \end{array}%
    \right)\rightarrow \left(%
   \begin{array}{c}
    V_0 \\
    \hline
    V_1 \\
    \end{array}%
    \right).
\ee
If to denote the basis of $V_0$ by $|\phi^a\rangle$ and the basis of $V_1$ by $|\psi^\a\rangle$ then the action of the generators is the following:
\be\label{definingrep}
  L^a_b|\phi^c\rangle &=& \delta_b^c|\phi^a\rangle-\frac 12\delta^a_b|\phi^c\rangle,\no\\
  \CL^\a_\b|\psi^\g\rangle &=& \delta_\b^\g|\psi^\a\rangle-\frac 12\delta^\a_\b|\psi^\g\rangle,\no\\
  Q^a_\b|\psi^\g\rangle &=& \delta_\b^\g |\phi^a\rangle,\no\\
  S^\a_b|\phi^c\rangle &=& \delta_c^b |\psi^\a\rangle\;.
\ee

The explicit action of $L^a_b$ is given by
\be
L^1_1=-L^2_2=\frac 12\mtwo 100{-1}
,\ \ L^1_2=\mtwo 0100,\ \ L^2_1=\mtwo 0010.
\ee
Commutation relations in the algebra are described as follows. $L^a_b$ and $\CL^\a_\b$ form the $su(2)\oplus su(2)$ bosonic subgroup and act on $Q,S$ by the fundamental representation for the correspondent index. The nonzero commutator\footnote{When dealing with superalgebras, by commutator we always understand the graded commutator.} between the fermionic generators is given by
\be
  [Q^a_\a,S^\b_b]=\delta^a_b \CL^{\b}_{\a}+\delta^\a_\b L^a_b+\delta^a_b \delta^\a_\b C,
\ee
where the central charge $C$ equals $1/2$ for the four-dimensional representation. Abstract definition of $su(2|2)$ has no restriction on $C$. For arbitrary value of $C$ the minimal dimension of the representation is $16$. The condition $C=1/2$ is also called a shortening condition since the four-dimensional representation has a smaller dimension than the minimal possible dimension for arbitrary $C$.

An exceptional feature of the $su(2|2)$ algebra is that its central extension  has two central charges in addition to $C$, therefore a complete centrally extended $su(2|2)$ algebra is $su(2|2)\rtimes\MR^2=psu(2|2)\rtimes\MR^3$. The centrally extended algebra possesses $SU(2)$ group of external automorphisms. These features are due to the fact that there exists an antisymmetric invariant bilinear form $\varepsilon_{ab}$ which allows raising and lowering of indexes. Due to this the adjoint actions of the bosonic subalgebra $su(2)\oplus su(2)$ on $Q$ and on $S$ are equivalent.
 To fix the notation we define
\be
  v_a=\e_{ab}v^b,\ \ \e^{ab}\e_{bc}=\delta^{a}_{c},\ \ \e_{12}=1.
\ee

 Let us introduce
\be
  Q_{a\a 1}=\varepsilon_{ab}Q^b_\a, \ \ Q_{a\a 2}=\varepsilon_{\a\b}S^{\b}_a\;.
\ee
Then the commutation relations of the fermionic generators are given by\footnote{This form of the commutator is related to the fact that the centrally extended $su(2|2)$ can be viewed as a ${\bf{a}}\to 0$ limit of the exceptional Lie superalgebra $D(2,1;{\bf{a}})$. The latter has $su(2)\oplus su(2)\oplus su(2)$ bosonic subalgebra. The generators of the bosonic subalgebra are identified with $L_{ab},\CL_{\a\b}$ and $\tilde \CC_{mn}=\CC_{mn}/{\bf{a}}$. Fermionic subspace is spanned by $Q_{a\a m}$. Adjoint action of $L,\CL$ and $\tilde \CC$ on $Q$-s is given by the fundamental representation related respectively to $a,\a$ and $m$ indices. The commutator of $Q$-s is given by
\be\label{QQo21ecommutator}
   [Q_{a\a m},Q_{b\b n}]=\varepsilon_{\a\b}\varepsilon_{mn}L_{ab}+(1-{\bf{a}})\,\varepsilon_{ab}\varepsilon_{mn}\CL_{\a\b}+{\bf{a}}\,\varepsilon_{ab}\varepsilon_{\a\b}\tilde \CC_{mn}.
\ee
For generic ${\bf{a}}$ the generators $\CC_{nm}$ do not commute with each other. But they do in the limit ${\bf{a}}\to 0$. In this limit we recover (\ref{QQsu22commutator}) from (\ref{QQo21ecommutator}).}
\be\label{QQsu22commutator}
  [Q_{a\a m},Q_{b\b n}]=\varepsilon_{\a\b}\varepsilon_{mn}L_{ab}+\varepsilon_{ab}\varepsilon_{mn}\CL_{\a\b}+\varepsilon_{ab}\varepsilon_{\a\b}\CC_{mn}.
\ee
$\CC_{mn}$ is a symmetric matrix of central charges. $\CC_{12}=\CC_{21}=-C$. In the notations of \cite{Beisert:2006qh} $\CC_{11}=-P$ and $\CC_{22}=-K$.

The $SU(2)$ outer automorphisms rotate the third index of $Q_{a\a m}$.
The matrix $\CC_{mn}$ form a symmetric three-dimensional representation of the external $SU(2)$ group with an invariant $-\det \CC=C^2-PK$.

The existence of a four dimensional representation is possible if a shortening condition $-\det \CC=1/4$ is satisfied. We will denote such representation by $\langle{\CC}\rangle\equiv\langle C,P,K\rangle$.

For $P K\neq 0$ the four dimensional representation is not faithful since $P\propto K\propto C\propto id$. The faithful representation with nonzero central charges is infinite dimensional. The faithful infinite dimensional representation is important for identification of the scattering matrix. Therefore we will now discuss this representation.

The basis in the vector space of the faithful representation is chosen to be the following: $|\phi^a Z^n\rangle$ and $|\psi^\a Z^n\rangle$, where $n$ is integer. The bosonic generators $L$ and $\CL$ do not act on $Z$. They act on $\phi^a$ and $\psi^\a$ according to (\ref{definingrep}).

The action of the other generators is the following
\be\label{exactsu22rep}
  Q^a_\b|\psi^\g Z^n\rangle &=&a\, \delta_\b^\g |\phi^a Z^n\rangle,\no\\
  Q^a_\b|\phi^c Z^n\rangle &=&b\, \e^{ac}\e_{\b\g} |\psi^\g Z^{n+1}\rangle,\no\\
  S^\a_b|\phi^c Z^n\rangle &=&d\, \delta_c^b |\psi^\a Z^n\rangle,\no\\
  S^\a_b|\psi^\g Z^n\rangle &=&c\, \e^{\a\g}\e_{bc}|\phi^c Z^{n-1}\rangle,\no\\
  P|\Upsilon Z^n\rangle &=& ab|\Upsilon Z^{n+1}\rangle,\no\\
  K|\Upsilon Z^n\rangle &=& cd|\Upsilon Z^{n-1}\rangle,\\
 \label{valueC}  C|\Upsilon Z^n\rangle &=& \frac{ad+bc}2|\Upsilon Z^n\rangle,
\ee
where $\Upsilon$ is either $\phi^a$ or $\psi^\a$.
\begin{figure}[t]
\centering
\includegraphics[width=6.52cm]{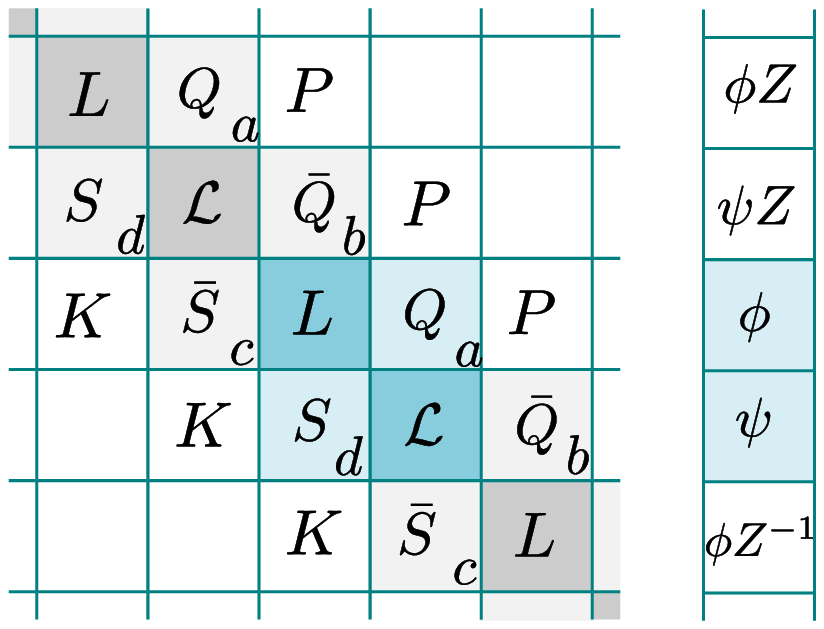}
\caption{\label{fig:su22infrep}Faithful representation of centrally extended $su(2|2)$.}
\end{figure}
The closure of the commutation relations requires that
\be\label{closurecondition}
  ad-bc=1.
\ee
The case (\ref{definingrep}) corresponds to $P=K=0$, $a=d=1$ and $b=c=0$. If we reduce the representation to four dimensions by identifying $|\Upsilon Z^n\rangle\equiv |\Upsilon Z^m\rangle$, it is easy to see that (\ref{closurecondition}) is equivalent to the shortening condition $C^2-PK=1/4$.


\subsection{Action of the algebra on 1-particle and two particle states}
According to the nested Bethe Ansatz construction which was explained in the first chapter, a generic one-particle state is given by
\be
  &&\hspace{7.5EM} n{\rm-th\ position}\no\\
  |\Upsilon[p]\rangle&=&\sum_n e^{i p n}|\ldots ZZZ\Upsilon ZZ\ldots\rangle.
\ee
The action of the symmetry algebra on this state is given by (\ref{exactsu22rep}) if to identify the field $Z$ with the markers $Z$ in (\ref{exactsu22rep}) and the field $\Upsilon$ with $\phi^a$ or $\psi^\a$ \cite{Beisert:2005tm}. Notice an unusual property of the symmetry algebra: its action changes the length of the spin chain.

In the basis $|\Upsilon[p]\rangle$ the representation (\ref{exactsu22rep}) rearranges into four-dimensional blocks. The action of the generators on $|\Upsilon[p]\rangle$ is the same as in (\ref{exactsu22rep}) except for the replacements:
\be\label{repl1}
  |\Upsilon Z^n\rangle\to |\Upsilon[p]\rangle,\ \ |Z^n\Upsilon \rangle\to \e^{-i p n}|\Upsilon[p]\rangle.
\ee
The second replacement plays an important role for the case of two and more particles and in particular leads to a nontrivial coproduct structure. Indeed, a generic two-particle state can be written as
\be
    &&\hspace{7.5EM} n_1{\rm-th\ position}\hspace{2.5EM} n_2{\rm-th\ position}\no\\
  |\Upsilon_1[p_1]\Upsilon_2[p_2]\rangle&=&\sum_{n_1<n_2} e^{i p_1 n_1+i p_2 n_2}|\ldots ZZZ\Upsilon_1 ZZ\ldots ZZ\Upsilon_2 ZZZ\ldots\rangle.
\ee
Therefore
\be
  |\Upsilon_1Z^n,\Upsilon_2 Z^m\rangle \to e^{-i p_2 n}|\Upsilon_1[p_1]\Upsilon_2[p_2]\rangle.
\ee
As a consequence, the action of the generators that change the number of $Z$-s contain an additional factor $e^{-ip_2}$. In particular, the co-product of $P$ is given by \cite{Plefka:2006ze,Beisert:2006qh}
\be
  \Delta P=1\otimes P[p_2]+P[p_1]\otimes e^{-ip_2}.
\ee
To put the coproduct in the symmetric form
\be
  \Delta P\propto (1-e^{-ip_1-ip_2})1\otimes 1
\ee
one should impose
\be
  P[p]=\hat g\;\a (1-e^{-ip}),
\ee
where $\hat g$ and $\a$ are two arbitrary parameters.

We see that the representation becomes dependent on the momentum $p$ of the particle.

Equivalently, the action of $K$ is given by $K[p]=\hat g\;\a^{-1}(1-e^{ip})$. From the shortening condition we obtain the expression for C:
\be\label{disper2}
  C=\frac 12\sqrt{1+16\;\hat g\;^2\sin^2\frac p2}.
\ee

In view of (\ref{Definition of C}) and  (\ref{definition of H}), the energy of the magnon should be identified with $2C-1$. Therefore (\ref{disper2}) can be seen as the dispersion relation.

The constant $\hat g$ cannot be fixed using the properties of the $psu(2|2)$ algebra. However, the dispersion relation (\ref{disper2}) was derived in \cite{Santambrogio:2002sb} using a different approach. Based on \cite{Santambrogio:2002sb} it is possible to identify $\hat g$ with the coupling $g$ defined by (\ref{gdefinition}).

\paragraph{Parametrization.}

Using rescaling of basis vectors in (\ref{exactsu22rep}) we can always adjust the constants $a,b,c,d$ to fulfill the following two criteria. First, we demand that $|\a|=1$. We can even put $\a=1$, however for the reasons explained below $\a$ should be kept explicitly. Second, we demand $a=d$. Using these restrictions, the constants $a,b,c,d$ are uniquely parameterized as:
\be\label{abcdcoef}
  a&=&\eta,\ b=g\frac{\a}{\eta}\(\ 1-e^{-ip}\ \),
  d=\eta,\ c=g\frac{\a^*}{\eta}\(\ 1-e^{ip}\ \)\;.
\ee
The parameter $\eta$ is fixed by the shortening condition (\ref{closurecondition}) which implies
\be\label{etacondition}
    \eta^4-\eta^2-4g^2\sin^2\frac p2=0\;.
\ee
We should choose the solution with the property: $\eta\to 1$ when $g\to 0$.

We see that the representation is uniquely defined by the momentum of the particle $p$ and the coupling constant $g$.

\paragraph{Hopf algebra picture.}

In the Hopf algebra picture, proposed for the AdS/CFT scattering matrix in \cite{Plefka:2006ze},  for each particle with momentum $p$ we associate the representation $\langle{\CC}\rangle$ with the coefficients $a,b,c,d$ given by (\ref{abcdcoef}). The coefficient $\a$ should be equal for all particles and may be put to $1$.

The coproduct is nontrivial for $P,K,S,Q$ and is given by
\be
  \Delta P&=&1\otimes P+P\otimes e^{-ip_2},\no\\
  \Delta K&=&1\otimes K+K\otimes e^{ip_2},\no\\
  \Delta Q&=&1\otimes Q+Q\otimes e^{-ip_2/2},\no\\
  \Delta S&=&1\otimes S+S\otimes e^{ip_2/2}.
\ee
The two-particle wave function is given by
\be\label{2particlewavefunction}
  \Psi_{12}=|\Upsilon_1[p_1]\Upsilon_2[p_2]\rangle+S_{12}|\Upsilon_1[p_2]\Upsilon_2[p_1]\rangle
\ee
where $S$-matrix should satisfy
\be
  [\Delta[J],S]=0.
\ee
In this notation $S$ matrix acts on two $4$-dimensional representations:
\be
S:\langle{\CC}[p_1]\rangle\otimes\langle{\CC}[p_2]\rangle\to \langle{\CC}[p_1]\rangle\otimes\langle{\CC}[p_2]\rangle.
\ee

\paragraph{Twisted picture}
The Hopf product becomes the standard one if we redefine the value of $\a$. Before scattering we will put $\a[p_1]=e^{-ip_2}$ and $\a[p_2]=1$. After scattering it will be opposite $\a[p_1]=1$, $\a[p_2]=e^{-ip_1}$. In this picture the $S$-matrix is invariant under the standard coproduct:
\be
[1\otimes J+J\otimes 1,S]=0,
\ee
but the representations of the particles before scattering are different from the representations of the particles after scattering:
\be
  S:\langle{\CC_1}\rangle\otimes\langle{\CC_2}\rangle&\to& \langle{\CC_1'}\rangle\otimes\langle{\CC_2'}\rangle,\no\\
  \CC_1&=&\langle C_1,e^{-ip_2}P_1,e^{ip_2}K_1\rangle,\no\\
  \CC_2&=&\langle C_2,P_2,K_2\rangle,\no\\
  \CC_1'&=&\langle C_1,P_1,K_1\rangle,\no\\
  \CC_2'&=&\langle C_2,e^{-ip_1}P_2,e^{ip_2}K_2\rangle.\\
\ee

\subsection{Zhukovsky parameterization}
Let us introduce two parameters $\xp$ and $\xm$ defined by
\be
   e^{ip}=\frac \xp\xm,\ \ \eta^2=-ig(\xp-\xm).
\ee
The shortening condition (\ref{etacondition}) is then expressed as
\be
   \left(\xp+\frac 1{\xp}\right)-\left(\xm+\frac 1\xm\right)=\frac ig.
\ee
This equation defines an algebraic curve which is a torus. A useful parameterization, not uniform however, for the torus is given in terms of the parameter $u$ defined by:
\be
    \xpm+\frac 1\xpm=\frac {u\pm \frac i2}g.
\ee
Let us define the Zhukovsky map $x[u]$ by the relation
\be\label{Juk}
  x+\frac 1x=\frac ug,\ \ x=\frac u{2g}\(1+\sqrt{1-\frac {4g^2}{u^2}}\).
\ee

Then the parameters $\xp$ and $\xm$ are the double-valued functions of $u$:
\be
    \xp=x\[u\pm\frac i2\].
\ee
A choice of the branches of these functions is constrained by the demand that the $g\to 0$ limit should be smooth. If momentum of the particle is real at all values of the coupling constant then smooth $g\to 0$ limit implies that $|\xpm|>1$. The region defined by $|\xpm|>1$ is usually called the physical region.

The parameter $u$ is an equivalent of the rapidity in the Heisenberg ferromagnet and for $g=0$ coincides with it. Let us show the coincidence on the example of the energy:
\be
 E[p]=\sqrt{1+16g^2\sin^2\frac p2}-1=\frac 2{\xp\xm-1}=2ig\left(\frac 1\xp-\frac 1\xm\right).
\ee
In the limit $g\to 0$ we get
\be
    E[p]\to\frac{2g^2}{u^2+\frac 14}
\ee
which is a correct expression for the energy of the XXX magnon (\ref{Energyformula}).

\subsection{Scattering matrix}
When central charges do not obey any shortening condition, a minimal dimension of an irrep of $psu(2|2)\rtimes\MR^3$ algebra is $16$. In general position the three central charges of the tensor product of two short irreps do not obey a shortening condition. Therefore the tensor product of two four-dimensional representations of the $psu(2|2)\rtimes\MR^3$ algebra is in general position an irreducible representation. Thus the $S$-matrix can be fixed up to a scalar factor from symmetry constraints.

The matrix elements of scattering matrix obtained in Hopf algebra formulation differ from those obtained in twisted picture by a phase. This is due to the different choice of the parameter $\a$ and therefore a basis redefinition. Different choices of $\a$ distinguish $S$-matrix obtained from the Hopf algebra formulation, $S$-matrix obtained from a Zamolodchikov-Faddeev algebra \cite{Arutyunov:2006yd}, and $R$-matrix of the Hubbard model \cite{Martins:2007hb}. In the following we deal with the $S$-matrix which is used for construction of a wave function \ref{2particlewavefunction}.

The structure of the $S$-matrix is quite involved. It can be found in the original work of Beisert \cite{Beisert:2005tm}. Here we will give the $sl(2|1)$ invariant formulation
of the scattering matrix presented in \cite{Beisert:2005tm}.

The infinite dimension representation (\ref{exactsu22rep}) is decomposed into four-dimensional blocks if we consider the action of the $sl(2|1)$ subalgebra. This subalgebra is constructed from the generators $L^1_1,\CL^\a_\b,Q^2_\a,S^\a_2$ and the central charge $C$. The commutation relations of $sl(2|1)$ algebra are recovered after the following identification with the generators of Sec.~\ref{sec:sl21algebra}:
\be
  B&=&C-L_1^1,\ \ h=\CL^1_1,\ \ e=\CL^1_2,\ \ f=\CL^2_1\;,\no\\
  Q_+&=&S^1_2,\ \ Q_-=S^2_2,\ \ S_+=-Q^2_2,\ \ S_-=Q^2_1\;.\no\\
  .
\ee
The subspace $\{\phi^2Z^n,\psi^1Z^n,\psi^2Z^n,\phi^1 Z^{n-1}\}$ is invariant under the action of the $sl(2|1)$ subalgebra and forms the typical irrep $\langle 1/2,b\rangle$ with $b=C=\frac {ad+bc}2$. The tensor product of two such irreps of $sl(2|1)$ is decomposed into direct sum of three representations
\be
  \langle 1/2,b\rangle\otimes\langle 1/2,b'\rangle= \langle 1/2,b+b'-1/2\rangle\oplus\langle 1,b+b'\rangle\oplus\langle 1/2,b+b'+1/2\rangle
\ee
of dimension $4$, 8, and 4 respectively.
\begin{figure}[t]
\centering
\includegraphics[width=6.52cm]{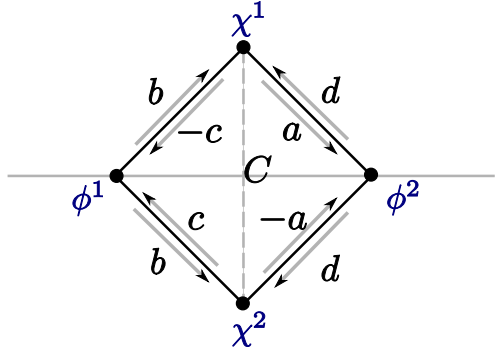}
\caption{\label{fig:weighttypicalSL21}Typical 4-dimensional representation of $SL(2|1)$. The normalization of the basis is chosen with respect to the imbedding to the $SU(2|2)\rtimes \MR^2$ algebra.}
\end{figure}
The scattering matrix is given by the sum of the projectors on these representations
\be\label{BSmat}
  S[\xpm,\ypm]=\frac{\yp-\xm}{\ym-\xp}{\rm proj}_1+{\rm proj}_2+\frac{\ym\xp}{\yp\xm}\frac{\yp-\xm}{\ym-\xp}{\rm proj}_3.
\ee
The scattering matrix has two special points. When $\xm=\yp$, the scattering matrix reduces to the ${\rm proj}_2$. 
When $\xp=\ym$ the scattering matrix reduces to ${\rm proj}_1+\frac{\ym\xp}{\yp\xm}{\rm proj}_3$. 
These properties of the scattering matrix allow performing the fusion procedure as it was in the case of rational R-matrices. In particular, the transfer matrices constructed from fused $S[\xpm[u],\ypm[v]]$ satisfy Hirota equation as a function of $u$ on a fat hook shape of $su(2|2)$ algebra.

\section{\label{Sec:BSasBA}Beisert-Staudacher asymptotic Bethe Ansatz}
The asymptotic Bethe Ansatz was proposed by Beisert and Staudacher \cite{Beisert:2005fw} before the full factorized $S$-matrix was found. Later it was rigorously derived based on the knowledge of the $S$-matrix (\ref{BSmat}) and using the coordinate nested Bethe Ansatz \cite{Beisert:2006qh,deLeeuw:2007uf} and the algebraic Bethe Ansatz \cite{Martins:2007hb}.
It is given by the following set of equations:
\be
  (-1)^\e&=&\prod_{j=1}^{K_2}\frac{u_{1,i}-u_{2,j}-\frac i2}{u_{1,i}-u_{2,j}+\frac i2}\prod_{j=1}^{K_4}\frac{1-\frac 1{x_{1,i}\xm_{4,j}}}{1-\frac 1{x_{1,i}\xp_{4,j}}}\sqrt{\frac{\xm_{4,j}}{\xp_{4,j}}}\;,\no\\
  -1&=&\prod_{j=1}^{K_2}\frac{u_{2,i}-u_{2,j}+i}{u_{2,i}-u_{2,j}-i}\prod_{j=1}^{K_1}\frac{u_{2,i}-u_{1,j}-\frac i2}{u_{2,i}-u_{1,j}-\frac i2}\prod_{j=1}^{K_3}\frac{u_{2,i}-u_{3,j}-\frac i2}{u_{2,i}-u_{3,j}+\frac i2}\;,\no\\
  (-1)^\e&=&\prod_{j=1}^{K_2}\frac{u_{3,i}-u_{2,j}-\frac i2}{u_{3,i}-u_{2,j}+\frac i2}\prod_{j=1}^{K_4}\frac{x_{3,i}-\xm_{4,j}}{x_{3,i}-\xp_{4,j}}\sqrt{\frac{\xp_{4,j}}{\xm_{4,j}}}\;,\no\\
  -\(\frac{\xp_4}{\xm_4}\)^L&=&\prod_{j=1}^{K_4}\frac{u_{4,i}-u_{4,j}+i}{u_{4,i}-u_{4,j}-i}\sigma^2[u_{4,i},u_{4,j}]\times\no\\&&\times\prod_{j=1}^{K_1} \frac{1-\frac 1{\xm_{4,i}x_{1,j}}}{1-\frac 1{\xp_{4,i}x_{1,j}}}\prod_{j=1}^{K_3}\frac{\xm_{4,i}-x_{3,j}}{\xp_{4,i}-x_{3,j}}\prod_{j=1}^{K_5}\frac{\xm_{4,i}-x_{5,j}}{\xp_{4,i}-x_{5,j}}\prod_{j=1}^{K_7}\frac{1-\frac 1{\xm_{4,i} x_{7,j}}}{1-\frac 1{\xp_{4,i}x_{7,j}}}\;,\no\\
  (-1)^{\e}&=&\prod_{j=1}^{K_6}\frac{u_{5,i}-u_{6,j}-\frac i2}{u_{5,i}-u_{6,j}+\frac i2}\prod_{j=1}^{K_4}\frac {x_{5,i}-\xm_{4,j}}{x_{5,i}-\xp_{4,j}}\sqrt{\frac{\xp_{4,j}}{\xm_{4,j}}}\;,\no\\
  -1&=&\prod_{j=1}^{K_6}\frac{u_{6,i}-u_{6,j}+i}{u_{6,i}-u_{6,j}-i}\prod_{j=1}^{K_7}\frac{u_{6,i}-u_{7,j}-\frac i2}{u_{6,i}-u_{7,j}+\frac i2}\prod_{j=1}^{K_5}\frac{u_{6,i}-u_{5,j}-\frac i2}{u_{6,i}-u_{5,j}+\frac i2}\;,\no\\
  (-1)^\e&=&\prod_{j=1}^{K_6}\frac{u_{7,i}-u_{6,j}-\frac i2}{u_{7,i}-u_{6,j}+\frac i2}\prod_{j=1}^{K_4}\frac {1-\frac 1{x_{7,i}\xm_{4,j}}}{1-\frac 1{x_{7,i}\xp_{4,j}}}\sqrt{\frac{\xm_{4,j}}{\xp_{4,j}}}\;.\label{BSaBA}
\ee

The notations here are the following:
\bn
    \item The parameter $K_a$, $a=\overline{1,7}$ counts the number of Bethe roots of the $a$-th type.

     The relation between $K_a$, $L$ and the charges $J_1,J_2,J_3=J$, $S_1,S_2$ defined by (\ref{ChargesDefinition}) is given by the following formulas:
     \be
        J_1&=&\frac 12(K_1+K_3-K_5-K_7)\;,\no\\
        J_2&=&K_4-\frac 12(K_7+K_5+K_1+K_3)\;,\no\\
        J_3&=&L-K_4+\frac 12(K_3-K_1)+\frac 12(K_5-K_7)\;,\no\\
        S_1&=&-K_2-K_6+\frac 12(K_7+K_5+K_1+K_3)\;,\no\\
        S_2&=&-K_2+K_6+\frac 12(K_1+K_3-K_5-K_7)\;.\label{charges}
     \ee
     \item The functions $x_{a,i}$, $\xpm_{a,i}$ are defined through the Zhukovsky map by
     \be
        x_{a,i}\equiv x[u_{a,i}],\ \ \xpm_{a,i}\equiv x[u_{a,i}\pm\frac i2]
     \ee
     supplemented by the condition $|x[u]|>1$. This condition is natural for the central roots $u_{4,j}$ since the energy is expressed through this roots by
     \be
        E=\sum_{j=1}^{K_4}=2ig\(\frac 1{\xp_{4,j}}-\frac 1{\xm_{4,j}}\)
     \ee
     and it should have a proper $g\to 0$ limit. However, the condition $|x[u]|>1$ is somehow artificial for the Bethe roots of type $1,3,5,7$. Indeed, up to the replacement $x_{1,i}\to 1/x_{3,i}$ the first equation coincides with the third one. Therefore it is possible to unite the equations for the roots of type $1$ and $3$ into one if to not impose constraints on the choice of the branch for $x[u]$. The same is true for the roots of type $5$ and $7$.
     \item The Bethe equations describe the spectrum of the single trace operators. Since the shift of all nodes by one does not change the single trace operator due to the cyclicity of the trace, the overall momentum of the spin chain should obey a constraint
         \be\label{conscond2}
            e^{i(p_1+p_2+\ldots+p_{K_4})}\equiv e^{iP}=\prod_{j=1}^{K_4}\frac{\xp_{4,j}}{\xm_{4,j}}=1.
         \ee
     \item The factor $(-1)^\e$ was introduced in \cite{Arutyunov:2007tc}. It is equal to $1$ when the fermionic excitations obey periodic boundary conditions and to $-1$ when the boundary conditions are antiperiodic. For the AdS/CFT case the boundary conditions are such that
         \be
            (-1)^{\e}e^{iP/2}=1\;.
         \ee
         We could in principle cancel out the factor $(-1)^{\e}$ and the terms $\sqrt{\xp/\xm}$ in the Bethe Ansatz. However we do not make this cancelation since the Bethe Ansatz equations in the form (\ref{BSaBA}) are valid even if the condition (\ref{conscond2}) is not satisfied.
     \item The explicit form of the scalar factor $\s^2$ will be given in the next section after the crossing equations that determine $\s^2$ are solved.
\en
To finish the description of the asymptotic Bethe Ansatz a map between the configuration of Bethe roots and single trace operators should be given. A graphical answer to this question is presented in Fig.1 of \cite{Rej:2007vm}. Here we will focus on the two simplest examples discussed in the previous chapter: $SU(2)$ and $SL(2)$ sectors of the theory.

The $SU(2)$ sector is spanned by the operators (\ref{forspan}). This sector is described by the Bethe equations (\ref{bethesu2sector}) which are recovered from (\ref{BSaBA}) by putting $K_{i\neq 4}=0$. Number $M$ of fields $X$ is equal to $K_4$, number of fields $Z$ is equal to $J$, $L=J+K_4$.

The $SL(2)$ sector is spanned by the operators of type (\ref{twistJ}). In this case the length of the spin chain is equal to the number of the fields $Z$, $L=J$. Number of covariant derivatives acting on one field $Z$ is interpreted as the number of excitation of the corresponding node. Each covariant derivative corresponds to a complex\footnote{This complex is not a string-like configuration. All Bethe roots are real.} of three Bethe roots: one bosonic root of type $4$, one fermionic root of type $3$, and one fermionic root of type $5$. Therefore the total number $M$ of excitations is equal to $K_4=K_3=K_5$, the parameters $K_1,K_2,K_6,K_7$ are equal to zero. In view of (\ref{charges}) the AdS charge $S\equiv S_1$ is equal to $M$.

\begin{figure}[t]
\centering
\includegraphics[width=12.00cm]{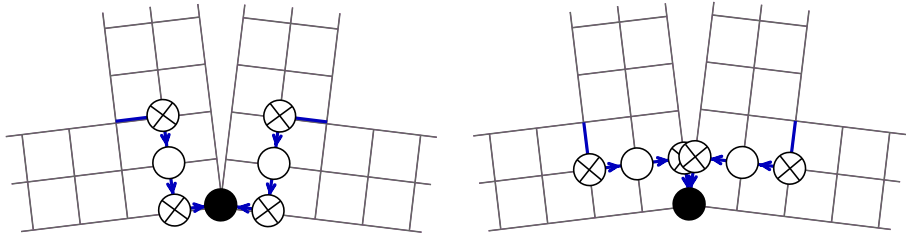}
\caption{\label{fig:dualitysu2sl2} To the left: The Dynkin diagram and the set of Backlund transforms which correspond to the set of Bethe equations (\ref{BSaBA}). To the right: The Dynkin diagram and the set of Backlund transforms which are obtained after duality transformation of fermionic nodes. Although for obtaining equation (\ref{bethesl2sector}) it is sufficient to transform nodes $3$ and $5$, all fermionic nodes are dual-transformed to preserve the bosonic type of nodes $2$ and $6$. Note that the depicted construction is not a T-hook but a pair of fat hooks that have the same source term denoted by the black node. The equation for $u_{4,k}$ which corresponds to the black node is not obtained from the Backlund transforms. Additional requirement of periodicity is needed to obtain this equation.}
\end{figure}
Instead of writing three types of Bethe equations for three types of Bethe roots it is possible to perform a duality transformation on the fermionic roots and get only one type of Bethe equations. For $g=0$ it is the same duality transformation that was discussed in section \ref{sec:DualityTransformation}. Its generalization exists for the case $g\neq 0$ \cite{Beisert:2006qh}. We depicted in Fig.~\ref{fig:dualitysu2sl2} the Dynkin diagrams which correspond to the Bethe equations (\ref{BSaBA}) and dual transformed Bethe equations.

After the duality transformation the $SL(2)$ sector is described by the following set of Bethe equations:
\be\label{bethesl2sector}
  \(\frac{\xp_{4,k}}{\xm_{4,k}}\)^L=-\prod_{j=1}^{M}\frac{u_{4,k}-u_{4,j}-i}{u_{4,k}-u_{4,j}+i}\left(\frac{1-\frac 1{\xp_{4,k}\xm_{4,j}}}{1-\frac 1{\xm_{4,k}\xp_{4,j}}}\right)^2\sigma[u_{4,k},u_{4,j}]^2,\ \ \ k=1,\ldots,M\;.
\ee

\section{\label{sec:crossingequations}Crossing equations}
\subsection{\label{subsec:Janikderivation}Sketch of derivation}
The scalar factor $\s$ which appears in the Beisert-Staudacher asymptotic Bethe Ansatz cannot be fixed from the continuous symmetries of the system. The same situation happens in the bootstrap approach for the relativistic theories (see chapter \ref{ch:twodimqft}). However, in relativistic theories there are also discrete symmetries which lead to the existence of antiparticles. From these symmetries additional functional equations (crossing equations) on the scattering matrix can be derived. These equations together with an assumption on the physical spectrum of the theory and unitarity condition unambiguously fix the scalar factor \cite{Zamolodchikov:1978xm}.

 Although the light-cone quantized string theory is not relativistically invariant and the presence of discrete symmetries is not evident, it was argued by Janik \cite{Janik:2006dc} that one can hope to have a kind of the crossing equations on the scattering matrix (\ref{BSmat}). Janik derived such crossing equations assuming that, written in a general algebraic form, they are the same as the ones for the relativistic theories. This general form is the following\footnote{Strictly speaking, (\ref{crossingabstract}) is not the crossing equations used in a relativistic theory. Equation (\ref{crossingabstract}) is derived from the crossing equations and the unitarity condition.}:
 \be\label{crossingabstract}
  (\CA\otimes 1)S=S^{-1},
 \ee
 where $S:g\otimes g\to\,g\otimes g$, $g$ is a symmetry algebra, and $\CA$ is the antihomomorphism of the symmetry algebra.

On the level of the representation $\pi$, the action of the antipode $\CA$ on an arbitrary generator $J\,\e\,g$ is realized as:
 \be\label{crossingrepresentation}
  \pi[\CA[J]]=\CC^{-1}\overline\pi[J]^{st}\CC,
 \ee
where $\overline\pi$ is a representation (for anitparticles) which is in general different from $\pi$, $\CC$ is a charge conjugation (intertwining) matrix and $^{st}$ means supertransposition\footnote{we need to perform transposition since $\CA$ is an antihomomorphism.}.

For the case of the $su(2|2)$ $S$-matrix (\ref{BSmat}) the representation $\pi$ is the four dimensional representation of the centrally extended $su(2|2)$ algebra defined by the parameters $a,b,c,d$.

The equation (\ref{crossingabstract}), realized on the representation $\pi\otimes\pi$, is a \textit{matrix} equation on the scattering matrix (\ref{BSmat}) in which only a single \textit{scalar} factor is not fixed. The demand that this equation holds fixes unambiguously $\overline\pi$ and $\CC$. This was done in \cite{Janik:2006dc} under the condition that $\CA[J]=-J$.

The representation $\overline\pi$ is parameterized by $\overline a,\overline b,\overline c,\overline d$ which are related to $a,b,c,d$ of $\pi$ via the substitution
\be\label{crossmeaning}
  \xpm\to\frac 1{\xpm}.
\ee
This substitution leads to the change of sign for the energy and the momentum. Note that the representation $\overline\pi$ is not contained in any tensor product of $\pi$. Therefore the antiparticles are not excitations of the asymptotic spin chain. This is an important difference with compact representations in the nonsupersymmetric case. This difference was discussed in chapter \ref{ch:susyspch}.

\begin{figure}[t]
\centering
\includegraphics[width=6.00cm]{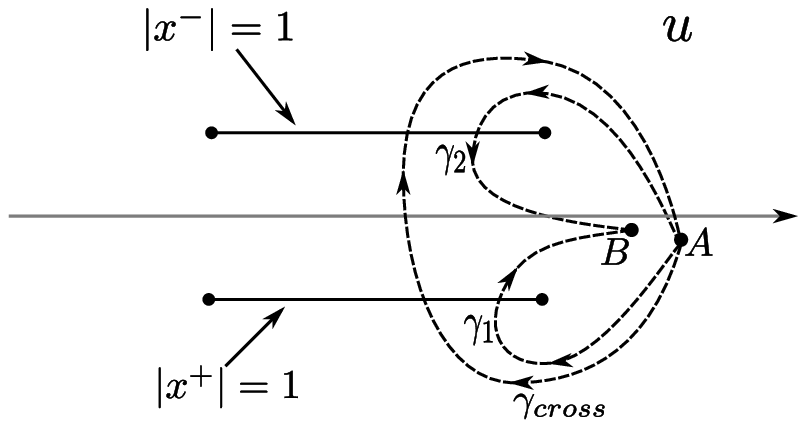}
\caption{\label{fig:crossing} The crossing equation (\ref{crossing}) relates $\s[u,v]$ at point $A$ and $\s[u,v]$ obtained by analytical continuation by the path $\g_{\rm cross}$. A simpler equation is obtained after analytical continuation by the path $\g_2$.}
\end{figure}

The existence of $\overline\pi$ and $\CC$ consistent with (\ref{crossingabstract}) is a nontrivial fact which supports the conjecture of the existence of the crossing equations for the scattering matrix (\ref{BSmat}). After identification of $\overline\pi$ and $\CC$ the crossing equations lead to the following equation on the scalar factor $\sigma$ in (\ref{BSaBA}) \cite{Janik:2006dc,Arutyunov:2006iu}:
\be\label{crossing}
  \sigma[\xpm,\ypm]\sigma^{cross}[\xpm ,\ypm]=\frac{\ym}{\yp}\frac{\xm-\yp}{\xp-\yp}\frac{1-\frac 1{\xm\ym}}{1-\frac 1{\xp\ym}}.
\ee
Let us explain the meaning of the superscript $^{cross}$. We consider the dressing factor $\sigma$ as a multivalued function of $u$. The spectral parameter $u$ defines $\xpm$, and therefore the parameters $a,b,c,d$ of the representation, by the relation $\xpm=x[u\pm i/2]$ (see (\ref{Juk})). The superscript $^{cross}$ means that the first particle in the $S$-matrix should be in the $\overline\pi$ representation, \textit{i.e.} we should use the replacement (\ref{crossmeaning}). This replacement is realized as the analytical continuation of $\sigma$ along the contour $\g_{\rm cross}$ in the $u$-plane\footnote{The monodromy over $\g_{\rm cross}^{-1}$ also leads to (\ref{crossmeaning}). But the solution of crossing equation with such monodromy seems to be pathological if one requires analytical properties of the dressing factor natural for the physical theory. The situation is opposite in the mirror theory. If to require analytical properties of the scattering matrix natural for the mirror theory, the crossing equation with the monodromy over $\g_{\rm cross}^{-1}$ is the right one. Its solution is the "improved" dressing phase of \cite{Arutyunov:2009kf} as we discuss in appendix \ref{app:mirror}.}. This contour is depicted in Fig.~\ref{fig:crossing}. It encircles the branch points $u=2g\pm i/2$. The dressing factor $\s$ has a nontrivial monodromy along the contour $\g_{\rm cross}$ as it is seen from the equations (\ref{crossing}).

In principle, it is possible to resolve both branch cuts $|\xp|=1$ and $|\xm|=1$ using an elliptic parameterization \cite{Janik:2006dc}. In terms of the elliptic variable $s$ used in \cite{KSV1} and related to the one of \cite{Janik:2006dc} by a Gauss-Landen transformation the crossing transformation is given by $\s^{cross}[s]=\s[s+2iK']$. However, as we will see below, the dressing phase has not two but infinitely many cuts. Therefore the elliptic parameterization for the dressing phase is not very useful. We will not use the elliptic parameterization in this work.

\subsection{Solution}
One can show that the solution of the crossing equation can be fixed uniquely by the following set of requirements on its analytical structure \cite{V3}:
\bi
\item $\sigma[u,v]$ can be  represented in the following form \cite{Arutyunov:2004vx,Arutyunov:2006iu}:
\be\label{param}
\s[u,v]=e^{i\theta[u,v]},\ \
\theta[u,v]=\chi[\xp,\ym]-\chi[\xm,\ym]-\chi[\xp,\yp]+\chi[\xm,\yp],
\ee
where $\chi[x,y]$ is antisymmetric with respect to the interchange of variables: $\chi[x,y]=-\chi[y,x]$.

In the following we will use the functions $\s_1[x,v]$ and $\s_2[x,y]$ defined by
\be
  \s_1[x,v]=e^{i\chi[x,\ym]-i\chi[x,\yp]},\ \
  \s_2[x,y]=e^{i\chi[x,y]}.
\ee

\item $\chi$ is an analytic single-valued function for the physical domain $|x|>1$. This domain is the analog of the physical strip $0<{\rm Im}[\theta]<\pi$ in relativistic models. The condition of analyticity can be seen as the minimality condition. 

\item $\chi$ as a function of $u$ does not have branch points except  those that are explicitly required by the crossing equation. The required branch points are of the square root type. This is a condition of compatibility of the analytical structure of the dressing phase and analytical structure of the Bethe equations. It can be compared with demand for the S-matrix in relativistic theories to be meromorphic function of the rapidity variable $\theta$.
\item $\chi[x,y]\to {\rm const}$ for $x\to\infty$. This asymptotics is compatible with the asymptotics of the strong coupling expansion of the dressing phase which is known from string theory calculations.
\ei

\begin{figure}[t]
\centering
\includegraphics[width=10.00cm]{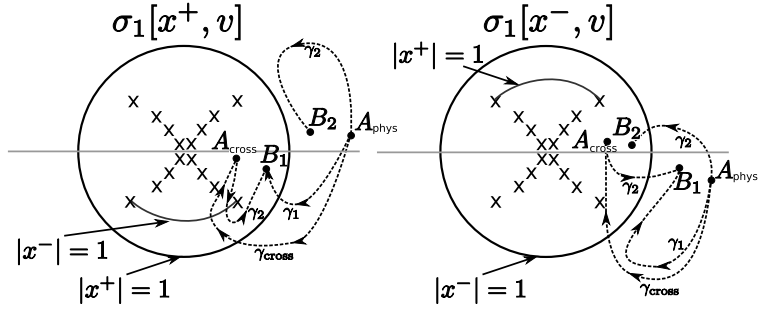}
\caption{\label{fig:AAx} Analytical structure of $\s_1[\xp,v]$ as a function of $\xp$ and $\s_1[\xm,v]$ as a function of $\xm$. Initial crossing equation relates two functions at points $A_{phys}$ and $A_{cross}$. Equation (\ref{mirrorcros}) relates two functions at points $B_1$ and $B_2$. The advantage of (\ref{mirrorcros}) is that at points $B_1$ and $B_2$ we are allowed to use (\ref{ABA}).}
\end{figure}

To solve the crossing equation let us first understand how the contour $\gamma_{\rm cross}$ looks like in the $x$ plane for the functions $\s_1[x^+,v]$ and $\s_1[x^-,v]$. This contour is shown if Fig.~\ref{fig:AAx}.

Since $\s_1[x,y]$ may not be single-valued in the domain $|x|<1$, it may happen (and it does) that $\s_1[x^+,v]$ as a function of $\xp$ and $\s_1[x^-,v]$ as a function of $\xm$ after the continuation $\gamma_{\rm cross}$ are on different Riemann sheets of the $x$ plane. To avoid this problem we analytically continue the crossing equation (\ref{crossing}) along the path $\gamma_2$, shown in Fig.~\ref{fig:crossing} and Fig.~\ref{fig:AAx}.

The crossing equation at the point $B$ is written as
\be\label{mirrorcros}
  \s^{\g_2}[u,v]\s^{\g_1}[u,v]=\frac{1-\frac 1{\xp\yp}}{1-\frac 1{\xm\ym}}\frac{1-\frac 1{\xm\yp}}{1-\frac 1{\xp\ym}},
\ee
where $\s^{\g_i}$ means the analytical continuation of $\s[u,v]$ from the point $A_{phys}$ to the point $B_i$ via the path $\g_i$.

The r.h.s. of (\ref{mirrorcros}) was obtained with the help of the property
\be
    \left(x-y\right)\left(1-\frac 1{xy}\right)=\frac{u-v}{g}.
\ee

At the point $B$, in contradistinction to the point $A$, we are allowed to write:
\be\label{ABA}
  \sigma^{\g_2}=\frac {\s_1[\xp,v]}{\s_1[1/\xm,v]},\ \
  \sigma^{\g_1}=\frac {\s_1[1/\xp,v]}{\s_1[\xm,v]}.
\ee
All four functions $\s_1$ which are used in (\ref{ABA}) are on the same Riemann sheet of the $x$ plane.

We can write (\ref{mirrorcros}) in terms of the shift operator as
\be\label{mirror}
  &&\(\ \s_1[x,v]\s_1[1/x,v]\ \)^{D-D^{-1}}=\(\frac{x-\frac 1{\yp}}{x-\frac 1{\ym}}\)^{\!\!D+D^{-1}}.
\ee
 The shift operator $D$ is not well defined inside the strip $|{\rm Re}[u]|\leq 2g$ since we can cross the cut of $x[u]$ and go to another sheet. To avoid this ambiguity, we will consider the crossing equation (\ref{mirror}) outside this strip, solve it, and then analytically continue the solution.

The function $\s_1[x,v]\s_1[1/x,v]$ as a function of $u$ does not have a branch cut $[-2g,2g]$. A solution of (\ref{mirror}) with this property is given by
\be\label{s1}
  \s_1[x,v]\s_1[1/x,v]=\(\frac{x-\frac 1{\yp}}{x-\frac 1\ym}\)^{-\frac{D^2}{1-D^2}+\frac {D^{-2}}{1-D^{-2}}},\ \ \frac{D^{\pm 2}}{1-D^{\pm 2}}=D^{\pm 2}+D^{\pm 4}+\ldots\ .\no
\ee
Strictly speaking, this expression should be regularized to have a precise meaning. However, the regulating terms will cancel for the complete dressing factor $\s[u,v]$.

The expression (\ref{s1}) can be further simplified if we use the functions $\s_2$ and the relation $\s_1[x,v]=\s_2[x,\yp]/\s_2[x,\ym]$:
\be
  \s_2[x,y]\s_2[1/x,y]=\(\frac{x-\frac 1y}{\sqrt{x}}\)^{-\frac{D^2}{1-D^2}+\frac {D^{-2}}{1-D^{-2}}}.
\ee
The multiplier $1/\sqrt{x}$ does not contribute to (\ref{s1}). It is needed for the consistency with the antisymmetry of $\chi[x,y]$ with respect to interchange $x\leftrightarrow y$. Indeed, a direct calculation shows that
\be\label{s2}
  \s_2[x,y]\s_2[1/x,y]\s_2[x,1/y]\s_2[1/x,1/y]=\no\\=(u-v)^{-\frac{D^2}{1-D^2}+\frac {D^{-2}}{1-D^{-2}}}=\frac{\Gamma[1-i(u-v)]}{\Gamma[1+i(u-v)]},
\ee
whose logarithm is antisymmetric with respect to $u\leftrightarrow v$ as it should.

By taking the logarithm of (\ref{s2}) we get a simple Riemann-Hilbert problem which is solved by
\be\label{s2solution}
  \chi[x,y]=-i\tilde{K}_u\tilde{K}_v\log\[\frac{\Gamma[1-i(u-v)]}{\Gamma[1+i(u-v)]}\],
\ee
with the kernel $\tilde K$ defined by
\be\label{Ktildedef}
  (\tilde K\cdot F)[u]\equiv\int_{-2g+i0}^{2g+i0}\frac{dw}{2\pi i}\frac {x-\frac 1x}{z-\frac 1{z}}\frac 1{w-u}F[w],\ \ \ \frac{w}{g}=z+\frac 1z.
\ee
The kernel $\tilde K$ is constructed to satisfy the following equation:
\be\label{Cauchy}
  (\tilde K\cdot F)[u+i0]+(\tilde K\cdot F)[u-i0]=F[u],\  u^2<4g^2.
\ee
The subscripts $u$ and $v$ in (\ref{s2solution}) refer to action of $\tilde K$ on $u$ and $v$ variables respectively.

This solution was chosen among the other possible solutions by the requirement that $\chi[x,y]$ should be analytic for $|x|>1$ and $\chi[x,y]\to{\rm const},\ x\to\infty$.
The kernel $\tilde K$ appears also in the asymptotic Bethe Ansatz. We will discuss this issue in Sec.~\ref{sec:Kernels}.

The expression (\ref{s2solution}) can be rewritten in the form proposed by Dorey, Hofman, and Maldacena \cite{Dorey:2007xn} if we rewrite the action of the kernel $\tilde K$ as the integral in the Zhukovsky plane
\be\label{toDHM}
  (\tilde K\cdot F)[u]=\int_{|z|=1}\hspace{-2.41EM}\circlearrowleft\hspace{1EM}\frac {dz}{2\pi i}\frac 1{x-z}F\[w\]-\int_{-2g+i0}^{2g+i0}\frac{dv}{2\pi i}\frac{g^{-1} F[v]}{y-\frac 1y}
\ee
and note that the second term does not contribute to the dressing phase.

\subsection{Analytical structure of the dressing factor}
We will discuss now the analytical structure of the factor $\s_2[u,v]$ as a function of $u$. The analytical properties of the whole dressing factor $\s$ can be then easily derived.
\begin{figure}[t]
\centering
\includegraphics[width=6.00cm]{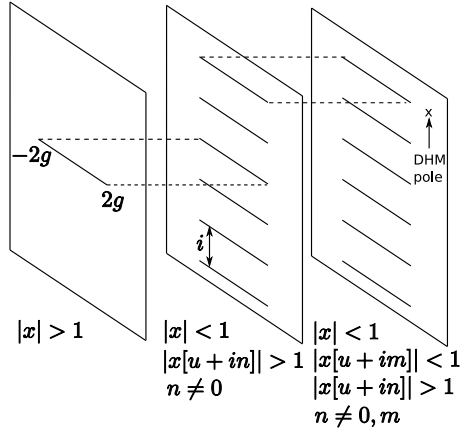}
\caption{\label{fig:dressinganalytics}Analytical structure of $\sigma_2[x,y]$ as a function of $u$.}
\end{figure}
On the physical sheet $|x|>1$ the function $\s_2[u,v]$ has only one square root cut $|x|=1$. Analytical structure of $\s_2[u,v]$ for $|x|<1$ can be seen from the formula (\ref{s1}):
\be\label{s3}
  \s_2[1/x,v]=\s_2[x,v]^{-1}\prod_{n\neq 0}\(\frac{x[u+in]-\frac 1y}{\sqrt{x[u+in]}}\)^{-\sign[n]}.
\ee
We see that in this region $\s_2[u,v]$ has infinitely many cuts $|x[u+in]|=1$.
In the product in the l.h.s. of (\ref{s3}) the branches for $x[u+in]$ are chosen in a way that $|x[u+in]|\geq 1$.

If we analytically continue $\s_2[u,v]$ through one of the cuts $|x[u+im]|=1$ we will arrive at a new Riemann sheet (the third one in Fig.~\ref{fig:dressinganalytics}) which contains, at the position $u=v-im$, pole if $m>0$ or zero if $m<0$. This pole/zero was first observed in \cite{Dorey:2007xn}. It was shown there that it does not correspond to any bound state but is due to the specific properties of Feynman diagrams in two-dimensional theories.

%
%
\section{\label{sec:Kernels}The kernel $\tilde K$ and a shorthand notation for the Bethe equations}
The kernels $\tilde K$ that had appeared in the derivation of the dressing phase appear also explicitly in the context of the Bethe Ansatz equations. Indeed, one can directly check that\footnote{A standard choice of branches is assumed in (\ref{Ktildeproperty}) for the region ${\rm Re}[v]>2g$, then analytical continuation is performed.}
\be\label{Ktildeproperty}
  \tilde K_{u}\cdot\log[v-u]=\log\left[\frac{y-\frac 1x}{\sqrt{y}}\right].
\ee
Therefore we get a shorthand notation for the building blocks of the Bethe equations:
\be
  \prod_{k}\frac{1-\frac 1{x\ym_k}}{1-\frac 1{x\yp_k}}\sqrt{\frac{\ym_k}{\yp_k}}&=&\prod_{k}(u-v_k)^{\tilde K_u (D-D^{-1})}=Q_{v}[u]^{\tilde K (D-D^{-1})},\no\\
  \prod_{k}\frac{x-\ym_k}{x-\yp_k}\sqrt{\frac{\yp_k}{\ym_k}}&=&Q_{v}[u]^{(1-\tilde K) (D-D^{-1})},\no\\
  \prod_{k}\frac{1-\frac 1{\xp y_k}}{1-\frac 1{\xm y_k}}&=&Q_{v}[u]^{(D-D^{-1})\tilde K},\no\\
  \prod_{k}\frac{\xp-y_k}{\xm-y_k}&=&Q_{v}[u]^{(D-D^{-1})(1-\tilde K)}\;.\label{blocksshorthand}
\ee

Using the property (\ref{Ktildeproperty}) and the representation for gamma functions used in (\ref{s2}) one can show that the dressing factor can be represented in the form
\be\label{dressingthroughshifts}
    \s[u,v]=(u-v)^{(D-D^{-1})\tilde K\left(\frac{D^2}{1-D^2}-\frac{D^{-2}}{1-D^{-2}}\right)\tilde K(D-D^{-1})},
\ee
where both the kernel $\tilde K$ and $D$-operator act on $u$ variable.

We should admit that in (\ref{blocksshorthand}) and (\ref{dressingthroughshifts}) we used the shift operator in the sense
\be\label{defD}
(D^{\pm 1} f)[u]\equiv f[u\pm i/2],
\ee
that is without performing analytical continuation. In opposite, in such formulas as (\ref{trescool}) we used the shift operator in the sense of analytical continuation. To avoid ambiguity, the following two rules of writing and interpreting formulas are applied. First, an expression if possible should be considered in a region where there is no ambiguity between two interpretations of the shift operator. For instance in the case (\ref{trescool}) the equation should be defined in the region ${\rm Im}[u]>1/2$, then it may be analytically continued. Second, if we cannot reach a region without ambiguities (for example we are required to stay on the real axis) then expression $D\, F$ is understood in the sense (\ref{defD}). To explicit the case of analytical continuation we use the notation $D\, F^{\pm 0}$ or $D\, F[u\pm i0]$, where $"+"$ corresponds to the analytic continuation from above the cut and $"-"$ corresponds to the analytic continuation from below the cut. As an example see (\ref{iequationON}) and (\ref{examplepm}).

Taking into account the remarks about the sense of (\ref{blocksshorthand}) and (\ref{dressingthroughshifts}), we can write down the Bethe Ansatz equations (\ref{BSaBA}) in a shorthand way. For this we introduce seven Baxter polynomials
\be
    Q_i[u]=\prod_{i=1}^{K_i}(u-u_k),\ \ \ i=1,\ldots, 7.
\ee
The Bethe equations (\ref{BSaBA}) will take the following form:
\be
    (-1)^\e&=&Q_2[u_{1,i}]^{-(D-D^{-1})}Q_4[u_{1,i}]^{\tilde K(D-D^{-1})},\no\\
    -1&=&Q_2[u_{2,i}]^{D^2-D^{-2}}Q_1[u_{2,i}]^{-(D-D^{-1})}Q_3[u_{2,i}]^{-(D-D^{-1})},\no\\
    (-1)^\e&=&Q_2[u_{3,i}]^{-(D-D^{-1})}Q_4[u_{3,i}]^{(1-\tilde K)(D-D^{-1})},\no\\
    -x^{L(D-D^{-1})}&=&Q_4[u_{4,i}]^{D^2-D^{-2}+2(D-D^{-1})\tilde K\left(\frac{D^2}{1-D^2}-\frac{D^{-2}}{1-D^{-2}}\right)\tilde K(D-D^{-1})}Q_1[u_{4,i}]^{-(D-D^{-1})\tilde K}
    \times\no\\&&\times
Q_3[u_{4,i}]^{-(D-D^{-1})(1-\tilde K)}
Q_7[u_{4,i}]^{-(D-D^{-1})(1-\tilde K)}Q_5[u_{4,i}]^{-(D-D^{-1})\tilde K},\no\\
(-1)^\e&=&Q_6[u_{5,i}]^{-(D-D^{-1})}Q_4[u_{5,i}]^{(1-\tilde K)(D-D^{-1})},\no\\
-1&=&Q_6[u_{6,i}]^{D^2-D^{-2}}Q_7[u_{6,i}]^{-(D-D^{-1})}Q_5[u_{6,i}]^{-(D-D^{-1})},\no\\
(-1)^\e&=&Q_6[u_{7,i}]^{-(D-D^{-1})}Q_4[u_{7,i}]^{\tilde K(D-D^{-1})}.\label{shorthandBethe}
\ee

The given notation is a generalization of (\ref{completeBetheAnsatz}), however without introducing string configurations. In the limit $g\to 0$ we have also $\tilde K\to 0$ and (\ref{completeBetheAnsatz}) is recovered from (\ref{shorthandBethe}).

We see that the Bethe equations (\ref{BSaBA}) can be expressed through the Baxter polynomials in the power of an operator which is a rational combination of $D$, $D^{-1}$, and $\tilde K$. To compare, analogical operator in the case of spin chain contains polynomial combination of $D$ and $D^{-1}$ and in the case of relativistic integrable models - rational combination of $D$ and $D^{-1}$.

\section{The kernel $K$ and functional equations.}
When we pass to the integral equations we should take logarithm and then derivative of the Bethe equations. For this purpose it is useful to introduce the kernel $K$ by
\be\label{kdef}
  (K\cdot F)[u]=\int_{-2g+i0}^{2g+i0}\frac{dw}{2\pi i}\frac {z-\frac 1z}{x-\frac 1{x}}\frac 1{w-u}F[w],\ \ \frac wg=z+\frac 1z.
\ee
The function $(K\cdot F)[u]$ also satisfies (\ref{Cauchy}). The relation between the kernels $K$ and $\tilde K$ is
\be
  \pd_u \tilde K F=K\pd_u F.
\ee
Indeed, using (\ref{toDHM}) we have

 \be
 \pd_u \tilde K F=\int_{|z|=1}\hspace{-2.41EM}\circlearrowleft\hspace{1EM}\frac {dz}{2\pi i}\(\pd_{u}\frac 1{x-z}\)F[w]=\int_{|z|=1}\hspace{-2.41EM}\circlearrowleft\hspace{1EM}\frac {dz}{2\pi i}\frac 1{x-z}\frac{\pd_u x}{\pd_w z}\pd_w F[w]=K\pd_u F.
 \ee

Therefore for example:
\be
  \pd_u\log\times Q[u]^{-\tilde K (D-D^{-1})}=K(D-D^{-1})R,\ \ R\equiv\pd_u\log\,Q[u].
\ee

Let us see how we can use the shorthand notations to derive a functional equation on the resolvent of Bethe roots from (\ref{bethesl2sector}) in the large $M$ limit. By taking the logarithm of (\ref{bethesl2sector}) we get
\be\label{presl2}
    -2\pi i\, n_k=L (D-D^{-1})\log x_{4,k}+(D^2-D^{-2})\log Q[u_{4,k}]-\tilde\CK\log Q[u_{4,k}],
\ee
where the derivative of $\tilde \CK$ can be read from (\ref{holll}).

Now we use the fact that in the large $M$ limit the approximation $\frac{dn_k}{du}=\rho[u]+\rho_h[u]$ is valid. Here
\be
    \rho[u]=-\frac 1{2\pi i}\left(R[u+i0]-R[u-i0]\right)
\ee
is the density of particles and $\rho_h$ is the density of holes (in the $SL(2)$ case holes are zeroes of the transfer matrix, see Sec.~\ref{sec:functionalequationinlog}). We can introduce the resolvent for the density of holes by
\be
    R_h=\int dv\frac{\rho_h[v]}{u-v}.
\ee
Then, in the large $M$ limit the derivative of (\ref{presl2}) is written as
\be\label{holll}
    R[u\!+\!i0]-R[u\!-\!i0]+R_h[u\!+\!i0]-R_h[u\!-\!i0]=L(D-D^{-1})\frac 1x\frac{dx}{du}+(D^2-D^{-2})R-\no\\
    -2\left(DKD-D^{-1}KD^{-1}+(D-D^{-1}) K\left(\frac {D^2}{1-D^2}-\frac {D^{-2}}{1-D^{-2}}\right) K(D-D^{-1})\right)R\;.\hspace{5EM}
\ee
This equation can be written in the form $G^+[u]=G^-[u]$, where the function $G^+$ is analytic in the upper half plane and the function $G^-$ is analytic in the lower half plane\footnote{Here we basically give a variation of  the procedure of holomorphic projection explained in Sec.~\ref{excitationsintheantiferro}.}. Therefore $G^+[u]=G^-[u]=0$. We therefore obtain an equation
\be\label{almostBESFRS}
    (1-D^2)R+R_h=-2D\,K\,D\,R-2D\,K\left(\frac {D^2}{1-D^2}-\frac {D^{-2}}{1-D^{-2}}\right) K(D-D^{-1})R
\ee
which is valid in the upper half plane. Equation in the lower half plane can be obtained from (\ref{almostBESFRS}) by replacement $D^{\pm 1}\to D^{\mp 1}$.

 Equation (\ref{almostBESFRS}) is almost a BES/FRS equation \cite{Beisert:2006ez,Freyhult:2007pz} which we discuss in chapter \ref{ch:BESFRS}. To obtain the BES/FRS equation we still have to perform a proper scaling of the resolvents as discussed in section \ref{sec:largejsolution}.

If the resolvent admit some parity properties, (\ref{almostBESFRS}) can be slightly simplified.
Let us decompose the resolvent into the symmetric and antisymmetric parts: $R=R_++R_-,\ R_\pm[-u]=\pm R_{\pm}$. The parity of the resolvent is opposite to the parity of the density.

We use
\be\label{kprop}
  K(D-D^{-1})R_\pm=\int_{-2g+i0}^{2g+i0}\frac{dv}{2\pi i}\frac{y-\frac 1y}{x-\frac 1x}\frac 1{v-u}\(R_\pm\[u+\frac i2\]\mp R_\pm\[-u+\frac i2\]\)\equiv K_\mp DR_\pm.\no\\
\ee
The kernels $K_\pm$ first appeared in \cite{KSV2} where they were derived using analytical transformations of the inverse Fourier transform of $K_0$ and $K_1$ of \cite{Beisert:2006ez}.

Using (\ref{kprop}) we can rewrite (\ref{almostBESFRS}) as
\be
(1-D^2)R+R_h=-2DKDR-2DK_-\frac {D^2}{1\!-\!D^2}K_+DR_--2DK_+\frac {D^2}{1\!-\!D^2}K_-DR_+.
\ee
In the case of an even density distribution ($R_+=0$) we get the BES/FRS equation, up to a proper rescaling of the resolvents, in the form used in
\cite{V1}.

\part{Functional form of integral Bethe Ansatz equations}
In order to compare the prediction of the Bethe Ansatz equations for the spectrum of the system with the field theoretical prediction one should often consider the thermodynamical limit in which the Bethe Ansatz equations reduce to the integral equation on the density of roots. Therefore it is important to be able to solve these integral equations.

In the simplest case the equations can be solved using the Fourier transform. For example such is the equation which describes the density of roots of the state that corresponds to the antiferromagnetic vacuum of the XXX spin chain:
\be\label{how1}
  \rho[u]=\frac {L}{2\pi}\frac{1}{u^2+\frac 14}-\intii \frac{dv}{\pi}\frac{1}{(u-v)^2+1}\rho[v].
\ee
The solution of this equation is
\be\label{rho1}
  \rho[u]=\frac{L}{2\cosh[\pi u]}.
\ee

In the first part of this thesis we discussed a slightly different approach which allows solving such equations. If we introduce the resolvent
\be
  R[u]=\intii dv\frac{\rho[v]}{u-v}
\ee
and integrate (\ref{how1}) with a Cauchy kernel $\intii dw\frac 1{w-u}\times\ldots$, then we will get the functional equation
\be
  R[w]=\frac{L}{w+i/2}-R[w+i],\ \ {\rm Im}[w]>0
\ee
and a complex conjugated equation for ${\rm Im}[w]<0$. This functional equation can be solved for ${\rm Im}[w]>0$ by
\be
  R[w]=\frac D{1+D^2}\frac {L}{w},\ \ D\equiv e^{\frac i2\pd_w},\ \ \frac 1{1+D^2}\equiv 1-D^2+D^4-D^6+\ldots.
\ee
Then the density function will be given by
\be\label{rho2}
  \rho[w]=-\frac 1{2\pi i}(R[w+i0]-R[w-i0])=-\frac 1{2\pi i}\sum_{n=-\infty}^\infty (-1)^n(D)^{2n+1}\frac Lw.
\ee
Comparing position of poles and the residues for (\ref{rho1}) and (\ref{rho2}) we conclude that (\ref{rho1}) and (\ref{rho2}) are equal.

A more complicated case is when the density of Bethe roots is defined on a compact support. For example, for the SU(2) XXX spin chain in the case when the number of Bethe roots $M$ is smaller than $L/2$ the equation (\ref{how1}) is replaced by (\ref{how2}):
\be\label{how2}
  \rho[u]+\rho_h[u]=\frac{L}{2\pi}\frac{1}{u^2+\frac 14}-\int_{-B}^B \frac{dv}{\pi}\frac{1}{(u-v)^2+1}\rho[v],
\ee
where the branch point $B$ is defined by $\int_{-B}^B du\rho[u]=M$. The function $\rho_h[u]$ ("density of holes") is zero on the interval $u^2<B^2$ and is defined outside it in such a way that the equation (\ref{how2}) is valid on the whole real axis.

 We cannot use directly the Fourier transform for solution of the equation (\ref{how2}). For finite values of $B$ such equations in general do not have explicit solution. However they can be solved perturbatively for small and large values of $B$.

We will consider a more interesting case of large values of $B$. In this case
it is important to consider two different regimes.

First, we can rescale our rapidity variable as $u= B\ \hat u$ and consider the limit $B\to\infty$ with $\hat u$ fixed. For this scaling we can perturbatively expand the shift operator as
\be
  D=e^{\frac i2\pd_u}=e^{\frac i{2B}\pd_{\hat u}}=1+\frac i{2B}\pd_{\hat u}-\frac 1{8B^2}\pd_{\hat u}^2+\ldots.
\ee
Using this expansion, we can reduce functional equations on the resolvents to a recursive tower of solvable Riemann-Hilbert problems.

Second, we can consider the vicinity of the branch point $u=B$. For this we should consider the double scaling limit
\be\label{refn}
  u,B\to\infty\ \ {\rm with}\ \ z\equiv 2(u-B)\ \ \ {\rm fixed.}
\ee
In this limit the equation (\ref{how2}) reduces to the integral equation on the semiinfinite axis. It can be solved by a Wiener-Hopf method \cite{WH}. This approach was used in many applications. We can mention \cite{Hasenfratz:1990zz,Hasenfratz:1990ab,Forgacs:1991rs} which are related to the integrability. We will discuss the solution in the double scaling limit (\ref{refn}) in details in Sec.~\ref{Sec:doublescaling}.

The large $B$ and the double scaling limit of (\ref{how2}) lead to the equations which allow in general many solutions. Only in the leading order the solution can be fixed uniquely by imposing correct analytical properties. We will show that in order to fix the solution in the subleading orders of $1/B$ expansion we have to consider solutions in both limits and demand their matching in the intermediate regime.

The search for the subleading orders of solutions of such equations has a number of different motivations.

In chapter \ref{ch:sl2} we discuss the integral equation which appears in a special limit of the $Sl(2)$ Heisenberg ferromagnet. As we will explain, the first three orders of the large $B$ expansion has a certain interest for the AdS/CFT correspondence. The large $B$ expansion of this system was investigated numerically in \cite{Beccaria:2008nf,V1}. Here we give an analytical derivation of this expansion. We also use this derivation to explain all the details of our method.

In chapter \ref{ch:BESFRS} we discuss the so called BES/FRS equation \cite{Eden:2006rx,Beisert:2006ez},\cite{Freyhult:2007pz} which appears in the context of the AdS/CFT correspondence. The first three orders of the strong coupling expansion of this equation in various regimes are required for the comparison with the perturbative results of string theory \cite{Roiban:2007dq,Roiban:2007ju}. This chapter is based on the works \cite{KSV1,KSV2,V1}.

In chapter \ref{ch:massgap} we discuss how to find the mass gap in sigma models. For instance, to fix the mass gap in the $O(n)$ sigma model \cite{Hasenfratz:1990zz} we need to find the first two (leading and subleading) orders of the large $B$ expansion. The higher orders allow studying the Borel summability properties of the model and in this way provide a check for the correctness of the Bethe Ansatz description of the model. This chapter is based on the author's work \cite{V2}.

\chapter{\label{ch:sl2}$Sl(2)$ Heisenberg magnet}
\section{Quantum mechanics reminder}
The quantum mechanics of one particle is described by the \Shrodinger equation
\be
  \(-{\hbar^2}\frac {d^2}{dx^2}+V[x]\)\psi[x]=E\ \psi[x].
\ee
To consider this equation in the quasiclassical limit it is convenient to introduce a new variable $p=\frac {\hbar}i\frac{\psi\prime}{\psi}$ in which the \Shrodinger equation reduces to the Ricatti equation
\be
p^2-i\hbar p'=E-V.
\ee
In the quasiclassical approximation which is valid for $p^2\gg \hbar p'$ the solution is simplified to $p=\sqrt{E-V}$. Then the wave function is given by the superposition of two functions
\be
  \psi[x]=C_1 e^{\frac i\hbar \int^x p dx}+C_2 e^{-\frac i\hbar \int^x p dx}.
\ee
For ${\rm Im}[x]\gg \hbar$ one of the exponents (say the second one) is exponentially suppressed. To glue solutions in two different limits one should perform in detail analysis in the vicinity of the turning points $x_0$ given by $E=V[x_0]$. This analysis leads to the Bohr-Sommerfeld quantization condition on the energy levels of the particle
\be
  \oint \sqrt{E-V}dx=2\pi\hbar (N+\mu),
\ee
where $\mu$ is a constant known as Maslow index. Its value depends on the analytical properties of the potential $V[x]$ near the turning points. In particular, if it is smooth then $\mu=\frac 12$, if it is a steep function then
$\mu=0$.

The Baxter equation (\ref{eq:Baxtersl2}) can be viewed as a discrete version of the \Shrodinger equation. We will see below that analogs of both smooth and steep potential $V[x]$ appear in the quasiclassical solution of the Baxter equation.

\section{Different regimes in the $Sl(2)$ Heisenberg magnet}
The $Sl(2)$ Heisenberg magnet (with spin zero) appeared in particular in the context of the integrable structures discovered in QCD \cite{Lipatov:1993yb,Faddeev:1994zg,Braun:1998id,Belitsky:2004cz}. It was intensively studied in \cite{Korchemsky:1995be,Kotanski:2005ci}. In the context of the AdS/CFT correspondence the spin 1/2 $Sl(2)$ Heisenberg magnet has a particular interest. It describes the one-loop approximation of the $SL(2)$ subsector of the AdS/CFT integrable system.

 The Bethe equations for a spin $1/2$ $Sl(2)$ magnet are written as follows:
\be\label{BetheBethe}
  \(\frac{u_k+\frac i2}{u_k-\frac i2}\)^L=-\prod_{j=1}^M \frac{u_k-u_j-i}{u_k-u_j+i}.
\ee
All the solutions of these equations are real. The Baxter equation is written as
\be\label{eq:Baxtersl2}
  \(u+\frac i2\)^L Q[u+i]+\(u-\frac i2\)^L Q[u-i]=Q[u]T[u].
\ee
The Baxter equation has a natural parameter which is a magnitude of the discrete shift. In the normalization chosen here this scale equals one. We will call it the Baxter scale.

Each solution has its own internal scales given by the typical distances between the Bethe roots. It is well known that solutions organize themselves into the cuts. We can distinguish two important scales: the distance between the roots inside the cut ($d$) and the distance between the cuts themselves ($a$).

In the thermodynamical limit one typically has a small parameter which plays the role of $\hbar$. The solution is supposed to be expanded in perturbative series within this parameter. It is natural to assume that for a sufficiently small $\hbar$ and inside one cut the typical displacement of the Bethe roots within the change of $\hbar$ does not exceed the distance between the Bethe roots. Under this assumption, the uniform expansion of the resolvent
\be
  R=\frac d{du}\log Q[u]
\ee
in the parameter $\hbar$ is possible at the distances from the cut larger than $d$.

The smallness of $\hbar$ usually means that the parameter $\eta=L+S$ is large. If it so, for ${\rm Im}[u]>0$ we have exponential suppression of the second term in the l.h.s. of (\ref{eq:Baxtersl2}). To estimate the magnitude of the suppression one can approximate
\be\label{eq:Bethesl2}
  \log \[\(\frac{u+\frac i2}{u-\frac i2}\)^L\frac{Q[u+i]}{Q[u-i}\]\simeq \frac{iL}{Re[u]}+\sum_{k=1}^M \frac {2i}{u-u_k}+L\frac{{ \rm Im}[u]}{{\rm Re}[u]^2}+\sum_{k=1}^M\frac {2{\rm Im}[u-u_k]}{{\rm Re}[u-u_k]^2}.
\ee
We see that the real part of the r.h.s. is positive. Its magnitude depends on the particular limit that we consider, but at any case for large $L$ and $|u|\lesssim\sqrt{L}$ the exponential suppression of the second term in the l.h.s. of (\ref{eq:Baxtersl2}) is granted. Usually the restriction on $u$ is weaker.

Under the assumption of the exponential suppression, the Baxter equations are approximated by
\be\label{Baxterprojected}
  \(u+\frac i2\)^L Q[u+i]=Q[u]T[u],\ \ {\rm Im}[u]>0,\no\\
  \(u-\frac i2\)^L Q[u-i]=Q[u]T[u],\ \ {\rm Im}[u]<0.
\ee
The equations (\ref{Baxterprojected}) are equivalent to the linearized equation for the counting function, see the discussion in section \ref{excitationsintheantiferro}.

The relation between the Baxter scale and $d$ can be different depending on the limit we take.

In the case $L\sim S\to\infty$, which was studied in \cite{Gromov:2005gp}, the typical distance between roots is much larger than the Baxter scale. Then the approximation (\ref{Baxterprojected}) is valid only at the large distances from the real axis. The perturbative $\hbar$ expansion includes performing the Taylor series of the type $R[u+i]=R[u]+i\partial_u R[u]+\ldots$. If we perform this expansion in
(\ref{Baxterprojected}) then in the leading order we get the equations that can be also obtained by the differentiation of (\ref{eq5}). This is not surprising since the limit $L\sim S\to\infty$ is the one studied in Sec.~\ref{sec:foldedandspinningstrings}.

In order to find the corrections to (\ref{Baxterprojected}) we need to study the vicinity of the branch point in the scaling limit which leads to the solution in terms of the Airy function \cite{Gromov:2005gp}. This type of limit is similar to the case of the quantum mechanical particle in a smooth potential.

In this chapter we consider another case which is analogical to the particle in a steep potential. In this case the distance between the Bethe roots is much smaller than the Baxter scale, the equations (\ref{Baxterprojected}) are valid for any value of $u$.

This case corresponds to the limit
\be\label{sudakovscaling}
  L,M\to\infty\ \ {\rm with }\ \ \frac{L}{\log M}=j \ \ {\rm fixed}
\ee
initially discussed in \cite{Belitsky:2006en}. We use the notation of \cite{Freyhult:2007pz}.

The scaling (\ref{sudakovscaling}) is possible because in the case of a noncompact spin chain the number of excitations $M$ can be greater than the length of the spin chain $L$.

\subsubsection{The root distribution near the branch cut}
 Let us take the logarithm of the Bethe equations (\ref{BetheBethe}). We obtain
 \be\label{eq:force}
  L\;F[2u_k]&+&\suml_{\substack{j=1 \\j\neq k}}^{M} F[u_k-u_j]=\sign[u_k]\;,\\
  F[u]&=&\frac 1{2\pi i}\log \frac{u+i}{u-i}\;.\no
 \ee
 The equation ($\ref{eq:force}$) can be interpreted as the equation for the equilibrium of forces in classical mechanics. For distances between the particles much larger than one, the interaction between the particles can be approximated with the Coulomb force $F[u_k-u_j]\simeq \frac 1{\pi}\frac 1{u_k-u_j}$. In this case the density of the particles is approximated by a square root cut in the leading order of the large spin limit. This case is realized in the scaling limit $L\sim S\to\infty$.

 For the scaling (\ref{sudakovscaling}) we have the opposite situation: the distances between the particles with the smallest absolute values of rapidities are much smaller than one. The Coulomb approximation is not applicable. To describe the distribution of roots in this limit it is better to represent $F[u]$ as
 \be
  F[u]=-\frac 1{\pi}\arctan[u]+\frac 12\sign[u]\;.
 \ee
 If we introduce the effective force $F_{\textrm{eff}}[u]=L\, F[2u]-\suml_{\substack{j=1 \\j\neq k}}^{M}\frac 1{\pi}\arctan[u-u_j]$, then for the positive roots equation of the equilibrium will be written as
 \be
    F_{\textrm{eff}}[u_{M/2+k}]=\frac 32-k.
 \ee
 Since $F_{\textrm{eff}}$ is a smooth function, we immediately get that in the vicinity of the branch point
 \be\label{behaviorroots}
  u_{M/2+k}-u_{M/2+1}\simeq \frac {1-k}{F'[u_{M/2+1}]}\simeq \frac{\pi}L(k-1).
 \ee
 The last estimation comes from the term $L\,F[2u]$ which is dominant in $F_{\textrm{eff}}$ for roots close to the origin. The expression (\ref{behaviorroots}) is consistent with the assumption of the small distance between the roots. It is valid for $u_{M/2+1}\ll L$ which is the case for any finite $j$. Equidistant distribution of the roots corresponds to a branch point of logarithmic type in the continuous limit.

\subsubsection{\label{sec:functionalequationinlog}Functional equation in the logarithmic scaling regime (\protect\ref{sudakovscaling}).}
Let us introduce the resolvent for holes\footnote{In opposite to the compact case, there are no accompanying roots (in the terminology of Sec.~\ref{sec:countingBethe}) in the r.h.s. of \ref{eq:Baxtersl2}. Due to this all zeroes of $T[u]$ are holes. For the same reason there is no string configurations.}:
\be
  R_h=\frac d{du}\log T[u].
\ee
Taking the derivative of the logarithm of (\ref{Baxterprojected}) we get
\be\label{somebaxter}
  D\frac L{u}&=&(1-D^2)R+R_h,\ \ {\rm Im}[u]>0,\no\\
  D^{-1}\frac L{u}&=&(1-D^{-2})R+R_h,\ \ {\rm Im}[u]<0.
\ee
We will consider the ground state in the limit (\ref{sudakovscaling}).
For the ground state all the holes except two are situated inside the interval $[-B,B]$ \cite{Korchemsky:1995be}. The two external holes are situated at the position $u^*=\pm M/\sqrt{2}+\CO(L)$. All the Bethe roots are situated between the interval $[-B,B]$ and external holes

Let us expand the equation (\ref{somebaxter}) for ${\rm Im}[u]>0$ at $|u|\gg B$. We will get
\be
  \frac Lu=-i\p_u R+\frac{L-2}{u}+\frac 1{u-u^*}+\frac 1{u+u^*}+\CO\(\frac 1{u^2}\).
\ee
Demanding that the resolvent decreases at infinity, we get:
\be\label{someir}
  iR=\log\[1-\frac{u^*}{u}\]+\log\[1+\frac{u^*}{u}\]+\CO(1/u)=2\log M+\CO\(\frac{\log[u]}{\log[M]}\)+\CO(1/u).
\ee
We see that the external holes define the behavior of the resolvent for the values of $u$ in the interval $M\gg u\gg 1$.

The energy can be calculated as
\be\label{someen}
  E=\int_{-\infty}^{\infty}du\frac{\rho[u]}{u^2+1/4}\ .
\ee

It was shown in \cite{Belitsky:2006en} that in the limit (\ref{sudakovds}) the energy scales logarithmically:
\be
  2g^2 E=f[j]\log M+\ldots\ ,
\ee
where we introduced the normalization of $f[j]$ which is used in the AdS/CFT correspondence.

We see from (\ref{someir}) and (\ref{someen}) that the Bethe roots with absolute value larger than $\log[M]$ do not contribute to $f[j]$. Therefore we can consider that the asymptotic behavior of the resolvent at infinity is given by (\ref{someir}).

\section{\label{sec:largejsolution}Large $j$ solution}
In view of the asymptotic behavior (\ref{someir}) it is useful to rescale the resolvents:
\be\label{resolventrescaling}
  R:=R\log M,\ \ R_h:=R_h\log M.
\ee
Then in the scaling limit (\ref{sudakovscaling}) we obtain the following functional equations:
\be\label{sudakoveq}
  (1-D^2)R+R_h&=&D\frac j{u},\ \ Im[u]>0,\no\\
  (1-D^{-2})R+R_h&=&D^{-1}\frac j{u},\ \ Im[u]<0,
\ee
with asymptotic behavior given by:
\be\label{sudassym}
  R[u+i0]&=&-2i+\CO\(\frac 1{u}\),\ \ u\to\infty,\no\\
  R_h[u]&=&\frac {j}{u}.
\ee
These equations are equivalent to the linear integral equation on the density of holes used in \cite{Freyhult:2007pz,Bombardelli:2008ah}.

The scaling function $f[j]$ can be found as
\be\label{sudenergy}
  f[j]=i2g^2(R[i/2]-R[-i/2])=2g^2\lim\limits_{u\to 0}\(\frac ju-D^{-1} R_{h}^{+0}[u]\),
\ee
where $D^{-1} R_{h}^{+0}[u]$ means that we analytically continue to the point $u-i/2$ from a point $u$ which has a  positive imaginary part.

We aim to solve (\ref{sudakoveq}) perturbatively for the case of large $j$. As we will see below, the position of the branch point, which we denote as $\BrP$, scales as $j$. Therefore the typical scale for the solution is $u\sim \BrP$ and the shift operator can be treated perturbatively:
\be
  D=e^{\frac i2\p_u}=1+\frac i2\p_u+\ldots\ .
\ee
Let us consider the equations (\ref{sudakoveq}) on the support $[-\BrP,\BrP]$ of the density of holes. After multiplying the first one by $D^{-1}$ and the second one by $D$ and taking the sum we will get
\be\label{sudakovrprm}
  D^{-1}R_h[u+i0]+D R_h[u-i0]=\frac j{u+i0}+\frac j{u-i0},\ \ u^2<B^2.
\ee
This equation can be rewritten in the following form:
\be\label{RH111}
  R_h[u+i0]+R_h[u-i0]&=&\frac {D-D^{-1}}{D+D^{-1}}(R_h[u+i0]-R_h[u-i0])+\no\\&&\hspace{1EM}+\frac 2{D+D^{-1}}\(\frac j{u+i0}+\frac j{u-i0}\).
\ee
Since $D-D^{-1}= i\p_u+\ldots=o(1)$, (\ref{RH111}) gives us the possibility to perform the perturbative expansion in the recursive manner. Of course, the solution of (\ref{RH111}) is not unique. We can always add to it a solution of the homogeneous equation
\be\label{RHhom}
  R_{h,hom}[u+i0]+R_{h,hom}[u-i0]=0.
\ee
The correct solution can be fixed from additional requirements, which we give below.

\subsubsection{Leading and subleading orders.}
Let us assume the following large $B$ expansion of the resolvent:
\be\label{rhexpansion}
  j^{-1}R_h=R_{h,1}+R_{h,2}+\ldots
\ee
 After performing the change of variables $u= B\ \hat u$, the term $R_{h,m}$ should scale as $B^{-m}$ times a polynomial of $\log[B]$.

A direct calculation gives the first two orders of the perturbative expansion:
\be
  R_{h,1}=\frac 1u,\ \
  R_{h,2}=\frac 12\frac{B}{u^3\sqrt{1-\frac{B^2}{u^2}}}.
\ee
The solution $R_{h,1}$ is obvious. For $R_{h,2}$ one can do a simple check:
\be
  R_{h,2}[u+i0]+R_{h,2}[u-i0]&=&\frac{B}{2}\(\(\frac 1{u^3\sqrt{1-\frac{B^2}{u^2}}}\)_++\(\frac 1{u^3\sqrt{1-\frac{B^2}{u^2}}}\)_-\)=\no\\
  &&\frac{-i}{2\sqrt{1-\frac{u^2}{B^2}}}\(\frac 1{(u+i0)^2}-\frac 1{(u-i0)^2}\).
\ee
The last expression is nonzero only for $u=0$, therefore we can replace $\sqrt{1-u^2/B^2}$ by $1$.

$R_{h,1}$ has the asymptotics at infinity required in (\ref{sudassym}). Therefore all the other terms in the expansion (\ref{rhexpansion}) should decrease at least as $1/u^3$ at infinity.

The solution of the homogeneous equation (\ref{RHhom}) can have singularities only at $u=\pm B$. The general solution of the homogeneous equation which decreases as $1/u^3$ at infinity is
\be\label{Rhhom}
  R_{h,hom}=\frac 1{u\sqrt{1-\frac {B^2}{u^2}}}\sum_{m=1}^\infty\sum_{n=1}^\infty\frac {c_{n,m}}{B^{m-n}(u^2-B^2)^n}.
\ee
The index $m$ cannot be equal $0$, since the first order in which $R_{h,hom}$ can appear is $R_{h,3}$. The reason is that the expansion (\ref{rhexpansion}) is an expansion of an integrable function, which implies that $R_{h,2}$ should be integrable.

To relate the boundary parameter $B$ with $j$ we have to use (\ref{sudakoveq}), which for ${\rm Im}[u]>0$ leads to
\be\label{RRh}
  R=\frac 1{1-D^2}\left(D \frac j{u}-R_h\right)=j\frac i{\p_u}\(-\frac i2\frac 1{u^2}-\frac 12\frac{B}{u^3\sqrt{1-\frac {B^2}{u^2}}}\)+\ldots.
\ee
The inverse differential operator means integration. The constant of integration should be fixed from the parity properties of $R$. The result for the leading order of R is:
\be
  j^{-1} R[u]=-\frac 1{2u}-\frac{i}{2B}\sqrt{1-\frac{B^2}{u^2}},\ \ {\rm Im}[u]>0.
\ee
Indeed, this function is odd if we take the cuts of the square root to be $(-\infty,-B]\cup[B,\infty)$ which is the support of the density of roots.

To have the correct asymptotics at infinity we have to demand $B=j/4$. This relation should be corrected at the next orders of the large $B$ expansion.

\section{\label{Sec:doublescaling}Solution in the double scaling limit}
Starting from $R_{h,3}$ we get a nontrivial contribution from (\ref{Rhhom}) and the solution of (\ref{RH111}) becomes ambiguous. To fix this ambiguity, and this is the key point for the perturbative solution of such kind of equations, we should consider the initial equations in the double scaling limit which is defined as follows. We introduce a new variable $z$
\be
  z=2(u-B)
\ee
and expand our equation in the powers of $1/B$ with $z$ fixed. We get the equations
\be\label{sudakovds}
  (1-D^2)R[z]+R_h[z]\simeq 0,\ \ {\rm Im}[z]>0,\no\\
  (1-D^{-2})R[z]+R_h[z]\simeq 0,\ \ {\rm Im}[z]<0,
\ee
where $R[z]$ has a discontinuity at $z\ \e\ [0,\infty)$, while $R_h[z]$ has a discontinuity at $z\ \e\ (-\infty,0]$. The weak equivalence to zero means that the $l.h.s.$ is proportional to a function $f[z]$ which is analytic at zero.

This equation can be solved explicitly after performing the inverse Laplace transform
\be
  \tilde R_h[s]=\int_{-i\infty+0}^{i \infty+0}\frac {dz}{2\pi i}e^{sz}R_h[z],\ \ \no
  \tilde R[s]=\int_{i \infty-0}^{-i \infty-0}\frac {dz}{2\pi i}e^{s z}R[s].
\ee
We should consider the equations (\ref{sudakovds}) order by order in the $1/B$ expansion. At each order the function $f[z]$ analytic at zero is just a polynomial. Its inverse Laplace transform is therefore a sum of delta function at $s=0$ and its derivatives. This sum does not affect the conclusions that we will make in this section. We therefore put $f[z]=0$.

Let us investigate analytic properties of the function $\tilde R_h[s]$.
\begin{figure}[t]
\centering
\includegraphics[height=4cm]{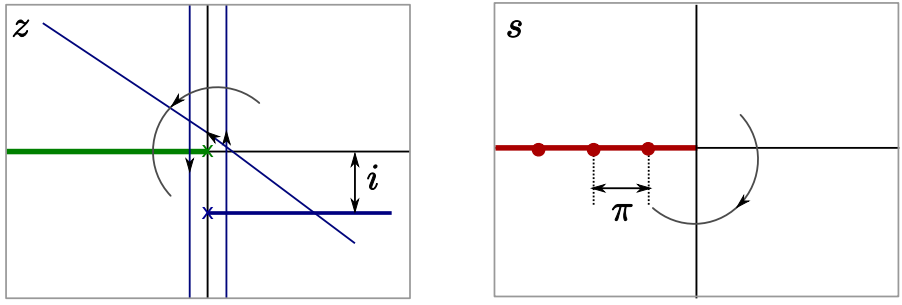}
\caption{\label{fig:laplaceandanalytics}Analytical continuation in $z$ and $s$ planes.}
\end{figure}
The resolvent $R_h[z]$ has only one branch cut on $z<0$ in $z$-plane. Therefore the Laplace transform should be defined for ${\rm Re}[s]>0$. We analytically continue the function $\tilde R_h[s]$ to the region with  ${\rm Re}[s]<0$ by changing the phase of $s$ and simultaneously rotating the contour of integration of the Laplace transform such that ${\rm Re}[z s]=0$. So, if we will change $s$ in the clockwise direction, the contour of integration should be rotated in the counterclockwise direction.

To get into the region ${\rm Re}[s]<0$ the contour of integration passes through the cut of $R_h[z]$. To get the analytical continuation of $R_h[z]$ to the other Riemann sheet we use the first equation in (\ref{sudakovds}) to reexpress $R_h[z]$ in terms of $R[z]$. We see for example that when we pass through the cut $z<0$ we encounter a branch point at $z=-i$.

Performing the Laplace transform of the first equation in (\ref{sudakovds}) we see that the analytical continuation of $\tilde R_h[s]$ by rotation $s\to s e^{-i\pi}$ is given by
\be
  \tilde R_h[s-i0]=(1-e^{-2is})\tilde R[s].
\ee
Since $\tilde R[s]$ is analytic for $s<0$ we deduce that $\tilde R_h[s]$ should have zeroes for $s=\pi n,\ n<0$.

Equivalently we can find that for $s<0$:
\be
  \tilde R_h[s+i0]=(1-e^{2is})\tilde R[s].
\ee

The conclusion is that $\tilde R_h[s]$ is a function analytic everywhere except on the axis $s<0$ where it satisfies the equation
\be\label{sudlaplace}
  e^{is}\tilde R_h[s-i0]+e^{-is}\tilde R_h[s+i0]=0.
\ee

We can fix the solution of this equation if we remark that $\tilde R_h[s]$ should be expanded in inverse powers of $s$ at infinity. Indeed, from the structure of the functional equations (\ref{sudakovds}) we conclude that $R_h[z]+R[z]$ should be analytic at $z=0$. Therefore the total density $\rho_h[z]+\rho[z]$ should be analytic as well. This implies that the expansion of $R_h[z]$ at $z=0$ is given by
\be
  R_h[z]=\log[z](\g_0+z\g_1+z^2\g_2+\ldots).
\ee
The inverse Laplace transform of this expansion is
\be
  \tilde R_h[s]=-\frac{\g_0}{s}+\frac{\g_1}{s^2}-\frac{2\g_2}{s^3}+\ldots,\ \ s\neq 0,
\ee
which is the stated behavior of $\tilde R_h[s]$ at infinity.

The solution of (\ref{sudlaplace}) with the required structure of the zeroes and behavior at infinity  is given by
\be\label{dssol}
  \tilde R_h[s]=\frac A s Q[s,B]\Phi[s],\ \ \Phi[s]=\sqrt{2s}\frac{e^{\frac s\pi\log\frac s\pi-\frac s\pi}}{\Gamma\[\frac s\pi+1\]},\ \
  Q[s,B]=\sum_{n,m=0}^\infty \frac{Q_{n,m}[\log B]}{B^{n+m}s^n}.
\ee
At each order of $1/B$ expansion the sum $Q[s,B]$ contains only a finite number of terms and therefore is well defined. For convenience we put the overall normalization $A$ such that $Q_{0,0}=1$.

The dependence of $Q[s,B]$ on $B$ of course cannot be understood from the equation (\ref{sudlaplace}). It will become clear when we will show how to find the coefficients $Q_{n,m}$.

\section{Higher orders in perturbative expansion}
The still unknown coefficients $Q_{n,m}$ in (\ref{Rhhom}) and $c_{n,m}$ in (\ref{dssol}) are fixed by demanding that the large $B$ expansion (\ref{rhexpansion}) and the expansion (\ref{dssol}) in the double scaling limit match in the intermediate regime. The expansion (\ref{rhexpansion}) in the double scaling regime (\ref{doublescaling}) organizes at each order of $1/B$ as a large $z$ expansion, therefore it should match with the Laplace transform of the small $s$ expansion of (\ref{dssol}).

Let us see how this works for $R_{h,3}$. The general solution of (\ref{RH111}) at this order is given by
\be
  R_{h,3}=-\frac 1{\sqrt{1-\frac{B^2}{u^2}}}\(\frac{3B}{4\pi}\frac{u^2-\frac 23B^2}{u^2-B^2}\frac{\log\frac{u-B}{u+B}}{u^4}+\frac 1{\pi u^3}-\frac{c_{1,1}}{u(u^2-B^2)}\)+\frac 1{4u^3}.
\ee
Reexpansion of $j^{-1}R_h$ in the double scaling regime is then given by
\be\label{lB1}
  R_{h,1}+R_{h,2}+R_{h,3}=\frac 1B-\frac z{2B^2}+\frac 1{(Bz)^{3/2}}\(\frac z2+c_{1,1}-\frac 1{4\pi}{\log\[\frac{z}{4B}\]}\)+\ldots
\ee
Correspondingly, the Laplace transform of the small $s$ expansion of (\ref{dssol}) is given by
\be\label{lB2}
\int_0^\infty ds e^{-sz}\tilde R_h[s]=\frac{2\sqrt{2\pi}A}{z^{3/2}}\(\frac z2+\frac {1-\log[4\pi z]}{4\pi}\).
\ee
The first two terms in (\ref{lB1}) are analytic at $z=0$. Demanding the weak equivalence between (\ref{lB1}) and (\ref{lB2}) in the sense of (\ref{sudakovds}) we find
\be
  A=\frac 1{2\sqrt{2\pi}B^{3/2}},\ \ c_{1,1}=\frac{1-\log[16\pi B]}{4\pi}.
\ee
To obtain the expansion (\ref{Eexpn}) we need also $R_{h,4}$. The solution at this order is given in appendix \ref{app:sl2}.

To relate the parameters $B$ and $j$ we find the resolvent $R$ from (\ref{RRh}) and demand that it has the correct asymptotic behavior (\ref{sudassym}). This leads to
\be
  B=\frac {j}4-\frac{1-\log[4\pi j]}{2\pi}+\frac {\frac 16-\frac 3{2\pi^2}}{j}+\CO(j^{-2}).
\ee
The scaling function $f[j]$ can be found from (\ref{sudenergy}). In terms of $B$ this is a rather complicated expression containing $\log[B]$ as well. But in terms of the parameter $j$ it significantly simplifies:
\be\label{Eexpn}
  f[j]=8g^2\(\frac 1j-\frac{8}{3\pi}\frac 1{j^2}+\frac 2{3}\frac 1{j^3}+\CO(j^{-4})\).
\ee
This expansion matches with numerical analysis in \cite{Beccaria:2008nf,V1}.

These three orders of large $j$ expansion lie in the BMN-like window of the AdS/CFT. More precisely, as we conclude from \cite{Frolov:2006qe} the large $j$ expansion of the generalized scaling function $f[g,j]$ has the form
\be\label{BMNlike}
  f[g,j]=\suml_{n\geq 1}\frac{g^{2n}}{j^{2n}}\suml_{m\geq 0}\frac{c_{nm}[g]}{j^{m-1}}.
\ee
In this expansion the coefficients $c_{10},c_{11},c_{12},c_{20},c_{21}$ do not depend on the coupling constant.
The coefficients $c_{10},c_{11},c_{12}$ define the expansion (\ref{Eexpn}). The coefficients $c_{20},c_{21}$ appear in the large $j$ expansion of the two-loop correction to the generalized scaling function and were not calculated here.

$c_{10}$ and $c_{11}$ can be found respectively from tree \cite{Gubser:2002tv,Frolov:2002av} and one loop \cite{Frolov:2006qe} calculations on the string side. In \cite{V1} $c_{12}$ was found from the strong coupling solution of the BES/FRS equation up to two loops. All findings are in agreement with (\ref{Eexpn}).

\chapter{\label{ch:BESFRS}Generalized scaling function at strong coupling}
\section{BES/FRS equation and different regimes at \\ strong coupling}
In the previous chapter we discussed the energy of the ground state of the $Sl(2)$ Heisenberg ferromagnet in various regimes of the thermodynamic limit. We saw that the answer depends strongly on the relation of two parameters - the twist $L$ and the number of magnons $M=S$. For $L\lesssim \log S$ the energy scales logarithmically with a prefactor $f[j]$, $j=L/\log S$. One expects \cite{Freyhult:2007pz} that the logarithmic scaling of the anomalous dimension
\be
  E-S=f[g,j]\log[S]+\CO(1)
\ee
in the limit with fixed $j$ takes place at any finite value of the coupling constant $g$. In section \ref{sec:cusp} we discussed how $f[g,j]$ can be used for the test of integrability in AdS/CFT. Now more technical details would be given.

The Heisenberg ferromagnet describes the spectrum of the dilatation operator only at one loop. The all-loop result may be obtained from the conjectured Beisert-Staudacher asymptotic Bethe Ansatz. In the $Sl(2)$ sector it reduces to the one integral equation known as the BES/FRS equation. Its derivation is completely analogous to the derivation of the one loop equation discussed above. This equation was initially derived in \cite{Freyhult:2007pz} and also in \cite{Bombardelli:2008ah}. In our notations it reads \cite{V1}:
\be\label{holobes}
(1-D^2)R_m+R_h&=&-D\CK D R_m + \frac 1{\e}D \frac {\ell}{x}\frac {dx}{du},\no\\
\CK&=&K_-+K_++2K_-D\frac 1{1-D^2}DK_+,
\ee
where
\be
\ell=\e j,\ \ \e=\frac 1{4g}.
\ee

 The kernels $K_\pm$ are defined by \cite{KSV2}
\be
  (K_\pm\cdot F)[u]=(K\cdot F)[u]\pm (K\cdot F)[-u],
\ee
where $K$ is given by (\ref{kdef}).

The asymptotic behavior of the resolvents at infinity is
\be\label{escaling}
  R_m&\rightarrow& \mp \frac {i}{\e}+\frac{\beta}u\,,\, u\rightarrow \infty\pm i0;\no\\
  R_h&\rightarrow& \frac{j}{u}\,,\,u\to\infty.
\ee

Based on the all-loop integral equation (\ref{holobes}), we can relate the generalized scaling function $f[g,j]$ and the parameter $\b$ in (\ref{escaling})\footnote{This property was initially remarked in \cite{Kotikov:2006ts}.}. Indeed, the anomalous dimension is given by the formula
\be\label{comp1}
  \g=\frac 1{2\e}\suml_{k}\(\frac i{\xp_k}-\frac i{\xm_k}\)&=&
  \frac {\log S}{2\pi\e}\intl_{\MR-i0} dv \frac 1{y}DR_m.
\ee
Expanding the BES/FRS equation (\ref{holobes}) at infinity, we obtain for the coefficients in front of $1/u^2$
\be\label{comp2}
\beta=\frac 1{2i\e}\intl_{-\infty}^{\infty}\frac {dv}{2\pi i}\frac 1y DR_m-\frac j2.
\ee
Comparing (\ref{comp1}) and (\ref{comp2}), we conclude that the generalized scaling function can be found as
\be\label{f2cj}
  f=-2\beta-j\;.
\ee
Note that the derivation of this relation includes expansion of the kernel $K$ at large $x$, therefore it is essentially based on the higher order corrections to the one-loop integral equation.

The resolvents $R_m$ and $R_h$ in the BES/FRS equation have two types of branch points. The first type of branch points has "kinematic" origin - these are branch points of the inverse Zhukovsky map $x[u]$ which is singular at $u=\pm 2g$. The "kinematic" branch points are located at $u=\pm 2g+(\MZ+1/2)i$. The branch points of the second type are located at $u=\pm a+ \MZ i$, where $a[g,j]$ is the end of the root distribution.

It is natural to assume that the only nonanalyticities in the generalized scaling function appear at the values of $g$ and $j$ for which the branch points of different types collide. For real finite $g$ and $j$ the collision is impossible. However, if we take into account complex values of these parameters, the collision is possible. Using this observation, we can determine in particular the radius of convergence of the weak coupling expansion of $2\G_{cusp}[g]$. Indeed, for $j=0$ we have $a=0$. The kinematic branch point touches the origin when $g=i(1/4+\MZ/2)$. Therefore, the radius of convergence equals $1/4$. This observation coincides with the numerical prediction in \cite{Beisert:2006ez}.

To investigate analytical properties of the generalized scaling function at strong coupling it is useful to introduce the rescaled coordinate $u_{rescaled}=u/2g$. At strong coupling  all the kinematic points condense into two points $u_{rescaled}=\pm 1$ that are situated on the real axis. Therefore we will potentially have different regimes depending where the branch point $a$ is situated: inside the interval $[-2g,2g]$, outside it, or at the finite distance from the point $u=\pm 2g$.

\paragraph{Finite-twist operators and small perturbations.}

This is a regime in which $a$ is situated inside the interval $[-2g,2g]$. In the particular case when $a=0$ (and therefore $j=0$) the generalized scaling function reduces to twice the cusp anomalous dimension
\be
  f[g,0]=2\Gamma_{cusp}[g]
\ee
To obtain a nonzero $a$ inside the interval $[-2g,2g]$ the parameter  $j$ should scale as
$j\simeq e^{\pi (a-2g)}$, so it is exponentially small. As was shown in the works of Alday and Maldacena \cite{Alday2007} and Basso and Korchemsky \cite{Basso:2008tx}, this regime is effectively described by the dynamics of the $O(6)$ sigma model. The quantity $2\G_{cusp}-j$ can be viewed as the vacuum energy and $f[g,0]-2\Gamma_{cusp}+j=\varepsilon_{O(6)}[j/m]$ is identified with the energy of the $O(6)$ sigma model. The physical motivation for this particular description was discussed in Sec.~\ref{sec:O6physical}. In Sec.~\ref{sec:O6} we give a derivation, alternative to the one in \cite{Basso:2008tx}, of the integral equation for the $O(6)$ sigma model from the BES/FRS integral equation.

\paragraph{The limit with $j\sim g$.}

In the large $g$ limit with $\ell=\e j$ fixed the branch point is situated outside the interval $[-2g,2g]$. The position of branch point at $\e=0$ is given by \cite{Casteill:2007ct,Gromov:2008en}:
\be\label{a2g}
    a[\e=0,\ell]=g\(b+\frac 1b\),\;\;\;\;b\equiv \sqrt{1+\ell^2}.
\ee
$a>2g$ for any positive $\ell$. Therefore if we perform the perturbation series as
\be\label{flexpansion}
  f[g,j]=\frac 1{\e}\left(f_0[\ell]+\e f_1[\ell]+\ldots\right).
\ee
each function $f_n[\ell]$ should be analytical function of $\ell$. This regime was a subject of investigation of \cite{V1}. In Sec.~\ref{sec:perturbativeregime} we discuss the results of this investigation.

\paragraph{Intermediate regime.}

Finally, let us consider the regime when the distance between the branch point $a$ and the Zhukovsky branch point $2g$ is of order of $1$. This limit should describe the transition between the $O(6)$ sigma model which is realized in the case $2g-a\gg 1$ and the string perturbative regime $a-2g\gg 1$. From the $O(6)$ sigma model we know that $j$ behaves as
\be\label{as1}
  j\propto g^{3/4}e^{\pi g(a-1)}+\ldots .
\ee
Performing the small $\ell$ expansion of the expression of (\ref{a2g}) we get
\be
  j=4\sqrt{2}g^{3/4}\(a-2g\)^{1/4}.
\ee
Therefore if we introduce the parameter $\delta=a-2g$ and consider the strong coupling limit with $\delta$ fixed we can expect that
\be\label{as2}
  j=g^{3/4}V[\delta]+o(g^{3/4})
\ee
with $V[\delta]$ interpolating smoothly between the asymptotics (\ref{as1}) and (\ref{as2}). This observation is further supported by the known analytic structure of the BES/FRS equation: we do not encounter new singular points while varying $\delta$.

The study of this regime was not done and may be the subject of a future research.

\section{\label{sec:cuspanomaly}Cusp anomalous dimension}
The BES/FRS equation for $j=0$ (and therefore $R_h=0$) reduces to the BES equation. This is the very equation in which the all-loop proposal for the BES/BHL dressing phase initially appeared. The cusp anomalous dimension which can be found from the solution of the BES equation is a nontrivial quantity and is  accessible up to four loops on the gauge side and up to two loops on the string side. The perturbative recursive solution of the BES equation at strong coupling was found, after a number of initial attempts \cite{Kotikov:2006ts},\cite{Benna:2006nd,Alday:2007qf},\cite{Beccaria:2007tk,KSV1},\cite{Casteill:2007ct,Belitsky:2007kf}, in \cite{Basso:2007wd},\cite{KSV2}. Here we will show that the method used for the solution is in fact the same as the one used in the previous chapter. Our analysis follows closely \cite{KSV2} with a few improvements.

The BES equation reads
\be
  (1-D^2)R_m=-D\CK DR_m,
\ee
with the asymptotic conditions on $R_m$ given by (\ref{escaling}). It turns out to be useful to introduce the resolvents\footnote{In \cite{KSV2} we used a shifted definition: $$R_{\pm,{\rm here}}=DR_{\pm,{\rm KSV2}}.$$} $R_+$ and $R_-$ in the upper half of the $u$-plane by the following equations valid for any value of the coupling constant:
\be
R_m&=&R_++R_-\,,\\
\label{S1x}R_+&=&-\frac{D}{1-D^2}K_+D(R_++R_-)\,,\\
\label{Sx}R_-&=&-\frac{D}{1-D^2}K_-D(-R_++R_-)\,.
\ee
There are combinations $G_\pm$ and $g_\pm$ of $R_+$ and $R_-$ with simple analytical properties. They are defined in the upper half plane by
\be
\label{Gpmgpm}
G_\pm&=&\frac{1\pm i}{2}(D\mp i D^{-1})(R_-\pm i R_+)\,,\\
g_\pm&=&\pm i(D-D^{-1})(R_-\pm i R_+)\,,
\ee
and further by analytical continuation\footnote{Similar linear combinations in the Fourier space were initially proposed in \cite{Basso:2007wd}.}.

The functions $G_\pm[u]$ and $g_\pm[u]$ have square root type branching at the points
$u=\pm 2g$ inherited from $R_+$ and $R_-$. If to choose the branch cut to be on $[-2g,2g]$ for $g_\pm$ and
on $\mathbb{R}\backslash(-2g,2g)$ for $G_\pm$ then one can show from the definitions of $g_\pm$ and $G_\pm$ and the property (\ref{analyticpropK}) of the kernels $K_\pm$ that $g_\pm$ and $G_\pm$ are analytic outside these cuts. Analytical continuation through the cuts reveals the branch points at $u=\pm 2g+i\mathbb{Z}$ on the next sheet.

The resolvents $G_\pm$ and $g_\pm$ are related in the upper half plane as
\be\label{Gg}
G_\pm=\frac{1\mp i}2\frac{D^2\mp i}{D^2-1}g_\pm.
\ee
Since $G_\pm$ and $g_\pm$ have discontinuities on the complementary cuts, (\ref{Gg}) is very similar to the integral equations which follows from the Baxter equations (for example like (\ref{somebaxter})).

The resolvents $G_\pm$ perturbatively at strong coupling, in the regime $g\to\infty$ with $u/g$ fixed, satisfy the equations \cite{KSV2}
\be\label{Gperturbative}
  G_\pm[u+i0]=\mp i G_\pm[u-i0],\ \ u^2<4g^2.
\ee
We see that the BES equation reduces to a very simple Riemann-Hilbert problem. Its most general solution is given by:
\be\label{solG}
  G_\pm=2i\e\sum_{n\e\MZ}\a_n[\e]\e^{|n|}\(\frac{u+2g}{u-2g}\)^{\pm n\pm \frac14}.
\ee
Using the formula (\ref{f2cj}) and the definitions of the resolvents, we derive that the scaling function can be expressed in terms of $\a_n[\e]$ as
\be
 f[g]=\sum_{n\e\MZ}(4n+1)\e^{|n|}\a_n[\e].
\ee
Similarly to the discussion in chapter~\ref{ch:sl2}, to fix the coefficients $\a_n$ we should solve (\ref{Gg}) in the double scaling limit
\be\label{BESdoublescaling}
\e\to 0,u\to 2g\hspace{2em}{\rm with}\hspace{2em} z=u-2g\ \ {\rm fixed}
\ee
and then demand the compatibility of this solution with (\ref{solG}). The demand that the double scaling limit exists explains the presence of $\e^{|n|}$ factor in (\ref{solG}). The coefficients $\a_n[\e]$ should be bounded at $\e=0$.

The solution in the double scaling limit can be most easily found if to apply the inverse Laplace transform:
\be
  \tilde g_\pm[s]=\int_{-i\infty+0}^{i\infty+0}\frac{dz}{2\pi i}e^{zs}g_\pm[z],\ \ \tilde G_\pm[s]=\int_{i\infty-0}^{-i\infty-0}\frac{dz}{2\pi i}e^{zs}G_\pm[z].
\ee
The study of the analytical properties of $\tilde g_\pm[s]$ and $\tilde G_\pm[s]$, see the discussion near Fig.~\ref{fig:laplaceandanalytics}, leads to the conclusion that $\tilde g_\pm[s]$ is analytic everywhere except on the negative real axis and $\tilde G_\pm[s]$ is analytic everywhere except on the positive real axis.

The inverse Laplace transform of (\ref{Gg}) has the following form
\be\label{Gglaplace}
\tilde G_\pm[s]=\frac{1\mp i}2\frac{e^{-is}\mp i}{e^{-is}-1}\tilde g_\pm[s-i0],\ \ s<0.
\ee
This equation and the equation for $\tilde g_\pm[s+i0]$, which can be similarly derived, imply that $\tilde g_\pm[s]$ and $\tilde G_\pm[s]$ have monodromy of order four around zero\footnote{We can see this also in the following way. The monodromy of $\tilde G_\pm[s]$ around zero is inherited from the monodromy of $G[z]$ around infinity. At large values of $z$ the solution in the double scaling limit should match with the perturbative solution (\ref{solG}). The monodromy of the latter is of order four, as it follows from (\ref{Gperturbative}).}.

Another consequence of (\ref{Gglaplace}) is the following. Let us analytically continue this equation to $s>0$ by the path that does not cross the real axis. Then, since $\tilde g_\pm[s]$ is analytic for $s>0$, $\tilde G_\pm[s]$ should have zeroes at $s=(2n+1\pm \frac 12)\pi,\ n\geq 0$ and may have poles only at $s=2\pi n,\ n\geq 0$.

Finally, $G_\pm[z]$ has a square root branch point at $z=0$, same as the square root branch point for $G_\pm[u]$ at finite $g$. This implies that at $s=\infty$ $\tilde G_\pm[s]$ is expanded in inverse half-integer powers of $s$.

The most general solution for $G_\pm[s]$ with stated above properties is given by
\be\label{solgs}
  \tilde G_\pm[s]=\frac{\(\frac s\e\)^{-1\pm \frac 14}}{\Gamma[\frac 12\pm\frac 14]}\!\!\!\!\!\!&&\!\!\!\!T_\pm[s]Q_\pm[s],\\
   T_\pm[s]=\frac{\Gamma\[1-\frac{s}{2\pi}\]\Gamma[\frac 12\pm\frac 14]}{\Gamma\[\frac 12-\frac s{2\pi}\pm \frac 14\]},&&\ Q_\pm[s]=\sum_{n=0}^\infty \b_n^\pm[\e]\left(\frac{s}{\e}\right)^{-n}.\no
\ee

The requirement of compatibility of the solutions (\ref{solG}) and (\ref{solgs}) fixes unambiguously all the coefficients $\a_n$ and $\b_n$. This requirement in particular implies the presence of an overall factor $\e^{\mp\frac 14}$ in (\ref{solgs}) and that $\b_n[\e]$ should be bounded at $\e=0$. The algorithmic procedure for calculation of $\a_n$ and $\b_n$, realized with the help of \textit{Mathematica}, is presented in appendix~\ref{app:cusp}.

As a result, at first three orders of the strong coupling expansion we get
\be\label{fgexpansion}
    f[g]=4g-\frac{3\log 2}{\pi}-\frac 1{4g}\frac{K}{\pi^2}-\ldots\ .
\ee
This result was initially obtained in \cite{Basso:2007wd}. The derivation there contained number of assumptions which we were able to prove in \cite{KSV2} confirming that result. The obtained expansion (\ref{fgexpansion}) coincides with the two-loop string prediction \cite{Roiban:2007jf,Roiban:2007dq}. This was the first two-loop test at strong coupling of the asymptotic Bethe Ansatz proposal.

\section{\label{sec:O6}$O(6)$ sigma model}
The appearance of the $O(6)$ sigma model in a special regime of the AdS/CFT system can be understood from the point of view of the string theory. If we consider a folded string solution, then the centrifugal forces make all the fluctuations on the AdS space, as well all the fermion excitations massive. Therefore the low-energy excitations in the theory should be described by the fluctuating modes that live on $S^5$, \textit{i.e.} by the $O(6)$ sigma model. The mass of the AdS fluctuating modes determines the energy scale below which the $O(6)$ sigma model approximation is applicable.

We can test the $O(6)$ sigma model regime if we consider the string which moves around equator of the $S^5$ in addition to the rotation in $AdS$, and therefore is described by the $AdS$ angular momentum $S$ and the $S^5$ angular momentum $J$. This is the configuration which is described by the $BES/FRS$ equation.

In order to stay in the $O(6)$ sigma model regime, we should put $J\ll S$. More precisely, $J$ should be at most of the order $\log S$. We again come back to a parameter $j=J/\log S$. As we will see below, the density\footnote{number per unit volume} of holes of the BES/FRS equation is proportional to $j$.

The $O(6)$ sigma model, through the mechanism of dimensional transmutation, acquires a nontrivial mass scale $m$ which should depend on the coupling constant by the law
\be
  m\propto g^{1/4}e^{-\pi g}.
\ee
A very nontrivial check of the integrability in the AdS/CFT correspondence would be to identify the $O(6)$ particles with holes of the BES/FRS equation as far as we stay at the energies much lower than the masses of fluctuations in AdS. This regime is possible when $j/m$ is finite. In this case we come back to the scaling law (\ref{as1}) for the strong coupling behavior of $j$.

The identification of the BES/FRS holes and the excitations of the $O(6)$ sigma model was done by Basso and Korchemsky in \cite{Basso:2008tx}. They showed that in the regime (\ref{as1}) the BES/FRS equation can be reduced to the integral equation obeyed by the particles of the $O(6)$ sigma model. Here we will give a different derivation of their result.

Let us consider in addition to the BES/FRS equation (\ref{holobes}) in the u.h.p, the BES/FRS equation in the l.h.p. It reads
\be\label{examplepm}
  R_h^{-0}=-(1-D^{-2})R_m^{-0}-D^{-1}(K_-^{-0}-K_+^{-0}+2K_-^{-0}\frac {D^2}{1-D^2}K_+)DR_m-D^{-1}\frac{2\ell/\e}{x-\frac 1x}.
\ee
For the meaning of notation $F^{\pm 0}$ see appendix \ref{app:shift}.

In our derivation we will rely on the properties of the kernels $K_\pm$
\be\label{analyticpropK}
  (K_\pm^{-0} F)[u]+(K_\pm^{+0}F)[u]=F[u+i0]\pm F[-u+i0]
\ee
and on the fact that $R_m$ and $R_h$ are odd functions.

Considering the BES/FRS equation for $u^2<4g^2$ we can make the following transformations:

\be
  K_-^{-0}DR_m&=&-K_-^{+0}DR_m+DR_m^{+0}+D^{-1}R_m^{-0},\no\\
  K_+^{-0}DR_m&=&-K_+^{+0}DR_m+DR_m^{+0}-D^{-1}R_m^{-0},\no\\
  2\(K_-^{-0}\frac {D^2}{1-D^2}K_+\)DR_m&=&-2\(K_-^{-0}\frac {D^2}{1-D^2}K_+\)DR_m+2\frac {D^2}{1-D^2}K_+^{+0}DR_m-\no\\&&-2\frac{D^{-2}}{1-D^{-2}}K_+^{-0}DR_m=\no\\
  -2\(K_-^{-0}\frac {D^2}{1-D^2}K_+\)DR_m\!\!&+&\!\!2\frac {D^2}{1-D^2}K_+^{+0}DR_m+2\frac{D^{-2}}{1-D^{-2}}K_+^{+0}DR_m-\no\\&&-2\frac {D^{-2}}{1-D^{-2}}(DR_m^{+0}-D^{-1}R_m^{-0}).
\ee
Let us introduce an additional notation
\be
\scP_A\equiv\sum_{n=-\infty}^\infty A^n.
\ee
Using this notation we can rewrite
\be
  2\frac{D^2}{1-D^2}F+2\frac{D^{-2}}{1-D^{-2}}F=2\scP_{D^2}F-F
\ee
Summarizing, we can rewrite the BES/FRS equation in the upper and the lower half planes as
\be\label{someBES}
  R_h^{+0}&=&-(1-D^{2})R_m^{+0}-DK^{+0}DR_m+\frac 1{\e}D\frac{2\ell}{x-\frac 1x},\ \ {\rm Im}[u]>0,\ \ -2g<{\rm Re}[u]<2g,\no\\
  R_h^{-0}&=&-(1-D^{-2})R_m^{-0}+D^{-1}\CK^{+0}DR_m-\frac 1{\e}D^{-1}\frac{2\ell}{x-\frac 1x}-2D^{-1}\scP_{D^2}(\CK_+^{+0}DR_m)+\no\\&&+2\frac{D^{-2}}{1-D^{-2}}(R_m^{+0}-R_m^{-0}),\ \ {\rm Im}[u]<0,\ \ -2g<{\rm Re}[u]<2g.
\ee
Let us consider the interval $u^2<a^2$. In this interval $R_m^{+0}=R_m^{-0}$. We therefore can derive
\be\label{eqblabla}
  D^{-1}R_h^{+0}+DR_h^{-0}=-2\scP_{D^2}(K_+^{+0}DR_m).
\ee
We see that remarkably $R_h$ is almost decoupled from $R_m$. In the l.h.s. of (\ref{eqblabla}) we recognize the l.h.s. of perturbative functional equation for the $O(6)$ sigma model (\ref{ONperturbative}).

The exact equation for the $O(6)$ sigma model is written as
\be
  \frac{1-D^{2}}{1+D^{4}}R_h^{+0}-\frac{1-D^{-2}}{1+D^{-4}}R_h^{-0}=-2\pi i \mass \cosh\[\frac{\pi u}2\].
\ee
Let us then take the l.h.s. of the last equation  and evaluate it using (\ref{someBES}). After some algebra we will get
\be
  \frac{1-D^{2}}{1+D^{4}}R_h^{+0}-\frac{1-D^{-2}}{1+D^{-4}}R_h^{-0}&=&\CA_1+\CA_2,\no\\
  \CA_1=2D^2\scP_{-D^4}R_m,\ \ \ \CA_2&=&\scP_{-D^4}(D^3-D)\(\CK^{+0}DR_m+\frac 1\e\frac{2\ell}{x-\frac 1x}\).
\ee
The functions $\CA_1$ and $\CA_2$ have the same periodicity as $e^{\frac{\pi u}2}$ and branch points at the positions $u=\pm 2g$. Therefore at the leading order we expect
\be
  \CA_1+\CA_2=-2\pi i k\frac{e^{\frac{\pi u}2-\pi g}+e^{-\frac{\pi u}2-\pi g}}2=-2\pi i k e^{-\pi g}\cosh\left[\frac{\pi u}{2}\right]
\ee
We see that this gives the correct integral equation for the $O(6)$ sigma model.

The calculation of the coefficient $k$ is tricky and was done explicitly in \cite{Basso:2008tx}. It is given by
\be
  k=g^{1/4}\frac{2^{3/4}\pi^{1/4}}{\Gamma[5/4]}\ .
\ee
Therefore we see that the induced mass of the $O(6)$ sigma model has a correct RG behavior (prefactor $g^{1/4}$).

\section{\label{sec:perturbativeregime}Connection to the perturbative string theory\\ calculations.}
We have already discussed the limits which lead $a<2g$ at strong coupling. The limit which leads to $a>2g$ at strong coupling is $g,j\to\infty$ with $\ell=j/{4g}$ fixed. The interest for this limit, which was studied in \cite{V1}, is motivated by the fact that naively the expansion (\ref{flexpansion}) can be directly compared with the string theory results. However, the order of limits in calculation of (\ref{flexpansion}) is different from that used on the string theory side. On the string theory side the logarithmic scaling regime is attained in the following two-step limiting procedure:
\be\label{stringlimit}
1)\;\;&& g\!\sim\! L\!\sim\! S\to\infty,\no\\
2)\;\;&& S/g\gg L/g\gg 1,\;\; \ell=\frac{L/g}{4\log [S/g]} \;\;\;\textrm{finite}.
\ee
To compare, (\ref{flexpansion}) is obtained using the following order of limits
\be\label{BESFRSlimit}
1)\;\;&&L,S\to\infty,\;\; j=J/\log S\;\;{\rm and}\;\; g\;\;{\rm finite,}\no\\
2)\;\;&&g,j\to\infty,\;\; \ell=\frac{j}{4g} \;\;\;{\rm finite.}
\ee

\begin{figure}[t]
\centering
\includegraphics[height=3cm]{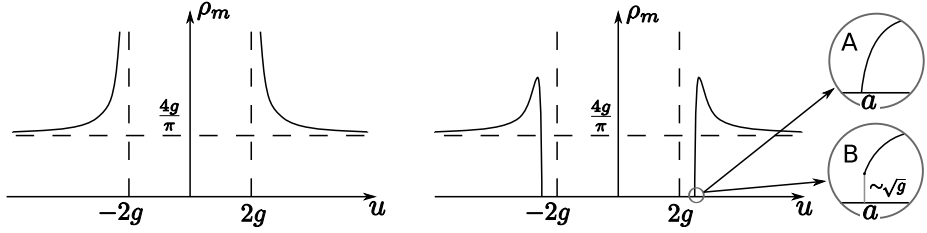}
\caption{\label{fig:densitybesfrs}The density of magnons (rescaled using (\ref{resolventrescaling}) at strong coupling). To the left: the case $j\ll g$. There is an exponential small with respect to $g$ number of magnons inside the interval $[-2g,2g]$. To the right: the case $j\sim g$. There are no magnons inside the interval $[-a,a]$, $a>2g$. The case A corresponds to the order of limits (\ref{stringlimit}). The branch point is of square root type. If we magnify up to the Baxter scale, we will see separate Bethe roots. The case B corresponds to the order of limits (\ref{BESFRSlimit}). The branch point is of logarithmic type. If we magnify up to the Baxter scale, there is to be still a dense distribution.}
\end{figure}

There is a qualitative difference between two limiting orders (\ref{stringlimit}) and (\ref{BESFRSlimit}), which may become important starting from two loops. It is shown at the r.h.s. of Fig.~\ref{fig:densitybesfrs}.

The limiting procedure (\ref{stringlimit}) is organized in a way that the distance between the Bethe roots in the vicinity of the branch point is larger than the Baxter scale. Therefore in the vicinity of the branch point we need both terms in the Baxter equation and approximation in which the BES/FRS equation was derived is inapplicable. This situation is similar to the WKB solution of quantum mechanics in a smooth potential.

In the limiting procedure (\ref{BESFRSlimit}) the distance between the Bethe roots in the vicinity of the branch point is much smaller than the Baxter scale. This is similar to the WKB solution in quantum mechanics with a steep function at the turning point.

This qualitative difference is absent in the case when we consider the strong coupling regime with $j\ll g$. As shown at the l.h.s. of Fig.~\ref{fig:densitybesfrs}, the density of roots is always large in this case.

In \cite{V1} the discrepancy at two-loop level was found between (\ref{flexpansion}) and the solution in \cite{Gromov:2008en} of the asymptotic Bethe Ansatz in the order of limits (\ref{stringlimit}). We however were not able to show explicitly that this discrepancy is due to the explained qualitative difference between two cases.

We see therefore that there are reasons to doubt the possibility of application of the BES/FRS equation, at least without modifications, for the comparison of the Bethe Ansatz and perturbative string results. The calculation in \cite{Gromov:2008en} was done using the order of limits (\ref{stringlimit}) and therefore should match the string theory prediction \cite{Roiban:2007ju}. However, there is a discrepancy at the two-loop level.

Another possibility to check the results of \cite{Gromov:2008en,Roiban:2007ju} was proposed in \cite{Bajnok:2008it} based on the $O(6)$ sigma model regime. The two-loop prediction of \cite{Bajnok:2008it} matched that of \cite{Gromov:2008en}. This matching should be however better understood, since the order of limits used in \cite{Bajnok:2008it} was different from (\ref{stringlimit}) (and from (\ref{BESFRSlimit})).


\chapter{\label{ch:massgap}Mass gap in integrable sigma models}

\section{Formulation of the problem}

Another important application of the techniques developed in previous chapters is the perturbative solution of integral equations which appear in two-dimensional integrable sigma models.

Two-dimensional nonlinear sigma models are often considered as toy models for QCD. They share with QCD the property of the asymptotic freedom \cite{Polyakov:1975rr}. Although these theories do not have dimensionful parameters on the classical level, they acquire a dynamical mass scale on the quantum level via the dimensional transmutation mechanism. We will call this mass scale $\LQCD$.

The symmetry group $G$ of the target space of the sigma model is the analog of the QCD gauge group. The sigma models with the target space being a coset manifold $G/H$ are classically integrable \cite{Zakharov:1973pp}. It is believed that when $H$ is simple the integrability is preserved on the quantum level. Examples of such quantum integrable theories are $O(N)$ vector sigma model and $SU(N)\times SU(N)$ principal chiral field (PCF) model. $CP(N)$ sigma model is an example of the theory which is integrable on the classical but not on the quantum level.

As in QCD, the sigma models can be studied perturbatively at scales much larger than the dynamical mass scale. In the infrared regime the perturbative description is not applicable. Instead, it is possible to solve these theories exactly. For example, large $N$ solutions of such models as $O(N)$ and $CP(N)$ sigma models were found in \cite{Bardeen:1976zh,Brezin:1976qa,D'Adda:1978uc,Witten:1978bc}. These solutions showed that the excitations are described on the quantum level by massive particles. Since $\LQCD$ is the only dimensionful parameter of the theory, the mass
of the particles should be proportional to it:
\be\label{mcL}
  m=c\ \LQCD.
\ee

If to assume that the description in terms of particles is appropriate for finite values of $N$ then quantum integrable sigma models can be solved by the asymptotic Bethe Ansatz in the infinite volume \cite{Zamolodchikov:1978xm}, as we discussed in the second chapter.

In this chapter we will explain how to determine the coefficient of proportionality $c$ in (\ref{mcL}) for the $O(N)$ sigma model at finite $N$. Although we will consider only this particular case, most of the statements can be applied for any sigma model solvable by the asymptotic Bethe Ansatz.

The idea of determination of the coefficient $c$ is not new. It goes back to \cite{Hasenfratz:1990ab,Hasenfratz:1990zz}. We refer also to \cite{Evans:1995dn} for an overview. The mass gap (\ref{mcL}) was found in \cite{Hasenfratz:1990ab,Hasenfratz:1990zz} numerically with a precision that allowed to guess the analytical expression. Our contribution is to give an analytical derivation of the mass gap\footnote{In \cite{Balog:1992cm} it was stated that Balog had provided such analytical derivation (see ref.[19] there). His result was however not published.}.

The determination of the mass gap is not the only issue of this chapter. We will also discuss the renorm-group behavior and Borel summability properties of the $O(N)$ sigma model. This will bring us an additional evidence that the Bethe Ansatz is a proper description of this QFT.

%

\subsubsection{Sigma-model in an external field}
Let us consider an asymptotically free model with a running coupling constant $\a$  that satisfies the Gell-Man--Low equation\footnote{We may choose $\a$ equal to $\sigmacoupling$ from (\ref{actionsofsigmamodels}). This is however not at all obligatory. In fact, below we will make another choice which is related to $\sigmacoupling$ via the finite renormalization (\ref{finiterenexact}).}:
\be\label{GLeq}
  \mu\frac{d\a}{d\mu}=\beta[\a]\equiv -\b_0 \a^2-\b_1 \a^3-\ldots\ .
\ee
The dynamically generated scale $\LQCD$ is the RG invariant which can be built from (\ref{GLeq}):
\be\label{LQCD}
  \LQCD=\mu\ e^{-\int^{\a[\mu]}\frac{d\a}{\beta[\a]}}=k\ \mu\ \a^{-\frac {\b_1}{\b_0^2}}e^{-\frac 1{\b_0\,\a}}(1+\CO(\a)).
\ee
The quantity $\b_1\b_0^{-2}$ is invariant under finite renormalizations of the coupling constant given by the transformations of the type
\be\label{finiterenormalization}
  \a\to \a(\k_0+\k_1 \a+\k_2 \a^2+\ldots).
\ee
This scheme independent object $\b_1\b_0^{-2}$ is an important prediction of the perturbation theory.

%

From (\ref{LQCD}) we can see that the mass gap (\ref{mcL}) cannot be found by perturbative means only. Therefore integrability, or another nonperturbative description, is needed. A typical way for identification of the coefficient $c$ is the comparison of the perturbative field theory prediction with the prediction of the asymptotic Bethe Ansatz. Since the theory is asymptotically free, we have to introduce a large scale in order to insure that we are in the perturbative regime. A strong external field $h$ coupled to one of the charges $Q$ of the system is often used for this purpose. $O(N)$ sigma model in the presence of the external field was initially considered in the works of Polyakov and Wiegmann \cite{Polyakov:1983tt,Wiegmann:1985jt}. The thermodynamics of the system, which we consider at zero temperature, can be described by the free energy which is a minimum with respect to the value of the charge $Q$ of
\be
  {\cal{F}}=E-h\ Q.
\ee
Here $E$ is the energy of the system.

$O(N)$ sigma model contains only one type of particles that belong to the vector multiplet of the $O(N)$ group. We chose a charge $Q$ which counts the number of particles polarized in a particular direction.
Then $h$ plays the role of the chemical potential. For $h$ larger than the mass gap a condensate of equally polarized particles is created.


In the limit of infinite volume $L\to\infty$ we should speak about intensive quantities: the density of the particles $\rho=Q/L$, the free energy density $f=F/L$, and the energy density $\varepsilon=E/L$.  These intensive quantities are related as
\be
  f[h]=\min\limits_\rho(\varepsilon[\rho]-h \rho).
\ee
The free energy density $f$ is a function of $h$. Instead, $\varepsilon$, which is the Legendre transform of $f$, is a function of $\rho$.

Perturbative field theory predicts the following asymptotic weak coupling expansion of the energy density:
\be\label{varepsginv}
  \varepsilon[\rho]=\rho^2\(\ \sum_{n=1}^\infty \chico_n \a^{n}+\CO\(\frac{\LQCD^2}{\rho^2}\)\),
\ee
where the running coupling constant $\a[\mu]$ is evaluated at the scale $\mu=h$. The ratio $\LQCD^2/\rho^2$ is exponentially small in terms of the coupling constant.

The dependence of $\rho$ on $h$ can be figured out from the condition $\pd\varepsilon/\pd\rho=h$ and decomposition (\ref{varepsginv}). It is given by the expansion
\be\label{h2rho}
    h=2\rho\,\chico_1(\a[h]+\CO(\a^2)).
\ee
It is useful to think about the coupling constant as the function of $\rho$ and to introduce the corresponding beta-function
\be
    \rho\frac{d\a[h[\rho]]}{d\rho}=\tilde\beta[\a]=-\tilde\beta_0\a^2-\tilde\beta_1\a^3-\ldots.
\ee
From (\ref{h2rho}) it follows that
\be
    \tilde\b_0=\b_0,\ \ \tilde\b_1=\b_1-\b_0^2.
\ee

The explicit value of the coefficients $\chico_n$ depends on the computation scheme which is used and on the definition of the coupling constant.
The energy of the $O(N)$ sigma model at tree level and its one-loop correction were calculated in \cite{Hasenfratz:1990ab,Hasenfratz:1990zz}\footnote{More precisely, the free energy was found. Then we can make the Legendre transform to obtain the expression for the energy.}. The two-loop correction was found in \cite{Bajnok:2008it}\footnote{This work contains a very good pedagogical explanation about how the perturbative calculations are done.}. The calculations were done in the $\overline{MS}$ scheme. With the coupling constant defined by
\be\label{Bajnokcoupling}
    \frac 1{\a}+(\Delta-1)\log\a=\log\left[\frac{\rho}{\Lambda_{\overline{MS}}}\right]+\log[2\pi\Delta],\ \ \Delta=\frac 1{N-2}
\ee
the first three coefficients $\chico_n$ were found to be
\be
    \chico_1=\pi\Delta,\ \ \chico_2=\frac 12\pi\Delta,\ \ \chico_3=\frac 12\pi\Delta^2.
\ee
The definition (\ref{Bajnokcoupling}) of the coupling constant is the same as in \cite{Bajnok:2008it}, where it was called $\tilde\a$. The reason to use such definition is that it leads to a particularly simple form of the coefficients $\chico_n$, as we discuss below.

We can rewrite the expansion (\ref{varepsginv}) in terms of the ratio $\rho/\LQCD$ only:
\be\label{energyft}
\varepsilon&=&\chico_1\tilde\beta_0\frac{\rho^2}{\log\frac\rho\LQCD}\left(1-(\tilde\b_1^{-1}\tilde\b_0^2)
\frac{\log\log\frac\rho\LQCD}{\log\frac\rho\LQCD}+\frac{r_\Lambda}{\log\frac\rho\LQCD}+\CO\left(\frac
{\log\log\frac\rho\LQCD}{\log\frac\rho\LQCD}\right)^2\right)+\CO(\LQCD^2),\no\\
&&\tilde\b_0=1,\ \ \tilde\b_1=\Delta-1,\ \ r_\Lambda=\frac 12-\log[2\pi\Delta].
\ee
 The idea of \cite{Hasenfratz:1990ab,Hasenfratz:1990zz} to calculate the mass gap (\ref{mcL}) is based on the fact that the energy of the considered system can be also found from the Bethe Ansatz. Instead of $\rho/\LQCD$, the solution should depend on the ratio of the density and the mass of the particles, since the mass of the particles serves as an input for the construction of the Bethe Ansatz. One should expect the expression of the following form
\be\label{energyba}
\varepsilon\!\!&=&\!\!\frac{\chico_1\tilde\beta_0\rho^2}{\log\frac{\rho}{m}}\left(1-(\tilde\b_1^{-1}\tilde\b_0^2)
\frac{\log\log\frac{\rho}{m}}{\log\frac{\rho}{m}}+\frac{r_m}{\log\frac{\rho}{m}}+\CO\!\!\left(\frac
{\log\log\frac{\rho}{m}}{\log\frac{\rho}{m}}\right)^{\!2}\right)+\CO(\m^2).
\ee
Comparison of (\ref{energyft}) and (\ref{energyba}) gives us a possibility to find $c$ in (\ref{mcL}). The expansions (\ref{energyft}) and (\ref{energyba}) contain also another valuable information. They contain the invariant ratio $\tilde\b_1\tilde\b_0^{-2}$. Therefore we can verify that the Bethe Ansatz reproduces a correct renorm-group dynamics of the model. It is a nontrivial check of the bootstrap approach. \textit{A priory}, the Bethe Ansatz does not contain coupling constant at all. But it turns out to be useful to introduce a coupling constant in order to rewrite (\ref{energyba}) in terms of the power series of the coupling constant (like (\ref{varepsginv})).

\subsubsection{Integral equation}
Since only the states with a certain polarization are excited, the nested levels of
the Bethe Ansatz are turned off. Therefore in the large volume limit the system with a given density of particles $\rho$ is described by the one integral equation
\be\label{iequationform}
  &&\chi[\theta]-\int_{-B}^{B}d\theta'K[\theta-\theta']\chi[\theta']=m \cosh[\theta],\no\\
  &&K[\theta]=\frac 1{2\pi i}\frac d{d\theta}\log S[\theta],\ \ S[\theta]=-\frac{\Gamma\[\frac 12+\frac{i\theta}{2\pi}\]\Gamma\[\Delta+\frac{i\theta}{2\pi}\]}{\Gamma\[1+\frac{i\theta}{2\pi}\]\Gamma\[\frac 12+\Delta+\frac{i\theta}{2\pi}\]}/\textrm{c.c}.
\ee
Here $\chi[\theta]$ is a density of the Bethe roots distribution. 
The boundary $B$ of the Bethe roots distribution is the value of rapidity up to which the Fermi sea is filled. This value should be found from the normalization condition
\be
  \rho=\int_{-B}^{B}\frac {d\theta}{2\pi}\chi[\theta].
\ee
The energy density of the system is given by
\be
  \varepsilon=m\int_{-B}^{B}\frac {d\theta}{2\pi} \cosh[\theta]\chi[\theta].
\ee
The energy density $\varepsilon$ depends on the particle density $\rho$ through the parameter $B$. Our goal is to find this parametric dependence for large $\rho$, or equivalently large $B$.

One can see (from numerical solution for example) that the density of roots grows when it approaches the points $\pm B$. Therefore we can use the methods explained in the previous chapters to perturbatively solve the integral equation. The discussed below solution is the extended version of the work \cite{V2}.

The $S$-matrix is given by ratio of the gamma functions. Using the formal equivalence\footnote{In (\ref{gamma}) we already used such notation. Moreover, there we defined the r.h.s. of (\ref{gammaxencore}) as the gamma function.}
\be\label{gammaxencore}
  \Gamma[x]\simeq \frac 1{x}\frac 1{x+1}\ldots=x^{-\frac 1{1-e^{\p_x}}}
\ee
and introducing the resolvent
\be
  R[\theta]=\int_{-B}^{B}dv\frac {\chi[v]}{\theta-v}
\ee
we can rewrite the integral equation (\ref{iequationform}) in an elegant form:
\be\label{iequationON}
  \frac {1-D^{4\Delta}}{1+D^2}R[\theta+i0]-\frac {1-D^{-4\Delta}}{1+D^{-2}}R[\theta+i0]=-2\pi i m\cosh[\theta],\ \ \theta^2<B^2.
\ee
Here the shift operator is defined by $D=e^{\frac i2\pi\p_\theta}$ and
\be
\Delta=\frac 1{N-2}.
\ee
%

\section{Perturbative expansion}
%

Let us act on (\ref{iequationON}) with the operator $1+D^2$, which is well defined operation. The r.h.s. of (\ref{iequationON}) will vanish and we are left with
\be
  (1-D^{4\Delta})(R[\theta+i0]+D^{-4\Delta+2}R[\theta-i0])=0.
\ee
The action of the shift operator in this notation is understood as an analytical continuation.

If to consider this equation perturbatively in the regime
\be
\theta\sim B\to\infty,
\ee
the overall operator $(1-D^{4\Delta})$ can be dropped in the perturbative expansion by the following reason. This operator is perturbatively understood as a power series in $\p_\theta$. Integrating the obtained equation in each order of the perturbation theory we get
\be\label{ONperturbative}
 D^{2\Delta-1}R[\theta+i0]+D^{-2\Delta+1}R[\theta-i0]=0,
\ee
up to a constant of integration. However, this constant is zero since $R$ is an odd function.

The equation (\ref{ONperturbative}) can be solved perturbatively. Its most general solution is given by
 \be\label{sol2}
  R[\theta]=\!\!\sum_{n,m=0}^\infty\sum_{k=0}^{m+n}\frac{\sqrt{B}\,c_{n,m,k}(\theta/B)^{\e[k]}}{B^{m-n}\(\theta^2-B^2\)^{n+1/2}}\log\[\frac{\theta\!-\!B}{\theta\!+\!B}\]^k\!\!\!,
 \ee
 where $\e[k]=k\ \textrm{mod}\ 2$. The perturbative meaning of the expansion (\ref{sol2}) is most easily seen in terms of the variable $u=\theta/B$.

 The coefficients $c_{n,m,k}$ are not all independent. The coefficients for the terms which contain logarithms can be defined uniquely through the other coefficients using equation (\ref{ONperturbative}). However, below we will show a more efficient way to fix them.

 The solution (\ref{sol2}) gives us the value for the particle density from the residue of the resolvent at infinity:
 \be
  \rho=\frac{\sqrt{B}}{2\pi}\(c_{0,0,0}+\sum_{m=1}^\infty \frac{c_{0,m,0}-2c_{0,m,1}}{B^m}\).
 \ee

\section{Solution in the double scaling limit}
To fix all the coefficients $c_{n,m,k}$, we follow exactly the same procedure as we did in the previous chapters. Namely, we consider a double scaling limit
\be\label{doublescaling}
  B,\theta\to\infty,\ \ z=2(\theta-B)\ \ {\rm fixed}
\ee
and perform the inverse Laplace transform defined by
\be\label{ONlaplacetransform}
 \hat R[s]=\int_{-i\infty+0}^{i\infty+0}\frac {dz}{2\pi i}e^{s z}R[z].
\ee
The double scaling limit of (\ref{iequationON}) and further its inverse Laplace transform give the following equation:
\be\label{iLTequation}
  &&K_H[s-i0]\hat R[s-i0]-K_H[-(s+i0)]\hat R[s+i0]=\frac m2 e^B\(\frac 1{s+\frac 12-i0}-\frac 1{s+\frac 12=i0}\),\no\\
  &&K_H[s]=\frac{1-e^{-4i\pi\Delta s}}{1+e^{-2i\pi s}}.
\ee
Here we neglected the exponentially small terms since our goal is to find the asymptotic large $B$ expansion. Neglecting these terms allows reexpressing the energy in terms of $\hat R[1/2]$. Indeed,
\be\label{e1}
  \frac{\varepsilon}{\mass}\simeq\int_{0}^B\!\! e^{\theta}\chi[\theta]\frac{d\theta}{2\pi}\simeq e^B\int_{-\infty}^0\!\!\!\! e^{\frac {z}{2}}\chi[z]\frac{dz}{4\pi}=\frac {e^B}{4\pi}\hat R[1/2].
\ee
The study of the analytical structure of the resolvent goes similarly to study in chapter \ref{ch:sl2} (note that the role of $R$ and $R_h$ are exchanged in these two cases). The conclusion is that:
\bi
  \item $\hat R[s]$ is analytic everywhere except on the negative real axis,
  \item For $s<0$ $\hat R[s]$ has a branch cut. The monodromy around the branch point $s=0$ is given by the equation:
      \be
        e^{\frac {i w\pi}2 s}R[s-i0]=-e^{-\frac {i w\pi}2 s}R[s+i0],
      \ee
  \item $\hat R[s]$ has an infinite set of poles and zeroes on the negative real axes. The position of poles and zeroes is such that all the poles and zeroes of $K_H[e^{-i\pi s}]$ for $s<0$ are canceled out, except for the pole at $s=-1/2$, which is a singularity of the r.h.s. of (\ref{iLTequation}).
  \item The asymptotic behavior of $\hat R[s]$ at large $s$ is given by
  \be
    \hat R[s]=\frac{\a_1}s+\frac{\a_2}{s^2}+\ldots\ .
  \ee
\ei
The solution of (\ref{iLTequation}) which satisfies these properties is given by
\be\label{DSsolution}
  \hat R[s]&=&\(\frac {1}{s+\frac 12}+Q[s]\)A\Phi[s],\ \  A=\frac{m}{4\Delta^{\Delta}}e^{-\frac 12+B+\Delta}\G[\Delta],\no\\
  Q[s]&=&\frac 1{Bs}\sum_{n,m=0}^\infty\frac{Q_{n,m}[\log B]}{B^{m+n}s^n},\no\\
  \Phi[s]&=&\frac 1{\sqrt{s}}e^{(1-2\Delta)s\log\[\frac se\]-2\Delta s\log[2\Delta]}\frac{\Gamma[2\Delta s+1]}{\Gamma\[s+\frac 12\]}.
\ee

The dependence of $Q[s]$ on $B$ is not a consequence of (\ref{iLTequation}). It comes from the comparison of (\ref{DSsolution}) and (\ref{sol2}). $Q[s]$ contains only finite number of terms at each order of $1/B$ expansion. Therefore the solution (\ref{DSsolution}) is well defined.

To fix all the coefficients $c_{n,m,k}$ and $Q_{n,m}$, we should consider (\ref{sol2}) in the double scaling limit. In this limit the expansion organizes at each order of $1/B$ as a $1/z$ expansion. This expansion should match with the Laplace transform of the small $s$ expansion of $\hat R[s]$. This requirement uniquely fixes all the coefficients. We used \textit{Mathematica} to obtain first 25 orders of the perturbative expansion. The routine for this calculation is given in appendix \ref{app:massgap}.

\section{Results for the mass gap coefficient and the RG dynamics.}
At the leading and subleading orders the energy and particle densities are given by
\be\label{rhoeefirsttwo}
    \varepsilon=\frac{\Delta A^2}{\pi}\left(1+\frac 1{4B}\right),\ \
    \rho=\frac{A}{\pi}\sqrt{B}\left(1-\frac {\frac 32+(1\!-\!2\Delta)\log \frac{8B}{\Delta}-\log\frac{2}{\Delta}}{4B}\right).
\ee
By explicit solution of this parametric dependence and comparison with (\ref{energyft}) we get the correct value of the mass gap known in the literature and given by \cite{Hasenfratz:1990ab,Hasenfratz:1990zz}
\be
  m=\(\frac{8}e\)^{\frac 1{N-2}}\frac 1{\Gamma\[1+\frac 1{N-2}\]}\Lambda.
\ee
Here the $\Lambda$-parameter is defined in the $\overline{MS}$ scheme.

We would like to note that the importance of the solution is not only in finding the coefficient of proportionality. The expression for the energy density in terms of the parameter $\rho$ contains both $\log \rho$ and $\log\log \rho$ terms. The fact that we can cancel all the double logarithm terms by introducing a proper coupling constant is very nontrivial. Moreover, this cancelation is only possible if the coupling constant has a beta-function with the correct ratio $\tilde\b_1\tilde\b_0^{-2}$ predicted by the quantum field theory. This shows that the Bethe Ansatz correctly reproduces the renorm-group dynamics of the model.

The presence of the CDD factors\footnote{The crossing and unitarity equations constrain the scalar factor in the scattering matrix up to a multiplier known as the CDD factor. This factor can be fixed if the complete particle content of the theory is known.} would change this renorm-group dynamics. Therefore this calculation also serves as a verification of the assumption about the particle content of the theory.

\section{Borel summability}
The recursive procedure allows us to find analytical expression for the first $25$ orders of the perturbation theory. We are able therefore to analyze the large order behavior of the coefficients of the asymptotic expansion. Comparison of the large order behavior that we get from the Bethe Ansatz with the prediction of the quantum field theory gives us an additional check of the validity of the Bethe Ansatz description of the $O(N)$ sigma model.

Let us briefly recall the basic facts about asymptotic series. Typically, a small coupling expansion in a quantum field theory is an asymptotic expansion with zero radius of convergence. Probably the only known example of the convergent series is when models are considered in the 't Hooft planar limit.

In opposite to the case with convergent series, there is no bijection between the function and the asymptotic series. Although the asymptotic expansion of the function is uniquely defined, the inverse is not true. It is often postulated that the function which corresponds to a given asymptotic expansion is given by the Borel ressumation of the asymptotic series.

Suppose that this series is given by
\be\label{asseries}
  f[g]=\sum_{n=0}^{\infty} f_n g^n.
\ee
The Borel sum of this series is defined by the integral
\be\label{Borel}
  f[g]_{\rm Borel}\equiv \int_0^{\infty}ds\, e^{-\frac s{g}}\sum_{n=0}^\infty \frac{f_n s^{n-1}}{\Gamma[n]}.
\ee
The integrand is given by the series which is convergent provided that $f_n$ grow at most factorially. In the most interesting cases $f_n$ grow exactly in factorial way. Therefore the series has a finite radius of convergence. The integrand in (\ref{Borel}) is defined using the analytical continuation of the function
\be
  h[s]=\sum_{n=0}^\infty \frac{f_n s^{n-1}}{\Gamma[n]},
\ee
which is well defined by its series expansion for a sufficiently small $s$.

\begin{figure}[t]
\centering
\includegraphics[height=3cm]{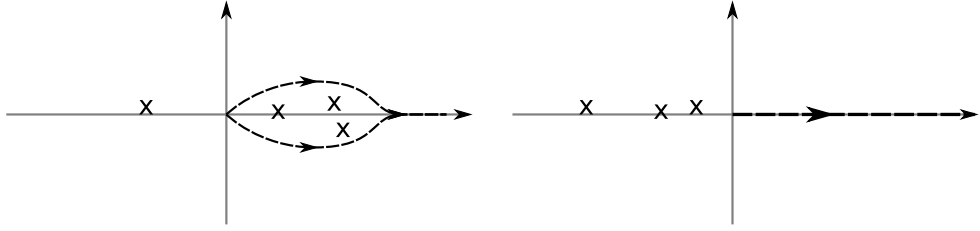}
\caption{\label{fig:Borelnonborel}Position of singularities in the Borel plane. On the left: non-Borel-summable function. Due to the presence of poles to the right of the imaginary axes we can chose the contour of integration in (\ref{Borel}) in different ways. On the right: Borel-summable function. The contour of integration is chosen uniquely.}
\end{figure}

Let us suppose that $h[s]$ has a singularity (we will consider only poles for simplicity) at $s=s_0$ with $Re[s_0]>0$. Then we can choose different contours of integration, as it shown in Fig.~\ref{fig:Borelnonborel}.
Different choices of the contours obviously lead to different functions $f[g]_{\rm Borel}$. However, these functions have the same asymptotic expansion. We see that the Borel transform is an ambiguous procedure in this case. A series (\ref{asseries}) with such properties is called non-Borel-summable.

When all the singularities of $h[s]$ are situated to the left from the imaginary axes, we can uniquely choose the contour of integration. In this case the Borel summation is the unambiguous procedure. An asymptotic series for which Borel transform is uniquely defined is called Borel-summable.


The large order behavior of the asymptotic expansion gives a valuable information about the existence/absence of singularities in the Borel plane. If the large order behavior of the coefficients $f_n$ is approximated by
\be
  f_n\simeq \Gamma[n]/A^n,
\ee
then $h[s]$ has a singularity at $s=A$. If the sign of $A$ is positive,  we get a non-Borel-summable theory. Therefore sign oscillation of the asymptotic expansion is a necessary condition for the Borel summability. Of course, this condition is not at all sufficient.

Let us now investigate what is the asymptotic behavior of the coefficients in the $O(N)$ sigma model. The large $n$ behavior for the coefficients $\chico_n$ is approximated by
\be
  \chico_n\simeq \frac{\Gamma[n]}{2^{n-1}}a_n[\Delta].
\ee
The qualitative behavior of the coefficients $a_n$ is different for large and small values of $N$. For $\Delta=0$, and therefore $N=\infty$, we have $a_n[0]=(-1)^{n-1}$, so we have a sign oscillating series. We will see below that this is in agreement with the observation that exactly at $N=\infty$ the asymptotic expansion is Borel-summable.

For sufficiently small $N$ the coefficients of the series become positive. So, $a_n\simeq 1.09$ for $\Delta=1$ and $a_{n+1}\simeq n^{-1}(2.09 - 0.43(n\ \rm{mod}\  2))$ for $\Delta=1/2$. We see that at least for sufficiently small $N$ the series is non-Borel-summable. The Borel ambiguity is of order ${\Lambda^2}/{\rho^2}$ as it should be from the field theory point of view (see (\ref{energyft})).

For other values of $\Delta$ the behavior of the coefficients is more complicated. It seems that they reflect the concurrence between two singularities at $s=\pm 2$ in the Borel plane.

For $\Delta >1$ the asymptotic behavior of $a_n$ is estimated by $a_n\simeq -\a_1 n^{\a_2} \Delta^{n-2}$, where $\a_1$ is positive. This shows that we have an additional singularity in the Borel plane at $s=2(N-2)$. This singularity determines the asymptotic behavior of the coefficients for $N<3$.
\begin{figure}[t]
\centering
\includegraphics[height=4cm]{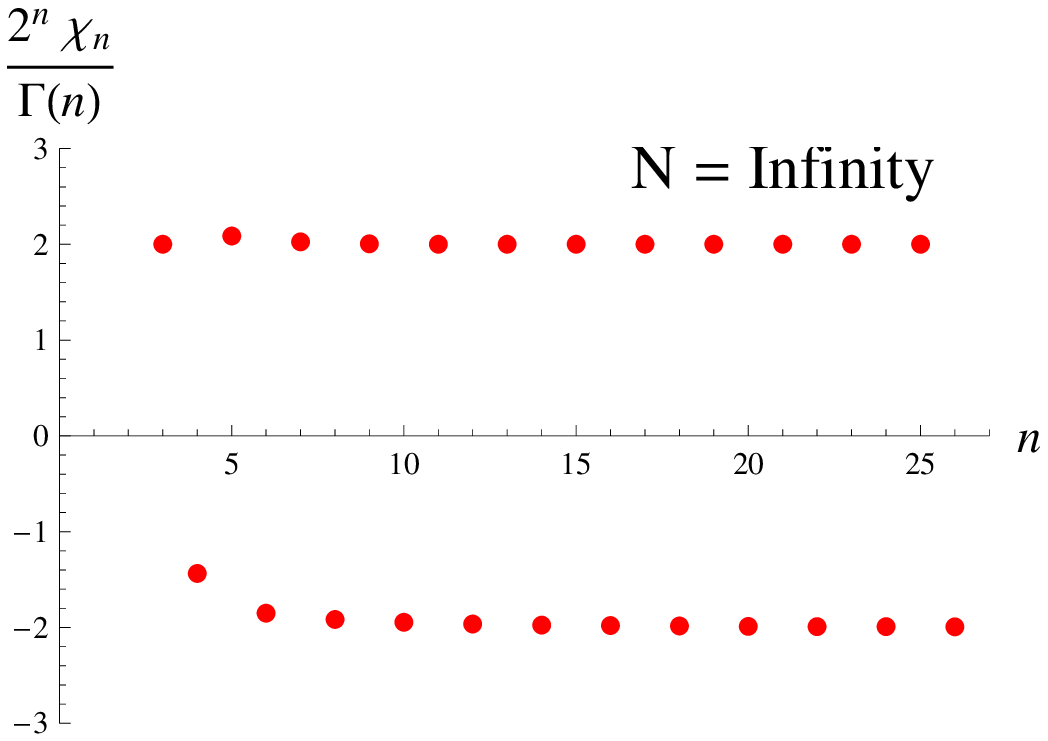}
\includegraphics[height=4cm]{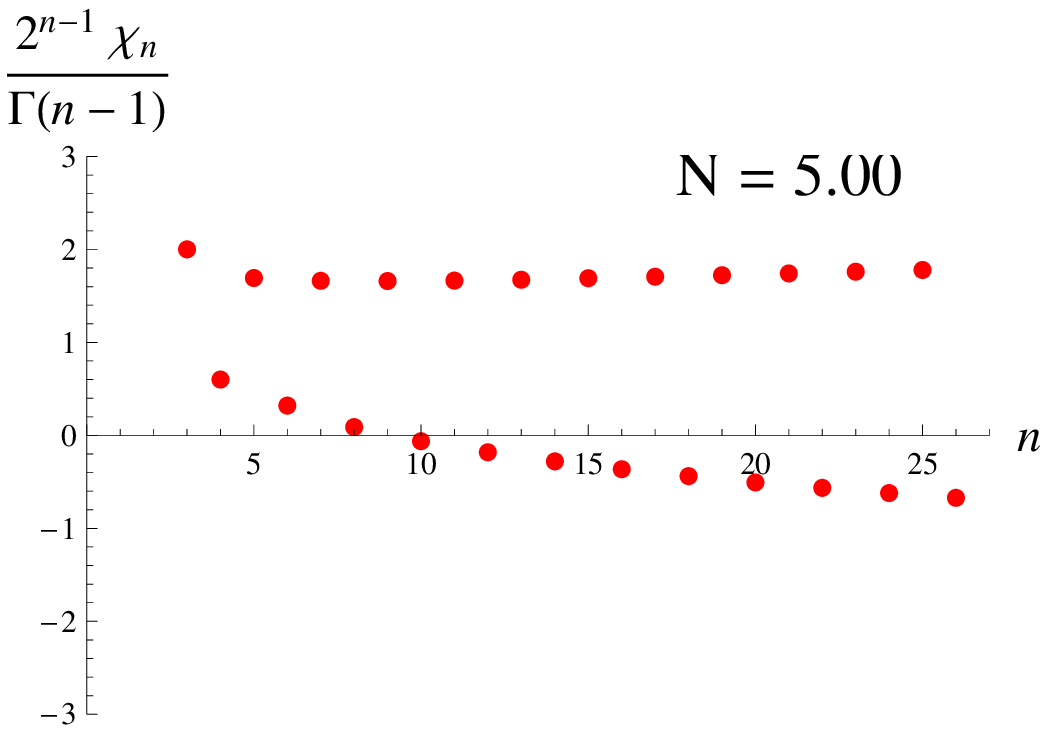}
\includegraphics[height=4cm]{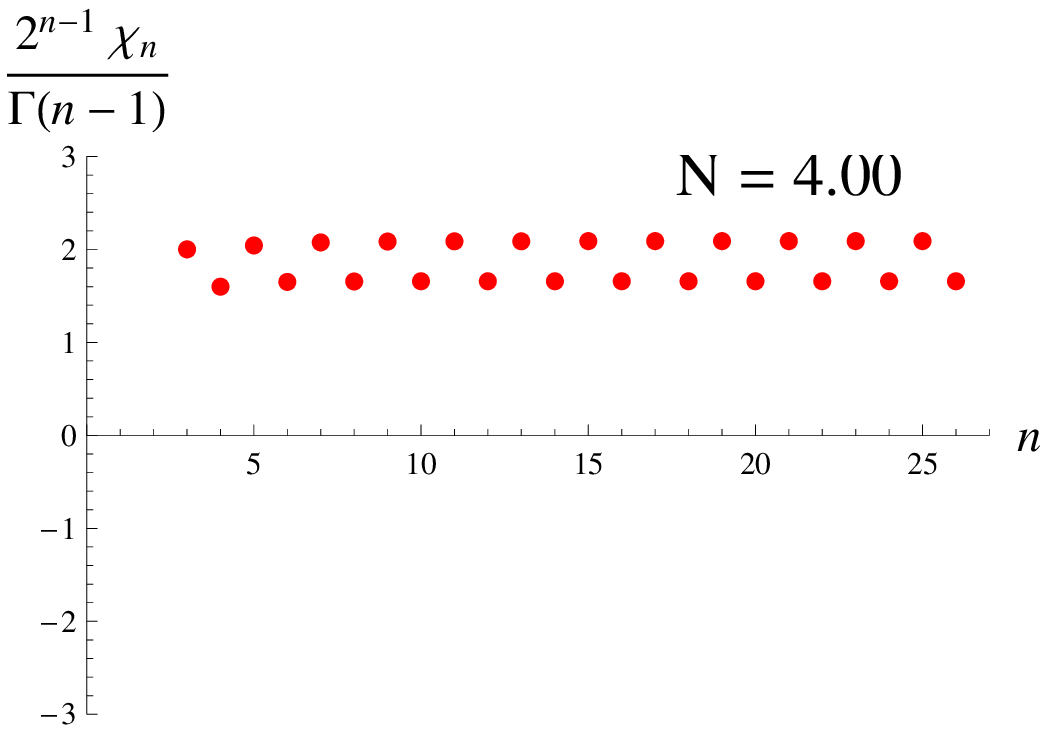}
\includegraphics[height=4cm]{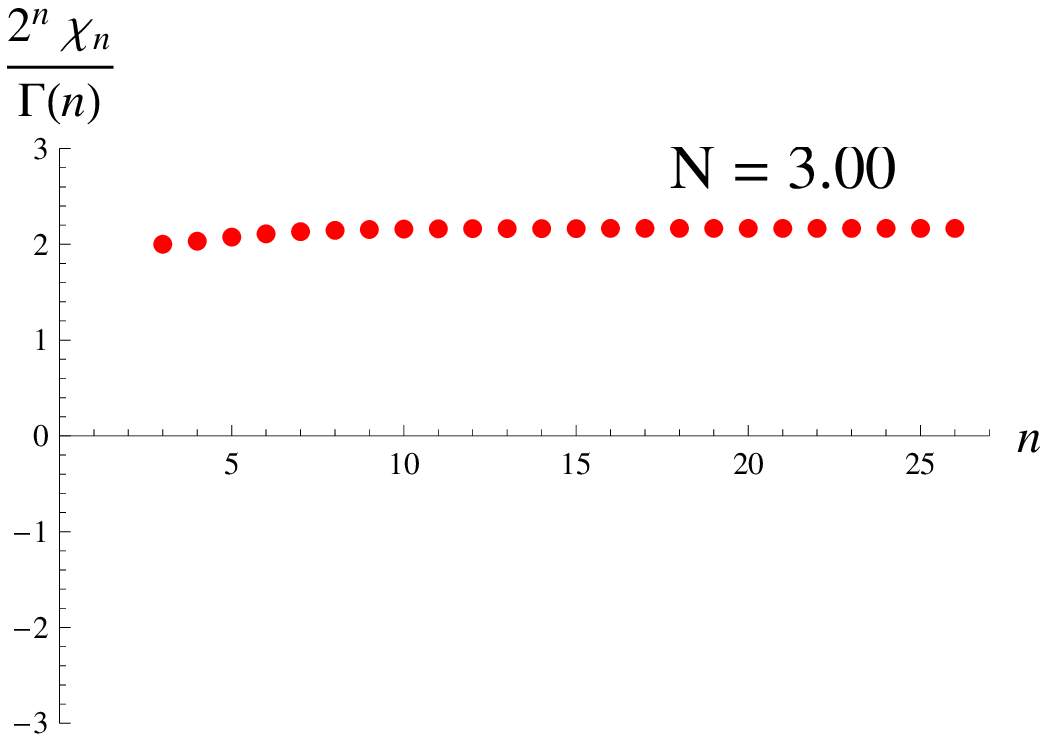}
\caption{\label{fig:ONbehavior}Behavior of the coefficients $\chi_n=\chico_n/\Delta$ at different values of $D$.}
\end{figure}

\begin{figure}[t]
\centering
\includegraphics[height=4cm]{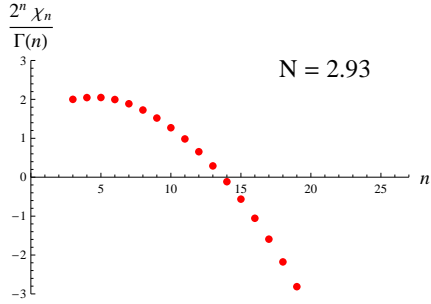}
\caption{\label{fig:ON3behavior}For $N<3$ the behavior of the coefficients signal to the presence of the pole in the Borel plane at position $A=2(N-2)$}
\end{figure}

\ \\

Let us compare our results with the predictions of the quantum field theory. In quantum field theory there are two sources of the singularities in the Borel plane: instantons and renormalons. Both of them are present in the $O(N)$ sigma model.

\paragraph{Instantons.} Instanton is defined as a solution of the equations of motion in the imaginary time with finite value of the Euclidian action. As was observed by Lipatov \cite{Lipatov:1977hj}, there are singularities in the Borel plane related to instanton solutions. We will apply the method of \cite{Lipatov:1977hj} to find the position of these singularities.

The relation between the coupling constant $\a$ that we use and $\sigmacoupling$ that was used in (\ref{actionsofsigmamodels}) is the following:
\be\label{finiterenexact}
    \sigmacoupling^2=\frac{\a}{2\beta_0[\l]}(1+\CO(\a^2))=2\pi\Delta\a(1+\CO(\a^2)).
\ee
For our discussion it would be sufficient only the leading order. Let us consider a physical quantity
\be
    I[\a]=\sum_{k} I_k\a^k=\int \CD n\ \CO[n] e^{-\frac 1{4\pi\Delta\a}S_E[n]}\ .
\ee
The $k$-th order of the asymptotic expansion of $I[\a]$ can be found from the correspondent residue under the functional integral
\be
    I_k=\int \CD n \oint \frac{d\a}{2\pi i}\CO[n]e^{-\frac 1{4\pi\Delta\a}S_E[n]-(k+1)\log[\a]}
\ee
Using the saddle point approximation at large values of $k$, both for $\a$ and $n$, we can estimate that
\be\label{kthcoef}
    I_k\propto k!\,\(\frac{4\pi\Delta}{S_{ins}}\)^k,
\ee
where $S_{ins}$ is the extremal value of the action. The smallest nonzero value of $S_{ins}$ is $8\pi$. A particular instanton solution which corresponds to this value is the following: $n^1[x],n^2[x]$, and $n^3[x]$ form the instanton solution for the $O(3)$ sigma model with topological charge equal one, $n^i[x]=0$ for $i>3$. Of course, we should integrate over all solutions of this type. This will lead to a common multiplier in (\ref{kthcoef}) that does not affect the position of the Borel singularity. Higher instanton solutions are built analogically. For the $n$-th instanton solution $S_{ins,n}=8\pi n$.

Substituting $S_{ins,n}=8\pi n$ to (\ref{kthcoef}) we obtain the positions of the Borel singularities related to the instantons:
\be
    A_{ins,n}=\frac{S_{ins,n}}{4\pi\Delta}=2(N-2)n.
\ee

The instanton solutions are not stable for $N>3$\footnote{To show this we can consider the one-parametric deformation $\overrightarrow{n}[t]=\overrightarrow{n}_{ins}\cos t+\overrightarrow{n}_{\perp}\sin t$, $t\,\e\,[0,\pi/2]$, where $\overrightarrow{n}_{\perp}$ is a constant unit vector which satisfies the condition $\overrightarrow{n}_{ins}\cdot \overrightarrow{n}_\perp=0$. The value of the action, $S[n[t]]\propto\cos^2t,$ interpolates monotonously  between $S_{ins}$ and $0$.}. Naively we could think that an unstable solution is not a local minimum of the action and therefore it does not lead to nonperturbative corrections. However, there are examples from the quantum mechanics which show that the situation is less trivial and that unstable instantons may lead to the singularities in the Borel plane.

From our explicit calculation we observed for $N<3$ a leading singularity in the Borel plane equal to $A_{ins,1}$. It is natural to identify it with the instanton solution. For $N>3$ this singularity, if present, is not a leading one and therefore cannot be seen from our analysis. Thus we cannot conclude about the presence or absence of nonperturbative corrections related to instantons for $N>3$.

\paragraph{Renormalons.}  Renormalon ambiguities do not correspond to any physical effects but reflect an ambiguity in the definition of the asymptotic perturbative expansion. It is believed that the ambiguities related to renormalons can be fixed by proper definition of the subleading terms in the operator product expansion. For a review of the subject see for example \cite{Beneke:1998ui}.

 The renormalons in the $O(N)$ sigma model were studied by David \cite{David:1982qv,David:1983gz}. For $N=\infty$ the $O(N)$ sigma model is just an $N$-component gaussian field theory. It contains only UV renormalons at positions $A_{{\rm UV},k}=2k$ with $k$ being negative integer. The $1/N$ corrections to the large $N$ solutions suggest that for finite $N$ both UV and IR renormalons are likely to be present. The IR renormalons are situated at positions $A_{{\rm IR},k}=2k$ with $k$ being positive integer.

In view of this field theory prediction we can naturally interpret our results in the following way. For $N=\infty$ we observe that the asymptotic expansion is governed by the leading UV renormalon singularity at $s=-2$. For finite $N$ there are two leading singularities: the UV renormalon at $s=-2$ and the IR renormalon at $s=2$. The concurrence of them is reflected in the nonsymmetrical with respect to the real axis oscillating behavior of the coefficients of the expansion.

We see that the results about the large order behavior obtained from the Bethe Ansatz are in accordance with the predictions of the field theory. We got a new confirmation that the Bethe Ansatz correctly describes the $O(N)$ sigma model.

\chapter*{Conclusions and discussions}
\addcontentsline{toc}{chapter}{Conclusions and discussion}

The study of the AdS/CFT correspondence gave us an example of integrable system with remarkable properties and posed new questions to answer in the field of quantum integrable systems. In this thesis we described the AdS/CFT integrable system in analogy with known integrable systems based on the rational $R$-matrix.

We saw that the symmetry group is always behind all the equations that describe rational integrable systems, whether these are Bethe Ansatz equations, fusion Hirota equations or integral equations for the resolvents of the density functions.

It seems that the deepest way to describe an integrable system is to consider the Hirota equations ($T$-system). Imposing different requirements on the analytical structure of the transfer matrices we can obtain different integrable systems, not necessarily based on the rational $R$-matrix. In this way the AdS/CFT system can be also obtained.

It is not always simple to deal with the Hirota equations due to their quadratic structure. We saw however that the $T$-system is defined, roughly speaking, by the "exponentiation" of the functional equations for the resolvents (compare (\ref{vcool}) and (\ref{prefY})). This resembles to the description of the Lie group via its Lie algebra. The functional equations for the resolvents are linear and are simple to investigate. Therefore we may consider them as an indicator of the possible relations among different integrable theories and then probably raise this relation to the level of $T$-systems.

For example, we saw that the structure of the functional equations for the resolvents in the $SU(N)$ XXX spin chain is simple and is defined by two deformed Cartan matrices (one for $A_{N-1}$ algebra and another for $A_{\infty}$ algebra) (\ref{vcool}). The excitations over the antiferromagnetic vacuum in such a spin chain in the thermodynamic limit describe a certain relativistic field theory. To see this we should perform a particle-hole transformation which is very simple in the language of functional equations. In this way we also see that an overall scalar factor of the scattering matrix is defined basically by the inverse deformed Cartan matrix of the $A_{N-1}$ algebra. The "exponentiation" of the functional equations through TBA leads to a $T$-system which remarkably coincides with the fusion relations for the transfer matrices of the spin chain. This leads to a conjecture that transfer matrices of a spin chain discretization coincide with the $T$-functions which appear in the TBA approach that describes the field theory.

Another example is the functional equations for the $GL(N|M)$ integrable spin chains. We showed that whatever is the choice of the Kac-Dynkin diagram, the functional equations are always organized into a fat hook structure (more generally,  $T$-hook structure). Then, through the TBA we get a $T$-system defined on such fat hook (or $T$-hook). One should then ask what is the meaning of the obtained $T$-functions. Provided by proper boundary condition, the $T$-system on the fat hook is solved by the Bethe Ansatz equations through the chain of Backlund transforms \cite{Kazakov:2007fy}; the $T$-functions have the meaning of the transfer matrices of the corresponding spin chain \cite{Kazakov:2007fy}. However the boundary conditions that allow Bethe Ansatz solution are not the ones that are obtained through TBA, therefore relation between $T$-functions of TBA and transfer matrices is not clear.

In the $T$-hook case the solution in terms of the Bethe equations is still an open problem, though some progress was made in \cite{Hegedus:2009ky}. The $T$-hook case is more involved also because we do not know how to identify the representations of the $gl(N|M)$ algebra with the points of $T$-hook.

\ \\

The AdS/CFT integrable system which is much more complicated from the first glance, in fact can be treated very similarly to the theories with the rational $R$-matrix. To show this in a more explicit way we may use the kernels $K$ and $\tilde K$ defined respectively by (\ref{kdef}) and (\ref{Ktildedef}). These kernels are specially designed to treat the presence of the square root cut. Using $\tilde K$, we can rewrite the Bethe equations in a more tractable form (see section \ref{sec:Kernels}). For example
\be
  \prod_{k}\frac{1-\frac 1{x\yp_k}}{1-\frac 1{x\ym_k}}\sqrt{\frac{\yp_k}{\ym_k}}&=&\prod_{k}(u-v_k)^{-\tilde K_u (D-D^{-1})}.
\ee
We see the appearance of the rational structure in such notation.

As we showed in \cite{V3}, the solution of the crossing equations can be also naturally formulated in terms of the kernel $\tilde K$ and the shift operator $D$. Using this representation for the dressing phase, we can write down the functional representation of the BES and the BES/FRS equations (as well as the other integral equations which follow from the Beisert-Staudacher asymptotic Bethe Ansatz). It is particularly simple to derive these functional equations directly from the Bethe Ansatz.

In the third part of the thesis we presented a way to perturbatively solve the functional equations at strong coupling. The usefulness of the functional representation is based first of all on the defining property of the kernel $K$
\be
   (K\cdot F)[u+i0]+(K\cdot F)[u-i0]=F[u],\ \ -2g<\,u\,<2g,
\ee
which allows performing various analytical transformations of the functional equations. Such transformations of the BES equation at strong coupling allow to reduce a problem to a simple functional equation (\ref{Gpmgpm}) which does not contain the kernel $K$. The solution of (\ref{Gpmgpm}) was found by considering of the related Riemann-Hilbert problems in two different scaling regimes and then identification of unknown constants by comparing these two regimes.

We applied the same strategy for the solution of the BES/FRS equation and for performing weak coupling asymptotic expansion of the energy of the $O(N)$ sigma model. When solving the latter problem we learned that our approach can be reformulated in terms of the Wiener-Hopf solution of the integral equations.

\ \\

To a great extent, the AdS/CFT integrable system was sufficiently well studied from the point of view of the asymptotic Bethe Ansatz. The asymptotic Bethe Ansatz allows computation of the energy in the infinite volume. Recent developments based on TBA allowed for study of finite size operators \cite{Ambjorn:2005wa,Arutyunov:2007tc,Arutyunov:2009zu}, \cite{Gromov:2009tv,Arutyunov:2009ur,Bomb,GromovKKV}, \cite{Bajnok:2008qj,Bajnok:2008bm}, \cite{Lukowski:2009ce,Fiamberti:2009zz}. The anomalous dimension of the Konishi operator, the simplest nontrivial finite-size operator, was found analytically at weak coupling up to five loops \cite{Bajnok:2009vm,Arutyunov:2010gb}.
This anomalous dimension was also computed numerically up to a sufficiently large values of the coupling constant \cite{Gromov:2009zb} which allowed predicting first two orders of the strong coupling expansion. Still, the analytic derivation at strong coupling expansion is missing.

There is one intriguing question that can be asked on the level of the asymptotic Bethe Ansatz. The integrable system as we know it now may be an effective one and there it may be a different, probably much simpler formulation. The one reason for this idea is that the integrable system was built around a fast spinning string solution and we would rather like to describe all possible states. This idea appeared soon after the discovery of the all-loop integrability. In particular, in \cite{Rej:2005qt} it was suggested that the $SU(2)$ sector can be described in terms of the Hubbard model. This description turned out to be incorrect starting only from four loops.

One probable answer to the question what is a simpler formulation is just a $T$-system constructed through TBA. Then we still should give an interpretation for the $T$-functions. We therefore come back to the question if there a spin chain behind $T$-hook. If such spin chain exists, we can think about it as a discretization of the string sigma model.

Another possibility is that the Zhukovsky cut appears effectively from the condensation of particles from a hidden level. This idea was formulated in \cite{Mann:2005ab,Gromov:2006dh}. In \cite{Gromov:2006cq} it was shown on the classical level of the sigma model that the central equation of the Beisert-Staudacher Bethe Ansatz can be indeed reproduced by this approach. The consideration of \cite{Gromov:2006cq} was not however generalized to the quantum level, also it is not evident how to include fermionic interactions in this approach. We hope that the intuition developed in our work can help to improve the study in this direction.

Let us finally recall recent promising formulation of the string sigma model as a Pohlmeyer reduced theory \cite{Roiban:2009vh,Hoare:2009rq,Hoare:2009fs}. This formulation is relativistically invariant and describes excitations around true vacuum. Therefore finding of the quantum integrable model behind this theory would potentially give a desired simpler description of the AdS/CFT system.

We see that despite a huge progress in recent years there are still many questions to answer and that the study of integrability has many directions to develop in the future.

\chapter*{Note added}
\addcontentsline{toc}{chapter}{Note added}
It is important to mention two papers that appeared after the defense of this thesis.

In \cite{Giombi:2010fa} the calculation of the generalized scaling function at two-loops from the string theory was done. The result coincides with the one of \cite{Gromov:2008en,Bajnok:2008it}. Before \cite{Giombi:2010fa} string theory calculation \cite{Roiban:2007ju} was in disagreement with \cite{Gromov:2008en,Bajnok:2008it} as we discussed in this thesis.

In \cite{Gromov:2010vb} certain infinite dimensional representations of $GL(N|M+M'|N')$ were identified with the nodes of the $T$-hook plotted in Fig.~\ref{fig:thookstrings}. Hence, the $T$-functions living on the $T$-hook can be interpreted as transfer matrices in these representations.

\appendix
\chapter{\label{app:shift}Conventions for action of the shift operator.}
Throughout this text we often use the shift operator $D$ which is defined in the following way:
\be\label{definition of D}
    (D\,F)[u]\equiv F[u+i/2].
\ee
A usage of this operator can potentially lead to a number of ambiguities. The goal of this appendix is to define the conventions which will give a precise meaning to all the expressions that contain the shift operator.

\paragraph{Analytical continuation.}
When the function $F$ is multivalued, we should define in (\ref{definition of D}) the path that connects points $u$ and $u+i/2$. The default definition is the following. For $F$ a default set of branch cuts should be given. Then the points $u$ and $u+i/2$ are to be connected by the path that do not intersect the cuts.

\begin{figure}[t]
\centering
\includegraphics[height=3.5cm]{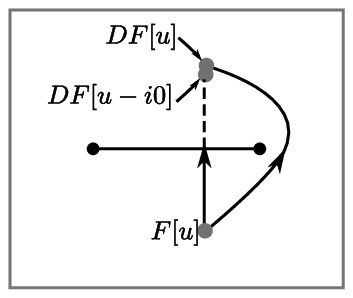}
\caption{\label{fig:analyticalcontinuations}Analytical continuations that correspond to two different usages of a shift .}
\end{figure}

Sometimes we need also an analytical continuation over the straight line irrespectively if it crosses any cut or not. To denote this case we use notations
\be\label{definition of D2}
    D\, F^{-0}\ \ \ \ {\rm or}\ \ \ D\,F[u-i0].
\ee
Correspondingly, if we want use the operator $D^{-1}$ to make a shift by $-i/2$ and the shift must be made by analytical continuation over the straight line, we use notations
\be
    D^{-1}\, F^{+0}\ \ \ \ {\rm or}\ \ \ D^{-1}\,F[u+i0].
\ee
The difference between (\ref{definition of D}) and (\ref{definition of D2}) is shown in Fig.~\ref{fig:analyticalcontinuations}

\paragraph{Function in the power of the operator.}
For a given operator $\CO$, by definition
\be\label{defpower}
    F[u]^{\CO}\equiv e^{\CO \log F[u]}.
\ee
For example
\be
    u^{D-D^{-1}}=e^{(D-D^{-1})\log[u]}=e^{\log\left[\frac{u+i/2}{u-i/2}\right]}=\frac{u+i/2}{u-i/2}.
\ee
In most cases a choice of the branch of the logarithm is not important since the logarithm is exponentiated. However, if an ambiguity may appear we take a standard definition of the branch cut of the logarithm and consider (\ref{defpower}) in the region of the variable $u$ where no crossing of the logarithm cut occurs.

\paragraph{Rational function of the shift operator}
Whenever the shift operator appears in the denominator, simultaneously only positive or only negative powers of $D$ are allowed. Then the denominator is understood as a power series assuming that $|D|$ or $|D^{-1}|$ is sufficiently small. For example:
\be\label{seriesDDD}
    \frac 1{1+D^2}=1-D^2+D^4-D^6+\ldots,\ \ \ \frac 1{1-D^{-2}}=1+D^{-2}+D^{-4}+\ldots\;.
\ee
The definition makes sense if the action of (\ref{seriesDDD}) on a function gives a convergent series.

Note that according to this definition for example
\be
    \frac 1{1+D}-\frac{D^{-1}}{1+D^{-1}}\neq 0.
\ee
Indeed,
\be
    \left(\frac 1{1+D}-\frac{D^{-1}}{1+D^{-1}}\right)\frac 1u\equiv \sum_{n\in\MZ}(-1)^n\frac 1{u+in}=\frac{\pi}{\sin\pi u}.
\ee

There is however one situation when we can use simultaneously $D$ and $D^{-1}$ in the denominator. That is when due to cancelations the actual expression is a polynomial. An example is a definition of a "D-number"
\be
    [s]_D\equiv \frac{D^{s}-D^{-s}}{D-D^{-1}}\equiv D^{s-1}+D^{s-3}+\ldots D^{1-s}
\ee
which is used for operations with string configurations (of Bethe roots) of the length $s$.

There is also one case when we formally deal with nonconvergent series. Namely, we define
\be\label{formalequality}
    u^{\frac 1{1-D^{\pm\a}}}\equiv\Gamma\left[\pm\frac{2u}{i\a}\right].
\ee
The reason for such definition is that the second derivative of the logarithm of the formal equality (\ref{formalequality}) gives a true equality.

For expressions of type
\be\label{uQD}
    u^{Q[D]}
\ee
where $Q[D]$ is a rational function we use the following definition. First, we represent $u^{Q[D]}$ as
\be
    Q[D]=\sum_{a}{c_a \frac{D^{b_a}}{1-D^a}}+\sum_{a'}{c_{a'} \frac{D^{-b_{a'}}}{1-D^{-a'}}}.
\ee
All the expressions used in the text allow such representation. The expression $\frac 1{1+D^\a}$ should be understood as $\frac{1}{1-D^{2\a}}-\frac{D^\a}{1-D^{2\a}}$.

Then we define (\ref{uQD}) by
\be\label{def145}
    u^{Q[D]}\equiv\prod_{a}u^{c_a \frac{D^{b_a}}{1-D^a}}\prod_{a'}u^{c_{a'} \frac{D^{-b_{a'}}}{1-D^{-a'}}}\equiv\prod_{a}\left(\Gamma\left[\frac{2u}{ia}\right]\right)^{c_aD^{b_a}}\prod_{a'}\left(\Gamma\left[-\frac{2u}{ia'}\right]\right)^{c_{a'}D^{b_{a'}}}.
\ee
From the definition (\ref{def145}) of (\ref{uQD}) it follows that in general
\be\label{notequal}
    u^{Q[D]}u^{Q'[D]}\neq u^{Q[D]+Q'[D]}.
\ee

However, since in all the cases presented in this text $Q[D]\frac 1u$ is a convergent series, l.h.s. and r.h.s. of (\ref{notequal}) may differ only by a constant:
\be
    u^{Q[D]}u^{Q'[D]}= c\;u^{Q[D]+Q'[D]}.
\ee
Throughout the text we track out this constant which is typically $\pm 1$. However, the essential information about algebraic structure of the expression is contained in the argument $Q[D]$ of the exponent and reader may not follow the preexponent constants. 
\chapter{Structure of the integral equations for the $gl(N|M)$ Bethe ansatz.}\label{app:susy}
Here we will derive the formula (\ref{fathookie}) and give an explicit form for $T_{M,N}$ in (\ref{TMN}).

Let us first consider the bosonic node. One can keep in mind the node 4 in Fig.~\ref{fig:stringfathook}. However all the arguments are applicable for any node. The Bethe equations are written as
\be\label{yaBAE}
  -1=\prod_{u_3}\frac{u_4-u_3-\frac i2}{u_4-u_3+\frac i2}\prod_{u_4'}\frac{u_4-u_4'+i}{u_4-u_4'-i}\prod_{u_5}\frac{u_4-u_5-\frac i2}{u_4-u_5+\frac i2}.
\ee
If there are roots $u_4'$ that are part of some stack configuration, then the  terms in (\ref{yaBAE}) which contain them will cancel against terms with $u_3$ and/or $u_5$ belonging to the same stack. Therefore what is left are the roots $u_4'$ which form string configurations themselves. These roots interact in a standard way with $u_3$ and $u_5$ and therefore (\ref{fathookie}) is correct for bosonic node. This is because (\ref{fathookie}) is the same equation that appeared in $gl(n)$ case. It also happens that part of $u_3$ roots belong to the end of the stack (type B in Fig.~\ref{fig:stringstacks} if $u_3$ is bosonic or type C if $u_3$ is fermionic). Such roots should be separated from other $u_3$ Bethe roots. The corresponding node for such roots is joined to $u_4$ by the colorless (black) line (in terms of Fig.~\ref{fig:stringfathook}. For $u_4$ this node is below the node 4).

\begin{figure}[t]
\centering
\includegraphics[height=4cm]{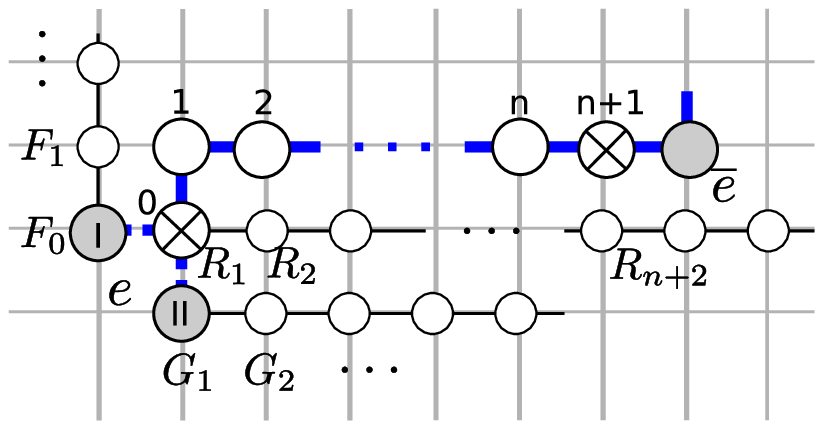}
\caption{\label{fig:generalfermion} The most general imbedding of the fermionic node 0 and stacks $R_s$ generated by it into the fat hook shape.}
\end{figure}

The fermionic Bethe roots generically form stacks of type B or C. Let us derive the integral equations for the resolvents corresponding to these stacks. The most general situation for the fermionic node is shown in Fig.~\ref{fig:generalfermion}. We will also adopt numeration from this figure.

We introduce the following notations for the resolvents:
\bi
  \item{$R_s$} - resolvent for s-stack generated by the 0 node. If $s\leq n+1$ this is the type B stack. If $s>n+1$ - type C.
  \item{$W_{(\a),s}$, $\a=\overline{1,n}$} - the resolvent of $s$-string for the node $n$.
  \item{$F_s$, $G_s$} - the resolvents shown on the picture. Their meaning is slightly different depending on whether node $e$ is bosonic or fermionic.

      If the node $e$ is fermionic (case I) then $F_s$ is the resolvent for  type C $(s+1)$-stacks formed by the nodes $e$ and $0$. $G_s$ is the resolvent for type C stacks terminated by the $s$-string at the node $e$.

      If the node $e$ is bosonic (case II) then $F_s, s\geq 1$ is the resolvent for type C stacks terminated by the $s$-string at the node $0$, $F_0$ is the resolvent for type B stack terminated at the node $e$, and $G_s$ is the resolvent for $s$-strings of the node $e$.

      The resolvents $F_s$ and $G_s$ and the corresponding Bethe roots enter the equations considered below in the same way independently of whether we deal with case I or II.
  \item{$W_{(n+1),s}$} - the resolvent for $s$-stacks originated by the node $n+1$.
  \item{$W_{(\overline{e}),s}$} - the resolvent for $s$-string if the node $\overline e$ is bosonic, otherwise it is the resolvent for $s$-stack originated at the node $\overline e$.
\ei
One have to prove the equations (\ref{fathookie}) for the $s$-stacks originated at the node 0. First we consider the case $1\leq s\leq n+1$ when the considered stacks are of B-type. Using the notations of Sec.~\ref{sec:countingBethe} one can write Bethe equations for the center of this stack as
\be
  -Q_e^{\CL_{s,1}}=Q_0^{\CL_{s-1,1}}Q_{s}^{\CL_{1,1}},
\ee
The corresponding integral equations are
\be\label{ie1}
  \(C^\infty\)^{-1}_{s,s'}(G_{s'}+F_{s'-1})=
  R_s+R_s^*+\(C^\infty\)^{-1}_{s,s'}(R_{s'}+F_{s'})
  +\(C^\infty\)^{-1}_{1,s'}(R_{s'+s}+W_{(s),s'}).
\ee
For $2\leq s\leq s$ one can apply $C_{ss'}$ operator on this equation. It is straightforward to check that we reproduce (\ref{fathookie}) if $W_{(s),s'}$ satisfies
\be\label{eq3}
  \(C^\infty\)^{-1}_{1,s'}\((D+D^{-1})W_{(s),s'}-W_{(s+1),s'}-W_{(s-1),s'}\)=-W_{(s),1}^*+DR_s.
\ee
This is indeed the case since the bosonic nodes satisfy (\ref{fathookie}).

The marginal case $s=1$ gives the  correct equations if we apply $C_{ss'}$
and (\ref{eq3}) and define $W_{(0),s'}\equiv F_{s'}$.

Now we consider the case $s> n+1$ (type C stacks). In this case the Bethe equation for the center of the $s$-stacks are written as
\be
  -Q_e^{\CL_{s,1}}=Q_0^{\CL_{s-1,1}}Q_{n+1}^{\CL_{s-n,1}}Q_{\overline{e}}^{-\CL_{s-n-1,1}}.
\ee
The corresponding integral equation is
\be\label{ie2}
  \(C^\infty\)^{-1}_{s,s'}(G_{s'}+F_{s'-1})&=&R_s+R_s^*+\(C^\infty\)^{-1}_{s,s'}(R_{s'}+F_{s'})
  +\no\\&&+\(C^\infty\)^{-1}_{s-n,s'}(R_{s'+n+1}+W_{(n+1),s'})-\no\\
  &&-\(C^\infty\)^{-1}_{s-n-1,s'}(W_{(n+1),s'+1}+W_{(\overline{e},s'}).
\ee
This equation is also valid for $s=n+1$ and coincides with (\ref{ie1}) for the same value of $s$. Therefore one can safely apply $C_{ss'}$ on (\ref{ie2}) for $s>n+1$. This application leads again to (\ref{fathookie}).

The case $s=n+1$ should be treated separately. In this case for the application of $C_{ss'}$ one need both equations (\ref{ie1}) and (\ref{ie2}).
The obtained integral equation will coincide with (\ref{fathookie}) if the following equation holds:
\be\label{ie3}
  W_{(n+1),1}^*+W_{(n+1),1}=\(C^\infty\)^{-1}_{s,s'}(R_{s'+n}+W_{(n),s'}-
  W_{(n+1),s'+1}-W_{(\overline e),s'}).
\ee
This equation is nothing than the integral equation for the simple Bethe roots of $(n+1)$ node.

We are left with the special case when $n=0,s=1$. In this case the same considerations show that we again obtain (\ref{fathookie}) if the following equation is correct
\be
  W_{(1),1}+W_{(1),1}^*=F_1+\(C^\infty\)^{-1}_{s,s'}(R_{s'}+F_{s'}-
  W_{(1),s'+1}-W_{(\overline e),s'}).
\ee
This is indeed the case.

Therefore we considered all possible cases and proved (\ref{fathookie}).

It is left to derive the integral equation for the corner point $\{N,M\}$.
If the terminating node of the Kac-Dynkin diagram is fermionic than this node will be exactly in the corner. One can think of it as the $(n+1)$ node in Fig.~\ref{fig:generalfermion} (the $\overline e$ node is absent in this case). Then it is easy to see that the integral equations will be
\be
  W_{(n+1),1}+W_{(n+1),1}^*=\(C^\infty\)^{-1}_{1,s'}(R_{s'+n}+W_{(n),s'}),
\ee
or in the notations $R_{a,s}$
\be\label{TNM1}
  R_{N,M}+R_{N,M}^*=\(C^\infty\)^{-1}_{1,s'}(R_{N-1,M+s'-1}+{R}^*_{N+s'-1,M-1})
\ee
The role of $R$ and $R^*$ can be interchanged depending on the situation.

If the terminating node of the Kac-Dynkin diagram is a bosonic one, then in the terms of Fig.~\ref{fig:generalfermion} there is no $(n+1)$ and $\overline e$ node. The corner resolvent will be $R_{n}$. The integral equation for it is written as
\be\label{TNM2}
  R_{n}+R_{n}^*=\(C^\infty\)^{-1}_{n,s'}G_{s'}+D^{n}\(C^\infty\)^{-1}_{1,s'}F_{s'}-\(C^\infty\)^{-1}_{n-1,s'}R_{s'}
\ee
We did not found a particularly nice equation that follows from the given one.

\chapter{\label{app:mirror}Solution of the crossing equations in the mirror theory.}
The goal of this appendix is to show how the solution \cite{V3} of the crossing equations can be modified to be applicable in the mirror theory. The mirror theory is the initial point for the construction of the thermodynamic Bethe Ansatz. It was formulated in \cite{Ambjorn:2005wa,Arutyunov:2007tc} as the analytical continuation of the physical theory. The dressing phase was also found there as the analytical continuation. We would use a different way and define the mirror theory from the bootstrap approach. This would give us mirror crossing equations solution of which leads to the mirror dressing phase.

While in the physical theory the physical region is defined by the condition $|x[u]|>1$, in the mirror theory the physical region is defined by the condition ${\rm Im}[x[u]]>0$ \cite{Arutyunov:2007tc}. Therefore let us define a mirror Jukowsky variable $x_m$
\be
    x_m=\frac 1{2g}\left(u+i\sqrt{4g^2-u^2}\right)
\ee
for which condition ${\rm Im}[x[u]]>0$ is satisfied\footnote{Mirror Zhukovsky variables where also used in \cite{GromovKKV}}. Mirror and physical Zhukovsky variables are related by
\be
    x[u]&=&x_m[u],\ \ Im[u]>0,\no\\
    x[u]&=&1/x_m[u],\ \ Im[u]<0.
\ee


The mirror theory is conjectured to be integrable. It has the same symmetry algebra as the physical theory. Therefore we can apply the bootstrap approach similarly as it was done for the physical theory. We would get then the Bethe Ansatz equations with the central node equation given by
\be\label{mirrorBAE}
    e^{ip_{mirr}}=\prod_{j\neq i}\frac{u_i-u_j+i}{u_i-u_j-i}\sigma_m[u,v]^2\ldots,
\ee
where dots stay for the interaction with the Bethe roots from nested levels. This interaction as well as the nested Bethe equations are exactly the same as in (\ref{BSaBA}) except for the replacement $x\to x_m$.

\begin{figure}[t]
\centering
\includegraphics[width=8.00cm]{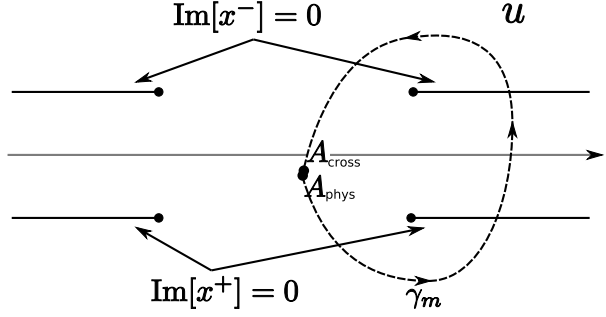}
\caption{\label{fig:mirrorgamma}Analytical continuation used in the crossing equation for the mirror theory.}
\end{figure}

The dressing phase $\sigma_m$ does not coincide with the dressing phase of the physical theory. We will now formulate the mirror crossing equations which this dressing phase should satisfy. 
The crossing equations can be derived by simple modification for the case of the mirror theory of the approach of Janik \cite{Janik:2006dc}. These equations now read
\be\label{mirrorcrossing}
  \sigma_m[\xpm_m,\ypm_m]\sigma^{cross}_m[\xpm_m ,\ypm_m]=\frac{\ym_m}{\yp_m}\frac{\xm_m-\yp_m}{\xp_m-\yp_m}\frac{1-\frac 1{\xm_m\ym_m}}{1-\frac 1{\xp_m\ym_m}},
\ee
where the $^{cross}$ stands for the analytical continuation over the contour $\g_m$ shown in Fig.~\ref{fig:mirrorgamma}. Note that $\g_m$ is not the same contour as was used for the crossing transformation in the physical theory. In fact, $\g_m=\g^{-1}$.

Solution of (\ref{mirrorcrossing}) undergoes the same steps as in the case of the physical theory. The only difference is that instead of the cut $|x[u]|=1$ ($u^2<4g^2$) we use the cut ${\rm Im}[x[u]]=0$ ($u^2>4g^2$). Therefore the solution of the mirror crossing equations is given by the following expression:
\be\label{dressingmirror}
  \sigma_m[u,v]=(u-v)^{(D-D^{-1})\tilde K_m\left(\frac{D^2}{1-D^2}-\frac{D^{-2}}{1-D^{-2}}\right)\tilde K_m(D-D^{-1})},
\ee
where $\tilde K_m$ is defined as
\be
  \tilde K_{m}\ F=\int_{\MR+i0\backslash[-2g,2g]}\frac {dv}{2\pi i}\frac {x_m-\frac 1{x_m}}{y_m-\frac 1{y_m}}\frac 1{v-u} F[v].
\ee

Let us find the relation between physical dressing factor $\sigma[u,v]$ and the mirror dressing factor $\sigma_m[u,v]$. For this we first need establish the relation between $(\tilde K\, F)[u]$ and $(\tilde K_m\, F)[u]$.  When $F$ has singularities only in the lower half plane then
\be\label{KR1}
  (\tilde K_m F)[u]&=&F[u]-(\tilde K F)[u],\ \ {\rm Im}[u]>0,\no\\
  (\tilde K_m F)[u]&=&(\tilde K F)[u],\ \ {\rm Im}[u]<0.
\ee
If $F$ has singularities only in the upper half plane then
\be
  (\tilde K_m F)[u]&=&(\tilde K F)[u],\ \ {\rm Im}[u]>0,\no\\
  (\tilde K_m F)[u]&=&F[u]-(\tilde K F)[u],\ \ {\rm Im}[u]<0.\label{KRR2}
\ee
If $F$ has singularities in the lower and the upper half planes then we have to represent $F$ as a sum of two functions and then apply (\ref{KR1}) and (\ref{KRR2}).

By application of transformations (\ref{KR1}) and (\ref{KRR2}) to the mirror dressing factor (\ref{dressingmirror}) we obtain in the region ${\rm Im}[u]>1/2 ,{\rm Im}[v]>1/2$:
\be
    \sigma_m[u,v]&=&(u-v)^{-(D-D^{-1})\tilde K\left(\frac{D^2}{1-D^2}-\frac{D^{-2}}{1-D^{-2}}\right)\tilde K(D-D^{-1})-D\tilde K D+D^{-1}\tilde K D^{-1}}=\no\\
    &=&\left(\sigma[u,v]\frac{1-\frac 1{\xp\ym}}{1-\frac 1{\xm\yp}}\sqrt{\frac{\ym}{\yp}}\right)^{-1}.
\ee
Therefore equation (\ref{mirrorBAE}) can be also written in the form
\be\label{mirrorbaeassl2}
    e^{-ip_{mirr}}=\prod_{j\neq i}\frac{u_i-u_j-i}{u_i-u_j+i}\left(\frac{1-\frac 1{\xp_i\xm_j}}{1-\frac 1{\xm_i\xp_j}}\sqrt{\frac{\xm_i}{\xp_i}}\sigma[u_i,u_j]\right)^2_{\rm an.cont.},
\ee
where "an.cont." means that expression in brackets is obtained by analytical continuation from the region ${\rm Im}[u_i]>1/2 ,{\rm Im}[u_j]>1/2$ to the real values of $u_i$ and $u_j$. Except for this analytical continuation, the r.h.s. of (\ref{mirrorbaeassl2}) coincides with the r.h.s. of the Bethe Ansatz equation (\ref{bethesl2sector}) for the $sl(2)$ sector of the physical theory.

We are now at the point to compare our considerations with the results obtained in the literature. The mirror theory was initially formulated as analytical continuation of the physical theory, the mirror dressing phase was defined not in the way that we define it here, but by the analytical continuation of $\sigma[u,v]$.

It was shown that the bound states of magnons in the mirror theory appear for the excitations on the AdS \cite{Arutyunov:2007tc} (in the physical theory analogical bound states appeared for excitations on $S^5$), therefore the $SL(2)$ sector was considered for the asymptotic Bethe Ansatz. The Bethe equations for this sector were given in the form (\ref{mirrorbaeassl2}) \cite{Arutyunov:2007tc}. Then after studying of the mirror dressing phase and the dressing phase for bound states obtained from fusion it was realized that the product of the mirror dressing phase and the BDS factor $\frac{1-\frac 1{\xp\ym}}{1-\frac 1{\xm\yp}}$ has much simple analytical properties \cite{GromovKKV,Arutyunov:2009kf,Arutyunov:2009ux}. The product of the mirror dressing phase and the BDS factor was called in \cite{Arutyunov:2009kf} the improved dressing phase. A representation analogical to (\ref{dressingmirror}) was first time derived in \cite{GromovKKV}.

In our approach we define the mirror theory purely by the bootstrap method and therefore without any reference to the physical theory. We see that what is called in \cite{Arutyunov:2009kf} the improved dressing phase is in fact a true dressing phase of the mirror theory which follows from the solution of the crossing equations (\ref{mirrorcrossing}).

Let us finally comment on the dressing factor for the magnon bound states. The magnon bound states from the point of view of the Bethe Ansatz are string-type configurations \cite{Dorey:2006dq}, exactly as in the case of XXX spin chain. Using the notations of section \ref{sec:countingBethe} we can write down the dressing factor for the interaction of $s$-string and $s'$-string as
\be
    \sigma^{ss'}_m[u,v]=\sigma[u,v]^{\frac{D^{s}_u-D^{-s'}_u}{D_u-D^{-1}_u}\frac{D^{s'}_v-D^{-s'}_v}{D_v-D^{-1}_v}}.
\ee
Then, using representation (\ref{dressingmirror}) we can immediately see that
\be\label{stringsigma}
    \sigma^{ss'}_m[u,v]=(u-v)^{(D^s-D^{-s})\tilde K_m\left(\frac{D^2}{1-D^2}-\frac{D^{-2}}{1-D^{-2}}\right)\tilde K_m(D^{s'}-D^{-s'})}.
\ee
In fact, the same derivation can be made in the physical theory. We have just to replace $\tilde K_m$ with $\tilde K$.

A word of caution should made here. Derivation of (\ref{stringsigma}) based on the string configurations is not valid in the strip where Zhukovsky cut is defined. This is a strip $|{\rm Im[u]}|<2g$ for the physical theory and $|{\rm Im[u]}|>2g$ for the mirror theory. However, once the dressing factor (\ref{stringsigma}) is established in the permitted region it can be smoothly continued to the forbidden strips.  
\chapter{\label{app:sl2}$Sl(2)$ Heisenberg magnet. Technical details}
The most general solution of (\ref{RH111}) in the $B^{-4}$ order is the following
\be
  R_{h,4}&=&\frac 1{4\pi^2\sqrt{1-\frac{B^2}{u^2}}}\(\frac{\pi^2 B}{u^5}+\frac{6-\frac{\pi^2}2}{Bu^3}+\frac{\log\[\frac{u-B}{u+B}\]}{(B^2-u^2)^2}\(\frac 92+\frac{6B^4}{u^4}-\frac{13B^2}{u^2}+\frac{3\log[16\pi B]}{2}\)+\right.\no\\&&+\left.\frac{\(\ \log\[\frac{u-B}{u+B}\]\)^2}{(B^2-u^2)^2}\(\frac{3B^5}{2u^5}-\frac{15B^3}{4u^3}+\frac {3B}u\)+\frac 1{uB}\frac{c[1,2]}{(u^2-B^2)}+\frac{B}{u}\frac{c[2,1]}{(u^2-B^2)^2}\).
\ee
To fix the coefficients $c[1,2]$ and $c[2,1]$ we should consider the following solution in the double scaling limit
\be\label{rhs}
  \tilde R_h[s]=\frac As\(1+\frac{Q_{0,1}}{B}+\frac{Q_{1,1}}{B s}\)\Phi[s].
\ee
Comparison of large $z$ expansion of $j^{-1}R_h$ and one that follows from (\ref{rhs}) fixes the coefficients $c[1,2]$ and $c[2,1]$ to be
\be
  c[1,2]=\frac{2\log[16\pi B]+\pi^2-15}{16\pi^2},\ \ c[2,1]=\frac{3(\log[16\pi B])^2-10\log[16\pi B]+\pi^2-5}{16\pi^2}.
\ee 
\chapter{Mathematica code for asymptotic expansions}
\section{\label{sec:derofgammafunctions}Derivatives of the gamma functions}
For the calculation of the asymptotic expansion of the cusp anomalous dimension and the energy density of the $O(N)$ sigma model we need to know the value of the following ratio:
\be\label{DG}
    DG[x,n]=\frac{\Gamma[x]^{(n)}}{\Gamma[x]}.
\ee
In principle we can evaluate this ratio using the code

\ \\
\verb"D[Gamma[x], {x, n}]/Gamma[x] // FunctionExpand // Expand"
\ \\

However, this evaluation at large values of $n$ turns out to be the most time consuming place in the whole recursive procedure. Therefore we will present here a quicker code.

The code is based on the equality
\be
    \frac{d\log\Gamma[x]}{dx}=-\g-\frac 1x+\sum_{k=1}^\infty\left(\frac 1k-\frac 1{x+k}\right)
\ee
and its consequence for $n\geq 2$
\be
    \frac{d^n\log\Gamma[x]}{dx^n}=(-1)^n\Gamma[n]\sum_{k=0}^\infty\frac 1{(x+k)^n}=(-1)^n\Gamma[n]\zeta[n,x].
\ee
Using the fact that
\be
   \frac{d^n\log\Gamma[x]}{dx^n}=\frac{\Gamma[x]^{(n)}}{\Gamma[x]}+\ldots,
\ee
where the dots stay for the polynomial in $\Gamma[x]^{(m)}/\Gamma[x]$ with $m<n$, we obtain the recursive procedure to calculate (\ref{DG}). The code for this recursive procedure is the following:
\vfil
\begin{minipage}{0.85\textwidth}
\footnotesize\sl{\begin{verbatim}
DG[a_,1] := DG[a,1] = D[Log[Gamma[a+x]],x]/. x->0//FunctionExpand;
DG[a_,n_/;n>1] := DG[a, n]=
(-1)^n Gamma[n]FunctionExpand@Zeta[n,a]-(D[Log[f[x]],{x,n}]/.f[x]->1
/.Derivative[n][f][x]->0/.Derivative[m_][f][x]:>DG[m])//Expand;
\end{verbatim}
}
\end{minipage}
\vfil
Below we will use a slightly modified version of this code.
\vfil
\vfil
\vfil
\vfil
\vfil
\vfil
\vfil
\vfil
\vfil
\vfil
\vfil
\vfil
\section{\label{app:cusp}Cusp anomalous dimension}
We reexpand the solution (\ref{solG}) in the double scaling limit (\ref{BESdoublescaling}):
\be
    G_\pm[z]&=&2i\e\a_0[\e]\left(\frac{\e z}{1+\e z}\right)^{\mp\frac 14}
    +\no\\
    &+&2i\e\sum_{n=1}^\infty\e^{|n|}\left(\a_n[\e]\left(\frac{\e z}{1+\e z}\right)^{\mp n\mp\frac 14}+\a_{-n}[\e]\left(\frac{\e z}{1+\e z}\right)^{\pm n\mp\frac 14}\right).
\ee
The inverse Laplace transform of this series is given by
\be\label{tildeG1}
    \tilde G_\pm[s]&=&\frac{2i}{\G[1\pm\frac 14]}\left(\frac \e s\right)^{1\mp \frac 14}\sum_{n=0}^\infty\sum_{k=0}^\infty\e^n\left(\frac \e s\right)^{k-n}A_{n,k}^\pm[\e],
\ee
\be\label{aArelation}
    A_{0,k}^\pm[\e]&=&\frac{\Gamma[1\pm\frac 14]^2}{\Gamma[k+1]\Gamma[1\pm\frac 14-k]\Gamma[\pm\frac 14-k]}\a_0[\e],\\
    A_{n,k}^\pm[\e]&=&\frac{\Gamma[1\pm\frac 14]}{\Gamma[n+1\pm\frac 14-k]\Gamma[n\pm\frac 14-k]}\left(\frac{\Gamma[n+1\pm\frac 14]}{\G[k+1]}\a_n[\e]+\frac{\Gamma[-n+1\pm\frac 14]}{\Gamma[k+1-2n]}\a_{\mp n}[\e]\).\no
\ee
From the other side, the solution in the double scaling limit $\tilde G_\pm[s]$ is given by (\ref{solgs}):
\be\label{tildeG2}
    \tilde G_\pm[s]=\frac{2i}{\G[1\pm\frac 14]}\left(\frac \e s\right)^{1\mp \frac 14}T_\pm[s]\sum_{n=0}^\infty\beta_n^\pm[\e]\left(\frac \e s\right)^n.
\ee
Comparing (\ref{tildeG1}) and (\ref{tildeG2}) we see that
\be\label{eqtosolve}
  \sum_{n=0}^\infty\beta_n^\pm[\e]\left(\frac \e s\right)^n=\frac 1{T_\pm[s]}\sum_{n=0}^\infty\sum_{k=0}^\infty\e^n\left(\frac \e s\right)^{k-n}A_{n,k}^\pm[\e].
\ee
The coefficients $\a,\b,A$ are expanded in the power series of $\e$\footnote{The recursion procedure that we build allows to express all the coefficients as $\a_{0}[0]$ times a polynomial in $\e$; $\a_{0}[0]=1$ due to (\ref{importantconstraint}).}:
\be\label{aA}
    \a_{n}[\e]=\sum_{k=0}^\infty{\a_{n,k}}\e^{k},\ \b_{n}[\e]=\sum_{k=0}^\infty{\b_{n,k}}\e^{k},\
    A_{n,k}^\pm[\e]=\sum_{m=0}^\infty{A_{n,k,m}^\pm}\e^{m}.
\ee
Using this expansion we can find the coefficient for the $\e^m$ term in (\ref{eqtosolve}):
\be\label{lhsof}
    \sum_{k=0}^m\frac{\b^\pm_{k,m-k}}{s^k}=\frac{1}{T_\pm[s]}\sum_{n=0}^\infty\sum_{k=0}^ms^{n-k}A^\pm_{n,k,m-k}=\sum_{n=-m}^\infty s^n\sum_{k=0}^m\sum_{p=0}^{k+n}\(\frac 1{T_\pm}\)_p A^\pm_{n+k-p,k,m-k},
\ee
where $\(1/{T_\pm}\)_p$ is the coefficient in the expansion
\be
    \frac 1{T_\pm[s]}=\sum_{p=0}^\infty\(\frac 1{T_\pm}\)_ps^n.
\ee
Since the l.h.s. of (\ref{lhsof}) contains only nonpositive powers of $s$, we conclude that
\be\label{constraint}
   \sum_{k=0}^m\sum_{p=0}^{k+n}\(\frac 1{T_\pm}\)_p A^\pm_{n+k-p,k,m-k}=0,\ \ n\geq 1.
\ee
This gives us the recursion relation:
\be
    A^\pm_{n,0,m}=-\sum_{p=1}^n \(\frac 1{T_\pm}\)_p A^\pm_{n-p,0,m}-\sum_{k=1}^m\sum_{p=0}^{k+n}\(\frac 1{T_\pm}\)_p A^\pm_{n+k-p,k,m-k}.
\ee
This relation expresses\footnote{$\a$ and $A^\pm$ are related by (\ref{aArelation}).} $\a_{n,m}$ in terms of $\a_{\tilde n,\tilde m}$ with $\tilde n+\tilde m<n+m$ and $\a_{n+k,m-k}$ with $k>0$. Since $\a_{n,m}=0$ for $m< 0$, the recursion finally allows to express $\a_{n,m}$ in terms of $\a_{0,m}$. The latter cannot be found from the condition (\ref{constraint}). It is fixed by the normalization condition (\ref{escaling}) which implies that the coefficients $\a_n[\e]$ satisfy the constraint
\be\label{importantconstraint}
    \sum_{n\e\MZ}\e^{|n|}\a_n[\e]=1.
\ee
The consequence of this constraint is that
\be
    \a_{0,0}=1,\ \ a_{0,k}=-\sum_{n=1}^k(\a_{n,k-n}+\a_{-n,k-n}).
\ee
This completes the recursive procedure and it can be implemented in \textit{Mathemtica}. The code is the following:
\vfil
\begin{minipage}{0.9\textwidth}
\footnotesize\sl{
\verb"G"\verb" = Gamma; ClearAll[a, F1, F2, F3, F4];"\\
\verb"a[0, 0] = 1; a[0, k_ /; k < 0] = 0;"\\
\verb"a[0, k_] := a[0, k] = - Sum[a[n, k - n] + a[-n, k - n], {n, 1, k}];"\\
\verb"F1[n_, k_, s_] :=  F1[n, k, s] ="\\
\verb"  FunctionExpand[("\verb"G"\verb"[1 + s/4]"\verb"G"\verb"[n+s/4])/("\verb"G"\verb"[1 + s/4 - k]"\verb"G"\verb"[s/4 - k])]/"\verb"G"\verb"[k + 1];"\\
\verb"F2[n_, p_, s_] :=  F2[n, p, s] ="\\
\verb"  FunctionExpand["\verb"G"\verb"[n + s/4]/"\verb"G"\verb"[n - p + s/4]];"\\
\verb"F3[n_, p_, k_, s_] :=  F3[n, p, k, s] ="\\
\verb"  FunctionExpand[("\verb"G"\verb"[n + s/4]"\verb"G"\verb"[n + k - p + 1 + s/4])/("\verb"G"\verb"[n + 1 + s/4 - p]"\\
\verb"                  "\verb"G"\verb"[n + s/4 - p])]/"\verb"G"\verb"[k + 1];"\\
\verb"F4[n_, p_, k_, s_] :=  F4[n, p, k, s] ="\\
\verb"  FunctionExpand[("\verb"G"\verb"[n + s/4]"\verb"G"\verb"[1 + s/4 - n - k + p])/("\verb"G"\verb"[n + 1 + s/4 - p]"\\
\verb"                  "\verb"G"\verb"[n + s/4 - p])]/"\verb"G"\verb"[2 p + 1 - 2 n - k];"
\begin{verbatim}
a[n_ /; n != 0, m_] :=  a[n, m] =  Block[{s = Sign[n], nn = Abs[n]},
  - Expand@Plus@@Expand/@{
  Sum[invT[s, nn + k] F1[nn, k, s] a[0, m - k], {k, 0, m}],
  Sum[invT[s, p] F2[nn, p, s] a[s (nn - p), m], {p, 1, nn - 1}],
  Sum[invT[s, p] F3[nn, p, k, s] a[s (nn+k-p),m-k],{k,1,m},{p,0,k+nn-1}],
  Sum[invT[s, p] F4[nn, p, k, s] a[-s(nn+k-p),m-k],{k,1,m},{p,0,k+nn-1}]}];
\end{verbatim}
}
\end{minipage}
\vfil
The function \verb"invT[s, p]" encodes $(1/T_\pm)_p$. A direct way to evaluate it is
\vfil
\begin{minipage}{0.9\textwidth}
\footnotesize\sl{
\verb"invT[s_, p_] :=  invT[s, p] =  Expand@FunctionExpand@SeriesCoefficient["\\
\verb" If[s == 1,"\verb"G"\verb"[3/4 - x/(2"\verb"\[Pi]"\verb")]/("\verb"G"\verb"[3/4]"\verb"G"\verb"[1-x/(2"\verb"\[Pi]"\verb")]),"\\
\verb"           "\verb"G"\verb"[1/4 - x/(2"\verb"\[Pi]"\verb")]/("\verb"G"\verb"[1/4]"\verb"G"\verb"[1-x/(2"\verb"\[Pi]"\verb")])], {x, 0, p}];"
}
\end{minipage}
\vfil

Now we can find the coefficients $f_n$ for the expansion of the scaling function:
\be\label{fgstrongcoupling}
    f[g]=\frac 1{\e}(f_0+f_1\,\e +f_2\,\e^2+f_3\,\e^3+\ldots).
\ee
\vfil
\begin{minipage}{0.9\textwidth}
\footnotesize\sl{
\verb"ClearAll[f]; f[0] = 4; f[k_] := f[k] ="\\
\verb"    Expand@Sum[4 m ( a[m, k - m] - a[-m, k - m]), {m, 1, k}]"
}
\end{minipage}
\vfil

\subsubsection{Structure of the result and improvement of the code}
The proposed above code for \verb"invT[s,p]" allows find the first 10 coefficients $f_n$ in about 20 seconds. The result obtained in this way is however not well structured, its further simplification is time consuming. Here we will give another code for \verb"invT[s,p]" that leads from the beginning to a nice representation of the result.

We evaluate the derivatives of the gamma functions as was explained in section \ref{sec:derofgammafunctions}. This evaluation is the recursive procedure with $(d^{n}\log\G[a])/(da^n)$ as the input. In our case $a=1,1/4,3/4$. For these particular values of $a$ we have the following explicit expressions:

\noindent for $n=1$
\be\label{deriv1}
  (\log\G[1])'&=&-\gamma,\ \ (\log\G[1/2\pm1/4])'=-\g\pm\frac{\pi}{2}-\log[8],
\ee
for $n\geq 2$
\be\label{derivn}
    (\log\G[1])^{(n)}&=&(-1)^n \G[n]\zeta[n],\no\\
    (\log\G[1/2\pm1/4])^{(n)}&=&(-1)^n \frac{4^n\G[n]}{2}((1-2^{-n})\zeta[n]\mp\beta[n]),
\ee
where $\zeta[n]$ and $\beta[n]$ are defined by
\be
    \zeta[n]=\sum_{k=0}^\infty k^{-n},\ \ \beta[n]=\sum_{k=0}^\infty\frac{(-1)^k}{(2k+1)^n}.
\ee
The ratios $\zeta[n]/\pi^n$ for even values of $n$ and $\beta[n]/\pi^n$ for odd vales of $n$ are rational numbers.

Let us look more carefully on the structure of the recursive procedure which takes (\ref{deriv1}) and (\ref{derivn}) as the input and gives $f_n$ as the output. Each coefficient $(1/T_\pm)_p$ has an overall multiplier $\pi^{-p}$ due to the expansion over the combination $s/2\pi$. Except for this multiplication all the operations which are involved in the recursive procedure reduce to the polynomial combinations, with {\it rational} coefficients, of (\ref{deriv1}) and (\ref{derivn}).

\paragraph{Cancelation of logarithms and $\g$.} As was initially observed in \cite{Basso:2007wd}, the coefficients of the expansion of the scaling function do not contain $\log[2]$ if to expand over the shifted constant $g'=g-\frac{3\log 2}{4\pi}$. The coefficients of the expansion also do not contain $\g$. We checked these observations up to the tenth order and will assume that they are true for arbitrary order. Using this assumption and the rational structure of the recursive procedure, we can put $\log[2]\to 0$ and $\g\to 0$ in (\ref{deriv1}).

\paragraph{The property of maximal transcendentality.} Let us assign the transcendentality $n$ to $\zeta[n]$ of odd argument and to $\beta[n]$ of even argument. We assign transcendentality 0 to $\pi$ and to rational numbers. As we checked up to the tenth order and assume to be true for any order, the coefficient $f_n$ of the expansion (\ref{fgstrongcoupling}) has the transcendentality $n$ which is the maximal possible transcendentality at this order. This was initially observed also in \cite{Basso:2007wd}. Due to this property we can leave only the terms with maximal transcendentality in (\ref{deriv1}) and (\ref{derivn}). We put $\pi\to 0$ in (\ref{deriv1}) and $\zeta[2n]\to 0,\beta[2n+1]\to 0$ in (\ref{derivn}).

After these simplifications we can put $\pi\to 1$ in the combination $\frac{s}{2\pi}$. Indeed, since $\frac{s}{2\pi}$ is now the only place where $\pi$ appears, $\pi$ becomes an overall multiplier and can be restored at the end of the calculation ($f_n$ should contain $\pi^{-n}$ as an overall multiplier).

Using these simplifications, we propose the following code for $invT[s,p]$:

\vfil
\begin{minipage}{0.9\textwidth}
\footnotesize\sl{
\verb""\verb"G"\verb" = Gamma;ClearAll[DG, DGp, D1G, invT];"\\
\verb"DG[1] = 0; DGp[1] = 0; DG[0] = 1; DGp[0] = 1; D1G[0] = 1;"\\
\verb"DG[n_ /; n > 1] := DG[n] = Expand["\\
\verb"  (-1)^n "\verb"G"\verb"[n] If[EvenQ[n],  0, "\verb"\[Zeta]"\verb"[n]] - (D[Log[h[x]], {x, n}]"\\
\verb"  /.h[x] -> 1/. Derivative[n][h][x] -> 0/. Derivative[m_][h][x] :> DG[m])];"\\
\verb"DGp[n_ /; n > 1] := DGp[n] =  Expand[(-1)^n 4^n/2 "\verb"G"\verb"[n]"\\
\verb"  If[EvenQ[n], -"\verb"\[Beta]"\verb"[n], (1 - 2^-n) "\verb"\[Zeta]"\verb"[n]] - (D[Log[h[x]], {x, n}]"\\
\verb"  /. h[x] -> 1/. Derivative[n][h][x] -> 0/. Derivative[m_][h][x] :> DGp[m])];"
\verb"D1G[n_ /; n > 0] := D1G[n] = Expand[D[1/h[x], {x, n}]"\\
\verb"  /. h[x] -> 1/. Derivative[m_][h][x] :> DG[m]];"\\
\ \\
\verb"invT[_, 0] = 1; invT[_, 1] = 0;"\\
\verb"invT[1, p_ /; p >= 2] := invT[1, p] ="\\
\verb"  Expand@Sum[(-1/2)^p 1/("\verb"G"\verb"[r + 1]"\verb"G"\verb"[p - r + 1]) DGp[r] D1G[p - r],{r, 0, p}];"\\
\verb"invT[-1, p_ /; p >= 2] := invT[-1, p] = "\\
\verb"  Expand@(invT[1, p] /. "\verb"\[Beta]"\verb"[x_] -> (-"\verb"\[Beta]"\verb"[x]));"
}
\end{minipage}
\vfil

Now we can evaluate $f_n$, using the improved definition of \verb"invT[s,p]". For example, the first five orders are given by
\vfil
\begin{minipage}{0.9\textwidth}
\footnotesize\sl{
\verb"In=   Table[f[n]/\[Pi]^n, {n, 0, 4}]//Expand"\\
}
\ \\
\verb"Out=  {4, 0, -("$\beta$\verb"[2]/\[Pi]^2), -((27 "$\zeta$\verb"[3])/32/\[Pi]^3),"\\
\verb"       -(2 "$\beta$\verb"[2]^2 + (21 "$\beta$\verb"[4])/4)/\[Pi]^4}"
\ \\
\end{minipage}
\vfil
The result coincides with the one given in \cite{Basso:2007wd}. This code allows to calculate first 10 orders in few seconds and first 30 orders in half of an hour.

\section{\label{app:massgap}Energy density of the $O(N)$ sigma model}
In chapter~\ref{ch:massgap} we gave a general form of the perturbative expansion of the resolvent in the large $B$ limit (\ref{sol2}):
\be\label{appsol2}
R[\theta]=\frac{2A\sqrt{B}}{\theta\sqrt{1-\frac{B^2}{\theta^2}}}\!\!\sum_{n,m=0}^\infty\sum_{k=0}^{m+n}\frac{c_{n,m,k}
(\theta/B)^{\e[k]}}{B^{m-n}\(\theta^2-B^2\)^{n}}
\log\[\frac{\theta\!-\!B}{\theta\!+\!B}\]^k\ .
\ee
We replace in (\ref{appsol2}) $\theta$ with the double scaling variable $z=2(\theta-B)$, perform large $B$ expansion, and make the inverse Laplace transform (\ref{ONlaplacetransform}). To do these steps it is useful to represent $(\log ...)^k$ as $\lim\limits_{x\to 0}\frac{d^k}{dx^k}e^{x ...}$. The result is:
\be\label{expansion1}
    \hat R[s]&=&\frac{A}{\sqrt{\pi s}}\sum_{m=0}^\infty\sum_{n=-m}^\infty\sum_{t=0}^{n+m}\frac{s^n(\log[4Bs])^t}{B^m}V_c[n,m,t],\no\\
    V_c[n,m,t]&=&\sum_{k=t}^{n+m}\sum_{r=Max[0,-n]}^m c[n+r,m-r,k]F[n,t,k,r],\no\\ F[n,t,k,r]&=&2^{1-2r}(-1)^k\frac{1}{\G[r+1]}\frac{\G[k+1]\G[1/2]}{\G[t+1]\G[k-t+1]}\times\no\\
    &&\times\lim_{x\to 0}\frac{d^{k-t}}{dx^{k-t}}\(\frac{\frac{\G[-n-x+1/2-r]}{\G[-n-x+1/2-2r]}+
    2r\e[k]\frac{\G[-n-x+1/2-r]}{\G[-n-x+3/2-2r]}}{\Gamma[n-x+1/2]}\).
\ee

This expression should be equal to the solution in the double scaling limit (\ref{DSsolution}):
\be\label{appDSsolution}
  \hat R[s]&=&\(\frac {1}{s+\frac 12}+Q[s]\)A\,\Phi[s],\ \ \ A=\frac{\mass}{4\Delta^{\Delta}}e^{-\frac 12+B+\Delta}\Gamma[\Delta],\no\\
  \Phi[s]&=&\frac 1{\sqrt{s}}e^{(1-2\Delta)s\log\[\frac se\]-2\Delta s\log[2\Delta]}\frac{\Gamma[2\Delta s+1]}{\Gamma\[s+\frac 12\]},\no\\
  Q[s]&=&\frac 1{Bs}\sum_{n,m=0}^\infty\frac{Q_{n,m}[\log B]}{B^{m+n}s^n}
\ee
To obtain the expression of type (\ref{expansion1}) we should perform a small $s$ expansion of (\ref{appDSsolution}). Before doing this let us rewrite the argument of the exponent in $\Phi[s]$ in the following way:
\be
   (1-2\Delta)s\log\[\frac se\]-2\Delta s\log[2\Delta]= (1-2\Delta)s(\log\[4 B s\]-\log[B/B_0]).
\ee
The asymptotic expansion of the energy density $\varepsilon$ can be expressed as $\rho^2$ times a power series in $1/B$ and $\log[B/B_0]$. Due to the expected renorm-group properties of the solution it is possible to introduce a such coupling constant $\a$ that this series
will be expressed as the power series of the coupling constant. Assuming this property, which we checked up to 10 first orders, we replace $\log[B/B_0]$ with zero. Below we give a more accurate explanation of this step.

 Note that to derive the correct expansion of $\varepsilon$ and $\rho$ in the powers of $1/B$ and $\log B$, given by (\ref{rhoeefirsttwo})\footnote{We leave the derivation of (\ref{rhoeefirsttwo}), which requires keeping the term $\log[B/B_0]$, as an exercise for a curious reader (see also formula (22) in \cite{V2}). The logic of the derivation remains the same, even if $\log[B/B_0]$ is present.}, we cannot do the replacement $\log[B/B_0]\to 0$. It is only admissible when our goal is to express $\varepsilon$ in terms of the coupling constant.

Within $\log[B/B_0]\to 0$, expansion of (\ref{appDSsolution}) at small $s$ leads to:
\be
    \hat R[s]&=&\frac{A}{\sqrt{\pi s}}\sum_{m=0}^\infty\sum_{n=-m}^\infty\sum_{t=0}^{n+m}\frac{s^n(\log[4Bs])^t}{B^m}V_Q[n,m,t],\no\\
    V_Q[n,0,t]&=&\frac{(1-2\Delta)^t\Gamma[1/2]}{\Gamma[t+1]\Gamma[n-t+1]}\lim_{x\to 0}\(\frac{\Gamma[1+2x\Delta]}{\Gamma[3/2+x]}\)^{(n-t)},\no\\
    V_Q[n,m,t]&=&\sum_{c=Max[0,t-n-1]}^{m-1}\Phi[n+c+1,t]Q[c,m-c-1],\ \ m>0,\no\\
    \Phi[a,b]&=&\frac{(1-2\Delta)^b\Gamma[1/2]}{\Gamma[b+1]\Gamma[a-b+1]}\lim_{x\to 0}\(\frac{\Gamma[1+2x\Delta]}{\Gamma[1/2+x]}\)^{(a-b)}.
\ee
The requirement
\be\label{requirement}
V_c[n,m,t]=V_Q[n,m,t]
\ee
unambiguously fixes the coefficients $c_{a,b,c}$ and $Q_{a,b}$ in a recursive way. To express $Q_{a,b}$ in terms of $Q$ and $c$ known from the earlier steps of recursion we need to consider (\ref{requirement}) with $n=-a-1$, $m=b+a+1$, $t=0$. To do the same for $c_{a,b,c}$ we need to consider (\ref{requirement}) with $n=a$, $m=b$, $t=c$. The recursion procedure allows to express $Q$ and $c$ in terms of $V_Q[n,0,t]$ which is explicitly known.

Among all the equations (\ref{requirement}) with $n<0$ we use only those for which $t=0$. All the other equations with $n<0$ can serve for the verification of the self-consistency of our solution.

The evaluation of the derivatives of the gamma-functions is made using the trick of Sec.~\ref{sec:derofgammafunctions}. As in the case of the cusp anomalous dimension, a special structure of the answer allows to simplify this calculation. As we observed up to tenth order, the coefficients of asymptotic expansion are given in terms of zeta-functions of odd argument and rational numbers only. Assuming that this property holds at any order, we perform the following replacements in $(\log\G[...])'$: $\g\to 0$, $\log 2\to 0$, $\pi\to 0$.

The \textit{Mathematica} code for the recursive procedure is the following:

\vfil
\begin{minipage}{0.85\textwidth}
\footnotesize\sl{\begin{verbatim}
(* Derivatives of the gamma functions *)
ClearAll[DG, G, subs]; G = Gamma;
subs = {_Log -> 0, EulerGamma -> 0, \[Pi] -> 0, Zeta -> \[Zeta]};
DG[_, 0] = 1;
DG[a_List, n_?IntegerQ /; n > 0] := DG[a, n] =
  Block[{f, y}, (Plus @@ FunctionExpand@ Flatten[
  {1, -1} D[Log@G@a, {x, n}] /. x -> 0] /. subs) -
  (D[Log[f[y]], {y, n}] /. f[y] -> 1 /. Derivative[n][f][y] -> 0
  /. Derivative[m_][f][y] :> DG[a, m]) // Expand]
\end{verbatim}
}
\end{minipage}
\vfil
\begin{minipage}{0.85\textwidth}
\footnotesize\sl{\begin{verbatim}
(* Definition of structure constants *)
ClearAll[F, \[CapitalPhi], Vc, VQ, Q, c];
F[n_, t_, k_, r_] := F[n, t, k, r] = ((-1)^k 2^(1-2 r))/G[r+1]Binomial[
  k, t](FunctionExpand[(G[1/2] G[-n+1/2-r])/G[n+1/2]/{
  G[-n+1/2-2r],G[-n+3/2-2r]}].{DG[{{-n+1/2-r},{-n+1/2-2r,n+1/2}}+x,k-t],
  2 r Mod[k, 2]DG[{{-n+1/2-r},{-n+3/2-2r,n+1/2}}+x,k-t]}) //  Expand;
\[CapitalPhi][a_, b_] := \[CapitalPhi][a, b] =
  (1 - 2 \[CapitalDelta])^b/(G[b + 1] G[a - b + 1])DG[
  {{1 + 2 x \[CapitalDelta]}, {1/2 + x}}, a - b] // Expand;
VQ[n_, 0, t_] := VQ[n, 0, t] =
  (2 (1 - 2 \[CapitalDelta])^t)/(G[t + 1] G[n - t + 1]) DG[
  {{1 + 2 x \[CapitalDelta]}, {3/2 + x}}, n - t] // Expand;
\end{verbatim}
}
\end{minipage}
\vfil
\begin{minipage}{0.85\textwidth}
\footnotesize\sl{\begin{verbatim}
(* Recursive procedure *)
Vc[n_, m_, t_] := Plus @@ Flatten[Table[
  pc[n + r, m - r, k] F[n, t, k, r], {k, t, n + m}, {r, Max[0, -n], m}]
  /. If[n >= 0, pc[n, m, t] -> Pc, 0 -> 0] /.  pc -> c] // Expand;
VQ[n_, m_, t_] := Plus @@ (
  Table[\[CapitalPhi][n+c+1,t]pQ[c,m-c-1],{c, Max[0, t - n - 1], m - 1}]
   /. If[n<0 && t==0, pQ[-n-1, m+n] -> PQ, 0 -> 0] /.pQ -> Q)//Expand;
c[n_, m_, t_] := c[n, m, t] =
   Pc /. Solve[VQ[n, m, t] == Vc[n, m, t], Pc][[1]] // Expand;
Q[a_, b_] := Q[a, b] =  PQ /. Solve[
   VQ[-a - 1, b + a + 1, 0] == Vc[-a - 1, b + a + 1, 0],PQ][[1]]//Expand
\end{verbatim}
}
\end{minipage}
\vfil

The energy density and the particle density represented in the form
\be
    \varepsilon=\frac{\Delta}{\pi}A^2\,n_{\varepsilon},\  \ \ \rho=\frac{\sqrt{B}}{\pi}A\, n_{\rho}
\ee
can be evaluated using
\vfil
\begin{minipage}{0.85\textwidth}
\footnotesize\sl{\begin{verbatim}
(* Energy and particle density *)
ClearAll[ne, nr];
ne[0] = nr[0] = 1;
ne[n_] := ne[n] = Expand@Sum[2^(s + 1) Q[s, n - s - 1], {s, 0, n - 1}];
nr[n_] := nr[n] = Expand[c[0, n, 0] - 2 c[0, n, 1]];
\end{verbatim}
}
\end{minipage}
\ \\

\ \\

\paragraph{Expression in terms of the coupling constant.} We would like now to express the dimensionless ratio
\be\label{exp33}
\frac{\varepsilon}{\pi\Delta\rho^2}=\frac 1{B}\frac{n_\varepsilon}{n_\rho}
\ee
in terms of the coupling constant defined in (\ref{Bajnokcoupling}). At this step we would like to better explain the substitution $\log[B/B_0]\to 0$, therefore at the moment we consider that the term $\log[B/B_0]$ is present. In terms of the parameter $B$, the coupling constant is defined as:
\be\label{eq335}
    \frac 1{\a}+(\Delta-1)\log\a=B-\frac 12+\Delta\log\[\frac{8}{\Delta}\]-\log 2+\frac 12\log B+\log n_\rho.
\ee
Solving this equation perturbatively at large $B$, we express $\a$ as a series in $1/B$ and $\log B$. Expression (\ref{exp33}) is also given by such series. We can treat $\log B$ as the independent (from $B$) parameter and reexpress (\ref{exp33}) in terms of $\a$. The expected renorm-group dynamics predicts that all the $\log B$ terms cancel out. Therefore we can put $\log B$ equal to any quantity. For simplicity reasons we choose $\log B=\log B_0$. Performing this replacement in  (\ref{eq335}) includes one step which should be done carefully:
\be
    \log\a=\log[\a B]-\log[B]\to\log[\a B]-\log[B_0].
\ee
Taking this into account, we get the following equation
\be\label{afinal}
    \frac 1\a+(\Delta-1)\log[B\a]=B-\Delta+\log n_\rho,
\ee
which is valid modulo the terms proportional to $\log[B/B_0]$\footnote{or to $\log[2]$ that also cancel out from the final expression}.

Substituting
\be
\frac 1B=\sum_{k=1}^\infty b_k\a^k
\ee
to (\ref{afinal}), we find recursively the coefficients $b_k$ which allows us to express (\ref{eq335}) in terms of $\a$ and to find the coefficients $\chico_n$ defined in (\ref{varepsginv}). The {\it Mathematica} code for this procedure is the following.

\vfil
\begin{minipage}{0.85\textwidth}
\footnotesize\sl{\begin{verbatim}
M = 10;(* Maximal desired order *)
a = \[Alpha]; ClearAll[b, Binv, Bdir, Bdird, Bdirreg];
b[1] = 1; b[2] = -\[CapitalDelta];
Binv[0] = 1; Bdirreg[0] = 1; Bdird[0] = 1;
Binv[1] = Sum[b[n] a^n, {n, 1, M}] + O[a]^(M + 1);
Bdir = Expand /@ (1/Binv[1]);
Bdirreg[1] = 1/a + \[CapitalDelta] - Bdir;
Bdird[1] = a Bdir - 1 + O[a]^(M - 1);
Binv[n_] := Binv[n] = (Binv[n - 1] + O[a]^M) (Binv[1] + O[a]^M)
Bdirreg[n_] := Bdirreg[n] =  Expand /@ (
  (Bdirreg[n - 1] + O[a]^(M - 2)) (Bdirreg[1] + O[a]^(M - 2)))
Bdird[n_] :=  Bdird[n] =  Expand /@ (
  Bdird[n - 1] + O[a]^(M - 2)) (Bdird[1] + O[a]^(M - 2));

left = Expand /@ Sum[nr[k] Binv[k], {k, 0, M - 2}] + O[a]^(M - 1);
right = Expand /@ (Sum[FunctionExpand[Gamma[\[CapitalDelta]]/(
     Gamma[k + 1] Gamma[\[CapitalDelta] - k])] Bdird[k], {k, 0, M - 2}]
     Sum[Bdirreg[n]/Gamma[n + 1], {n, 0, M - 2}]);

Do[b[n] =  b[n] /. Expand@ Solve[left[[3, n - 1]] == right[[3, n - 1]],
  b[n]][[1]], {n, 3, M}]

nra = Expand /@ Sum[nr[k] Expand /@ Binv[k], {k, 0, M}];
nea = Expand /@ Sum[ne[k] Expand /@ Binv[k], {k, 0, M}];
answer = Expand /@ (Binv[1] nea/nra^2);
\end{verbatim}
}
\end{minipage}
\vfil

The first few coefficients $\chico_n$ are given by:
\vfil
\begin{minipage}{0.85\textwidth}
\footnotesize\sl{\begin{verbatim}
chi[n_] := answer[[3, n]] // Expand
Table[chi[n],{n,3,6}]
\end{verbatim}
}
\end{minipage}
\vfil

With this code $\chico_{10}$ can be found in less then one minute using a single core at 2Ghz. To get the expression for $\chico_{26}$ approximately 20 hours is needed.

%
%


%

%
%
%

 \bibliography{phdbib}        

\providecommand{\href}[2]{#2}\begingroup\raggedright\begin{thebibliography}{10%
0}

\bibitem{Bethe:1931hc}
H.~Bethe, ``{On the theory of metals. 1. Eigenvalues and eigenfunctions for the
  linear atomic chain},'' {\em Z. Phys.} {\bf 71} (1931) 205--226.

\bibitem{Maldacena:1997re}
J.~M. Maldacena, ``{The large N limit of superconformal field theories and
  supergravity},'' {\em Adv. Theor. Math. Phys.} {\bf 2} (1998) 231--252,
  \href{http://xxx.lanl.gov/abs/hep-th/9711200}{{\tt hep-th/9711200}}.

\bibitem{Gubser:1998bc}
S.~S. Gubser, I.~R. Klebanov, and A.~M. Polyakov, ``{Gauge theory correlators
  from non-critical string theory},'' {\em Phys. Lett.} {\bf B428} (1998)
  105--114, \href{http://xxx.lanl.gov/abs/hep-th/9802109}{{\tt
  hep-th/9802109}}.

\bibitem{Witten:1998qj}
E.~Witten, ``{Anti-de Sitter space and holography},'' {\em Adv. Theor. Math.
  Phys.} {\bf 2} (1998) 253--291,
  \href{http://xxx.lanl.gov/abs/hep-th/9802150}{{\tt hep-th/9802150}}.

\bibitem{Minahan:2002ve}
J.~A. Minahan and K.~Zarembo, ``{The Bethe-ansatz for N = 4 super
  Yang-Mills},'' {\em JHEP} {\bf 03} (2003) 013,
  \href{http://xxx.lanl.gov/abs/hep-th/0212208}{{\tt hep-th/0212208}}.

\bibitem{Beisert:2003tq}
N.~Beisert, C.~Kristjansen, and M.~Staudacher, ``{The dilatation operator of N
  = 4 super Yang-Mills theory},'' {\em Nucl. Phys.} {\bf B664} (2003) 131--184,
  \href{http://xxx.lanl.gov/abs/hep-th/0303060}{{\tt hep-th/0303060}}.

\bibitem{Zamolodchikov:1978xm}
A.~B. Zamolodchikov and A.~B. Zamolodchikov, ``{Factorized S-matrices in two
  dimensions as the exact solutions of certain relativistic quantum field
  models},'' {\em Annals Phys.} {\bf 120} (1979) 253--291.

\bibitem{Staudacher:2004tk}
M.~Staudacher, ``{The factorized S-matrix of CFT/AdS},'' {\em JHEP} {\bf 05}
  (2005) 054, \href{http://xxx.lanl.gov/abs/hep-th/0412188}{{\tt
  hep-th/0412188}}.

\bibitem{Beisert:2006qh}
N.~Beisert, ``{The Analytic Bethe Ansatz for a Chain with Centrally Extended
  $su(2|2)$ Symmetry},'' {\em J. Stat. Mech.} {\bf 0701} (2007) P017,
  \href{http://xxx.lanl.gov/abs/nlin/0610017}{{\tt nlin/0610017}}.

\bibitem{Santambrogio:2002sb}
A.~Santambrogio and D.~Zanon, ``{Exact anomalous dimensions of N = 4 Yang-Mills
  operators with large R charge},'' {\em Phys. Lett.} {\bf B545} (2002)
  425--429, \href{http://xxx.lanl.gov/abs/hep-th/0206079}{{\tt
  hep-th/0206079}}.

\bibitem{Beisert:2005fw}
N.~Beisert and M.~Staudacher, ``{Long-range $PSU(2,2|4)$ Bethe Ansaetze for
  gauge theory and strings},'' {\em Nucl. Phys.} {\bf B727} (2005) 1--62,
  \href{http://xxx.lanl.gov/abs/hep-th/0504190}{{\tt hep-th/0504190}}.

\bibitem{Metsaev:1998it}
R.~R. Metsaev and A.~A. Tseytlin, ``{Type IIB superstring action in AdS(5) x
  S(5) background},'' {\em Nucl. Phys.} {\bf B533} (1998) 109--126,
  \href{http://xxx.lanl.gov/abs/hep-th/9805028}{{\tt hep-th/9805028}}.

\bibitem{Bena:2003wd}
I.~Bena, J.~Polchinski, and R.~Roiban, ``{Hidden symmetries of the AdS(5) x
  S**5 superstring},'' {\em Phys. Rev.} {\bf D69} (2004) 046002,
  \href{http://xxx.lanl.gov/abs/hep-th/0305116}{{\tt hep-th/0305116}}.

\bibitem{Hofman:2006xt}
D.~M. Hofman and J.~M. Maldacena, ``{Giant magnons},'' {\em J. Phys.} {\bf A39}
  (2006) 13095--13118, \href{http://xxx.lanl.gov/abs/hep-th/0604135}{{\tt
  hep-th/0604135}}.

\bibitem{Arutyunov:2006yd}
G.~Arutyunov, S.~Frolov, and M.~Zamaklar, ``{The Zamolodchikov-Faddeev algebra
  for AdS(5) x S**5 superstring},'' {\em JHEP} {\bf 04} (2007) 002,
  \href{http://xxx.lanl.gov/abs/hep-th/0612229}{{\tt hep-th/0612229}}.

\bibitem{Janik:2006dc}
R.~A. Janik, ``{The AdS$_5$xS$^5$ superstring worldsheet S-matrix and crossing
  symmetry},'' {\em Phys. Rev.} {\bf D73} (2006) 086006,
  \href{http://xxx.lanl.gov/abs/hep-th/0603038}{{\tt hep-th/0603038}}.

\bibitem{Arutyunov:2004vx}
G.~Arutyunov, S.~Frolov, and M.~Staudacher, ``{Bethe ansatz for quantum
  strings},'' {\em JHEP} {\bf 10} (2004) 016,
  \href{http://xxx.lanl.gov/abs/hep-th/0406256}{{\tt hep-th/0406256}}.

\bibitem{Hernandez:2006tk}
R.~Hernandez and E.~Lopez, ``{Quantum corrections to the string Bethe
  ansatz},'' {\em JHEP} {\bf 07} (2006) 004,
  \href{http://xxx.lanl.gov/abs/hep-th/0603204}{{\tt hep-th/0603204}}.

\bibitem{Kazakov:2004qf}
V.~A. Kazakov, A.~Marshakov, J.~A. Minahan, and K.~Zarembo, ``{Classical /
  quantum integrability in AdS/CFT},'' {\em JHEP} {\bf 05} (2004) 024,
  \href{http://xxx.lanl.gov/abs/hep-th/0402207}{{\tt hep-th/0402207}}.

\bibitem{Beisert:2006ib}
N.~Beisert, R.~Hernandez, and E.~Lopez, ``{A crossing-symmetric phase for
  AdS(5) x S$^5$ strings},'' {\em JHEP} {\bf 11} (2006) 070,
  \href{http://xxx.lanl.gov/abs/hep-th/0609044}{{\tt hep-th/0609044}}.

\bibitem{Beisert:2006ez}
N.~Beisert, B.~Eden, and M.~Staudacher, ``{Transcendentality and crossing},''
  {\em J. Stat. Mech.} {\bf 0701} (2007) P021,
  \href{http://xxx.lanl.gov/abs/hep-th/0610251}{{\tt hep-th/0610251}}.

\bibitem{Eden:2006rx}
B.~Eden and M.~Staudacher, ``Integrability and transcendentality,'' {\em J.
  Stat. Mech.} {\bf 0611} (2006) P014,
  \href{http://xxx.lanl.gov/abs/hep-th/0603157}{{\tt hep-th/0603157}}.

\bibitem{Bern:2006ew}
Z.~Bern, M.~Czakon, L.~J. Dixon, D.~A. Kosower, and V.~A. Smirnov, ``{The
  Four-Loop Planar Amplitude and Cusp Anomalous Dimension in Maximally
  Supersymmetric Yang-Mills Theory},'' {\em Phys. Rev.} {\bf D75} (2007)
  085010, \href{http://xxx.lanl.gov/abs/hep-th/0610248}{{\tt hep-th/0610248}}.

\bibitem{Freyhult:2007pz}
L.~Freyhult, A.~Rej, and M.~Staudacher, ``{A Generalized Scaling Function for
  AdS/CFT},'' {\em J. Stat. Mech.} {\bf 0807} (2008) P07015,
  \href{http://xxx.lanl.gov/abs/0712.2743}{{\tt 0712.2743}}.

\bibitem{Bombardelli:2008ah}
D.~Bombardelli, D.~Fioravanti, and M.~Rossi, ``{Large spin corrections in
  ${\cal N}=4$ SYM sl(2): still a linear integral equation},'' {\em Nucl.
  Phys.} {\bf B810} (2009) 460--490,
  \href{http://xxx.lanl.gov/abs/0802.0027}{{\tt 0802.0027}}.

\bibitem{Kotikov:2006ts}
A.~V. Kotikov and L.~N. Lipatov, ``{On the highest transcendentality in N = 4
  SUSY},'' {\em Nucl. Phys.} {\bf B769} (2007) 217--255,
  \href{http://xxx.lanl.gov/abs/hep-th/0611204}{{\tt hep-th/0611204}}.

\bibitem{Alday:2007qf}
L.~F. Alday, G.~Arutyunov, M.~K. Benna, B.~Eden, and I.~R. Klebanov, ``{On the
  strong coupling scaling dimension of high spin operators},'' {\em JHEP} {\bf
  04} (2007) 082, \href{http://xxx.lanl.gov/abs/hep-th/0702028}{{\tt
  hep-th/0702028}}.

\bibitem{Beccaria:2007tk}
M.~Beccaria, G.~F. De~Angelis, and V.~Forini, ``{The scaling function at strong
  coupling from the quantum string Bethe equations},'' {\em JHEP} {\bf 04}
  (2007) 066, \href{http://xxx.lanl.gov/abs/hep-th/0703131}{{\tt
  hep-th/0703131}}.

\bibitem{Benna:2006nd}
M.~K. Benna, S.~Benvenuti, I.~R. Klebanov, and A.~Scardicchio, ``{A test of the
  AdS/CFT correspondence using high-spin operators},'' {\em Phys. Rev. Lett.}
  {\bf 98} (2007) 131603, \href{http://xxx.lanl.gov/abs/hep-th/0611135}{{\tt
  hep-th/0611135}}.

\bibitem{Casteill:2007ct}
P.~Y. Casteill and C.~Kristjansen, ``{The Strong Coupling Limit of the Scaling
  Function from the Quantum String Bethe Ansatz},'' {\em Nucl. Phys.} {\bf
  B785} (2007) 1--18, \href{http://xxx.lanl.gov/abs/0705.0890}{{\tt
  0705.0890}}.

\bibitem{Belitsky:2007kf}
A.~V. Belitsky, ``{Strong coupling expansion of Baxter equation in N=4 SYM},''
  {\em Phys. Lett.} {\bf B659} (2008) 732--740,
  \href{http://xxx.lanl.gov/abs/0710.2294}{{\tt 0710.2294}}.

\bibitem{Basso:2007wd}
B.~Basso, G.~P. Korchemsky, and J.~Kotanski, ``{Cusp anomalous dimension in
  maximally supersymmetric Yang- Mills theory at strong coupling},'' {\em Phys.
  Rev. Lett.} {\bf 100} (2008) 091601,
  \href{http://xxx.lanl.gov/abs/0708.3933}{{\tt 0708.3933}}.

\bibitem{Basso:2009gh}
B.~Basso and G.~P. Korchemsky, ``{Nonperturbative scales in AdS/CFT},'' {\em J.
  Phys.} {\bf A42} (2009) 254005, \href{http://xxx.lanl.gov/abs/0901.4945}{{\tt
  0901.4945}}.

\bibitem{Basso:2008tx}
B.~Basso and G.~P. Korchemsky, ``{Embedding nonlinear O(6) sigma model into N=4
  super-Yang- Mills theory},'' {\em Nucl. Phys.} {\bf B807} (2009) 397--423,
  \href{http://xxx.lanl.gov/abs/0805.4194}{{\tt 0805.4194}}.

\bibitem{Fioravanti:2008ak}
D.~Fioravanti, P.~Grinza, and M.~Rossi, ``{The generalised scaling function: a
  note},'' \href{http://xxx.lanl.gov/abs/0805.4407}{{\tt 0805.4407}}.

\bibitem{Fioravanti:2008bh}
D.~Fioravanti, P.~Grinza, and M.~Rossi, ``{The generalised scaling function: a
  systematic study},'' \href{http://xxx.lanl.gov/abs/0808.1886}{{\tt
  0808.1886}}.

\bibitem{Bajnok:2008it}
Z.~Bajnok, J.~Balog, B.~Basso, G.~P. Korchemsky, and L.~Palla, ``{Scaling
  function in AdS/CFT from the O(6) sigma model},'' {\em Nucl. Phys.} {\bf
  B811} (2009) 438--462, \href{http://xxx.lanl.gov/abs/0809.4952}{{\tt
  0809.4952}}.

\bibitem{Hasenfratz:1990ab}
P.~Hasenfratz and F.~Niedermayer, ``{The Exact mass gap of the O(N) sigma model
  for arbitrary N $\geq$ 3 in d = 2},'' {\em Phys. Lett.} {\bf B245} (1990)
  529--532.

\bibitem{Hasenfratz:1990zz}
P.~Hasenfratz, M.~Maggiore, and F.~Niedermayer, ``{The Exact mass gap of the
  O(3) and O(4) nonlinear sigma models in d = 2},'' {\em Phys. Lett.} {\bf
  B245} (1990) 522--528.

\bibitem{Kazakov:2007fy}
V.~Kazakov, A.~Sorin, and A.~Zabrodin, ``{Supersymmetric Bethe ansatz and
  Baxter equations from discrete Hirota dynamics},'' {\em Nucl. Phys.} {\bf
  B790} (2008) 345--413, \href{http://xxx.lanl.gov/abs/hep-th/0703147}{{\tt
  hep-th/0703147}}.

\bibitem{Ambjorn:2005wa}
J.~Ambjorn, R.~A. Janik, and C.~Kristjansen, ``{Wrapping interactions and a new
  source of corrections to the spin-chain / string duality},'' {\em Nucl.
  Phys.} {\bf B736} (2006) 288--301,
  \href{http://xxx.lanl.gov/abs/hep-th/0510171}{{\tt hep-th/0510171}}.

\bibitem{Arutyunov:2007tc}
G.~Arutyunov and S.~Frolov, ``{On String S-matrix, Bound States and TBA},''
  {\em JHEP} {\bf 12} (2007) 024, \href{http://xxx.lanl.gov/abs/0710.1568}{{\tt
  0710.1568}}.

\bibitem{Arutyunov:2009zu}
G.~Arutyunov and S.~Frolov, ``{String hypothesis for the AdS$_5$xS$^5$
  mirror},'' {\em JHEP} {\bf 03} (2009) 152,
  \href{http://xxx.lanl.gov/abs/0901.1417}{{\tt 0901.1417}}.

\bibitem{Gromov:2009tv}
N.~Gromov, V.~Kazakov, and P.~Vieira, ``{Integrability for the Full Spectrum of
  Planar AdS/CFT},'' \href{http://xxx.lanl.gov/abs/0901.3753}{{\tt 0901.3753}}.

\bibitem{Arutyunov:2009ur}
G.~Arutyunov and S.~Frolov, ``{Thermodynamic Bethe Ansatz for the AdS$_5$xS$^5$
  Mirror Model},'' {\em JHEP} {\bf 05} (2009) 068,
  \href{http://xxx.lanl.gov/abs/0903.0141}{{\tt 0903.0141}}.

\bibitem{Bomb}
D.~Bombardelli, D.~Fioravanti, and R.~Tateo, ``{Thermodynamic Bethe Ansatz for
  planar AdS/CFT: a proposal},'' \href{http://xxx.lanl.gov/abs/0902.3930}{{\tt
  0902.3930}}.

\bibitem{GromovKKV}
N.~Gromov, V.~Kazakov, A.~Kozak, and P.~Vieira, ``{Integrability for the Full
  Spectrum of Planar AdS/CFT II},''
  \href{http://xxx.lanl.gov/abs/0902.4458}{{\tt 0902.4458}}.

\bibitem{Gromov:2008en}
N.~Gromov, ``{Generalized Scaling Function at Strong Coupling},'' {\em JHEP}
  {\bf 11} (2008) 085, \href{http://xxx.lanl.gov/abs/0805.4615}{{\tt
  0805.4615}}.

\bibitem{Alday2007}
L.~F. Alday and J.~M. Maldacena, ``{Comments on operators with large spin},''
  {\em JHEP} {\bf 11} (2007) 019, \href{http://xxx.lanl.gov/abs/0708.0672}{{\tt
  0708.0672}}.

\bibitem{Gaudin}
M.~Gaudin, {\em La fonction d'onde de Bethe}.
\newblock Masson, 1983.

\bibitem{Faddeev:1996iy}
L.~D. Faddeev, ``{How Algebraic Bethe Ansatz works for integrable model},''
  \href{http://xxx.lanl.gov/abs/hep-th/9605187}{{\tt hep-th/9605187}}.

\bibitem{IntroBethe1}
M.~Karbach and G.~Muller, ``{Introduction to the Bethe ansatz I},'' {\em
  Computers in Physics} {\bf 11} (1997) 36--43,
  \href{http://xxx.lanl.gov/abs/cond-mat/9809163}{{\tt cond-mat/9809163}}.

\bibitem{IntroBethe2}
M.~Karbach, K.~Hu, and G.~Muller, ``{Introduction to the Bethe ansatz II},''
  {\em Computers in Physics} {\bf 12} (1998) 563--573,
  \href{http://xxx.lanl.gov/abs/cond-mat/9809162}{{\tt cond-mat/9809162}}.

\bibitem{IntroBethe3}
M.~Karbach, K.~Hu, and G.~Muller, ``{Introduction to the Bethe ansatz III},''
  \href{http://xxx.lanl.gov/abs/cond-mat/0008018}{{\tt cond-mat/0008018}}.

\bibitem{PZJ}
P.~Zinn-Justin, ``{Quelques applications de l'Ansatz de Bethe},''
  \href{http://xxx.lanl.gov/abs/solv-int/9810007}{{\tt solv-int/9810007}}.

\bibitem{CompletenessSU2}
A.~Kirillov, ``Combinatorial identities, and completeness of eigenstates of the
  heisenberg magnet,'' {\em Journal of Mathematical Sciences} {\bf 30} (August,
  1985) 2298--2310.

\bibitem{Faddeev:1979gh}
L.~D. Faddeev, E.~K. Sklyanin, and L.~A. Takhtajan, ``{The Quantum Inverse
  Problem Method. 1},'' {\em Theor. Math. Phys.} {\bf 40} (1980) 688--706.

\bibitem{Faddeev:1981ft}
L.~D. Faddeev and L.~A. Takhtajan, ``{Spectrum and scattering of excitations in
  the one- dimensional isotropic Heisenberg model},'' {\em J. Sov. Math.} {\bf
  24} (1984) 241--267.

\bibitem{Kulish:1983rd}
P.~P. Kulish and N.~Y. Reshetikhin, ``{Diagonalization of Gl(N) invariant
  transfer matrices and quantum N wave system (Lee model)},'' {\em J. Phys.}
  {\bf A16} (1983) L591--L596.

\bibitem{BetheCompleteness}
A.~Kirrilov, ``Completeness of states of the generalised heisenberg magnet,''
  {\em Journal of Mathematical Sciences} {\bf 36} (January, 1987) 115--128.

\bibitem{Ogievetsky:1986hu}
E.~Ogievetsky and P.~Wiegmann, ``{Factorized S matrix and the Bethe Ansatz for
  simple Lie groups},'' {\em Phys. Lett.} {\bf B168} (1986) 360.

\bibitem{Isler:1993fc}
K.~Isler and M.~B. Paranjape, ``{Violations of the string hypothesis in the
  solutions of the Bethe ansatz equations in the XXX Heisenberg model},'' {\em
  Phys. Lett.} {\bf B319} (1993) 209--214,
  \href{http://xxx.lanl.gov/abs/hep-th/9304078}{{\tt hep-th/9304078}}.

\bibitem{Ilakovac:1999pe}
A.~Ilakovac, M.~Kolanovic, S.~Pallua, and P.~Prester, ``{Violation of the
  string hypothesis and Heisenberg XXZ spin chain},'' {\em Phys. Rev.} {\bf
  B60} (1999) 7271, \href{http://xxx.lanl.gov/abs/hep-th/9907103}{{\tt
  hep-th/9907103}}.

\bibitem{Antipov2006}
A.~Antipov and K.~I.V., ``The isotropic heisenberg chain of arbitrary spin by
  direct solution of the baxter equation,'' {\em Physica D.} {\bf 221} (2006)
  101--109.

\bibitem{Bargheer:2008kj}
T.~Bargheer, N.~Beisert, and N.~Gromov, ``{Quantum Stability for the Heisenberg
  Ferromagnet},'' {\em New J. Phys.} {\bf 10} (2008) 103023,
  \href{http://xxx.lanl.gov/abs/0804.0324}{{\tt 0804.0324}}.

\bibitem{Zabrodin:1996vm}
A.~Zabrodin, ``{Discrete Hirota's equation in quantum integrable models},''
  \href{http://xxx.lanl.gov/abs/hep-th/9610039}{{\tt hep-th/9610039}}.

\bibitem{Gross:1974jv}
D.~J. Gross and A.~Neveu, ``{Dynamical Symmetry Breaking in Asymptotically Free
  Field Theories},'' {\em Phys. Rev.} {\bf D10} (1974) 3235.

\bibitem{Polyakov:1983tt}
A.~M. Polyakov and P.~B. Wiegmann, ``{Theory of nonabelian Goldstone bosons in
  two dimensions},'' {\em Phys. Lett.} {\bf B131} (1983) 121--126.

\bibitem{Polyakov:1984et}
A.~M. Polyakov and P.~B. Wiegmann, ``{Goldstone Fields in Two-Dimensions with
  Multivalued Actions},'' {\em Phys. Lett.} {\bf B141} (1984) 223--228.

\bibitem{Fateev:1994ai}
V.~A. Fateev, V.~A. Kazakov, and P.~B. Wiegmann, ``{Principal chiral field at
  large N},'' {\em Nucl. Phys.} {\bf B424} (1994) 505--520,
  \href{http://xxx.lanl.gov/abs/hep-th/9403099}{{\tt hep-th/9403099}}.

\bibitem{Bardeen:1976zh}
W.~A. Bardeen, B.~W. Lee, and R.~E. Shrock, ``{Phase Transition in the
  Nonlinear Sigma Model in Two + Epsilon Dimensional Continuum},'' {\em Phys.
  Rev.} {\bf D14} (1976) 985.

\bibitem{Brezin:1976qa}
E.~Brezin and J.~Zinn-Justin, ``{Spontaneous Breakdown of Continuous Symmetries
  Near Two- Dimensions},'' {\em Phys. Rev.} {\bf B14} (1976) 3110.

\bibitem{Forgacs:1991nk}
P.~Forgacs, S.~Naik, and F.~Niedermayer, ``{The Exact mass gap of the chiral
  Gross-Neveu model},'' {\em Phys. Lett.} {\bf B283} (1992) 282--286.

\bibitem{Iagolnitzer:1977sw}
D.~Iagolnitzer, ``{Factorization of the Multiparticle s Matrix in Two-
  Dimensional Space-Time Models},'' {\em Phys. Rev.} {\bf D18} (1978) 1275.

\bibitem{Chew}
G.~Chew, {\em The analytic S matrix (a basis for nuclear democracy)}.
\newblock Benjamin, New York, 1966.

\bibitem{Faddeev:1981ip}
L.~D. Faddeev and L.~A. Takhtajan, ``{What is the spin of a spin wave?},'' {\em
  Phys. Lett.} {\bf A85} (1981) 375--377.

\bibitem{Takhtajan:1982zz}
L.~Takhtajan, ``{The picture of low-lying excitations in the isotropic
  Heisenberg chain of arbitrary spins},'' {\em Phys. Lett.} {\bf A87} (1982)
  479--482.

\bibitem{Faddeev:1984ft}
L.~D. Faddeev and L.~A. Takhtajan, ``{Spectrum and scattering of excitations in
  the one- dimensional isotropic Heisenberg model},'' {\em J. Sov. Math.} {\bf
  24} (1984) 241--267.

\bibitem{Korepin:1979qq}
V.~E. Korepin, ``{Direct calculation of the S matrix in the massive Thirring
  model},'' {\em Theor. Math. Phys.} {\bf 41} (1979) 953--967.

\bibitem{Ogievetsky:1987vv}
E.~Ogievetsky, P.~Wiegmann, and N.~Reshetikhin, ``{The Principal Chiral Field
  in Two-Dimensions on Classical Lie Algebras: The Bethe Ansatz Solution and
  Factorized Theory of Scattering},'' {\em Nucl. Phys.} {\bf B280} (1987)
  45--96.

\bibitem{Faddeev:1985qu}
L.~D. Faddeev and N.~Y. Reshetikhin, ``{Integrability of the Principal Chiral
  Field model in (1+1)-dimension},'' {\em Ann. Phys.} {\bf 167} (1986) 227.

\bibitem{Yang:1968rm}
C.-N. Yang and C.~P. Yang, ``{Thermodynamics of a one-dimensional system of
  bosons with repulsive delta-function interaction},'' {\em J. Math. Phys.}
  {\bf 10} (1969) 1115--1122.

\bibitem{Zamolodchikov:1989cf}
A.~B. Zamolodchikov, ``{Thermodynamic Bethe Ansatz in relativistic models.
  Scaling three state Potts and Lee-Yang models},'' {\em Nucl. Phys.} {\bf
  B342} (1990) 695--720.

\bibitem{SKR}
F.~Spill, P.~Koroteev, and A.~Rej, ``Classification of yangians of lie
  superalgebras and their r-matrices,'' June, 2009.
\newblock presented at "Integrability in Gauge and String Theory" conference,
  June.

\bibitem{Bazhanov:1996aq}
V.~V. Bazhanov, S.~L. Lukyanov, and A.~B. Zamolodchikov, ``{Quantum field
  theories in finite volume: Excited state energies},'' {\em Nucl. Phys.} {\bf
  B489} (1997) 487--531, \href{http://xxx.lanl.gov/abs/hep-th/9607099}{{\tt
  hep-th/9607099}}.

\bibitem{Dorey:1996re}
P.~Dorey and R.~Tateo, ``{Excited states by analytic continuation of TBA
  equations},'' {\em Nucl. Phys.} {\bf B482} (1996) 639--659,
  \href{http://xxx.lanl.gov/abs/hep-th/9607167}{{\tt hep-th/9607167}}.

\bibitem{Gromov:2008gj}
N.~Gromov, V.~Kazakov, and P.~Vieira, ``{Finite Volume Spectrum of 2D Field
  Theories from Hirota Dynamics},''
  \href{http://xxx.lanl.gov/abs/0812.5091}{{\tt 0812.5091}}.

\bibitem{Kac:1977em}
V.~G. Kac, ``{Lie Superalgebras},'' {\em Adv. Math.} {\bf 26} (1977) 8--96.

\bibitem{Kac:1977qb}
V.~G. Kac, ``{A Sketch of Lie Superalgebra Theory},'' {\em Commun. Math. Phys.}
  {\bf 53} (1977) 31--64.

\bibitem{Balantekin:1980qy}
A.~Baha~Balantekin and I.~Bars, ``{Dimension and Character Formulas for Lie
  Supergroups},'' {\em J. Math. Phys.} {\bf 22} (1981) 1149.

\bibitem{Balantekin:1980pp}
A.~Baha~Balantekin and I.~Bars, ``{Representations of Supergroups},'' {\em J.
  Math. Phys.} {\bf 22} (1981) 1810.

\bibitem{Bars:1982se}
I.~Bars, B.~Morel, and H.~Ruegg, ``{Kac-Dynkin diagrams and supertableaux},''
  {\em J. Math. Phys.} {\bf 24} (1983) 2253.

\bibitem{Bazhanov:1989yk}
V.~Bazhanov and N.~Reshetikhin, ``{Restricted solid on solid models connected
  with simply based algebras and conformal field theory},'' {\em J. Phys.} {\bf
  A23} (1990) 1477.

\bibitem{Tsuboi:1997iq}
Z.~Tsuboi, ``{Analytic Bethe ansatz and functional equations for Lie
  superalgebra sl(r+1|s+1)},'' {\em J. Phys.} {\bf A30} (1997) 7975--7991.

\bibitem{Tsuboi:1998ne}
Z.~Tsuboi, ``{Analytic Bethe Ansatz And Functional Equations Associated With
  Any Simple Root Systems Of The Lie Superalgebra SL(r+1|s+1)},'' {\em Physica}
  {\bf A252} (1998) 565--585.

\bibitem{Kazakov:2007na}
V.~Kazakov and P.~Vieira, ``{From Characters to Quantum (Super)Spin Chains via
  Fusion},'' {\em JHEP} (2008) 050,
  \href{http://xxx.lanl.gov/abs/0711.2470}{{\tt 0711.2470}}.

\bibitem{Ragoucy:2007kg}
E.~Ragoucy and G.~Satta, ``{Analytical Bethe Ansatz for closed and open
  $gl(M|N)$ super-spin chains in arbitrary representations and for any Dynkin
  diagram},'' {\em JHEP} {\bf 09} (2007) 001,
  \href{http://xxx.lanl.gov/abs/0706.3327}{{\tt 0706.3327}}.

\bibitem{Scheunert:1976wj}
M.~Scheunert, W.~Nahm, and V.~Rittenberg, ``{Irreducible Representations of the
  OSP(2,1) and SPL(2,1) Graded Lie Algebras},'' {\em J. Math. Phys.} {\bf 18}
  (1977) 155.

\bibitem{Marcu:1979se}
M.~Marcu, ``{The representations of spl(2,1): an example of representations of
  basic superalgebras},'' {\em J. Math. Phys.} {\bf 21} (1980) 1277.

\bibitem{Gotz:2005jz}
G.~Gotz, T.~Quella, and V.~Schomerus, ``{Representation theory of $sl(2|1)$},''
  {\em J. Algebra} {\bf 312} (2007) 829--848,
  \href{http://xxx.lanl.gov/abs/hep-th/0504234}{{\tt hep-th/0504234}}.

\bibitem{Essler:2005ag}
F.~H.~L. Essler, H.~Frahm, and H.~Saleur, ``{Continuum Limit of the Integrable
  $sl(2|1)$ 3-$\bar{3}$ Superspin Chain},'' {\em Nucl. Phys.} {\bf B712} (2005)
  513--572, \href{http://xxx.lanl.gov/abs/cond-mat/0501197}{{\tt
  cond-mat/0501197}}.

\bibitem{Saleur:2006tf}
H.~Saleur and V.~Schomerus, ``{On the $SU(2|1)$ WZNW model and its statistical
  mechanics applications},'' {\em Nucl. Phys.} {\bf B775} (2007) 312--340,
  \href{http://xxx.lanl.gov/abs/hep-th/0611147}{{\tt hep-th/0611147}}.

\bibitem{Takahashi:1972}
M.~Takahashi, ``{One-Dimensional Hubbard model at finite temperature},'' {\em
  Prog. Theor. Phys.} {\bf 47} (1972) 69.

\bibitem{Saleur:1999cx}
H.~Saleur, ``{The continuum limit of $sl(N|K)$ integrable super spin chains},''
  {\em Nucl. Phys.} {\bf B578} (2000) 552--576,
  \href{http://xxx.lanl.gov/abs/solv-int/9905007}{{\tt solv-int/9905007}}.

\bibitem{Woynarovich}
F.~Woynarovich, ``{Low-energy excited states in a Hubbard chain with on-site
  attraction},'' {\em J. Phys. C: Solid State Phys} {\bf 16} (1983) 6593.

\bibitem{tJmodel}
P.-A. Bares, J.~M.~P. Carmelo, J.~Ferrer, and P.~Horsch, ``{Charge-spin
  recombination in the one-dimensional supersymmetric t-J model},'' {\em Phys.
  Rev. B} {\bf 46} (1992) 14624–14654.

\bibitem{GohmannSeel}
F.~Gohmann and A.~Seel, ``{A note on the Bethe ansatz solution of the
  supersymmetric t-J model},'' \href{http://xxx.lanl.gov/abs/0309138}{{\tt
  0309138}}.

\bibitem{Hegedus:2009ky}
A.~Hegedus, ``{Discrete Hirota dynamics for AdS/CFT},''
  \href{http://xxx.lanl.gov/abs/0906.2546}{{\tt 0906.2546}}.

\bibitem{Klebanov:2000me}
I.~R. Klebanov, ``{TASI lectures: Introduction to the AdS/CFT
  correspondence},'' \href{http://xxx.lanl.gov/abs/hep-th/0009139}{{\tt
  hep-th/0009139}}.

\bibitem{Maldacena:2003nj}
J.~M. Maldacena, ``{Lectures on AdS/CFT},''
  \href{http://xxx.lanl.gov/abs/hep-th/0309246}{{\tt hep-th/0309246}}.

\bibitem{Aharony:1999ti}
O.~Aharony, S.~S. Gubser, J.~M. Maldacena, H.~Ooguri, and Y.~Oz, ``{Large N
  field theories, string theory and gravity},'' {\em Phys. Rept.} {\bf 323}
  (2000) 183--386, \href{http://xxx.lanl.gov/abs/hep-th/9905111}{{\tt
  hep-th/9905111}}.

\bibitem{D'Hoker:2002aw}
E.~D'Hoker and D.~Z. Freedman, ``{Supersymmetric gauge theories and the AdS/CFT
  correspondence},'' \href{http://xxx.lanl.gov/abs/hep-th/0201253}{{\tt
  hep-th/0201253}}.

\bibitem{Beisert:2004ry}
N.~Beisert, ``{The dilatation operator of N = 4 super Yang-Mills theory and
  integrability},'' {\em Phys. Rept.} {\bf 405} (2005) 1--202,
  \href{http://xxx.lanl.gov/abs/hep-th/0407277}{{\tt hep-th/0407277}}.

\bibitem{Rej:2009je}
A.~Rej, ``{Integrability and the AdS/CFT correspondence},'' {\em J. Phys.} {\bf
  A42} (2009) 254002, \href{http://xxx.lanl.gov/abs/0907.3468}{{\tt
  0907.3468}}.

\bibitem{SerbanMemoire}
D.~Serban, ``{Integrability and the AdS/CFT correspondence},''
  \href{http://xxx.lanl.gov/abs/1003.4214}{{\tt 1003.4214}}.

\bibitem{Tseytlin:2003ii}
A.~A. Tseytlin, ``{Spinning strings and AdS/CFT duality},''
  \href{http://xxx.lanl.gov/abs/hep-th/0311139}{{\tt hep-th/0311139}}.

\bibitem{Vicedo:2008jk}
B.~Vicedo, ``{Finite-g Strings},''
  \href{http://xxx.lanl.gov/abs/0810.3402}{{\tt 0810.3402}}.

\bibitem{GromovThese}
N.~Gromov, ``{Integrability in AdS/CFT correspondence: quasi-classical
  analysis},'' {\em J. Phys.} {\bf A42} (2009) 254004.

\bibitem{Arutyunov:2009ga}
G.~Arutyunov and S.~Frolov, ``{Foundations of the AdS$_5$xS$^5$ Superstring.
  Part I},'' {\em J. Phys.} {\bf A42} (2009) 254003,
  \href{http://xxx.lanl.gov/abs/0901.4937}{{\tt 0901.4937}}.

\bibitem{Okamura:2008jm}
K.~Okamura, ``{Aspects of Integrability in AdS/CFT Duality},''
  \href{http://xxx.lanl.gov/abs/0803.3999}{{\tt 0803.3999}}.

\bibitem{VieiraThese}
P.~Vieira, ``{Integrability in AdS/CFT},'' {\em PhD thesis} (2008).

\bibitem{DoreyReview}
N.~Dorey, ``Notes on integrability in gauge theory and string theory,'' {\em J.
  Phys.} {\bf A42} (2009) 254001.

\bibitem{Alday:2008yw}
L.~F. Alday and R.~Roiban, ``{Scattering Amplitudes, Wilson Loops and the
  String/Gauge Theory Correspondence},'' {\em Phys. Rept.} {\bf 468} (2008)
  153--211, \href{http://xxx.lanl.gov/abs/0807.1889}{{\tt 0807.1889}}.

\bibitem{Alday:2008zz}
L.~F. Alday and R.~Roiban, ``{Scattering amplitudes at weak and strong coupling
  in N=4 super-Yang-Mills theory},'' {\em Acta Phys. Polon.} {\bf B39} (2008)
  2979--3046.

\bibitem{tHooft:1973jz}
G.~'t~Hooft, ``{A planar diagram theory for strong interactions},'' {\em Nucl.
  Phys.} {\bf B72} (1974) 461.

\bibitem{Kazakov:1985ds}
V.~A. Kazakov, ``{Bilocal Regularization of Models of Random Surfaces},'' {\em
  Phys. Lett.} {\bf B150} (1985) 282--284.

\bibitem{David:1985nj}
F.~David, ``{A Model of Random Surfaces with Nontrivial Critical Behavior},''
  {\em Nucl. Phys.} {\bf B257} (1985) 543.

\bibitem{Ambjorn:1985az}
J.~Ambjorn, B.~Durhuus, and J.~Frohlich, ``{Diseases of Triangulated Random
  Surface Models, and Possible Cures},'' {\em Nucl. Phys.} {\bf B257} (1985)
  433.

\bibitem{Kazakov:1985ea}
V.~A. Kazakov, A.~A. Migdal, and I.~K. Kostov, ``{Critical Properties of
  Randomly Triangulated Planar Random Surfaces},'' {\em Phys. Lett.} {\bf B157}
  (1985) 295--300.

\bibitem{DiFrancesco:2004qj}
P.~Di~Francesco, ``{2D quantum gravity, matrix models and graph
  combinatorics},'' \href{http://xxx.lanl.gov/abs/math-ph/0406013}{{\tt
  math-ph/0406013}}.

\bibitem{Gross:1992tu}
D.~J. Gross, ``{Two-dimensional QCD as a string theory},'' {\em Nucl. Phys.}
  {\bf B400} (1993) 161--180,
  \href{http://xxx.lanl.gov/abs/hep-th/9212149}{{\tt hep-th/9212149}}.

\bibitem{Gross:1993hu}
D.~J. Gross and W.~Taylor, ``{Two-dimensional QCD is a string theory},'' {\em
  Nucl. Phys.} {\bf B400} (1993) 181--210,
  \href{http://xxx.lanl.gov/abs/hep-th/9301068}{{\tt hep-th/9301068}}.

\bibitem{Minahan:1992sk}
J.~A. Minahan, ``{Summing over inequivalent maps in the string theory
  interpretation of two-dimensional QCD},'' {\em Phys. Rev.} {\bf D47} (1993)
  3430--3436, \href{http://xxx.lanl.gov/abs/hep-th/9301003}{{\tt
  hep-th/9301003}}.

\bibitem{Polchinski:1996na}
J.~Polchinski, ``{TASI Lectures on D-branes},''
  \href{http://xxx.lanl.gov/abs/hep-th/9611050}{{\tt hep-th/9611050}}.

\bibitem{Witten:1995im}
E.~Witten, ``{Bound states of strings and p-branes},'' {\em Nucl. Phys.} {\bf
  B460} (1996) 335--350, \href{http://xxx.lanl.gov/abs/hep-th/9510135}{{\tt
  hep-th/9510135}}.

\bibitem{Aharony:2008ug}
O.~Aharony, O.~Bergman, D.~L. Jafferis, and J.~Maldacena, ``{N=6 superconformal
  Chern-Simons-matter theories, M2-branes and their gravity duals},'' {\em
  JHEP} {\bf 10} (2008) 091, \href{http://xxx.lanl.gov/abs/0806.1218}{{\tt
  0806.1218}}.

\bibitem{Erlich:2009me}
J.~Erlich, ``{How Well Does AdS/QCD Describe QCD?},''
  \href{http://xxx.lanl.gov/abs/0908.0312}{{\tt 0908.0312}}.

\bibitem{Brink:1982wv}
L.~Brink, O.~Lindgren, and B.~E.~W. Nilsson, ``{The Ultraviolet Finiteness of
  the N=4 Yang-Mills Theory},'' {\em Phys. Lett.} {\bf B123} (1983) 323.

\bibitem{Novikov:1983uc}
V.~A. Novikov, M.~A. Shifman, A.~I. Vainshtein, and V.~I. Zakharov, ``{Exact
  Gell-Mann-Low Function of Supersymmetric Yang-Mills Theories from Instanton
  Calculus},'' {\em Nucl. Phys.} {\bf B229} (1983) 381.

\bibitem{Howe:1983sr}
P.~S. Howe, K.~S. Stelle, and P.~K. Townsend, ``{Miraculous Ultraviolet
  Cancellations in Supersymmetry Made Manifest},'' {\em Nucl. Phys.} {\bf B236}
  (1984) 125.

\bibitem{Dobrev:1985qv}
V.~K. Dobrev and V.~B. Petkova, ``{All Positive Energy Unitary Irreducible
  Representations of Extended Conformal Supersymmetry},'' {\em Phys. Lett.}
  {\bf B162} (1985) 127--132.

\bibitem{Dobrev:1985vh}
V.~K. Dobrev and V.~B. Petkova, ``{On the group theoretical approach to
  extended conformal supersymmetry: classification of multiplets},'' {\em Lett.
  Math. Phys.} {\bf 9} (1985) 287--298.

\bibitem{Lipatov:1993yb}
L.~N. Lipatov, ``{High-energy asymptotics of multicolor QCD and exactly
  solvable lattice models},''
  \href{http://xxx.lanl.gov/abs/hep-th/9311037}{{\tt hep-th/9311037}}.

\bibitem{Faddeev:1994zg}
L.~D. Faddeev and G.~P. Korchemsky, ``{High-energy QCD as a completely
  integrable model},'' {\em Phys. Lett.} {\bf B342} (1995) 311--322,
  \href{http://xxx.lanl.gov/abs/hep-th/9404173}{{\tt hep-th/9404173}}.

\bibitem{Braun:1998id}
V.~M. Braun, S.~E. Derkachov, and A.~N. Manashov, ``{Integrability of
  three-particle evolution equations in {QCD}},'' {\em Phys. Rev. Lett.} {\bf
  81} (1998) 2020--2023, \href{http://xxx.lanl.gov/abs/hep-ph/9805225}{{\tt
  hep-ph/9805225}}.

\bibitem{Braun:1999te}
V.~M. Braun, S.~E. Derkachov, G.~P. Korchemsky, and A.~N. Manashov, ``{Baryon
  distribution amplitudes in {QCD}},'' {\em Nucl. Phys.} {\bf B553} (1999)
  355--426, \href{http://xxx.lanl.gov/abs/hep-ph/9902375}{{\tt
  hep-ph/9902375}}.

\bibitem{Belitsky:1999ru}
A.~V. Belitsky, ``{Integrability and WKB solution of twist-three evolution
  equations},'' {\em Nucl. Phys.} {\bf B558} (1999) 259--284,
  \href{http://xxx.lanl.gov/abs/hep-ph/9903512}{{\tt hep-ph/9903512}}.

\bibitem{Belitsky:1999bf}
A.~V. Belitsky, ``{Renormalization of twist-three operators and integrable
  lattice models},'' {\em Nucl. Phys.} {\bf B574} (2000) 407--447,
  \href{http://xxx.lanl.gov/abs/hep-ph/9907420}{{\tt hep-ph/9907420}}.

\bibitem{Derkachov:1999ze}
S.~E. Derkachov, G.~P. Korchemsky, and A.~N. Manashov, ``{Evolution equations
  for quark gluon distributions in multi-color QCD and open spin chains},''
  {\em Nucl. Phys.} {\bf B566} (2000) 203--251,
  \href{http://xxx.lanl.gov/abs/hep-ph/9909539}{{\tt hep-ph/9909539}}.

\bibitem{Beisert:2003jj}
N.~Beisert, ``{The complete one-loop dilatation operator of N = 4 super
  Yang-Mills theory},'' {\em Nucl. Phys.} {\bf B676} (2004) 3--42,
  \href{http://xxx.lanl.gov/abs/hep-th/0307015}{{\tt hep-th/0307015}}.

\bibitem{Beisert:2003ys}
N.~Beisert, ``{The su(2|3) dynamic spin chain},'' {\em Nucl. Phys.} {\bf B682}
  (2004) 487--520, \href{http://xxx.lanl.gov/abs/hep-th/0310252}{{\tt
  hep-th/0310252}}.

\bibitem{Berenstein:2003gb}
D.~E. Berenstein, J.~M. Maldacena, and H.~S. Nastase, ``{Strings in flat space
  and pp waves from N=4 Super Yang Mills},'' {\em AIP Conf. Proc.} {\bf 646}
  (2003) 3--14.

\bibitem{Gross:2002su}
D.~J. Gross, A.~Mikhailov, and R.~Roiban, ``{Operators with large R charge in N
  = 4 Yang-Mills theory},'' {\em Annals Phys.} {\bf 301} (2002) 31--52,
  \href{http://xxx.lanl.gov/abs/hep-th/0205066}{{\tt hep-th/0205066}}.

\bibitem{Serban:2004jf}
D.~Serban and M.~Staudacher, ``{Planar N = 4 gauge theory and the Inozemtsev
  long range spin chain},'' {\em JHEP} {\bf 06} (2004) 001,
  \href{http://xxx.lanl.gov/abs/hep-th/0401057}{{\tt hep-th/0401057}}.

\bibitem{Rej:2005qt}
A.~Rej, D.~Serban, and M.~Staudacher, ``{Planar N = 4 gauge theory and the
  Hubbard model},'' {\em JHEP} {\bf 03} (2006) 018,
  \href{http://xxx.lanl.gov/abs/hep-th/0512077}{{\tt hep-th/0512077}}.

\bibitem{Beisert:2007hz}
N.~Beisert, T.~McLoughlin, and R.~Roiban, ``{The Four-Loop Dressing Phase of
  N=4 SYM},'' {\em Phys. Rev.} {\bf D76} (2007) 046002,
  \href{http://xxx.lanl.gov/abs/0705.0321}{{\tt 0705.0321}}.

\bibitem{Bajnok:2008bm}
Z.~Bajnok and R.~A. Janik, ``{Four-loop perturbative Konishi from strings and
  finite size effects for multiparticle states},'' {\em Nucl. Phys.} {\bf B807}
  (2009) 625--650, \href{http://xxx.lanl.gov/abs/0807.0399}{{\tt 0807.0399}}.

\bibitem{Bajnok:2008qj}
Z.~Bajnok, R.~A. Janik, and T.~Lukowski, ``{Four loop twist two, BFKL, wrapping
  and strings},'' {\em Nucl. Phys.} {\bf B816} (2009) 376--398,
  \href{http://xxx.lanl.gov/abs/0811.4448}{{\tt 0811.4448}}.

\bibitem{Bajnok:2009vm}
Z.~Bajnok, A.~Hegedus, R.~A. Janik, and T.~Lukowski, ``{Five loop Konishi from
  AdS/CFT},'' \href{http://xxx.lanl.gov/abs/0906.4062}{{\tt 0906.4062}}.

\bibitem{Fiamberti:2009jw}
F.~Fiamberti, A.~Santambrogio, and C.~Sieg, ``{Five-loop anomalous dimension at
  critical wrapping order in N=4 SYM},''
  \href{http://xxx.lanl.gov/abs/0908.0234}{{\tt 0908.0234}}.

\bibitem{Lukowski:2009ce}
T.~Lukowski, A.~Rej, and V.~N. Velizhanin, ``{Five-Loop Anomalous Dimension of
  Twist-Two Operators},'' \href{http://xxx.lanl.gov/abs/0912.1624}{{\tt
  0912.1624}}.

\bibitem{Bargheer:2008jt}
T.~Bargheer, N.~Beisert, and F.~Loebbert, ``{Boosting Nearest-Neighbour to
  Long-Range Integrable Spin Chains},'' {\em J. Stat. Mech.} {\bf 0811} (2008)
  L11001, \href{http://xxx.lanl.gov/abs/0807.5081}{{\tt 0807.5081}}.

\bibitem{Bargheer:2009xy}
T.~Bargheer, N.~Beisert, and F.~Loebbert, ``{Long-Range Deformations for
  Integrable Spin Chains},'' {\em J. Phys.} {\bf A42} (2009) 285205,
  \href{http://xxx.lanl.gov/abs/0902.0956}{{\tt 0902.0956}}.

\bibitem{Kotikov:2007cy}
A.~V. Kotikov, L.~N. Lipatov, A.~Rej, M.~Staudacher, and V.~N. Velizhanin,
  ``{Dressing and Wrapping},'' {\em J. Stat. Mech.} {\bf 0710} (2007) P10003,
  \href{http://xxx.lanl.gov/abs/0704.3586}{{\tt 0704.3586}}.

\bibitem{Beisert:2005tm}
N.~Beisert, ``{The $su(2|2)$ dynamic S-matrix},'' {\em Adv. Theor. Math. Phys.}
  {\bf 12} (2008) 945, \href{http://xxx.lanl.gov/abs/hep-th/0511082}{{\tt
  hep-th/0511082}}.

\bibitem{Martins:2007hb}
M.~J. Martins and C.~S. Melo, ``{The Bethe ansatz approach for factorizable
  centrally extended S-matrices},'' {\em Nucl. Phys.} {\bf B785} (2007)
  246--262, \href{http://xxx.lanl.gov/abs/hep-th/0703086}{{\tt
  hep-th/0703086}}.

\bibitem{Beisert:2004hm}
N.~Beisert, V.~Dippel, and M.~Staudacher, ``{A novel long range spin chain and
  planar N = 4 super Yang- Mills},'' {\em JHEP} {\bf 07} (2004) 075,
  \href{http://xxx.lanl.gov/abs/hep-th/0405001}{{\tt hep-th/0405001}}.

\bibitem{Zakharov:1973pp}
V.~E. Zakharov and A.~V. Mikhailov, ``{Relativistically Invariant
  Two-Dimensional Models in Field Theory Integrable by the Inverse Problem
  Technique. (In Russian)},'' {\em Sov. Phys. JETP} {\bf 47} (1978) 1017--1027.

\bibitem{Dorey:2006mx}
N.~Dorey and B.~Vicedo, ``{A symplectic structure for string theory on
  integrable backgrounds},'' {\em JHEP} {\bf 03} (2007) 045,
  \href{http://xxx.lanl.gov/abs/hep-th/0606287}{{\tt hep-th/0606287}}.

\bibitem{Kluson:2007vw}
J.~Kluson, ``{Current Algebra and Integrability of Principal Chiral Model on
  the World-sheet with General Metric},'' {\em JHEP} {\bf 04} (2007) 040,
  \href{http://xxx.lanl.gov/abs/hep-th/0703003}{{\tt hep-th/0703003}}.

\bibitem{Kluson:2007md}
J.~Kluson, ``{Note About Integrability and Gauge Fixing for Bosonic String on
  AdS(5)xS(5)},'' {\em JHEP} {\bf 07} (2007) 015,
  \href{http://xxx.lanl.gov/abs/0705.2858}{{\tt 0705.2858}}.

\bibitem{Magro:2008dv}
M.~Magro, ``{The Classical Exchange Algebra of AdS5 x S5},'' {\em JHEP} {\bf
  01} (2009) 021, \href{http://xxx.lanl.gov/abs/0810.4136}{{\tt 0810.4136}}.

\bibitem{Gubser:2002tv}
S.~S. Gubser, I.~R. Klebanov, and A.~M. Polyakov, ``{A semi-classical limit of
  the gauge/string correspondence},'' {\em Nucl. Phys.} {\bf B636} (2002)
  99--114, \href{http://xxx.lanl.gov/abs/hep-th/0204051}{{\tt hep-th/0204051}}.

\bibitem{Frolov:2002av}
S.~Frolov and A.~A. Tseytlin, ``{Semiclassical quantization of rotating
  superstring in AdS(5) x S(5)},'' {\em JHEP} {\bf 06} (2002) 007,
  \href{http://xxx.lanl.gov/abs/hep-th/0204226}{{\tt hep-th/0204226}}.

\bibitem{Minahan:2002rc}
J.~A. Minahan, ``{Circular semiclassical string solutions on AdS(5) x S**5},''
  {\em Nucl. Phys.} {\bf B648} (2003) 203--214,
  \href{http://xxx.lanl.gov/abs/hep-th/0209047}{{\tt hep-th/0209047}}.

\bibitem{Frolov:2003qc}
S.~Frolov and A.~A. Tseytlin, ``{Multi-spin string solutions in AdS(5) x
  S**5},'' {\em Nucl. Phys.} {\bf B668} (2003) 77--110,
  \href{http://xxx.lanl.gov/abs/hep-th/0304255}{{\tt hep-th/0304255}}.

\bibitem{Arutyunov:2003uj}
G.~Arutyunov, S.~Frolov, J.~Russo, and A.~A. Tseytlin, ``{Spinning strings in
  AdS(5) x S**5 and integrable systems},'' {\em Nucl. Phys.} {\bf B671} (2003)
  3--50, \href{http://xxx.lanl.gov/abs/hep-th/0307191}{{\tt hep-th/0307191}}.

\bibitem{Beisert:2003ea}
N.~Beisert, S.~Frolov, M.~Staudacher, and A.~A. Tseytlin, ``{Precision
  spectroscopy of AdS/CFT},'' {\em JHEP} {\bf 10} (2003) 037,
  \href{http://xxx.lanl.gov/abs/hep-th/0308117}{{\tt hep-th/0308117}}.

\bibitem{Frolov:2003xy}
S.~Frolov and A.~A. Tseytlin, ``{Rotating string solutions: AdS/CFT duality in
  non- supersymmetric sectors},'' {\em Phys. Lett.} {\bf B570} (2003) 96--104,
  \href{http://xxx.lanl.gov/abs/hep-th/0306143}{{\tt hep-th/0306143}}.

\bibitem{Beisert:2005bm}
N.~Beisert, V.~A. Kazakov, K.~Sakai, and K.~Zarembo, ``{The algebraic curve of
  classical superstrings on AdS(5) x S**5},'' {\em Commun. Math. Phys.} {\bf
  263} (2006) 659--710, \href{http://xxx.lanl.gov/abs/hep-th/0502226}{{\tt
  hep-th/0502226}}.

\bibitem{Beisert:2004ag}
N.~Beisert, V.~A. Kazakov, and K.~Sakai, ``{Algebraic curve for the SO(6)
  sector of AdS/CFT},'' {\em Commun. Math. Phys.} {\bf 263} (2006) 611--657,
  \href{http://xxx.lanl.gov/abs/hep-th/0410253}{{\tt hep-th/0410253}}.

\bibitem{Dorey:2006zj}
N.~Dorey and B.~Vicedo, ``{On the dynamics of finite-gap solutions in classical
  string theory},'' {\em JHEP} {\bf 07} (2006) 014,
  \href{http://xxx.lanl.gov/abs/hep-th/0601194}{{\tt hep-th/0601194}}.

\bibitem{Gromov:2007aq}
N.~Gromov and P.~Vieira, ``{The AdS(5) x S$^5$ superstring quantum spectrum
  from the algebraic curve},'' {\em Nucl. Phys.} {\bf B789} (2008) 175--208,
  \href{http://xxx.lanl.gov/abs/hep-th/0703191}{{\tt hep-th/0703191}}.

\bibitem{Vicedo:2008jy}
B.~Vicedo, ``{Semiclassical Quantisation of Finite-Gap Strings},'' {\em JHEP}
  {\bf 06} (2008) 086, \href{http://xxx.lanl.gov/abs/0803.1605}{{\tt
  0803.1605}}.

\bibitem{Gromov:2007ky}
N.~Gromov and P.~Vieira, ``{Complete 1-loop test of AdS/CFT},'' {\em JHEP} {\bf
  04} (2008) 046, \href{http://xxx.lanl.gov/abs/0709.3487}{{\tt 0709.3487}}.

\bibitem{Roiban:2007dq}
R.~Roiban and A.~A. Tseytlin, ``{Strong-coupling expansion of cusp anomaly from
  quantum superstring},'' {\em JHEP} {\bf 11} (2007) 016,
  \href{http://xxx.lanl.gov/abs/0709.0681}{{\tt 0709.0681}}.

\bibitem{Roiban:2007ju}
R.~Roiban and A.~A. Tseytlin, ``{Spinning superstrings at two loops:
  strong-coupling corrections to dimensions of large-twist SYM operators},''
  {\em Phys. Rev.} {\bf D77} (2008) 066006,
  \href{http://xxx.lanl.gov/abs/0712.2479}{{\tt 0712.2479}}.

\bibitem{Roiban:2009aa}
R.~Roiban and A.~A. Tseytlin, ``{Quantum strings in AdS$_5$ x S$^5$:
  strong-coupling corrections to dimension of Konishi operator},'' {\em JHEP}
  {\bf 11} (2009) 013, \href{http://xxx.lanl.gov/abs/0906.4294}{{\tt
  0906.4294}}.

\bibitem{Penros}
R.~Penrose, ``{Any space-time has a plane wave as a limit},'' in {\em
  Differential geometryand relativity}.
\newblock Reidel, Dordrecht, 1976, pp. 271–275.

\bibitem{Metsaev:2001bj}
R.~R. Metsaev, ``{Type IIB Green-Schwarz superstring in plane wave Ramond-
  Ramond background},'' {\em Nucl. Phys.} {\bf B625} (2002) 70--96,
  \href{http://xxx.lanl.gov/abs/hep-th/0112044}{{\tt hep-th/0112044}}.

\bibitem{Pohlmeyer:1975nb}
K.~Pohlmeyer, ``{Integrable Hamiltonian Systems and Interactions Through
  Quadratic Constraints},'' {\em Commun. Math. Phys.} {\bf 46} (1976) 207--221.

\bibitem{Maldacena:2006rv}
J.~M. Maldacena and I.~Swanson, ``{Connecting giant magnons to the pp-wave: An
  interpolating limit of AdS$_5$xS$^5$},'' {\em Phys. Rev.} {\bf D76} (2007)
  026002, \href{http://xxx.lanl.gov/abs/hep-th/0612079}{{\tt hep-th/0612079}}.

\bibitem{Beisert:2003xu}
N.~Beisert, J.~A. Minahan, M.~Staudacher, and K.~Zarembo, ``{Stringing spins
  and spinning strings},'' {\em JHEP} {\bf 09} (2003) 010,
  \href{http://xxx.lanl.gov/abs/hep-th/0306139}{{\tt hep-th/0306139}}.

\bibitem{Brezin:1977sv}
E.~Brezin, C.~Itzykson, G.~Parisi, and J.~B. Zuber, ``{Planar Diagrams},'' {\em
  Commun. Math. Phys.} {\bf 59} (1978) 35.

\bibitem{Kostov:1988fy}
I.~K. Kostov, ``{O(n) vector model on a planar random lattice: spectrum of
  anomalous dimensions},'' {\em Mod. Phys. Lett.} {\bf A4} (1989) 217.

\bibitem{Beisert:2005cw}
N.~Beisert and A.~A. Tseytlin, ``{On quantum corrections to spinning strings
  and Bethe equations},'' {\em Phys. Lett.} {\bf B629} (2005) 102--110,
  \href{http://xxx.lanl.gov/abs/hep-th/0509084}{{\tt hep-th/0509084}}.

\bibitem{Freyhult:2006vr}
L.~Freyhult and C.~Kristjansen, ``{A universality test of the quantum string
  Bethe ansatz},'' {\em Phys. Lett.} {\bf B638} (2006) 258--264,
  \href{http://xxx.lanl.gov/abs/hep-th/0604069}{{\tt hep-th/0604069}}.

\bibitem{Gromov:2007cd}
N.~Gromov and P.~Vieira, ``{Constructing the AdS/CFT dressing factor},'' {\em
  Nucl. Phys.} {\bf B790} (2008) 72--88,
  \href{http://xxx.lanl.gov/abs/hep-th/0703266}{{\tt hep-th/0703266}}.

\bibitem{Belitsky:2006en}
A.~V. Belitsky, A.~S. Gorsky, and G.~P. Korchemsky, ``{Logarithmic scaling in
  gauge / string correspondence},'' {\em Nucl. Phys.} {\bf B748} (2006) 24--59,
  \href{http://xxx.lanl.gov/abs/hep-th/0601112}{{\tt hep-th/0601112}}.

\bibitem{Korchemsky:1988si}
G.~P. Korchemsky, ``{Asymptotics of the Altarelli-Parisi-Lipatov Evolution
  Kernels of Parton Distributions},'' {\em Mod. Phys. Lett.} {\bf A4} (1989)
  1257--1276.

\bibitem{Korchemsky:1992xv}
G.~P. Korchemsky and G.~Marchesini, ``{Structure function for large x and
  renormalization of Wilson loop},'' {\em Nucl. Phys.} {\bf B406} (1993)
  225--258, \href{http://xxx.lanl.gov/abs/hep-ph/9210281}{{\tt
  hep-ph/9210281}}.

\bibitem{Korchemsky:1985xj}
G.~P. Korchemsky and A.~V. Radyushkin, ``{Loop space formalism and
  renormalization group for the infrared asymptotics of QCD},'' {\em Phys.
  Lett.} {\bf B171} (1986) 459--467.

\bibitem{Korchemsky:1987wg}
G.~P. Korchemsky and A.~V. Radyushkin, ``{Renormalization of the Wilson Loops
  Beyond the Leading Order},'' {\em Nucl. Phys.} {\bf B283} (1987) 342--364.

\bibitem{Bern:2005iz}
Z.~Bern, L.~J. Dixon, and V.~A. Smirnov, ``{Iteration of planar amplitudes in
  maximally supersymmetric Yang-Mills theory at three loops and beyond},'' {\em
  Phys. Rev.} {\bf D72} (2005) 085001,
  \href{http://xxx.lanl.gov/abs/hep-th/0505205}{{\tt hep-th/0505205}}.

\bibitem{Vogt:2004gi}
A.~Vogt, S.~Moch, and J.~Vermaseren, ``{The three-loop splitting functions in
  QCD},'' {\em Nucl. Phys. Proc. Suppl.} {\bf 152} (2006) 110--115,
  \href{http://xxx.lanl.gov/abs/hep-ph/0407321}{{\tt hep-ph/0407321}}.

\bibitem{Kotikov:2004er}
A.~V. Kotikov, L.~N. Lipatov, A.~I. Onishchenko, and V.~N. Velizhanin,
  ``{Three-loop universal anomalous dimension of the Wilson operators in N = 4
  SUSY Yang-Mills model},'' {\em Phys. Lett.} {\bf B595} (2004) 521--529,
  \href{http://xxx.lanl.gov/abs/hep-th/0404092}{{\tt hep-th/0404092}}.

\bibitem{Cachazo:2006az}
F.~Cachazo, M.~Spradlin, and A.~Volovich, ``{Four-Loop Cusp Anomalous Dimension
  From Obstructions},'' {\em Phys. Rev.} {\bf D75} (2007) 105011,
  \href{http://xxx.lanl.gov/abs/hep-th/0612309}{{\tt hep-th/0612309}}.

\bibitem{Dorey:2008vp}
N.~Dorey and M.~Losi, ``{Spiky Strings and Spin Chains},''
  \href{http://xxx.lanl.gov/abs/0812.1704}{{\tt 0812.1704}}.

\bibitem{Frolov:2006qe}
S.~Frolov, A.~Tirziu, and A.~A. Tseytlin, ``{Logarithmic corrections to higher
  twist scaling at strong coupling from AdS/CFT},'' {\em Nucl. Phys.} {\bf
  B766} (2007) 232--245, \href{http://xxx.lanl.gov/abs/hep-th/0611269}{{\tt
  hep-th/0611269}}.

\bibitem{Fioravanti:2008rv}
D.~Fioravanti, P.~Grinza, and M.~Rossi, ``{Strong coupling for planar ${\cal
  N}=4$ SYM theory: an all-order result},'' {\em Nucl. Phys.} {\bf B810} (2009)
  563--574, \href{http://xxx.lanl.gov/abs/0804.2893}{{\tt 0804.2893}}.

\bibitem{Arutyunov:2006ak}
G.~Arutyunov, S.~Frolov, J.~Plefka, and M.~Zamaklar, ``{The off-shell symmetry
  algebra of the light-cone AdS(5) x S$^5$ superstring},'' {\em J. Phys.} {\bf
  A40} (2007) 3583--3606, \href{http://xxx.lanl.gov/abs/hep-th/0609157}{{\tt
  hep-th/0609157}}.

\bibitem{deLeeuw:2007uf}
M.~de~Leeuw, ``{Coordinate Bethe Ansatz for the String S-Matrix},'' {\em J.
  Phys.} {\bf A40} (2007) 14413--14432,
  \href{http://xxx.lanl.gov/abs/0705.2369}{{\tt 0705.2369}}.

\bibitem{Arutyunov:2009kf}
G.~Arutyunov and S.~Frolov, ``{The Dressing Factor and Crossing Equations},''
  \href{http://xxx.lanl.gov/abs/0904.4575}{{\tt 0904.4575}}.

\bibitem{Plefka:2006ze}
J.~Plefka, F.~Spill, and A.~Torrielli, ``{On the Hopf algebra structure of the
  AdS/CFT S-matrix},'' {\em Phys. Rev.} {\bf D74} (2006) 066008,
  \href{http://xxx.lanl.gov/abs/hep-th/0608038}{{\tt hep-th/0608038}}.

\bibitem{Rej:2007vm}
A.~Rej, M.~Staudacher, and S.~Zieme, ``{Nesting and dressing},'' {\em J. Stat.
  Mech.} {\bf 0708} (2007) P08006,
  \href{http://xxx.lanl.gov/abs/hep-th/0702151}{{\tt hep-th/0702151}}.

\bibitem{Arutyunov:2006iu}
G.~Arutyunov and S.~Frolov, ``{On AdS(5) x S$^5$ string S-matrix},'' {\em Phys.
  Lett.} {\bf B639} (2006) 378--382,
  \href{http://xxx.lanl.gov/abs/hep-th/0604043}{{\tt hep-th/0604043}}.

\bibitem{Dorey:2007xn}
N.~Dorey, D.~M. Hofman, and J.~M. Maldacena, ``{On the singularities of the
  magnon S-matrix},'' {\em Phys. Rev.} {\bf D76} (2007) 025011,
  \href{http://xxx.lanl.gov/abs/hep-th/0703104}{{\tt hep-th/0703104}}.

\bibitem{WH}
B.~Noble, {\em {Methods Based on the Wiener-Hopf Technique for the Solution of
  Partial Differential Equations}}.
\newblock Pergamon, London, 1958.

\bibitem{Forgacs:1991rs}
P.~Forgacs, F.~Niedermayer, and P.~Weisz, ``{The Exact mass gap of the
  Gross-Neveu model. 1. The Thermodynamic Bethe ansatz},'' {\em Nucl. Phys.}
  {\bf B367} (1991) 123--143.

\bibitem{Beccaria:2008nf}
M.~Beccaria, ``{The generalized scaling function of AdS/CFT and semiclassical
  string theory},'' {\em JHEP} {\bf 07} (2008) 082,
  \href{http://xxx.lanl.gov/abs/0806.3704}{{\tt 0806.3704}}.

\bibitem{Belitsky:2004cz}
A.~V. Belitsky, V.~M. Braun, A.~S. Gorsky, and G.~P. Korchemsky,
  ``{Integrability in QCD and beyond},'' {\em Int. J. Mod. Phys.} {\bf A19}
  (2004) 4715--4788, \href{http://xxx.lanl.gov/abs/hep-th/0407232}{{\tt
  hep-th/0407232}}.

\bibitem{Korchemsky:1995be}
G.~P. Korchemsky, ``{Quasiclassical QCD pomeron},'' {\em Nucl. Phys.} {\bf
  B462} (1996) 333--388, \href{http://xxx.lanl.gov/abs/hep-th/9508025}{{\tt
  hep-th/9508025}}.

\bibitem{Kotanski:2005ci}
J.~Kotanski, ``{Reggeized gluon states in quantum chromodynamics},''
  \href{http://xxx.lanl.gov/abs/hep-th/0511279}{{\tt hep-th/0511279}}.

\bibitem{Gromov:2005gp}
N.~Gromov and V.~Kazakov, ``{Double scaling and finite size corrections in
  sl(2) spin chain},'' {\em Nucl. Phys.} {\bf B736} (2006) 199--224,
  \href{http://xxx.lanl.gov/abs/hep-th/0510194}{{\tt hep-th/0510194}}.

\bibitem{Roiban:2007jf}
R.~Roiban, A.~Tirziu, and A.~A. Tseytlin, ``{Two-loop world-sheet corrections
  in AdS$_5\times$S$^5$ superstring},'' {\em JHEP} {\bf 07} (2007) 056,
  \href{http://xxx.lanl.gov/abs/0704.3638}{{\tt 0704.3638}}.

\bibitem{Polyakov:1975rr}
A.~M. Polyakov, ``{Interaction of Goldstone Particles in Two-Dimensions.
  Applications to Ferromagnets and Massive Yang-Mills Fields},'' {\em Phys.
  Lett.} {\bf B59} (1975) 79--81.

\bibitem{D'Adda:1978uc}
A.~D'Adda, M.~Luscher, and P.~Di~Vecchia, ``{A 1/n Expandable Series of
  Nonlinear Sigma Models with Instantons},'' {\em Nucl. Phys.} {\bf B146}
  (1978) 63--76.

\bibitem{Witten:1978bc}
E.~Witten, ``{Instantons, the Quark Model, and the 1/n Expansion},'' {\em Nucl.
  Phys.} {\bf B149} (1979) 285.

\bibitem{Evans:1995dn}
J.~M. Evans and T.~J. Hollowood, ``{Exact results for integrable asymptotically
  - free field theories},'' {\em Nucl. Phys. Proc. Suppl.} {\bf 45A} (1996)
  130--139, \href{http://xxx.lanl.gov/abs/hep-th/9508141}{{\tt
  hep-th/9508141}}.

\bibitem{Balog:1992cm}
J.~Balog, S.~Naik, F.~Niedermayer, and P.~Weisz, ``{The Exact mass gap of the
  chiral SU(n) x SU(n) model},'' {\em Phys. Rev. Lett.} {\bf 69} (1992)
  873--876.

\bibitem{Wiegmann:1985jt}
P.~B. Wiegmann, ``{Exact solution of the $O(3)$ nonlinear sigma model},'' {\em
  Phys. Lett.} {\bf B152} (1985) 209--214.

\bibitem{Lipatov:1977hj}
L.~N. Lipatov, ``{Divergence of Perturbation Series and Pseudoparticles},''
  {\em JETP Lett.} {\bf 25} (1977) 104--107.

\bibitem{Beneke:1998ui}
M.~Beneke, ``{Renormalons},'' {\em Phys. Rept.} {\bf 317} (1999) 1--142,
  \href{http://xxx.lanl.gov/abs/hep-ph/9807443}{{\tt hep-ph/9807443}}.

\bibitem{David:1982qv}
F.~David, ``{Nonperturbative effects and infrared renormalons within the 1/N
  expansion of the O(N) nonlinear sigma model},'' {\em Nucl. Phys.} {\bf B209}
  (1982) 433--460.

\bibitem{David:1983gz}
F.~David, ``{On the Ambiguity of Composite Operators, IR Renormalons and the
  Status of the Operator Product Expansion},'' {\em Nucl. Phys.} {\bf B234}
  (1984) 237--251.

\bibitem{Fiamberti:2009zz}
F.~Fiamberti, A.~Santambrogio, C.~Sieg, and D.~Zanon, ``{Wrapping at four loops
  in N=4 SYM theory},'' {\em Nucl. Phys. Proc. Suppl.} {\bf 192-193} (2009)
  187--189.

\bibitem{Arutyunov:2010gb}
G.~Arutyunov, S.~Frolov, and R.~Suzuki, ``{Five-loop Konishi from the Mirror
  TBA},'' \href{http://xxx.lanl.gov/abs/1002.1711}{{\tt 1002.1711}}.

\bibitem{Gromov:2009zb}
N.~Gromov, V.~Kazakov, and P.~Vieira, ``{Exact AdS/CFT spectrum: Konishi
  dimension at any coupling},'' \href{http://xxx.lanl.gov/abs/0906.4240}{{\tt
  0906.4240}}.

\bibitem{Mann:2005ab}
N.~Mann and J.~Polchinski, ``{Bethe Ansatz for a Quantum Supercoset Sigma
  Model},'' {\em Phys. Rev.} {\bf D72} (2005) 086002,
  \href{http://xxx.lanl.gov/abs/hep-th/0508232}{{\tt hep-th/0508232}}.

\bibitem{Gromov:2006dh}
N.~Gromov, V.~Kazakov, K.~Sakai, and P.~Vieira, ``{Strings as multi-particle
  states of quantum sigma- models},'' {\em Nucl. Phys.} {\bf B764} (2007)
  15--61, \href{http://xxx.lanl.gov/abs/hep-th/0603043}{{\tt hep-th/0603043}}.

\bibitem{Gromov:2006cq}
N.~Gromov and V.~Kazakov, ``{Asymptotic Bethe ansatz from string sigma model on
  S$^3$ x R},'' {\em Nucl. Phys.} {\bf B780} (2007) 143--160,
  \href{http://xxx.lanl.gov/abs/hep-th/0605026}{{\tt hep-th/0605026}}.

\bibitem{Roiban:2009vh}
R.~Roiban and A.~A. Tseytlin, ``{UV finiteness of Pohlmeyer-reduced form of the
  AdS$_5$xS$^5$ superstring theory},'' {\em JHEP} {\bf 04} (2009) 078,
  \href{http://xxx.lanl.gov/abs/0902.2489}{{\tt 0902.2489}}.

\bibitem{Hoare:2009rq}
B.~Hoare, Y.~Iwashita, and A.~A. Tseytlin, ``{Pohlmeyer-reduced form of string
  theory in AdS$_5$xS$^5$: semiclassical expansion},'' {\em J. Phys.} {\bf A42}
  (2009) 375204, \href{http://xxx.lanl.gov/abs/0906.3800}{{\tt 0906.3800}}.

\bibitem{Hoare:2009fs}
B.~Hoare and A.~A. Tseytlin, ``{Tree-level S-matrix of Pohlmeyer reduced form
  of AdS$_5$xS$^5$ superstring theory},''
  \href{http://xxx.lanl.gov/abs/0912.2958}{{\tt 0912.2958}}.

\bibitem{Giombi:2010fa}
S.~Giombi, R.~Ricci, R.~Roiban, A.~A. Tseytlin, and C.~Vergu, ``{Generalized
  scaling function from light-cone gauge AdS$_5$xS$^5$ superstring},''
  \href{http://xxx.lanl.gov/abs/1002.0018}{{\tt 1002.0018}}.

\bibitem{Gromov:2010vb}
N.~Gromov, V.~Kazakov, and Z.~Tsuboi, ``{PSU(2,2|4) Character of Quasiclassical
  AdS/CFT},'' \href{http://xxx.lanl.gov/abs/1002.3981}{{\tt 1002.3981}}.

\bibitem{Arutyunov:2009ux}
G.~Arutyunov and S.~Frolov, ``{Simplified TBA equations of the AdS$_5$xS$^5$
  mirror model},'' {\em JHEP} {\bf 11} (2009) 019,
  \href{http://xxx.lanl.gov/abs/0907.2647}{{\tt 0907.2647}}.

\bibitem{Dorey:2006dq}
N.~Dorey, ``{Magnon bound states and the AdS/CFT correspondence},'' {\em J.
  Phys.} {\bf A39} (2006) 13119--13128,
  \href{http://xxx.lanl.gov/abs/hep-th/0604175}{{\tt hep-th/0604175}}.

\end{thebibliography}\endgroup


\begin{thebibliography}{KSV1}
\bibitem[KSV1]{KSV1}
  I.~Kostov, D.~Serban and D.~Volin,
  \textit{``Strong coupling limit of Bethe Ansatz equations,''}
  Nucl.\ Phys.\  B {\bf 789}, 413 (2008)
  [arXiv:hep-th/0703031].

\bibitem[KSV2]{KSV2}
  I.~Kostov, D.~Serban and D.~Volin,
  \textit{``Functional BES equation,''}
  JHEP {\bf 0808}, 101 (2008)
  [arXiv:0801.2542 [hep-th]].

\bibitem[V1]{V1}
  D.~Volin,
  \textit{``The 2-loop generalized scaling function from the BES/FRS equation,''}
  [arXiv:0812.4407 [hep-th]].

\bibitem[V2]{V2}
  D.~Volin,
  \textit{``From the mass gap in O(N) to the non-Borel-summability in O(3) and O(4) sigma-models,''}
  [arXiv:0904.2744 [hep-th]].

\bibitem[V3]{V3}
  D.~Volin,
  \textit{``Minimal solution of the AdS/CFT crossing equation,''}
  J.\ Phys.\ A  {\bf 42}, 372001 (2009)
  [arXiv:0904.4929 [hep-th]].

\end{thebibliography}
 \bibliographystyle{utphys}  
\end{document}